\documentclass[traditabstract,longauth]{aa}

\usepackage[nonamebreak]{natbib}
\usepackage[stable]{footmisc}
\usepackage{amsmath}
\usepackage{txfonts}
\usepackage{placeins}
\usepackage{natbib}
\bibpunct{(}{)}{;}{a}{}{,} % to follow A&A style
\usepackage{graphicx}
\usepackage{epstopdf}
\usepackage{ifthen}
\usepackage[breaklinks, colorlinks, citecolor=blue]{hyperref}
\usepackage{subfigure}
\usepackage[utf8]{inputenc}

\usepackage[table,usenames,dvipsnames]{xcolor}

\defcitealias{planck2013-p11}{PCP13}
\defcitealias{planck2013-p08}{PPL13}
\defcitealias{pb2015}{BKP}

\providecommand{\sorthelp}[1]{}

\newcommand{\paramsI}{\citetalias{planck2013-p11}}
\newcommand{\likeI}{\citetalias{planck2013-p08}}
\newcommand{\BKP}{\citetalias{pb2015}}

\def\WMAP{{WMAP}}
\def\COBE{{COBE}}

\newcommand{\onesig}[1]{(68\%, \text{#1})}
\newcommand{\twosig}[1]{(95\%, \text{#1})}
\newcommand{\twoonesig}[3]{
\begin{equation}
\left.
 \begin{aligned}
#1 \\ #2
 \end{aligned}
\ \right\} \ \ \mbox{\text{#3}}
\end{equation}
}
\newcommand{\twotwosig}[3]{
\begin{equation}
\left.
 \begin{aligned}
#1 \\ #2
 \end{aligned}
\ \right\} \ \ \mbox{95\%, \text{#3}}
\end{equation}
}

\newcommand{\camspec}{{\tt CamSpec}}
\newcommand{\plik}{{\tt Plik}}

\newcommand{\mspec}{{\tt MSPEC}}
\newcommand{\smica}{{\tt SMICA}}

\newcommand{\CFHTLENS}{CFHTLenS}

\newcommand{\mksym}[1]{\ifmmode {\rm #1}\else #1\fi}
\newcommand{\dataplus}{{+}}
\newcommand{\WP}{\mksym{WP}}
\newcommand{\highL}{\mksym{highL}}
\newcommand{\BAO}{\mksym{BAO}}
\newcommand{\lensing}{\mksym{lensing}}
\newcommand{\ext}{\mksym{ext}}
\newcommand{\planckonly}{\planck}
\newcommand{\TT}{\mksym{TT}}
\newcommand{\TTTEEE}{\mksym{TT,TE,EE}}
\newcommand{\planckTTonly}{\planck\ \TT}
\newcommand{\planckTTTEEEonly}{\planck\ \TTTEEE}
\newcommand{\lowTEB}{\mksym{lowP}}
\newcommand{\lowEB}{\mksym{lowP}}
\newcommand{\WMAPTEB}{\lowTEB\dataplus\mksym{WP}}
\newcommand{\lowTP}{\mksym{lowT,P}}
\newcommand{\planckTT}{\planckTTonly\dataplus\lowTEB}
\newcommand{\planckall}{\planckTTTEEEonly\dataplus\lowTEB}
\newcommand{\planckTTBAO}{\planckTT\dataplus\BAO}
\newcommand{\planckTTlensing}{\planckTT\dataplus\lensing}
\newcommand{\planckallBAO}{\planckall\dataplus\BAO}
\newcommand{\planckalllensing}{\planckall\dataplus\lensing}
\newcommand{\planckTTlensext}{\planckTT\dataplus\lensing\dataplus\ext}
\newcommand{\datalabel}[1]{#1}

\newcommand{\shortTT}{\TT\dataplus\lowTEB}
\newcommand{\shortall}{\TTTEEE\dataplus\lowTEB}

\newcommand{\As}{A_{\rm s}}

\newcommand{\nt}{n_{\rm t}}
\newcommand{\ns}{n_{\rm s}}
\newcommand{\lcdm}{{$\rm{\Lambda CDM}$}}

\newcommand{\rzerotwo}{r_{0.002}}
\newcommand{\Alens}{A_{\rm L}}
\newcommand{\Aphiphi}{A_{\rm L}^{\phi\phi}}
\newcommand{\omegak}{\Omega_K}

\newcommand{\alphaiso}{\alpha}

\newcommand{\thetaMC}{\theta_{\rm MC}}
\newcommand{\nrun}{d \ns / d\ln k}
\newcommand{\nrunfrac}{\frac{d \ns}{d\ln k}}
\newcommand{\zre}{z_{\text{re}}}
\newcommand{\yhe}{Y_{\text{P}}}
\newcommand{\ypbbn}{Y_{\text{P}}^{\rm BBN}}
\newcommand{\zeq}{z_{\text{eq}}}
\newcommand{\nnu}{N_{\rm eff}}
\newcommand{\neff}{\nnu}
\newcommand{\mnu}{\sum m_\nu}
\newcommand{\sumnu}{\sum m_\nu}
\newcommand{\mnusterile}{m_{\nu,\, \mathrm{sterile}}^{\mathrm{eff}}}
\newcommand{\meffsterile}{\mnusterile}

\newcommand{\Tsterile}{T_{\mathrm{s}}}
\newcommand{\msthermal}{m_{\rm sterile}^{\rm thermal}}
\newcommand{\msDW}{m_{\rm sterile}^{\rm DW}}

\newcommand{\rdrag}{r_{\rm drag}}

\newcommand{\rstar}{r_{\ast}}
\newcommand{\rs}{r_{\rm s}}
\newcommand{\thetastar}{\theta_{\ast}}
\newcommand{\DAstar}{D_{\rm A}}
\newcommand{\keq}{k_{\rm eq}}

\newcommand{\DVBAO}{D_{\rm V}}
\newcommand{\DA}{D_{\rm A}}

\providecommand{\Planck}{\textit{Planck}}
\providecommand{\planck}{\Planck}

\providecommand{\text}[1]{\rm{#1}}

\newcommand{\Mpc}{\text{Mpc}}

\newcommand{\grad}{\nabla}

\newcommand{\Hunit}{~\text{km}~\text{s}^{-1} \Mpc^{-1}}

\providecommand{\muK}{\mu\rm{K}}

\newcommand{\mpl}{m_{\text{Pl}}}
\newcommand{\eV}{\,\text{eV}}
\newcommand{\MeV}{\,\text{MeV}}

\providecommand{\Omk}{\Omega_K}
\providecommand{\Oml}{\Omega_{\Lambda}}

\providecommand{\Omb}{\Omega_{\mathrm{b}}}
\providecommand{\Omc}{\Omega_{\mathrm{c}}}
\providecommand{\Omm}{\Omega_{\mathrm{m}}}
\providecommand{\omb}{\omega_{\mathrm{b}}}
\providecommand{\omc}{\omega_{\mathrm{c}}}
\providecommand{\omm}{\omega_{\mathrm{m}}}

\providecommand{\CAMB}{{\tt camb}}
\providecommand{\COSMOMC}{{\tt CosmoMC}}

\providecommand{\CLASS}{{\tt class}}

\providecommand{\LCDM}{{$\rm{\Lambda CDM}$}}
\providecommand{\COSMOREC}{{\tt CosmoRec}}
\providecommand{\HYREC}{{\tt HyRec}}
\providecommand{\RECFAST}{{\tt recfast}}
\providecommand{\HALOFIT}{{\tt halofit}}

\providecommand{\MontePython}{{\tt Monte Python}}
\providecommand{\HEALpix}{{\tt HEALpix}}

\newcommand{\begm}{\begin{pmatrix}}
\newcommand{\enm}{\end{pmatrix}}

\newcommand\ba{\begin{eqnarray}}
\newcommand\ea{\end{eqnarray}}
\newcommand\bea{\begin{eqnarray}}
\newcommand\eea{\end{eqnarray}}

\newcommand\be{\begin{equation}}
\newcommand\ee{\end{equation}}

\newcommand{\clo}{\mathcal{O}}

\def\pmb#1{\setbox0=\hbox{#1}%
    \kern-.025em\copy0\kern-\wd0
    \kern.05em\copy0\kern-\wd0
    \kern-.025em\raise.0433em\box0}

\def\muk{$(\mu{\rm K})^2$ }

\def\p2Y{\;_2Y}
\def\m2Y{\;_{-2}Y}
\def\beglet{
  \addtocounter{equation}{1}%
  \setcounter{parentequation}{\value{equation}}%
  \setcounter{equation}{0}%
  \def\theequation{\arabic{parentequation}\alph{equation}}%
  \ignorespaces
}
\def\endlet{
  \setcounter{equation}{\value{parentequation}}%
  \def\theequation{\arabic{equation}}%
}
\providecommand{\beglet}{\begin{subequations}}
\providecommand{\endlet}{\end{subequations}}

\newcommand{\neutron}{{\rm n}}

\def\setsymbol#1#2{\expandafter\def\csname #1\endcsname{#2}}
\def\getsymbol#1{\csname #1\endcsname}

\def\Planck{\textit{Planck}}

\newbox\tablebox    \newdimen\tablewidth
\def\leaderfil{\leaders\hbox to 5pt{\hss.\hss}\hfil}
\def\endPlancktable{\tablewidth=\columnwidth 
    $$\hss\copy\tablebox\hss$$
    \vskip-\lastskip\vskip -2pt}
\def\endPlancktablewide{\tablewidth=\textwidth 
    $$\hss\copy\tablebox\hss$$
    \vskip-\lastskip\vskip -2pt}
\def\tablenote#1 #2\par{\begingroup \parindent=0.8em
    \abovedisplayshortskip=0pt\belowdisplayshortskip=0pt
    \noindent
    $$\hss\vbox{\hsize\tablewidth \hangindent=\parindent \hangafter=1 \noindent
    \hbox to \parindent{$^#1$\hss}\strut#2\strut\par}\hss$$
    \endgroup}
\def\doubleline{\vskip 3pt\hrule \vskip 1.5pt \hrule \vskip 5pt}

\def\L2{\ifmmode L_2\else $L_2$\fi}

\def\DeltaT{\ifmmode \Delta T\else $\Delta T$\fi}
\def\deltat{\ifmmode \Delta t\else $\Delta t$\fi}
\def\fknee{\ifmmode f_{\rm knee}\else $f_{\rm knee}$\fi}
\def\Fmax{\ifmmode F_{\rm max}\else $F_{\rm max}$\fi}
\def\solar{\ifmmode{\rm M}_{\mathord\odot}\else${\rm M}_{\mathord\odot}$\fi}
\def\Msolar{\ifmmode{\rm M}_{\mathord\odot}\else${\rm M}_{\mathord\odot}$\fi}
\def\Lsolar{\ifmmode{\rm L}_{\mathord\odot}\else${\rm L}_{\mathord\odot}$\fi}
\def\inv{\ifmmode^{-1}\else$^{-1}$\fi}
\def\mo{\ifmmode^{-1}\else$^{-1}$\fi}
\def\sup#1{\ifmmode ^{\rm #1}\else $^{\rm #1}$\fi}
\def\expo#1{\ifmmode \times 10^{#1}\else $\times 10^{#1}$\fi}
\def\,{\thinspace}
\def\lsim{\mathrel{\raise .4ex\hbox{\rlap{$<$}\lower 1.2ex\hbox{$\sim$}}}}
\def\gsim{\mathrel{\raise .4ex\hbox{\rlap{$>$}\lower 1.2ex\hbox{$\sim$}}}}

\def\simprop{\mathrel{\raise .4ex\hbox{\rlap{$\propto$}\lower 1.2ex\hbox{$\sim$}}}}
\def\deg{\ifmmode^\circ\else$^\circ$\fi}
\def\pdeg{\ifmmode $\setbox0=\hbox{$^{\circ}$}\rlap{\hskip.11\wd0 .}$^{\circ}
          \else \setbox0=\hbox{$^{\circ}$}\rlap{\hskip.11\wd0 .}$^{\circ}$\fi}
\def\arcs{\ifmmode {^{\scriptstyle\prime\prime}}
          \else $^{\scriptstyle\prime\prime}$\fi}
\def\arcm{\ifmmode {^{\scriptstyle\prime}}
          \else $^{\scriptstyle\prime}$\fi}
\newdimen\sa  \newdimen\sb
\def\parcs{\sa=.07em \sb=.03em
     \ifmmode \hbox{\rlap{.}}^{\scriptstyle\prime\kern -\sb\prime}\hbox{\kern -\sa}
     \else \rlap{.}$^{\scriptstyle\prime\kern -\sb\prime}$\kern -\sa\fi}
\def\parcm{\sa=.08em \sb=.03em
     \ifmmode \hbox{\rlap{.}\kern\sa}^{\scriptstyle\prime}\hbox{\kern-\sb}
     \else \rlap{.}\kern\sa$^{\scriptstyle\prime}$\kern-\sb\fi}
\def\ra[#1 #2 #3.#4]{#1\sup{h}#2\sup{m}#3\sup{s}\llap.#4}
\def\dec[#1 #2 #3.#4]{#1\deg#2\arcm#3\arcs\llap.#4}
\def\deco[#1 #2 #3]{#1\deg#2\arcm#3\arcs}
\def\rra[#1 #2]{#1\sup{h}#2\sup{m}}

\def\dots{\relax\ifmmode \ldots\else $\ldots$\fi}
\def\WHzsr{\ifmmode $W\,Hz\mo\,sr\mo$\else W\,Hz\mo\,sr\mo\fi}
\def\mHz{\ifmmode $\,mHz$\else \,mHz\fi}
\def\GHz{\ifmmode $\,GHz$\else \,GHz\fi}
\def\mKs{\ifmmode $\,mK\,s$^{1/2}\else \,mK\,s$^{1/2}$\fi}
\def\muKs{\ifmmode \,\mu$K\,s$^{1/2}\else \,$\mu$K\,s$^{1/2}$\fi}
\def\muKRJs{\ifmmode \,\mu$K$_{\rm RJ}$\,s$^{1/2}\else \,$\mu$K$_{\rm RJ}$\,s$^{1/2}$\fi}
\def\muKHz{\ifmmode \,\mu$K\,Hz$^{-1/2}\else \,$\mu$K\,Hz$^{-1/2}$\fi}
\def\MJysr{\ifmmode \,$MJy\,sr\mo$\else \,MJy\,sr\mo\fi}
\def\MJysrmK{\ifmmode \,$MJy\,sr\mo$\,mK$_{\rm CMB}\mo\else \,MJy\,sr\mo\,mK$_{\rm CMB}\mo$\fi}
\def\microns{\ifmmode \,\mu$m$\else \,$\mu$m\fi}

\def\muK{\ifmmode \,\mu$K$\else \,$\mu$\hbox{K}\fi}
\def\microK{\ifmmode \,\mu$K$\else \,$\mu$\hbox{K}\fi}
\def\muW{\ifmmode \,\mu$W$\else \,$\mu$\hbox{W}\fi}
\def\kms{\ifmmode $\,km\,s$^{-1}\else \,km\,s$^{-1}$\fi}
\def\kmsMpc{\ifmmode $\,\kms\,Mpc\mo$\else \,\kms\,Mpc\mo\fi}
\providecommand{\sorthelp}[1]{}

\pdfmapfile{+txfonts.map}

\begin{document}

\title{\vglue -10mm\textit{Planck} 2015 results. XIII. Cosmological parameters}
\titlerunning{\textit{Planck} cosmological parameters}
\authorrunning{Planck Collaboration}
\author{The Planck team}
%\institute{L2 \and Earth}
%\date{Received 1 January 2013/Accepted 1 January 2013}

%This author list corresponds to \title{Author list for A15\_Cosmological\_parameters}
%Prepared by M. Lopez-Caniego (Marcos.Lopez.Caniego@sciops.esa.int), ESAC/ESA
%This version is from Thu Mar 17 17:52:44 2016 CET
%\subtitle{There are 261 co-authors in this list}
\author{\small
Planck Collaboration: P.~A.~R.~Ade\inst{105}
\and
N.~Aghanim\inst{71}
\and
M.~Arnaud\inst{87}
\and
M.~Ashdown\inst{83, 7}
\and
J.~Aumont\inst{71}
\and
C.~Baccigalupi\inst{103}
\and
A.~J.~Banday\inst{117, 12}
\and
R.~B.~Barreiro\inst{78}
\and
J.~G.~Bartlett\inst{1, 80}
\and
N.~Bartolo\inst{38, 79}
% ...
}
\institute{\small
APC, AstroParticule et Cosmologie, Universit\'{e} Paris Diderot, CNRS/IN2P3, CEA/lrfu, Observatoire de Paris, Sorbonne Paris Cit\'{e}, 10, rue Alice Domon et L\'{e}onie Duquet, 75205 Paris Cedex 13, France\goodbreak
\and
Aalto University Mets\"{a}hovi Radio Observatory and Dept of Radio Science and Engineering, P.O. Box 13000, FI-00076 AALTO, Finland\goodbreak
\and
African Institute for Mathematical Sciences, 6-8 Melrose Road, Muizenberg, Cape Town, South Africa\goodbreak
\and
Agenzia Spaziale Italiana Science Data Center, Via del Politecnico snc, 00133, Roma, Italy\goodbreak
\and
Aix Marseille Universit\'{e}, CNRS, LAM (Laboratoire d'Astrophysique de Marseille) UMR 7326, 13388, Marseille, France\goodbreak
\and
Aix Marseille Universit\'{e}, Centre de Physique Th\'{e}orique, 163 Avenue de Luminy, 13288, Marseille, France\goodbreak
\and
Astrophysics Group, Cavendish Laboratory, University of Cambridge, J J Thomson Avenue, Cambridge CB3 0HE, U.K.\goodbreak
\and
Astrophysics \& Cosmology Research Unit, School of Mathematics, Statistics \& Computer Science, University of KwaZulu-Natal, Westville Campus, Private Bag X54001, Durban 4000, South Africa\goodbreak
\and
Atacama Large Millimeter/submillimeter Array, ALMA Santiago Central Offices, Alonso de Cordova 3107, Vitacura, Casilla 763 0355, Santiago, Chile\goodbreak
\and
CGEE, SCS Qd 9, Lote C, Torre C, 4$^{\circ}$ andar, Ed. Parque Cidade Corporate, CEP 70308-200, Bras\'{i}lia, DF, Brazil\goodbreak
\and
CITA, University of Toronto, 60 St. George St., Toronto, ON M5S 3H8, Canada\goodbreak
\and
CNRS, IRAP, 9 Av. colonel Roche, BP 44346, F-31028 Toulouse cedex 4, France\goodbreak
\and
CRANN, Trinity College, Dublin, Ireland\goodbreak
\and
California Institute of Technology, Pasadena, California, U.S.A.\goodbreak
\and
Centre for Theoretical Cosmology, DAMTP, University of Cambridge, Wilberforce Road, Cambridge CB3 0WA, U.K.\goodbreak
\and
Centro de Estudios de F\'{i}sica del Cosmos de Arag\'{o}n (CEFCA), Plaza San Juan, 1, planta 2, E-44001, Teruel, Spain\goodbreak
\and
Computational Cosmology Center, Lawrence Berkeley National Laboratory, Berkeley, California, U.S.A.\goodbreak
\and
Consejo Superior de Investigaciones Cient\'{\i}ficas (CSIC), Madrid, Spain\goodbreak
\and
DSM/Irfu/SPP, CEA-Saclay, F-91191 Gif-sur-Yvette Cedex, France\goodbreak
\and
DTU Space, National Space Institute, Technical University of Denmark, Elektrovej 327, DK-2800 Kgs. Lyngby, Denmark\goodbreak
\and
D\'{e}partement de Physique Th\'{e}orique, Universit\'{e} de Gen\`{e}ve, 24, Quai E. Ansermet,1211 Gen\`{e}ve 4, Switzerland\goodbreak
\and
Departamento de Astrof\'{i}sica, Universidad de La Laguna (ULL), E-38206 La Laguna, Tenerife, Spain\goodbreak
\and
Departamento de F\'{\i}sica, Universidad de Oviedo, Avda. Calvo Sotelo s/n, Oviedo, Spain\goodbreak
\and
Department of Astronomy and Astrophysics, University of Toronto, 50 Saint George Street, Toronto, Ontario, Canada\goodbreak
\and
Department of Astrophysics/IMAPP, Radboud University Nijmegen, P.O. Box 9010, 6500 GL Nijmegen, The Netherlands\goodbreak
\and
Department of Physics \& Astronomy, University of British Columbia, 6224 Agricultural Road, Vancouver, British Columbia, Canada\goodbreak
\and
Department of Physics and Astronomy, Dana and David Dornsife College of Letter, Arts and Sciences, University of Southern California, Los Angeles, CA 90089, U.S.A.\goodbreak
\and
Department of Physics and Astronomy, Johns Hopkins University, Bloomberg Center 435, 3400 N. Charles St., Baltimore, MD 21218, U.S.A.\goodbreak
\and
Department of Physics and Astronomy, University College London, London WC1E 6BT, U.K.\goodbreak
\and
Department of Physics and Astronomy, University of Sussex, Brighton BN1 9QH, U.K.\goodbreak
\and
Department of Physics, Florida State University, Keen Physics Building, 77 Chieftan Way, Tallahassee, Florida, U.S.A.\goodbreak
\and
Department of Physics, Gustaf H\"{a}llstr\"{o}min katu 2a, University of Helsinki, Helsinki, Finland\goodbreak
\and
Department of Physics, Princeton University, Princeton, New Jersey, U.S.A.\goodbreak
\and
Department of Physics, University of California, Berkeley, California, U.S.A.\goodbreak
\and
Department of Physics, University of California, One Shields Avenue, Davis, California, U.S.A.\goodbreak
\and
Department of Physics, University of California, Santa Barbara, California, U.S.A.\goodbreak
\and
Department of Physics, University of Illinois at Urbana-Champaign, 1110 West Green Street, Urbana, Illinois, U.S.A.\goodbreak
\and
Dipartimento di Fisica e Astronomia G. Galilei, Universit\`{a} degli Studi di Padova, via Marzolo 8, 35131 Padova, Italy\goodbreak
\and
Dipartimento di Fisica e Scienze della Terra, Universit\`{a} di Ferrara, Via Saragat 1, 44122 Ferrara, Italy\goodbreak
\and
Dipartimento di Fisica, Universit\`{a} La Sapienza, P. le A. Moro 2, Roma, Italy\goodbreak
\and
Dipartimento di Fisica, Universit\`{a} degli Studi di Milano, Via Celoria, 16, Milano, Italy\goodbreak
\and
Dipartimento di Fisica, Universit\`{a} degli Studi di Trieste, via A. Valerio 2, Trieste, Italy\goodbreak
\and
Dipartimento di Matematica, Universit\`{a} di Roma Tor Vergata, Via della Ricerca Scientifica, 1, Roma, Italy\goodbreak
\and
Discovery Center, Niels Bohr Institute, Blegdamsvej 17, Copenhagen, Denmark\goodbreak
\and
Discovery Center, Niels Bohr Institute, Copenhagen University, Blegdamsvej 17, Copenhagen, Denmark\goodbreak
\and
European Southern Observatory, ESO Vitacura, Alonso de Cordova 3107, Vitacura, Casilla 19001, Santiago, Chile\goodbreak
\and
European Space Agency, ESAC, Planck Science Office, Camino bajo del Castillo, s/n, Urbanizaci\'{o}n Villafranca del Castillo, Villanueva de la Ca\~{n}ada, Madrid, Spain\goodbreak
\and
European Space Agency, ESTEC, Keplerlaan 1, 2201 AZ Noordwijk, The Netherlands\goodbreak
\and
Gran Sasso Science Institute, INFN, viale F. Crispi 7, 67100 L'Aquila, Italy\goodbreak
\and
HGSFP and University of Heidelberg, Theoretical Physics Department, Philosophenweg 16, 69120, Heidelberg, Germany\goodbreak
\and
Haverford College Astronomy Department, 370 Lancaster Avenue, Haverford, Pennsylvania, U.S.A.\goodbreak
\and
Helsinki Institute of Physics, Gustaf H\"{a}llstr\"{o}min katu 2, University of Helsinki, Helsinki, Finland\goodbreak
\and
INAF - Osservatorio Astrofisico di Catania, Via S. Sofia 78, Catania, Italy\goodbreak
\and
INAF - Osservatorio Astronomico di Padova, Vicolo dell'Osservatorio 5, Padova, Italy\goodbreak
\and
INAF - Osservatorio Astronomico di Roma, via di Frascati 33, Monte Porzio Catone, Italy\goodbreak
\and
INAF - Osservatorio Astronomico di Trieste, Via G.B. Tiepolo 11, Trieste, Italy\goodbreak
\and
INAF/IASF Bologna, Via Gobetti 101, Bologna, Italy\goodbreak
\and
INAF/IASF Milano, Via E. Bassini 15, Milano, Italy\goodbreak
\and
INFN, Sezione di Bologna, viale Berti Pichat 6/2, 40127 Bologna, Italy\goodbreak
\and
INFN, Sezione di Ferrara, Via Saragat 1, 44122 Ferrara, Italy\goodbreak
\and
INFN, Sezione di Roma 1, Universit\`{a} di Roma Sapienza, Piazzale Aldo Moro 2, 00185, Roma, Italy\goodbreak
\and
INFN, Sezione di Roma 2, Universit\`{a} di Roma Tor Vergata, Via della Ricerca Scientifica, 1, Roma, Italy\goodbreak
\and
INFN/National Institute for Nuclear Physics, Via Valerio 2, I-34127 Trieste, Italy\goodbreak
\and
IPAG: Institut de Plan\'{e}tologie et d'Astrophysique de Grenoble, Universit\'{e} Grenoble Alpes, IPAG, F-38000 Grenoble, France, CNRS, IPAG, F-38000 Grenoble, France\goodbreak
\and
ISDC, Department of Astronomy, University of Geneva, ch. d'Ecogia 16, 1290 Versoix, Switzerland\goodbreak
\and
IUCAA, Post Bag 4, Ganeshkhind, Pune University Campus, Pune 411 007, India\goodbreak
\and
Imperial College London, Astrophysics group, Blackett Laboratory, Prince Consort Road, London, SW7 2AZ, U.K.\goodbreak
\and
Infrared Processing and Analysis Center, California Institute of Technology, Pasadena, CA 91125, U.S.A.\goodbreak
\and
Institut N\'{e}el, CNRS, Universit\'{e} Joseph Fourier Grenoble I, 25 rue des Martyrs, Grenoble, France\goodbreak
\and
Institut Universitaire de France, 103, bd Saint-Michel, 75005, Paris, France\goodbreak
\and
Institut d'Astrophysique Spatiale, CNRS, Univ. Paris-Sud, Universit\'{e} Paris-Saclay, B\^{a}t. 121, 91405 Orsay cedex, France\goodbreak
\and
Institut d'Astrophysique de Paris, CNRS (UMR7095), 98 bis Boulevard Arago, F-75014, Paris, France\goodbreak
\and
Institut f\"ur Theoretische Teilchenphysik und Kosmologie, RWTH Aachen University, D-52056 Aachen, Germany\goodbreak
\and
Institute for Space Sciences, Bucharest-Magurale, Romania\goodbreak
\and
Institute of Astronomy, University of Cambridge, Madingley Road, Cambridge CB3 0HA, U.K.\goodbreak
\and
Institute of Theoretical Astrophysics, University of Oslo, Blindern, Oslo, Norway\goodbreak
\and
Instituto de Astrof\'{\i}sica de Canarias, C/V\'{\i}a L\'{a}ctea s/n, La Laguna, Tenerife, Spain\goodbreak
\and
Instituto de F\'{\i}sica de Cantabria (CSIC-Universidad de Cantabria), Avda. de los Castros s/n, Santander, Spain\goodbreak
\and
Istituto Nazionale di Fisica Nucleare, Sezione di Padova, via Marzolo 8, I-35131 Padova, Italy\goodbreak
\and
Jet Propulsion Laboratory, California Institute of Technology, 4800 Oak Grove Drive, Pasadena, California, U.S.A.\goodbreak
\and
Jodrell Bank Centre for Astrophysics, Alan Turing Building, School of Physics and Astronomy, The University of Manchester, Oxford Road, Manchester, M13 9PL, U.K.\goodbreak
\and
Kavli Institute for Cosmological Physics, University of Chicago, Chicago, IL 60637, USA\goodbreak
\and
Kavli Institute for Cosmology Cambridge, Madingley Road, Cambridge, CB3 0HA, U.K.\goodbreak
\and
Kazan Federal University, 18 Kremlyovskaya St., Kazan, 420008, Russia\goodbreak
\and
LAL, Universit\'{e} Paris-Sud, CNRS/IN2P3, Orsay, France\goodbreak
\and
LERMA, CNRS, Observatoire de Paris, 61 Avenue de l'Observatoire, Paris, France\goodbreak
\and
Laboratoire AIM, IRFU/Service d'Astrophysique - CEA/DSM - CNRS - Universit\'{e} Paris Diderot, B\^{a}t. 709, CEA-Saclay, F-91191 Gif-sur-Yvette Cedex, France\goodbreak
\and
Laboratoire Traitement et Communication de l'Information, CNRS (UMR 5141) and T\'{e}l\'{e}com ParisTech, 46 rue Barrault F-75634 Paris Cedex 13, France\goodbreak
\and
Laboratoire de Physique Subatomique et Cosmologie, Universit\'{e} Grenoble-Alpes, CNRS/IN2P3, 53, rue des Martyrs, 38026 Grenoble Cedex, France\goodbreak
\and
Laboratoire de Physique Th\'{e}orique, Universit\'{e} Paris-Sud 11 \& CNRS, B\^{a}timent 210, 91405 Orsay, France\goodbreak
\and
Lawrence Berkeley National Laboratory, Berkeley, California, U.S.A.\goodbreak
\and
Lebedev Physical Institute of the Russian Academy of Sciences, Astro Space Centre, 84/32 Profsoyuznaya st., Moscow, GSP-7, 117997, Russia\goodbreak
\and
Leung Center for Cosmology and Particle Astrophysics, National Taiwan University, Taipei 10617, Taiwan\goodbreak
\and
Max-Planck-Institut f\"{u}r Astrophysik, Karl-Schwarzschild-Str. 1, 85741 Garching, Germany\goodbreak
\and
McGill Physics, Ernest Rutherford Physics Building, McGill University, 3600 rue University, Montr\'{e}al, QC, H3A 2T8, Canada\goodbreak
\and
National University of Ireland, Department of Experimental Physics, Maynooth, Co. Kildare, Ireland\goodbreak
\and
Nicolaus Copernicus Astronomical Center, Bartycka 18, 00-716 Warsaw, Poland\goodbreak
\and
Niels Bohr Institute, Blegdamsvej 17, Copenhagen, Denmark\goodbreak
\and
Niels Bohr Institute, Copenhagen University, Blegdamsvej 17, Copenhagen, Denmark\goodbreak
\and
Nordita (Nordic Institute for Theoretical Physics), Roslagstullsbacken 23, SE-106 91 Stockholm, Sweden\goodbreak
\and
Optical Science Laboratory, University College London, Gower Street, London, U.K.\goodbreak
\and
Physics Department, Shahid Beheshti University, Tehran, Iran\goodbreak
\and
SISSA, Astrophysics Sector, via Bonomea 265, 34136, Trieste, Italy\goodbreak
\and
SMARTEST Research Centre, Universit\`{a} degli Studi e-Campus, Via Isimbardi 10, Novedrate (CO), 22060, Italy\goodbreak
\and
School of Physics and Astronomy, Cardiff University, Queens Buildings, The Parade, Cardiff, CF24 3AA, U.K.\goodbreak
\and
School of Physics and Astronomy, University of Nottingham, Nottingham NG7 2RD, U.K.\goodbreak
\and
Sorbonne Universit\'{e}-UPMC, UMR7095, Institut d'Astrophysique de Paris, 98 bis Boulevard Arago, F-75014, Paris, France\goodbreak
\and
Space Research Institute (IKI), Russian Academy of Sciences, Profsoyuznaya Str, 84/32, Moscow, 117997, Russia\goodbreak
\and
Space Sciences Laboratory, University of California, Berkeley, California, U.S.A.\goodbreak
\and
Special Astrophysical Observatory, Russian Academy of Sciences, Nizhnij Arkhyz, Zelenchukskiy region, Karachai-Cherkessian Republic, 369167, Russia\goodbreak
\and
Stanford University, Dept of Physics, Varian Physics Bldg, 382 Via Pueblo Mall, Stanford, California, U.S.A.\goodbreak
\and
Sub-Department of Astrophysics, University of Oxford, Keble Road, Oxford OX1 3RH, U.K.\goodbreak
\and
Sydney Institute for Astronomy, School of Physics A28, University of Sydney, NSW 2006, Australia\goodbreak
\and
The Oskar Klein Centre for Cosmoparticle Physics, Department of Physics,Stockholm University, AlbaNova, SE-106 91 Stockholm, Sweden\goodbreak
\and
Theory Division, PH-TH, CERN, CH-1211, Geneva 23, Switzerland\goodbreak
\and
UPMC Univ Paris 06, UMR7095, 98 bis Boulevard Arago, F-75014, Paris, France\goodbreak
\and
Universit\'{e} de Toulouse, UPS-OMP, IRAP, F-31028 Toulouse cedex 4, France\goodbreak
% ...
}

\abstract{\vspace{-2.5mm}This paper presents cosmological results based on
full-mission \Planck\ observations of temperature and polarization
anisotropies of the cosmic microwave background (CMB) radiation. Our
results are in very good agreement with the 2013 analysis of
the \Planck\ nominal-mission temperature data, but with increased
precision. The temperature and polarization power spectra are
consistent with the standard spatially-flat 6-parameter \LCDM\
cosmology with a power-law spectrum of adiabatic scalar perturbations
(denoted ``base \LCDM'' in this paper). From the \Planck\ temperature
data combined with \Planck\ lensing, for this cosmology we find a
Hubble constant, $H_0= (67.8 \pm 0.9)\,\Hunit$, a matter density
parameter $\Omega_{\rm m} = 0.308 \pm 0.012$, and a tilted scalar
spectral index with $n_{\rm s} = 0.968 \pm 0.006$, consistent with the
2013 analysis. Note that in this abstract we quote 68\,\% confidence limits on
measured parameters and 95\,\% upper limits on other parameters.  We
present the first results of polarization measurements with the Low
Frequency Instrument at large angular scales.  Combined with the \Planck\
temperature and lensing data,  these measurements give a reionization optical
depth of $\tau = 0.066 \pm 0.016$, corresponding to a reionization redshift of
$z_{\rm re}=8.8^{+1.7}_{-1.4}$.
These results are consistent with those from WMAP polarization
measurements cleaned for dust emission using 353-GHz polarization
maps from the High Frequency Instrument.  We find no evidence for any
departure from base \LCDM\ in the neutrino sector of the theory; for
example, combining \Planck\ observations with other astrophysical data
we find $N_{\rm eff} = 3.15 \pm 0.23$ for the effective number of relativistic
degrees of freedom, consistent with the value $N_{\rm eff} = 3.046$ of
the Standard Model of particle physics. The sum of
neutrino masses is constrained to $\sum m_{\nu} < 0.23\,{\rm eV}$.
The spatial curvature of our Universe is found to be very close to
zero, with $\vert \Omega_{K}\vert < 0.005$. Adding a tensor component as a
single-parameter extension to base \LCDM\ we find an upper limit on the
tensor-to-scalar ratio 
of $r_{0.002} <0.11$, consistent with the \Planck\ 2013 results and
consistent with the $B$-mode polarization constraints from a joint
analysis of BICEP2, \textit{Keck Array}, and \Planck\ (BKP)
data. Adding the BKP $B$-mode data to our analysis leads to a tighter
constraint of $r_{0.002} < 0.09$ and disfavours inflationary models
with a $V(\phi) \propto \phi^2$ potential. The addition of \Planck\
polarization data leads to strong constraints on deviations from a
purely adiabatic spectrum of fluctuations. We find no evidence for any
contribution from isocurvature perturbations or from cosmic
defects. Combining \Planck\ data with other astrophysical data,
including Type Ia supernovae, the equation of state of dark energy is
constrained to $w = -1.006 \pm 0.045$, consistent with the expected
value for a cosmological constant. The standard big bang
nucleosynthesis predictions for the helium and deuterium abundances
for the best-fit \Planck\ base \LCDM\ cosmology are in excellent
agreement with observations. We also analyse constraints on
annihilating dark matter and on possible deviations from the standard
recombination history. In neither case do we find no evidence for new
physics.  The \Planck\ results for base \LCDM\ are in good agreement
with baryon acoustic oscillation data and with the JLA sample of Type
Ia  supernovae.  However, as in the 2013 analysis, the amplitude of the
fluctuation spectrum is found to be higher than inferred from some
analyses of rich cluster counts and weak gravitational lensing. We
show that these tensions cannot easily be resolved with simple
modifications of the base \LCDM\ cosmology. Apart from these tensions,
the base \LCDM\ cosmology provides an excellent description of
the \Planck\ CMB observations and many other astrophysical data sets.
}


\date{\vspace{-0.23in} \today}
\titlerunning{Cosmological parameters}
\maketitle

%\tableofcontents
%\listoftables
%\listoffigures

\section{Introduction} \label{sec:introduction}

The cosmic microwave background (CMB) radiation offers an extremely
powerful way of testing the origin of fluctuations and of constraining
the matter content, geometry, and late-time evolution of the
Universe. Following the discovery of anisotropies in the CMB by the
\COBE\ satellite \citep{Smoot:92}, ground-based, sub-orbital
experiments and notably the \WMAP\ satellite \citep{Bennett:03,
  bennett2012} have mapped the CMB anisotropies with increasingly high
precision, providing a wealth of new information on cosmology.

\Planck\footnote{\Planck\ (\url{http://www.esa.int/Planck}) is a project of the European Space Agency  (ESA) with instruments provided by two scientific consortia funded by ESA member states and led by Principal Investigators from France and Italy, telescope reflectors provided through a collaboration between ESA and a scientific consortium led and funded by Denmark, and additional contributions from NASA (USA).}
is the third-generation space mission, following \COBE\ and
\WMAP, dedicated to measurements of the CMB anisotropies.  The first
cosmological results from \Planck\ were reported in a series of papers
\citep[for an overview see][and references therein] {planck2013-p01}
together with a public release of the first 15.5 months of temperature
data (which we will refer to as the nominal mission data).
Constraints on cosmological parameters from \Planck\ were reported in
\cite{planck2013-p11}.\footnote{This paper refers extensively to the
  earlier 2013 \Planck\ cosmological parameters paper and CMB power
  spectra and likelihood paper \citep{planck2013-p11, planck2013-p08}.
  To simplify the presentation, these papers will henceforth be
  referred to as \paramsI\ and \likeI, respectively.}  The
\Planck\ 2013 analysis showed that the temperature power spectrum from
\Planck\ was remarkably consistent with a spatially flat
\LCDM\ cosmology specified by six parameters, which we will refer to
as the base \LCDM\ model. However, the cosmological parameters of this
model were found to be in tension, typically at the 2--3$\,\sigma$
level, with some other astronomical measurements, most notably direct
estimates of the Hubble constant \citep{Riess:11}, the matter density
determined from distant supernovae \citep{Conley:2011, Rest:2014}, and
estimates of the amplitude of the fluctuation spectrum from weak
gravitational lensing \citep{Heymans:2013, Mandelbaum:13} and the
abundance of rich clusters of galaxies \citep{planck2013-p15,
  Benson:13, Hasselfield:2013wf}.  As reported in the revised version of
{\paramsI}, and discussed further in Sect.~\ref{sec:datasets}, some of
these tensions have been resolved with the acquisition of more
astrophysical data, while other new tensions have emerged.

The primary goal of this paper is to present the results from the full
\Planck\ mission, including a first analysis of the
\Planck\ polarization data.  In addition, this paper introduces some
refinements in data analysis and addresses the effects of small
instrumental systematics discovered (or better understood) since
{\paramsI} appeared.

The \Planck\ 2013 data were not entirely free of systematic effects.  The
\Planck\ instruments and analysis chains are complex and our understanding of
systematics has improved since {\paramsI}.  The most important of these
was the incomplete removal of line-like features in the
power spectrum of the time-ordered data, caused by interference of the 4-K
cooler electronics with the bolometer readout electronics.
This resulted in correlated systematics across detectors, leading to a small
``dip'' in the power spectra at multipoles $\ell\approx1800$ at 217\,GHz,
which is most noticeable in the first sky survey.
Various tests were presented in {\paramsI} that
suggested that this systematic caused only small shifts to
cosmological parameters. Further analyses, based on the full mission data from
the HFI (29 months, 4.8 sky surveys) are consistent with this conclusion (see
Sect.~\ref{sec:planck_only}).  In addition, we discovered
a minor error in the beam transfer functions applied to the 2013 217-GHz
spectra, which had negligible impact on the scientific results.  Another
feature of the \Planck\ data, not fully understood at the time of the 2013 data
release, was a $2.6\,\%$ calibration offset (in power) between \Planck\ and
\WMAP\ \citep[reported in {\paramsI}, see also][]{planck2013-p01a}.
As discussed in Appendix A of {\paramsI}, the
2013 \Planck\ and \WMAP\ power spectra agree to high precision if this
multiplicative factor is taken into account and it has no significant
impact on cosmological parameters apart from a rescaling of the
amplitude of the primordial fluctuation spectrum. The reasons for the 2013
calibration offsets are now largely understood and in the 2015 release the
calibrations of both \Planck\ instruments and \WMAP\ are consistent to
within about  $0.3\,\%$ in power \citep[see][for further details]{planck2014-a01}. In addition, the \Planck\ beams have been
characterized more accurately in the 2015 data release and there have
been minor modifications to the low-level data processing.

The layout of this paper is as follows.  Section~\ref{sec:model}
summarizes a number of small changes to the parameter estimation
methodology since {\paramsI}. The full mission temperature and
polarization power spectra are presented in
Sect.~\ref{sec:planck_only}.  The first subsection
(Sect.~\ref{subsec:planck_only1}) discusses the changes in the
cosmological parameters of the base \LCDM\ cosmology compared to those
presented in 2013.  Section~\ref{subsec:planck_only2} presents an
assessment of the impact of foreground cleaning (using the 545-GHz
maps) on the cosmological parameters of the base \LCDM\ model.  The
power spectra and associated likelihoods are presented in
Sect.~\ref{subsec:planck_only3}.  This subsection also discusses the
internal consistency of the \Planck\ $TT$, $TE$, and $EE$ spectra. The
agreement of $TE$ and $EE$ with the $TT$ spectra provides an important
additional test of the accuracy of our foreground corrections to the
$TT$ spectra at high multipoles.

{\paramsI} used the WMAP polarization likelihood at low multipoles to
constrain the reionization optical depth parameter $\tau$.  The 2015
analysis replaces the WMAP likelihood with polarization data from the
\Planck\ Low Frequency Instrument \citep[LFI,][]{planck2014-a03}.
The impact of this change on
$\tau$ is discussed in Sect.~\ref{subsec:tau}, which also presents an
alternative (and competitive) constraint on $\tau$ based on combining
the \Planck\ $TT$ spectrum with the power spectrum of the lensing
potential measured by \Planck. We also compare the LFI polarization
constraints with the WMAP polarization data cleaned with the
\Planck\ HFI 353-GHz maps.

Section~\ref{sec:actspt} compares the \Planck\ power spectra with the
power spectra from high-resolution ground-based CMB data from the
Atacama Cosmology Telescope \citep[ACT,][]{Das:2014} and the South
Pole Telescope \citep[SPT,][]{George:2014}.  This section applies a
Gibbs sampling technique to sample over foreground and other
``nuisance'' parameters to recover the underlying CMB power spectrum
at high multipoles \citep{Dunkley:2013,Calabrese:2013}.  Unlike
{\paramsI}, in which we combined the likelihoods of the high-resolution
experiments with the \Planck\ temperature likelihood, in
this paper we use the high-resolution experiments mainly to check the
consistency of the ``damping tail'' in the \Planck\ power spectrum at
multipoles $\ga 2000$.

Section~\ref{sec:datasets} introduces additional data, including the
\Planck\ lensing likelihood \citep[described in detail in][]{planck2014-a17}
and other astrophysical data sets. As in
{\paramsI}, we are highly selective in the
astrophysical data sets that we combine with
\Planck. As mentioned above, the main purpose of this
paper is to describe what the \Planck\ data have to
say about cosmology. It is not our purpose to present
an exhaustive discussion of what happens when the
\Planck\ data are combined with a wide range of
astrophysical data. This can be done by others, using
the publicly released
\Planck\ likelihood. Nevertheless, some cosmological
parameter combinations are highly degenerate using
CMB power spectrum measurements alone, the most
severe being the ``geometrical degeneracy'' that
opens up when spatial curvature is allowed to
vary. Baryon acoustic oscillation (BAO) measurements
are a particularly important astrophysical data
set. Since BAO surveys involve a simple geometrical
measurement, these data are less prone to systematic
errors than most other astrophysical data. As in
{\paramsI}, BAO measurements are used as a primary
astrophysical data set in combination with
\Planck\ to break parameter degeneracies.  It is
worth mentioning explicitly our approach to
interpreting tensions between \Planck\ and other
astrophysical data sets. Tensions may be indicators of
new physics beyond that assumed in the base
\LCDM\ model. However, they may also be caused by
systematic errors in the data. Our primary goal is to
report {\it whether the \Planck\ data support any
evidence for new physics.} If evidence for new
physics is driven primarily by astrophysical data,
but not by \Planck, then the emphasis must
necessarily shift to establishing whether the
astrophysical data are free of systematics. This type
of assessment is beyond the scope of this paper, but
sets a course for future research.

Extensions to the base \LCDM\ cosmology are discussed in
Sect.~\ref{sec:grid}, which explores a large grid of possibilities.
In addition to these models, we also explore
constraints on big bang nucleosynthesis, dark matter annihilation,
cosmic defects, and departures from the standard recombination
history.  As in {\paramsI}, we find no convincing evidence for a
departure from the base \LCDM\ model.  As far as we can tell, a simple
inflationary model with a slightly tilted, purely adiabatic, scalar
fluctuation spectrum fits the \Planck\ data and most other precision
astrophysical data.  There are some ``anomalies'' in this picture,
including the poor fit to the CMB temperature fluctuation spectrum at
low multipoles, as reported by WMAP \citep{Bennett:03} and in
{\paramsI}, suggestions of departures from statistical isotropy at low
multipoles \citep[as reviewed in][]{planck2013-p09,planck2014-a18}, and hints
of a discrepancy with the amplitude of the matter fluctuation spectrum at
low redshifts (see Sect.~\ref{sec:additional_data}).  However, none of
these anomalies are of decisive statistical significance at this stage.

One of the most interesting developments since the appearance of
{\paramsI} was the detection by the BICEP2 team of a $B$-mode
polarization anisotropy \citep{BicepDetection}, apparently in conflict with
the 95\,\% upper limit on the tensor-to-scalar ratio,
$r_{0.002} < 0.11$,\footnote{The subscript on $r$ refers to the pivot scale
in ${\rm Mpc}^{-1}$ used to define the tensor-to-scalar ratio. For
\Planck\ we usually quote $r_{0.002}$,
since a pivot scale of $0.002\,{\rm Mpc}^{-1}$
is close to the scale at which there is some sensitivity to tensor modes
in the large-angle temperature power spectrum.
For a scalar spectrum with no running and a scalar
spectral index of $\ns = 0.965$, $r_{0.05} \approx 1.12
r_{0.002}$ for small $r$. For  $r\approx 0.1$, assuming the
inflationary consistency relation, we have instead
$r_{0.05} \approx 1.08 r_{0.002}$.} reported in {\paramsI}.
Clearly, the detection of
$B$-mode signal from primordial gravitational waves would have
profound consequences for cosmology and inflationary theory. However,
a number of studies, in particular an analysis of \Planck\ 353-GHz
polarization data, suggested that polarized dust emission might
contribute a significant part of the BICEP2 signal
\citep{planck2014-XXX, Mortonson:2014, Flauger:2014}.  The situation
is now clearer following the joint analysis of BICEP2, Keck Array, and \Planck\
data \citep[][hereafter {\BKP}]{pb2015}; this increases the
signal-to-noise ratio on polarized dust emission primarily by directly
cross-correlating the BICEP2 and Keck Array data at 150\,GHz with the
\Planck\ polarization data at 353\,GHz. The results of {\BKP} give a
95\,\% upper limit on the tensor-to-scalar ratio of $r_{0.05} <
0.12$,
with no statistically significant evidence for a primordial gravitational
wave signal. Section~\ref{subsec:params_early} presents a brief discussion of
this result and how it fits in with the indirect constraints on $r$ derived
from the \Planck\ 2015 data.

Our conclusions are summarized in Sect.~\ref{sec:conclusions}.

\section{Model, parameters, and methodology} \label{sec:model}

The notation, definitions and methodology used in this paper largely
follow those described in {\paramsI},
and so will not be repeated here.  For completeness, we list some
derived parameters of interest in Sect.~\ref{subsec:derived}.
We have made a small number of modifications
to the methodology, as described in Sect.~\ref{subsec:Model}. We have also
made some minor changes to the model of unresolved foregrounds
and nuisance parameters used in the high-$\ell$ likelihood. These are
described in detail in \cite{planck2014-a13}, but to make this paper
more self-contained, these changes are summarized in
Sect.~\ref{subsec:foreground}.

%\subsection{Model and parameters}

\subsection{Theoretical model}
\label{subsec:Model}

We adopt the same general methodology as described in {\paramsI}, with small
modifications.  Our main results are now based on the lensed CMB power spectra
computed with the updated January 2015 version of the
\CAMB\footnote{\url{http://camb.info}} Boltzmann code~\citep{Lewis:1999bs},
and parameter constraints are based on the January 2015 version of \COSMOMC\
~\citep{Lewis:2002ah,Lewis:2013hha}. Changes in our physical
modelling are as follows.
\begin{itemize}
\item For each model in which the fraction of baryonic mass in helium $\yhe$ is
{\it not\/} varied independently of other parameters, it is now set from the big bang
nucleosynthesis (BBN) prediction by interpolation from a recent fitting
formula based on results from the {\tt PArthENoPE} BBN code
\citep{Pisanti:2007hk}.
We now use a fixed fiducial neutron decay constant of
$\tau_\neutron = 880.3\,{\rm s}$, and also account for the small difference
between the mass-fraction ratio $\yhe$ and the nucleon-based fraction
$Y_{\rm P}^{\rm BBN}$. These modifications result in changes of about  1\,\%
to  the inferred value of $\yhe$ compared to \paramsI, giving best-fit values
$\yhe \approx 0.2453$ ($\ypbbn \approx 0.2467$) in \lcdm.
See Sect.~\ref{sec:bbn} for a detailed discussion of the impact of
uncertainties arising from variations of $\tau_\neutron$ and nuclear reaction
rates; however, these uncertainties have minimal impact on our main results.
Section~\ref{sec:bbn} also corrects a small error arising from how the difference
between $\nnu=3.046$ and $\nnu=3$ was handled in the BBN fitting formula.

\item We have corrected a missing source term in the dark energy
modelling for $w\ne -1$. 
The correction of this error has very little impact on our science results,
since it is only important for values of $w$ far from $-1$.

\item To model the small-scale matter power spectrum, we use the \HALOFIT\
approach \citep{Smith:2002dz}, with the updates of \cite{Takahashi:2012em},
as in {\paramsI}, but with revised fitting parameters for
massive neutrino models.\footnote{Results for neutrino models with galaxy
and CMB lensing alone use
the \CAMB\ Jan 2015 version of \HALOFIT\ to avoid problems at large $\Omm$;
other results use the previous (April 2014) \HALOFIT\ version.}
 We also now include the
\HALOFIT\ corrections when calculating the lensed CMB power spectra.
\end{itemize}

As in {\paramsI} we adopt a Bayesian framework for testing theoretical models.
Tests using the ``profile likelihood'' method, described in
\cite{planck2013-XVI}, show excellent agreement for the mean
values of the cosmological parameters and their errors, for both the base
\lcdm\ model and its $N_{\rm eff}$
extension. Tests have also been carried out using the \CLASS\ Boltzmann
code~\citep{Lesgourgues:2011re} and the \MontePython\ MCMC code
\citep{Audren:2012wb} in place of \CAMB\ and \COSMOMC, respectively.
Again, for flat models
we find excellent agreement with the baseline
choices used in this paper.

\subsection{Derived parameters}
\label{subsec:derived}
Our base parameters are defined as in {\paramsI}, and we also calculate the
same derived parameters. In addition we now compute:
\begin{itemize}
\item the helium nucleon fraction defined by
 $Y_\mathrm{P}^\mathrm{BBN} \equiv 4 n_\mathrm{He}/n_\mathrm{b}$;
\item where standard BBN is assumed, the mid-value deuterium ratio predicted
 by BBN, $y_\mathrm{DP} \equiv 10^{5} n_\mathrm{D}/n_\mathrm{H}$, using a fit
 from the {\tt PArthENoPE} BBN code \citep{Pisanti:2007hk};
\item the comoving wavenumber of the perturbation mode that entered the Hubble radius at matter-radiation equality $\zeq$, where this redshift is calculated
 approximating all neutrinos as relativistic at that time, i.e.,
 $\keq \equiv a(\zeq)H(\zeq)$;
\item the comoving angular diameter distance to last scattering,
 $\DAstar(z_*)$;
\item the angular scale of the sound horizon at matter-radiation equality,
 $\theta_{\rm s, eq} \equiv \rs(\zeq)/\DAstar(z_*)$, where $\rs$ is the sound
 horizon and $z_*$ is the redshift of last scattering;
\item the amplitude of the CMB power spectrum $\mathcal{D}_\ell \equiv
\ell(\ell+1)C_\ell/2\pi$ in $\muK^2$, for
 $\ell = 40$, 220, 810, 1520, and 2000;
\item the primordial spectral index of the curvature perturbations at wavenumber
 $k=0.002\,\Mpc^{-1}$, $n_{{\rm s}, 0.002}$ (as in {\paramsI}, our default pivot
 scale is $k=0.05\, \Mpc^{-1}$, so that $\ns \equiv n_{{\rm s}, 0.05}$);
\item parameter combinations close to those probed by galaxy and CMB lensing
 (and other external data), specifically $\sigma_8 \Omm^{0.5}$ and
 $\sigma_8 \Omm^{0.25}$;
\item various quantities reported by BAO and redshift-space distortion
 measurements, as described in Sects.~\ref{sec:BAO} and \ref{subsec:RSD}.
\end{itemize}

\subsection{Changes to the foreground model}
\label{subsec:foreground}

Unresolved foregrounds contribute to the temperature power spectrum
and must be modelled to extract accurate cosmological parameters.
{\likeI} and {\paramsI}
used a parametric approach to modelling foregrounds,
similar to the approach adopted in the analysis of the SPT and ACT
experiments \citep{Reichardt:2012,Dunkley:2013}. The unresolved foregrounds
are described by a set of power spectrum
templates together with nuisance parameters, which are sampled via MCMC
along with the cosmological parameters.\footnote{Our treatment of Galactic
dust emission also differs from that used in {\likeI} and {\paramsI}.
Here we describe changes to the extragalactic model and our treatment
of errors in the  \Planck\ absolute
calibration, deferring a discussion of Galactic dust modelling in temperature and polarization to
Sect.~\ref{sec:planck_only}.} The components of the extragalactic foreground
model consist of:
\begin{itemize}
\item the shot noise from Poisson fluctuations in the number density of point sources;
\item the power due to clustering of point sources (loosely referred to as the CIB component);
\item a thermal Sunyaev-Zeldovich (tSZ) component;
\item a kinetic Sunyaev-Zeldovich (kSZ) component;
\item the cross-correlation between tSZ and CIB.
\end{itemize}

In addition, the likelihood includes a number of other nuisance parameters,
such as relative calibrations between frequencies, and beam eigenmode
amplitudes.
We use the same templates for the tSZ, kSZ, and tSZ/CIB cross-correlation
as in the 2013 papers. However, we have made a number of changes to the CIB
modelling and the priors adopted for the SZ effects, which
we now describe in detail.

\subsubsection{CIB}

In the 2013 papers, the CIB anisotropies were modelled as a power law:
\begin{equation}
{\cal D}_\ell^{\nu_1\times \nu_2} = A^{\rm CIB}_{\nu_1 \times \nu_2}
 \left ( {\ell \over 3000} \right)^{\gamma_{\rm CIB}}.  \label{NCIB1}
\end{equation}
\Planck\ data alone provide a  constraint on $A^{\rm CIB}_{217\times 217}$
and very weak constraints on the CIB amplitudes at lower frequencies.
{\paramsI} reported typical values of
$A^{\rm CIB}_{217\times 217} = (29 \pm 6)\,\mu{\rm K}^2$ and
$\gamma^{\rm CIB} = 0.40 \pm 0.15$, fitted over the range
$500 \le \ell \le 2500$. The addition of the ACT and SPT data (``highL'')
led to solutions with steeper values of $\gamma_{\rm CIB}$, closer to $0.8$,
suggesting that the CIB component was not well fit by a power law.

\Planck\ results on the CIB, using \ion{H}{i} as a tracer of Galactic dust,
are discussed in detail in  \citet{planck2013-pip56}.
In that paper, a model with 1-halo and 2-halo contributions was
developed that provides an accurate description of the \Planck and IRAS CIB
spectra from 217\,GHz through to 3000\,GHz.  At high
multipoles, $\ell \ga 3000$, the halo-model spectra are reasonably well approximated by power laws, with a slope $\gamma_{\rm CIB} \approx 0.8$ (though see Sect.~\ref{sec:actspt}). At multipoles
in the range $500 \la \ell \la 2000$, corresponding to the
transition from the 2-halo term dominating the clustering power to the 1-halo term dominating, the \citet{planck2013-pip56} templates have a shallower slope, consistent with
the results of {\paramsI}. The amplitudes of these templates at
$\ell=3000$ are
\begin{align}
 A^{\rm CIB}_{217 \times 217} &= 63.6\,\mu{\rm K}^2,
 &A^{\rm CIB}_{143 \times 217} &= 19.1,\mu{\rm K}^2, \nonumber \\
 A^{\rm CIB}_{143 \times 143} &= 5.9\,\mu{\rm K}^2,
 &A^{\rm CIB}_{100 \times 100} &= 1.4\,\mu{\rm K}^2.   \label{NCIB2}
\end{align}
Note that in {\paramsI}, the CIB amplitude of the $143\times217$ spectrum
was characterized by a correlation coefficient
\begin{equation}
A^{\rm CIB}_{143 \times 217}
 = r^{\rm CIB}_{143\times217}\sqrt{A^{\rm CIB}_{217 \times 217}
  A^{\rm CIB}_{143 \times 143}}.  \label{NCIB3}
\end{equation}
The combined \Planck+highL solutions in {\paramsI} always give a high
correlation coefficient with a $95\,\%$ lower limit of
$r^{\rm CIB}_{143\times217} \ga 0.85$, consistent with the model of
Eq.~\eqref{NCIB2}, which has $r^{\rm CIB}_{143\times217} \approx 1$.
In the 2015 analysis, we use the \citet{planck2013-pip56} templates,
fixing the relative amplitudes at $100\times100$,
$143\times143$, and $143\times217$ to the amplitude of the
$217\times217$ spectrum. Thus, the CIB model used in this paper
is specified by only one amplitude, $A^{\rm CIB}_{217\times217}$,
which is assigned a uniform prior in the range
0--200\,$\mu{\rm K}^2$.

In {\paramsI} we solved for the CIB amplitudes
{\it at the CMB effective frequencies\/} of $217$ and
$143$\,GHz, and so we included colour corrections in the amplitudes
$A^{\rm CIB}_{217\times217}$ and $A^{\rm CIB}_{143\times143}$
(there was no CIB component in the $100\times100$ spectrum).
In the 2015 \Planck\ analysis, we do not include
a colour term since we define $A^{\rm CIB}_{217\times217}$ to be the {\it actual\/}
CIB amplitude measured in the \Planck\ $217$-GHz
band. This is higher by a factor of about $1.33$ compared to the amplitude
at the CMB effective frequency of the
\Planck\ 217-GHz band.  This should be borne in mind by readers
comparing $2015$ and $2013$ CIB amplitudes measured by \Planck.

\subsubsection{Thermal and kinetic SZ amplitudes}

In the 2013 papers we assumed template shapes for the thermal (tSZ)
and kinetic (kSZ) spectra characterized by two amplitudes, $A^{\rm tSZ}$
and $A^{\rm kSZ}$, defined in equations~(26) and (27) of {\paramsI}. These
amplitudes were assigned uniform priors in the range 0--10\,\muk. We
used the \citet{Trac:2011} kSZ template spectrum and the
$\epsilon=0.5$ tSZ template from \citet{Efstathiou:2012}.  We adopt
the same templates for the 2015 \Planck\ analysis, since, for example, the tSZ
template is actually a good match to the results from the recent numerical
simulations of \citet{McCarthy:2014}.  In addition, we previously included a template
from \cite{Addison:2012} to model the cross-correlation between the CIB and tSZ emission from clusters of galaxies. The
amplitude of this template was characterized by a dimensionless
correlation coefficient, $\xi^{\rm tSZ\times CIB}$, which was assigned
a uniform prior in the range 0--1.
The three parameters $A^{\rm tSZ}$, $A^{\rm kSZ}$,
and $\xi^{\rm tSZ\times CIB}$, are not well constrained by \Planck\ alone.
Even when combined with ACT and SPT, the three parameters are highly
correlated with each other.
Marginalizing over $\xi^{\rm tSZ\times CIB}$, \citet{Reichardt:2012}
find that SPT spectra constrain the linear combination
\begin{equation}
A^{\rm kSZ} + 1.55\,A^{\rm tSZ} = (9.2 \pm 1.3)\,\mu{\rm K}^2. \label{NSZ1}
\end{equation}
The slight differences in the coefficients compared to the
formula given in \citet{Reichardt:2012} come from the different effective
frequencies used to define the \Planck\ amplitudes $A^{\rm kSZ}$
and $A^{\rm tSZ}$.
An investigation of the 2013 \Planck+highL solutions show a similar degeneracy
direction, which is almost independent of cosmology, even for extensions
to the base \lcdm\ model:
\begin{equation}
A^{\rm SZ} = A^{\rm kSZ} + 1.6\,A^{\rm tSZ}  =  (9.4\pm 1.4)\,\mu {\rm K}^2
 \label{NSZ2}
\end{equation}
for \planck+\WP+\highL,
which is very close to the degeneracy direction (Eq.~\ref{NSZ1}) measured by
SPT.  In the 2015 \Planck\ analysis, we impose a conservative Gaussian prior
for $A^{\rm SZ}$, as defined in Eq.~\eqref{NSZ2}, with a mean of $9.5\,\mu {\rm K}^2$ and a dispersion $3 \mu{\rm K}^2$ \citep[i.e., somewhat broader than
the dispersion measured by][]{Reichardt:2012}. The purpose of imposing
this prior on $A^{\rm SZ}$ is to prevent the parameters
$A^{\rm kSZ}$ and $A^{\rm tSZ}$ from wandering into unphysical
regions of parameter space when using \Planck\ data alone. We retain the
uniform prior of [0,1] for $\xi^{\rm tSZ\times CIB}$. As this paper was being
written, results from the complete 2540\,${\rm deg}^2$ SPT-SZ survey area
appeared \citep{George:2014}. These are consistent with Eq.~\eqref{NSZ2}
and in addition constrain the correlation parameter to low values,
$\xi^{\rm tSZ\times CIB} = 0.113^{+0.057}_{-0.054}$. The looser priors on
these parameters adopted in this paper are, however, sufficient to eliminate
any significant sensitivity of cosmological parameters derived from
\Planck\ to the modelling of the SZ components.

\subsubsection{Absolute \Planck\ calibration}

In {\paramsI}, we treated the calibrations of the 100 and 217-GHz channels
relative to $143\,{\rm GHz}$ as nuisance parameters.
This was an approximate way of dealing with small differences in relative
calibrations between different detectors at high multipoles, caused by
bolometer time-transfer function corrections and intermediate and far
sidelobes of the \Planck\ beams.
In other words, we approximated these effects as a purely multiplicative
correction to the power spectra over the multipole
range $\ell=50$--$2500$.  The absolute
calibration of the 2013 \Planck\ power spectra was therefore fixed, by
construction, to the absolute calibration of the 143-5 bolometer.
Any error in the absolute calibration of this reference bolometer was not
propagated into errors on cosmological parameters.
For the 2015 \Planck\ likelihoods we use an identical relative calibration
scheme between $100$, $143$, and $217$\,GHz, but we
now include an absolute calibration parameter $y_{\rm p}$, at the map level,
for the 143-GHz reference frequency. We adopt a Gaussian prior on $y_{\rm p}$ centred on unity
with a (conservative) dispersion of $0.25\,\%$.
This overall calibration uncertainty is then propagated through to cosmological
parameters such as $\As$ and $\sigma_8$. A discussion of the
consistency of the absolute calibrations across the nine \Planck\ frequency
bands is given in \citet{planck2014-a01}.

\section{ Constraints on the parameters of the base $\boldsymbol{\Lambda}$CDM
cosmology from \planck}
\label{sec:planck_only}

\subsection{Changes in the base \lcdm\ parameters compared to the 2013 data
release}
\label{subsec:planck_only1}

\begin{table*}[tb]
\begingroup
\newdimen\tblskip \tblskip=5pt
\caption{%
Parameters of the base \lcdm\ cosmology (as defined in
  {\paramsI}) determined from the publicly released
  nominal-mission \camspec\ DetSet  likelihood [2013N(DS)] and the
  2013 full-mission \camspec\ DetSet and cross-yearly (${\rm Y}1\times
  {\rm Y}2$) likelihoods with the extended sky coverage [2013F(DS) and
  2013F(CY)]. These three likelihoods are combined with the \WMAP\
  polarization likelihood to constrain $\tau$. The column labelled
  2015F(CHM) lists parameters for a \camspec\ cross-half-mission likelihood
  constructed from the 2015 maps using similar sky coverage to the
  2013F(CY) likelihood (but greater sky coverage at $217$\,GHz and
  different point source masks, as discussed in the text). The column
  labelled 2015F(CHM) (\plik) lists parameters for the \plik\
  cross-half-mission likelihood that uses identical
  sky coverage to the \camspec\ likelihood. The 2015
  temperature likelihoods are combined with the \Planck\ \lowTEB\
  likelihood to constrain $\tau$. The last two columns list the
  deviations of the \plik\ parameters from those of the nominal-mission and
  the \camspec\ 2015(CHM) likelihoods.  To help refer to specific columns,
  we have numbered the first six  explicitly. The high-$\ell$
  likelihoods used here include only $TT$ spectra.  $H_0$ is given in the
  usual units of ${\rm km}\,{\rm s}^{-1}\,{\rm Mpc}^{-1}$.
}
\label{paramtab1}
\nointerlineskip
\vskip -3mm
%\footnotesize
\scriptsize
\setbox\tablebox=\vbox{
   \newdimen\digitwidth
   \setbox0=\hbox{\rm 0}
   \digitwidth=\wd0
   \catcode`*=\active
   \def*{\kern\digitwidth}
   \newdimen\signwidth
   \setbox0=\hbox{+}
   \signwidth=\wd0
   \catcode`!=\active
   \def!{\kern\signwidth}
\halign{\hbox to 1.0in{#\leaderfil}\tabskip 1.2em&
\hfil#\hfil&
\hfil#\hfil&
\hfil#\hfil&
\hfil#\hfil&
\hfil#\hfil&
\hfil#\hfil&
\hfil#\hfil\tabskip 0pt\cr                            % Template goes here.
\noalign{\doubleline}
\omit\hfil[1] Parameter\hfil&[2] 2013N(DS)&[3] 2013F(DS)&[4] 2013F(CY)&[5] 2015F(CHM)&[6] 2015F(CHM) (\plik)&$([2]-[6])/\sigma_{[6]}$&$([5]-[6])/\sigma_{[5]}$\cr  % Table headings go here.
\noalign{\vskip 3pt\hrule\vskip 5pt}
                                    % Body of table goes here.
$100\theta_{\rm MC}$&$1.04131\pm0.00063$&$1.04126 \pm 0.00047$&$1.04121 \pm 0.00048$&$1.04094 \pm 0.00048$&$1.04086 \pm 0.00048$&$!0.71$&$!0.17$\cr
$\Omega_bh^2$&$0.02205\pm0.00028$&$0.02234 \pm 0.00023$&$0.02230 \pm 0.00023$&$0.02225 \pm 0.00023$&$0.02222 \pm 0.00023$&$-0.61$&$!0.13$\cr
$\Omega_ch^2$&$0.1199\pm0.0027$&$0.1189 \pm 0.0022$&$0.1188\pm0.0022$&$0.1194 \pm 0.0022$&$0.1199 \pm 0.0022$&$!0.00$&$-0.23$\cr
$H_0$&$67.3 \pm 1.2*$&$67.8 \pm 1.0*$&$67.8 \pm 1.0*$&$67.48 \pm 0.98*$&$67.26 \pm 0.98*$&$!0.03$&$!0.22$\cr
$\ns$&$0.9603 \pm 0.0073$&$0.9665 \pm 0.0062$&$0.9655 \pm0.0062$&  $0.9682 \pm 0.0062$&$0.9652 \pm 0.0062$&$-0.67$&$!0.48$\cr
$\Omm$&$0.315 \pm 0.017$&$0.308 \pm 0.013$&$0.308\pm0.013$& $0.313 \pm 0.013$&$0.316 \pm 0.014$&$-0.06$&$-0.23$\cr
$\sigma_8$&$0.829 \pm 0.012$&$0.831 \pm 0.011$&$0.828\pm0.012$&$0.829 \pm 0.015$&$0.830 \pm 0.015$&$-0.08$&$-0.07$\cr
$\tau$&$0.089 \pm 0.013$&$0.096 \pm 0.013$&$0.094\pm0.013$&$0.079 \pm 0.019$&$0.078\pm0.019$&$!0.85$&$!0.05$\cr
$10^9\As e^{-2\tau}$&$1.836 \pm 0.013$&$1.833 \pm 0.011$&$1.831\pm0.011$&$1.875 \pm 0.014$& $1.881\pm0.014$& $-3.46$& $-0.42$\cr
\noalign{\vskip 5pt\hrule\vskip 3pt}}}
\endPlancktablewide                 % ends two-column \halign
\endgroup
\end{table*}

The principal conclusion of \paramsI\ was the excellent agreement of
the base \LCDM\ model with the temperature power spectra measured by
\planck. In this subsection, we compare the parameters of the base
\lcdm\ model reported in \paramsI\ with those measured from the
full-mission 2015 data. Here we restrict the comparison to the high
multipole temperature ($TT$) likelihood (plus low-$\ell$
polarization), postponing a discussion of the $TE$ and $EE$ likelihood
blocks to Sect.~\ref{subsec:planck_only2}. The main differences
between the 2013 and 2015 analyses are as follows.

\paragraph{(1)}
There have been a number of changes to the low-level \Planck\ data
processing, as discussed in \cite{planck2014-a03} and
\cite{planck2014-a08}.  These include: changes to the filtering
applied to remove ``4-K'' cooler lines from the time-ordered data
(TOD); changes to the deglitching algorithm used to correct the TOD
for cosmic ray hits; improved absolute calibration based on the
spacecraft orbital dipole and more accurate models of the beams,
accounting for the intermediate and far sidelobes. These revisions
largely eliminate the calibration difference between \planck-2013 and
WMAP reported in {\paramsI} and \cite{planck2013-p01a},
leading to upward shifts of the HFI and LFI \planck\ power spectra of
approximately 2.0\,\% and 1.7\,\%, respectively.
In addition, the mapmaking used for 2015
data processing utilizes ``polarization destriping'' for the polarized
HFI detectors \citep{planck2014-a09}.

\paragraph{(2)}
The 2013 papers used WMAP polarization measurements
\citep{bennett2012} at multipoles $\ell \le 23$ to constrain the
optical depth parameter $\tau$; this likelihood was denoted ``WP'' in
the 2013 papers.  In the 2015 analysis, the WMAP polarization
likelihood is replaced by a \Planck\ polarization likelihood
constructed from low-resolution maps of $Q$ and $U$ polarization
measured by LFI at $70$\,GHz,
foreground cleaned using the LFI 30-GHz and HFI 353-GHz maps as
polarized synchrotron and dust templates, respectively, as described in
\cite{planck2014-a13}. After a comprehensive analysis of survey-to-survey
null tests, we found possible low-level residual systematics in Surveys 2 and 4, likely related to the unfavourable alignment of the CMB dipole in those
two surveys \citep[for details see][]{planck2014-a03}. We therefore conservatively use only six of the eight LFI 70-GHz full-sky surveys, excluding Surveys
2 and 4, The
foreground-cleaned LFI 70-GHz polarization maps are used over 46\,\%
of the sky, together with the temperature map from the {\tt Commander}
component-separation algorithm over 94\,\% of the sky \citep[see][for
further details]{planck2014-a11}, to form a low-$\ell$
\planck\ temperature+polarization pixel-based likelihood that extends
up to multipole $\ell=29$.  Use of the polarization information in this likelihood is denoted as
``lowP'' in this paper The optical depth inferred from the lowP
likelihood combined with the \Planck\ $TT$ likelihood
is
typically $\tau \approx 0.07$, and is about $1\,\sigma$ lower than the
typical values of $\tau \approx 0.09$ inferred from the WMAP polarization
likelihood (see Sect.~\ref{subsec:tau}) used in the 2013 papers.  As
discussed in Sect.~\ref{subsec:tau}
\citep[and in more detail in][]{planck2014-a13} the LFI 70-GHz and WMAP polarization maps are
consistent when both are cleaned with the HFI 353-GHz polarization maps.\footnote{%
Throughout this paper, we adopt the following labels for likelihoods:
(i) \planckTTonly\ denotes the combination of the $TT$ likelihood at multipoles $\ell \geq 30$ and a low-$\ell$ temperature-only likelihood based on the CMB map recovered with {\tt Commander}; (ii) \planckTT\ further includes the \planck\ polarization data in the low-$\ell$ likelihood, as described in the main text; (iii) labels such as {\planckonly}\ TE+\lowTEB\ denote the $TE$ likelihood at $\ell \geq 30$ plus the polarization-only component of the map-based low-$\ell$ \planck\ likelihood; and (iv) \planckall\ denotes the combination of the likelihood at $\ell \geq 30$ using $TT$, $TE$, and $EE$ spectra and the low-$\ell$ temperature+polarization likelihood. We make occasional use of combinations of the polarization likelihoods at $\ell \geq 30$ and the temperature+polarization data at low-$\ell$, which we denote with labels such as {\planckonly}\ TE+\lowTP.
}

\paragraph{(3)}
In the 2013 papers, the \planck\ temperature likelihood was a hybrid:
over the multipole range $\ell\,{=}\,2$--49, the likelihood was based
on the {\tt Commander} algorithm applied to 87\,\% of the sky computed
using a Blackwell-Rao estimatorl the likelihood at higher multipoles
($\ell\,{=}50$--2500) was constructed from cross-spectra over the
frequency range 100--217\,GHz using the \camspec\ software
\citep{planck2013-p08}, which is based on the methodology developed in
\citet{Efstathiou:2004} and \citet{Efstathiou:2006}. At each of the
\planck\ HFI frequencies, the sky is observed by a number of
detectors. For example, at 217\,GHz the sky is observed by four
unpolarized spider-web bolometers (SWBs) and eight polarization
sensitive bolometers (PSBs). The TOD from the 12 bolometers can be
combined to produce a single map at 217\,GHz for any given period of
time. Thus, we can produce 217-GHz maps for individual sky surveys
(denoted S1, S2, S3, etc.), or by year (Y1, Y2), or split by
half-mission (HM1, HM2).  We can also produce a temperature map from
each SWB and a temperature and polarization map from quadruplets of
PSBs. For example, at 217\,GHz we produce four temperature and two
temperature+polarization maps.  We refer to these maps as
detectors-set maps (or ``DetSets'' for short); note that the DetSet
maps can also be produced for any arbitrary time period. The high
multipole likelihood used in the 2013 papers was computed by
cross-correlating HFI DetSet maps for the ``nominal'' \Planck\ mission
extending over 15.5 months.\footnote{Although we analysed a
  \Planck\ full-mission temperature likelihood extensively, prior to
  the release of the 2013 papers.} For the 2015 papers we use the
full-mission \Planck\ data, extending over 29 months for the HFI and 48
months for the LFI. In the \Planck\ 2015 analysis, we have produced
cross-year and cross-half-mission likelihoods in addition to a DetSet
likelihood. The baseline 2015 \Planck\ temperature-polarization
likelihood is also a hybrid, matching the high-multipole likelihood at
$\ell = 30$ to the \Planck\ pixel-based likelihood at lower multipoles.

\paragraph{(4)}
The sky coverage used in the $2013$ \camspec\ likelihood was
intentionally conservative, retaining effectively 49\,\% of the sky at 100\,GHz
and 31\,\% of the sky at 143 and 217\,GHz.\footnote{These quantities are
explicitly the apodized effective $f^{\rm eff}_{\rm sky}$, calculated as the
average of the square of the apodized mask values (see Eq.~\ref{plik1}).}
This was done to ensure
that on the first exposure of \Planck\ cosmological results to the
community, corrections for Galactic dust emission were demonstrably
small and had negligible impact on cosmological parameters.  In the 2015
analysis we make more aggressive use of the sky at each of these frequencies.
We have also tuned the point-source masks to each frequency,
rather than using a single point-source mask constructed from the
union of the point source catalogues at 100, 143, 217, and 353\,GHz.
This results in many fewer point source holes in the 2015
analysis compared to the 2013 analysis.

\paragraph{(5)}
Most of the results in this paper are derived from a revised
\plik\ likelihood, based on cross-half-mission spectra. The
\plik\ likelihood has been modified since 2013 so that it is now
similar to the \camspec\ likelihood used in {\paramsI}. Both likelihoods
use similar approximations to compute the covariance matrices. The
main difference is in the treatment of Galactic dust corrections in
the analysis of the polarization spectra.  The two likelihoods have
been written independently and give similar (but not identical)
results, as discussed further below.  The \plik\ likelihood is
discussed in \cite{planck2014-a13}. The \camspec\ likelihood is
discussed in a separate paper (Efstathiou et al.\ in preparation).

\paragraph{(6)}
We have made minor changes to the foreground modelling and to the
priors on some of the foreground parameters, as discussed in
Sect.~\ref{subsec:foreground} and \cite{planck2014-a13}.
\smallskip

Given these changes to data processing, mission length, sky coverage, etc.,
it is reasonable to ask whether the base \LCDM\ parameters have changed
significantly compared to the 2013 numbers. In fact, the parameter shifts
are relatively small.  The situation is summarized in Table~\ref{paramtab1}.
The second column of this table lists the \Planck+WP parameters, as given in
table~5 of {\paramsI}. Since these numbers are based on
the 2013 processing of the nominal mission and computed via a DetSet
\camspec\ likelihood, the column is labelled
2013N(DS). We now make a number of specific remarks about these comparisons.

\paragraph{(1) 4-K cooler line systematics.} After the submission of
{\paramsI} we found strong evidence that a residual in the
$217\times217$ DetSet spectrum at $\ell \approx 1800$ was a systematic
caused by electromagnetic interference between the Joule-Thomson 4-K
cooler electronics and the bolometer readout electronics. This
interference leads to a set of time-variable narrow lines in the power spectrum
of the TOD. The data processing pipelines apply a filter to remove these
lines; however, the filtering failed to reduce their impact on the power spectra
to negligible levels. Incomplete removal of the 4-K cooler lines
affects primarily the $217\times217$ PSB$\times$PSB cross-spectrum in
Survey~1. The
presence of this systematic was reported in the revised versions of
2013 \Planck\ papers. Using simulations and also comparison with the 2013
full-mission likelihood (in which the $217\times217$ power spectrum ``dip'' is
strongly diluted by the additional sky surveys) we assessed that the 4-K line systematic was causing shifts in cosmological parameters
of less than $0.5\,\sigma$.\footnote{The revised version of \paramsI\ also reported
an error in the ordering of the beam-transfer functions applied to
some of the 2013 $217 \times 217$ DetSet cross-spectra, leading to an offset
of a few $(\mu{\rm K})^2$ in the coadded $217 \times 217$ spectrum. As
discussed in \paramsI, this offset is largely absorbed
by the foreground model and has negligible impact on the 2013 cosmological
parameters.}  Column~3 in Table~\ref{paramtab1}
lists the DetSet parameters for the full-mission 2013
data. This full-mission likelihood uses more extensive sky coverage
than the nominal mission likelihood (effectively 39\,\% of sky at 217\,GHz,
55\,\% of sky at 143\,GHz, and 63\,\% of sky at 100\,GHz);
otherwise the methodology
and foreground model are identical to the \camspec\ likelihood
described in \likeI. The parameter shifts are
relatively small and consistent with the improvement in
signal-to-noise of the full-mission spectra and the systematic shifts
caused by the $217\times217$ dip in the nominal mission (for example,
raising $H_0$ and $\ns$, as discussed in appendix~C4 of {\paramsI}).

\paragraph{(2) DetSets versus cross-surveys.} In a reanalysis of the publicly
released \Planck\ maps, \cite{Spergel:2013} constructed cross-survey
(${\rm S}1\times {\rm S}2$) likelihoods and found cosmological
parameters for the base \LCDM\ model that were close to (within
approximately $1\,\sigma$) the nominal mission parameters listed in
Table~\ref{paramtab1}. The \cite{Spergel:2013} analysis differs
substantially in sky coverage and foreground modelling compared to the
2013 \Planck\ analysis and so it is encouraging that they find no
major differences with the results presented by the
\Planck\ collaboration.  On the other hand, they did not identify the
reasons for the roughly $1\,\sigma$ parameter shifts. They argue that
foreground modelling and the $\ell\,{=}\,1800$ dip in the
$217\times217$ DetSet spectrum can contribute towards some of the
differences but cannot produce $1\,\sigma$ shifts, in agreement with
the conclusions of {\paramsI}.
The 2013F(DS) likelihood disfavours the
\cite{Spergel:2013} cosmology (with parameters listed in their
table~3) by $\Delta \chi^2 = 11$, i.e., by about $2\,\sigma$, and
almost all of the $\Delta \chi^2$ is contributed by the multipole
range 1000--1500, so the parameter shifts are not driven by
cotemporal systematics resulting in correlated noise biases at high
multipoles.
However, as discussed in \likeI\ and \cite{planck2014-a13}, low-level
correlated noise in
the DetSet spectra affects {\it all\/} HFI channels at high multipoles
where the spectra are noise dominated.  The impact of this
correlated noise on cosmological parameters is relatively small. This is
illustrated by column~4 of Table~\ref{paramtab1} (labelled
``2013F(CY)''), which lists the parameters of a 2013
\camspec\ cross-year likelihood using the same sky coverage and
foreground model as the DetSet likelihood used for column~3. The
parameters from these two likelihoods are in good agreement (better
than $0.2\,\sigma$), illustrating
that cotemporal systematics in the DetSets are at sufficiently low
levels that there is very little effect on cosmological
parameters. Nevertheless, in the 2015 likelihood analysis we apply
corrections for correlated noise to the DetSet cross-spectra, as
discussed in \cite{planck2014-a13}, and typically find agreement in
cosmological parameters between DetSet, cross-year, and
cross-half-mission likelihoods to better than $0.5\,\sigma$ accuracy
for a fixed likelihood code (and to better than $0.2\,\sigma$ accuracy for
base \LCDM).

\begin{figure*}
\begin{center}
\includegraphics[width=190mm,angle=0]{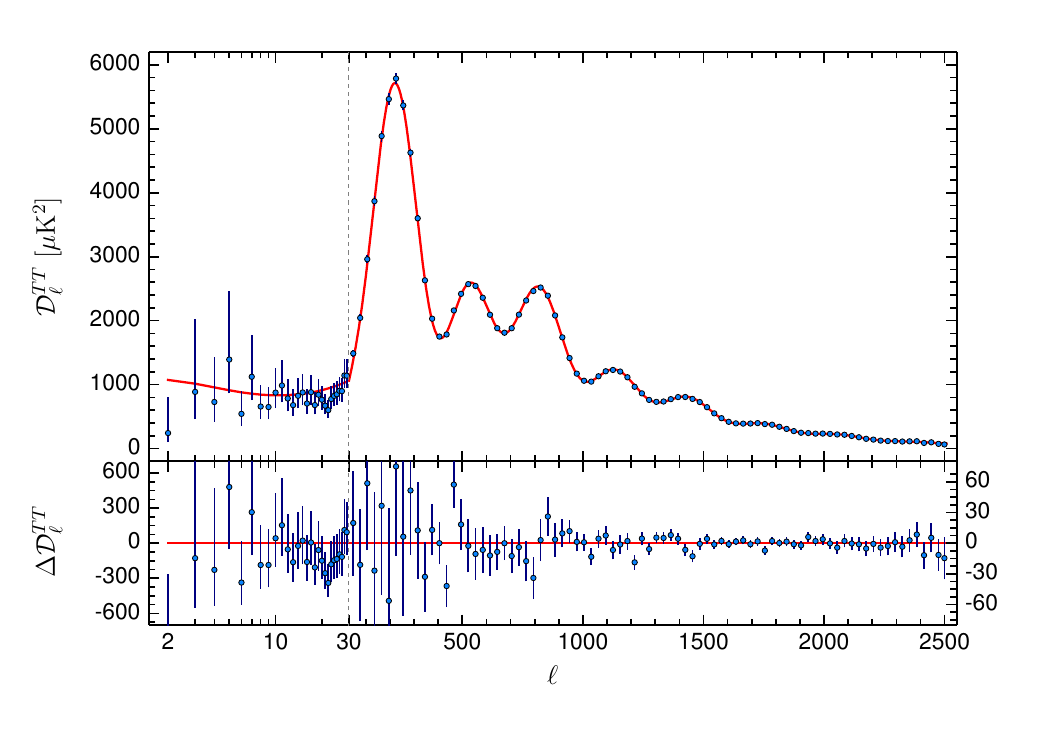}
\end{center}
\vspace{-10mm}
\caption {\Planck\ 2015 temperature power spectrum. At multipoles $\ell\ge30$
we show the maximum likelihood frequency-averaged temperature spectrum
computed from the \plik\ cross-half-mission likelihood, with foreground and other nuisance parameters
determined from the MCMC analysis of the base \LCDM\ cosmology. In the multipole range $2 \le \ell \le 29$, we plot the
power spectrum estimates from the {\tt Commander} component-separation algorithm, computed over 94\,\% of the sky.
The best-fit base \LCDM\ theoretical spectrum fitted to the \planckTT\ likelihood is plotted in the
upper panel. Residuals with respect to this model are shown in the lower panel.
The error bars show $\pm 1\,\sigma$ uncertainties.}
\label{pgTT_final}
\end{figure*}

\paragraph{(3) 2015 versus 2013 processing.}
Column~5 (labelled ``2015F(CHM)'') lists the parameters
computed from the \camspec\ cross-half-mission likelihood using the
HFI 2015 data with revised absolute calibration and beam-transfer functions.
We also replace the WP likelihood of the 2013 analysis with the \planck\
lowP likelihood. The 2015F(CHM) likelihood uses slightly more sky coverage
(60\,\%) at 217\,GHz, compared to the 2013F(CY) likelihood and also uses
revised point source masks.  Despite these changes, the base \lcdm\ parameters
derived from the 2015 \camspec\ likelihood are within $\approx 0.4\,\sigma$
of the 2013F(CY) parameters, with the exception of $\theta_{\rm MC}$,
which is lower by $0.67\,\sigma$,
$\tau$, which is lower by $1\,\sigma$, and $\As e^{-2\tau}$, which is
higher by about $4\,\sigma$ . The change in $\tau$
simply reflects the preference for a lower value of $\tau$ from the \planck\
LFI polarization data compared to the WMAP polarization likelihood in the form
delivered by the WMAP team (see Sect.~\ref{subsec:tau} for further discussion).
The large upward shift in $\As e^{-2\tau}$ reflects the change in the
absolute calibration of the HFI. As noted in Sect.~\ref{subsec:foreground},
the 2013 analysis did not propagate an error on the \Planck\ absolute
calibration through to cosmological parameters. Coincidentally, the
changes to the absolute calibration compensate for the downward change
in $\tau$ and variations in the other cosmological parameters to keep the
parameter $\sigma_8$ largely unchanged from the 2013 value.  This will be
important when we come to discuss possible tensions between the amplitude of
the matter fluctuations at low redshift estimated from various astrophysical
data sets and the \planck\ CMB values for the base \LCDM\ cosmology
(see Sect.~\ref{subsec:data_summary}).

\paragraph{(4) Likelihoods.}
Constructing a high-multipole likelihood for \planck, particularly
with $TE$ and $EE$ spectra, is complicated and difficult to check at the
sub-$\sigma$ level against numerical simulations because the simulations
cannot model the foregrounds, noise properties, and low-level
data processing of the real \Planck\ data to sufficiently high accuracy. Within the \Planck\
collaboration, we have tested the sensitivity of the results to the likelihood
methodology by developing several independent analysis pipelines.
Some of these are described in \citet{planck2014-a13}. The most highly
developed of them are the \camspec\ and revised \plik\ pipelines. For
the 2015 \planck\ papers,
 the \plik\ pipeline was chosen as the baseline.  Column~6 of Table~\ref{paramtab1}
lists the cosmological parameters for base \LCDM\
determined from the \plik\ cross-half-mission likelihood, together with the
lowP likelihood, applied to the 2015 full-mission data. The sky
coverage used in this likelihood is identical to that used
for the \camspec\ 2015F(CHM) likelihood. However, the two likelihoods differ in
the modelling of instrumental noise, Galactic dust, treatment of relative
calibrations, and multipole limits applied to each spectrum.

As summarized in column~8 of Table~\ref{paramtab1}, the
\plik\ and \camspec\ parameters agree to within $0.2\,\sigma$, except
for $\ns$, which differs by nearly
$0.5\,\sigma$. The difference in $\ns$ is perhaps not surprising, since this
parameter is sensitive to small differences in the foreground modelling.
Differences in $\ns$ between \plik\ and \camspec\ are systematic and persist throughout the grid of
extended \LCDM\ models discussed in Sect.~\ref{sec:grid}.
We emphasize that the \camspec\ and \plik\ likelihoods have been written
independently, though they are based on the same theoretical framework.
None of the conclusions in this paper (including those based
on the full ``TT{,}TE{,}EE'' likelihoods) would differ in any substantive
way had we chosen to use the \camspec\ likelihood in place of \plik.
The overall shifts of parameters between the \plik\ 2015 likelihood and the
published 2013 nominal mission parameters are summarized in column~7 of
Table~\ref{paramtab1}. These shifts are within $0.7\,\sigma$ except for
the parameters $\tau$ and $\As e^{-2\tau}$, which are sensitive to the
low-multipole polarization likelihood and absolute calibration.

In summary, the \planck\ 2013 cosmological parameters were pulled
slightly towards lower $H_0$ and $\ns$ by the $\ell\approx 1800$ 4-K
line systematic in the $217\times217$ cross-spectrum, but the net
effect of this systematic is relatively small, leading to shifts of $0.5\,\sigma$ or less
in cosmological parameters. Changes to the low-level data processing, beams,
sky coverage, etc., as well as the likelihood code also produce shifts of typically
$0.5\,\sigma$ or less. The combined effect of these changes is to introduce
parameter shifts relative to {\paramsI} of less than $0.7\,\sigma$,
with the exception of $\tau$ and $\As e^{-2\tau}$.
{\it The main scientific conclusions of {\paramsI} are therefore
consistent with the 2015 \Planck\ analysis.}

Parameters for the base \lcdm\ cosmology derived from full-mission
DetSet, cross-year, or cross-half-mission spectra are in extremely
good agreement, demonstrating that residual (i.e., uncorrected)
 cotemporal systematics are at low
levels. This is also true for the extensions of the \lcdm\ model
discussed in Sect.~\ref{sec:grid}.  It is therefore worth explaining
why we have adopted the cross-half-mission likelihood as the baseline
for this and other 2015 \Planck\ papers. The cross-half-mission
likelihood has lower signal-to-noise than the full-mission DetSet
likelihood; however, the errors on the cosmological parameters from the
two likelihoods are almost identical, as can be seen from the entries
in Table~\ref{paramtab1}.  This is also true for extended
\LCDM\ models. However, for more complicated tests, such as searches
for localized features in the power spectra \citep{planck2014-a24}, residual
4-K line systematic effects and residual uncorrected correlated noise at high
multipoles in the DetSet likelihood can produce
results suggestive of new physics (though not at a high significance level).
We have therefore decided to adopt the cross-half-mission
likelihood as the baseline for the 2015 analysis, sacrificing some
signal-to-noise in favour of reduced systematics. For almost all
of the models considered in this paper, the \Planck\ results are
limited by small systematics of various types, including systematic errors
in modelling foregrounds, rather than by signal-to-noise.

The foreground-subtracted, frequency-averaged, cross-half-mission
spectrum is plotted in Fig.~\ref{pgTT_final}, together with the
{\tt Commander} power spectrum at multipoles $\ell \le 29$.  The high
multipole spectrum plotted in this figure is an approximate maximum
likelihood solution based on equations~(A24) and (A25) of \likeI, with the
foregrounds and nuisance parameters for each spectrum fixed to the
best-fit values of the base \LCDM\ solution.  Note that a different way of
solving for the \Planck\ CMB spectrum, by marginalizing over foreground
and nuisance parameters, is presented in Sect.~\ref{sec:actspt}.  The
best-fit base \LCDM\ model is plotted in the upper panel, while residuals with
respect to this model are plotted in the lower panel.
In this plot, there are only four bandpowers at $\ell\ge 30$ that differ from the best-fit
model by more than $2\,\sigma$. These are: $\ell\,{=}\,434$ ($-2.0\,\sigma$);
$\ell\,{=}\,465$ ($2.5\,\sigma$); $\ell\,{=}\,1214$ ($-2.5\,\sigma$);
and $\ell\,{=}\,1455$ ($-2.1\,\sigma$). The $\chi^2$ of the coadded
$TT$ spectrum plotted in Fig.~\ref{pgTT_final} relative to the
best-fit base \LCDM\ model is $2547$ for $2479$ degrees of freedom
($30 \le \ell \le 2500$), which is a $0.96\,\sigma$ fluctuation
(PTE$\,{=}\,16.8\,\%$).
These numbers confirm the extremely good fit of the base \LCDM\ cosmology to
the \Planck\ $TT$ data at high multipoles. The consistency of the
\Planck\ polarization spectra with base \LCDM\ is discussed
in Sect.~\ref{subsec:planck_only3}.

{\paramsI} noted some mild internal tensions within the \Planck\ data,
for example, the preference of the phenomenological lensing parameter
$A_{\rm L}$ (see Sect.~\ref{sec:lensing}) towards values greater than unity and
a preference for a negative running of the scalar spectral index (see
Sect.~\ref{subsubsec:index}). These tensions were partly caused by the poor fit of
base \LCDM\ model to the temperature spectrum at multipoles below about 50.
As noted by the WMAP team \citep{hinshaw2003b}, the temperature spectrum
has a low quadrupole amplitude and a glitch in the multipole range $20 \la
\ell \la 30$. These features can be seen in the \Planck\ 2015 spectrum of Fig.~\ref{pgTT_final}. They have a similar (though slightly reduced)
effect on cosmological parameters to those described in {\paramsI}.

\subsection{545-GHz-cleaned spectra}
\label{subsec:planck_only2}

As discussed in {\paramsI}, unresolved extragalactic foregrounds (principally Poisson point
sources and the clustered component of the CIB)
contribute to the \Planck\ $TT$ spectra at high multipoles. The approach to
modelling these foreground contributions in {\paramsI}
is similar to that used by the ACT and SPT teams
\citep{Reichardt:2012,Dunkley:2013} in that the foregrounds are modelled
by a set of physically motivated power spectrum
template shapes with an associated set of
adjustable nuisance parameters. This approach
has been adopted as the baseline for the \Planck\ 2015 analysis. The foreground
model has been adjusted for this new analysis, in relatively minor ways,
as summarized in Sect.~\ref{subsec:foreground} and described in further detail
in \cite{planck2014-a14}. Galactic dust
emission also contributes to the temperature and polarization power spectra and must be subtracted
from the spectra used to form the \Planck\ likelihood.
Unlike the extragalactic foregrounds, Galactic dust emission is anisotropic
and so its impact can be reduced by appropriate masking of the
sky. In {\paramsI}, we intentionally adopted conservative masks, tuned for
each of the frequencies used to form the likelihood, to keep dust emission
at low levels. The results in {\paramsI} were therefore insensitive to the
modelling of residual dust contamination.

\begin{figure*}
\begin{center}
\includegraphics[width=68.5mm,angle=0]{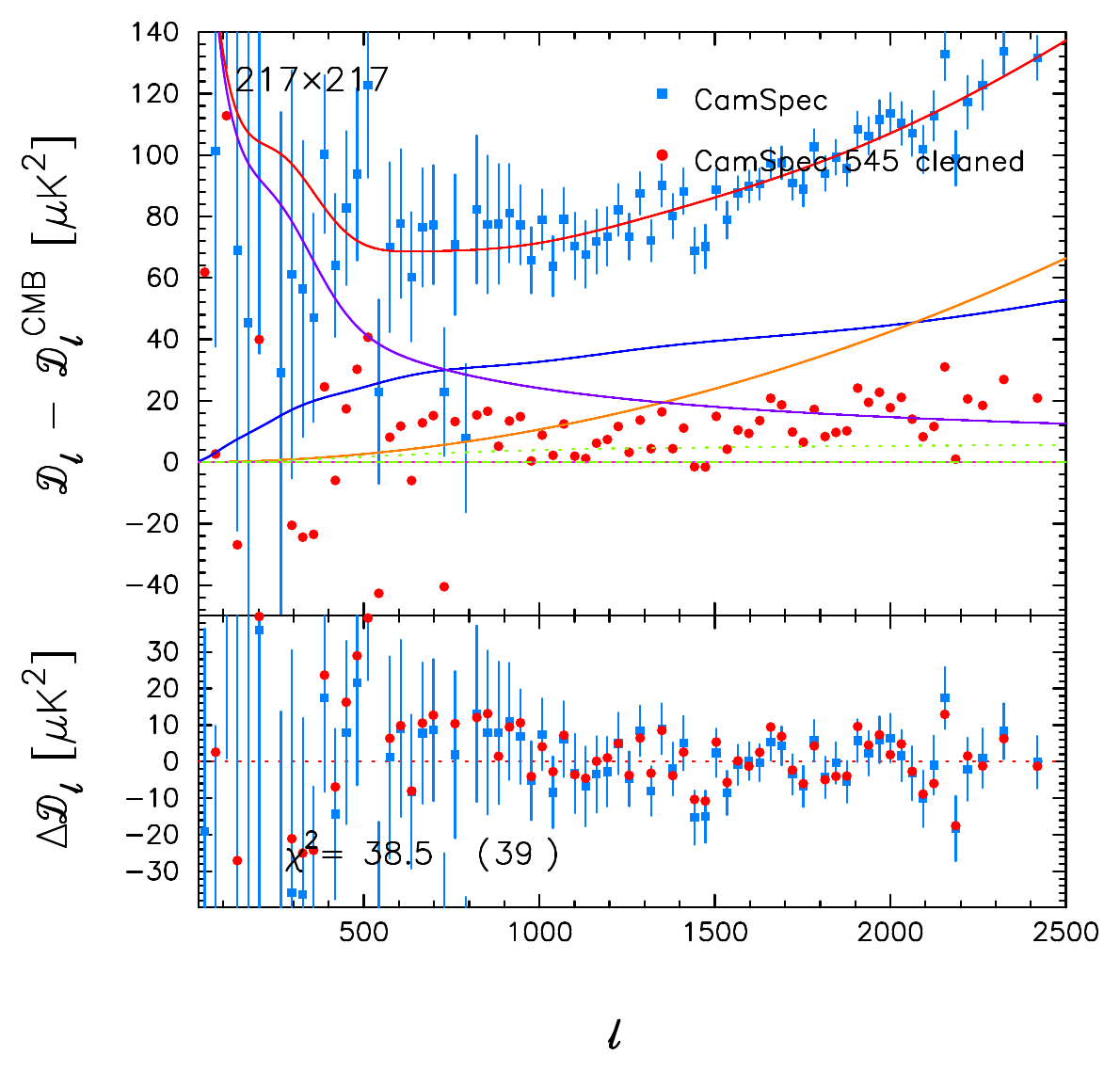}\hspace*{-2mm}
\includegraphics[width=58mm,angle=0]{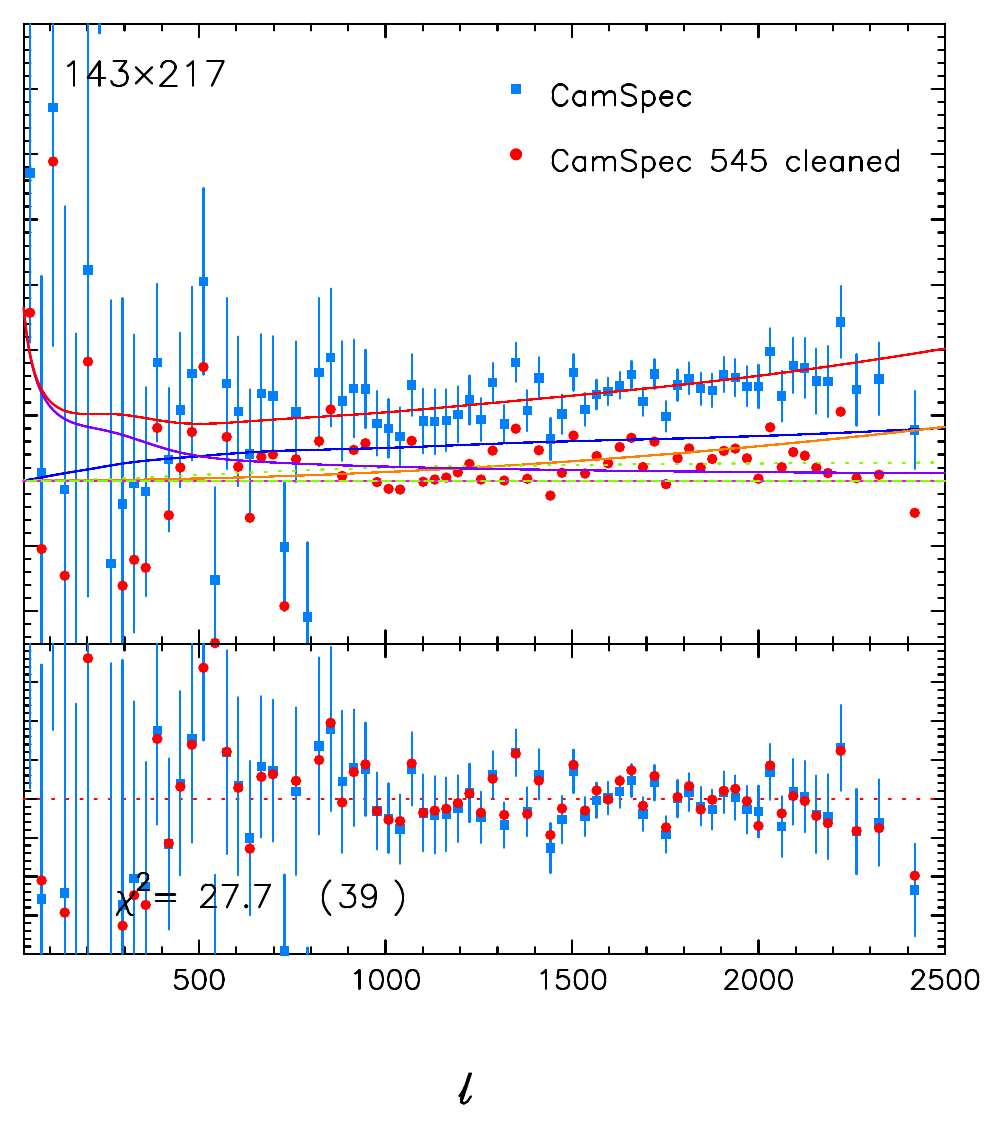}\hspace*{-2mm}
\includegraphics[width=58mm,angle=0]{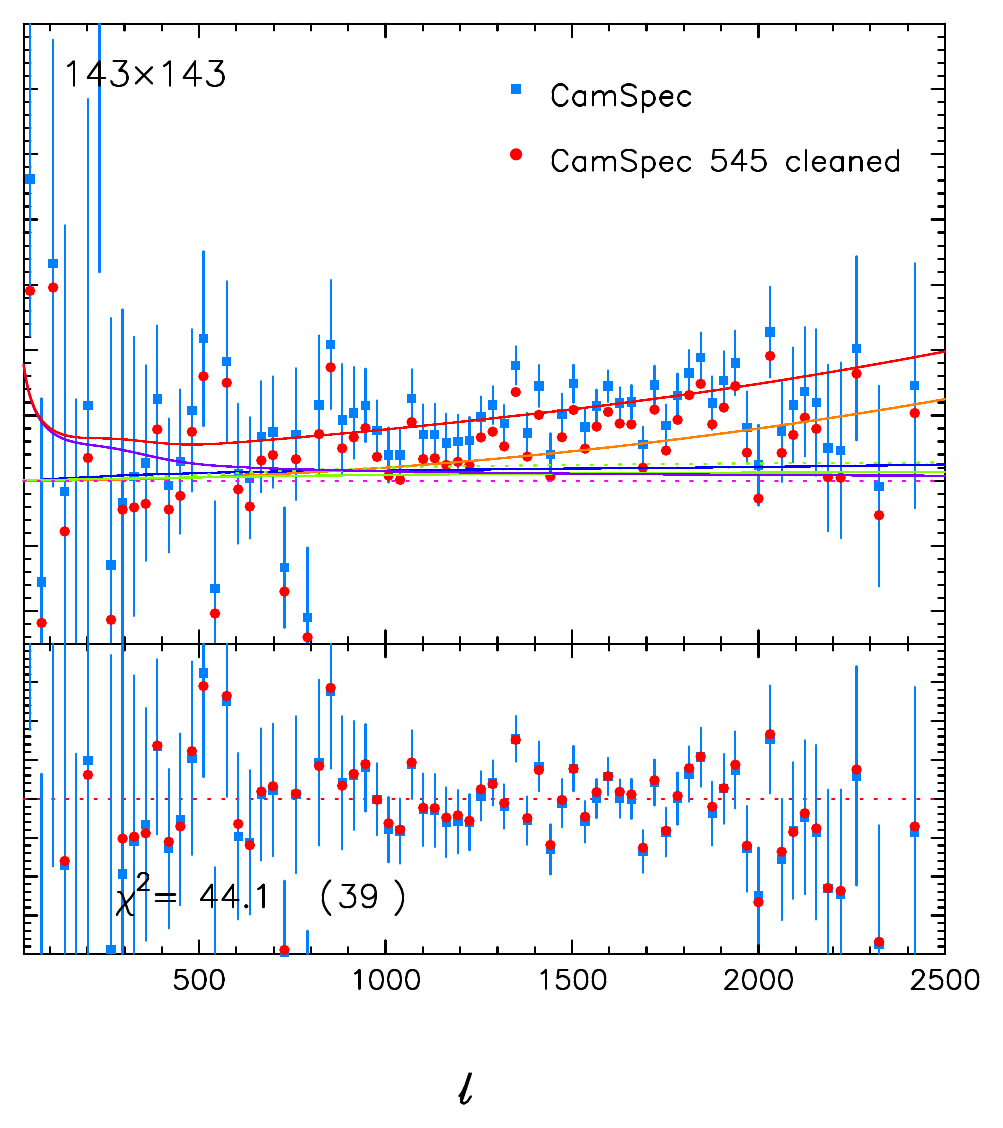}
\end{center}
\vspace{-3mm}
\caption {Residual plots illustrating the accuracy of the foreground modelling.
The blue points in the upper panels show the \camspec\ 2015(CHM) spectra
after subtraction of the best-fit \LCDM\ spectrum.  The residuals in the
upper panel should be accurately described by the foreground model. Major
foreground components are shown by the solid lines, colour coded as follows:
total foreground spectrum (red); Poisson point sources (orange);
clustered CIB (blue); thermal SZ (green); and Galactic dust
(purple). Minor foreground components are shown by the dotted lines,
colour-coded as follows: kinetic SZ (green); and tSZ$\times$CIB
cross-correlation (purple). The red points in the upper panels show
the 545-GHz-cleaned spectra (minus the best-fit CMB as subtracted from the uncleaned spectra)
that are fitted to a power-law residual foreground model, as discussed in the text.
The lower panels show the spectra after subtraction of the best-fit
foreground models. These agree to within a few $(\mu {\rm K})^2$.
The $\chi^2$ values of the residuals of the blue points, and the number of
bandpowers, are listed in the lower panels.}
\label{pgclean}
\end{figure*}

In the 2015 analysis, we have extended the sky coverage at each of 100,
143, and 217\,GHz, and so in addition to testing the
accuracy of the extragalactic foreground model, it is important to test
the accuracy of the Galactic dust model. As described in
\likeI\ and \cite{planck2014-a14} the Galactic dust templates used in the
\camspec\ and \plik\ likelihoods are derived by fitting
the 545-GHz mask-differenced power spectra. Mask differencing isolates the
anisotropic contribution of Galactic dust from the
isotropic extragalactic components. For the extended sky coverage used
in the 2015 likelihoods, the Galactic dust contributions
are a significant fraction of the extragalactic foreground contribution in
the $217\times217$ temperature spectrum at high multipoles, as illustrated in
Fig.~\ref{pgclean}. Galactic dust dominates over all other foregrounds
at multipoles $\ell \la 500$ at HFI frequencies.

A simple and direct test of the parametric foreground modelling used in the
\camspec\ and \plik\ likelihoods is to compare results with a completely
different approach in which the low-frequency maps are ``cleaned'' using
higher frequency maps as foreground templates \citep[see, e.g.,][]{Lueker:2010}.
In a similar approach to \cite{Spergel:2013},
we can form cleaned maps at lower frequencies $\nu$ by subtracting
a 545-GHz map as a template,
\begin{equation}
M^{T_\nu {\rm clean}} = (1 + \alpha^{T_\nu}) M^{T_\nu}
 - \alpha^{T_\nu} M^{T_{\nu_{\rm t}}}, \label{MC1}
\end{equation}
where $\nu_{\rm t}$ is the frequency of the template map
$M^{T_{\nu_{\rm t}}}$ and $\alpha^{T_\nu}$ is the cleaning
coefficient.  Since the maps have different beams, the
subtraction is actually done in the power spectrum domain:
\begin{eqnarray}
\hat C^{T_{\nu_1}T_{\nu_2} {\rm clean}} &=&
 (1 + \alpha^{T_{\nu_1}})(1 + \alpha^{T_{\nu_2}})
 \hat C^{T_{\nu_1} T_{\nu_2}} \nonumber \\
 & &  - (1 + \alpha^{T_{\nu_1}}) \alpha^{T_{\nu_2}}
 \hat C^{T_{\nu_2}T_{\nu_{\rm t}}} \nonumber \\
 & &  - (1 + \alpha^{T_{\nu_2}}) \alpha^{T_{\nu_1}}
 \hat C^{T_{\nu_1}T_{\nu_{\rm t}}}
 + \alpha^{T_{\nu_1}} \alpha^{T_{\nu_2}}
 \hat C^{T_{\nu_{\rm t}}T_{\nu_{\rm t}}},  \label{MC2}
\end{eqnarray}
where $\hat C^{T_{\nu_1}T_{\nu_2}}$ etc.\ are the mask-deconvolved
beam-corrected power spectra. The coefficients $\alpha^{T_{\nu_i}}$
are determined by minimizing
\begin{equation}
\sum_{\ell =\ell_{\rm min}}^{\ell_{\rm max}}
 \sum_{\ell^\prime =\ell_{\rm min}}^{\ell_{\rm max}}
 \hat{C}^{T_{\nu_i}T_{\nu_i}{\rm clean}}_\ell
 \left(
  \hat{\tens{M}}^{T_{\nu_i}T_{\nu_i}}_{\ell \ell^\prime}
 \right)^{-1}
 \hat{C}^{T_{\nu_i}T_{\nu_i} {\rm clean}}_{\ell^\prime},
\end{equation}
where $\hat{\tens{M}}^{T_{\nu_i}T_{\nu_i}}$ is the covariance
matrix of the estimates $\hat C^{T_{\nu_i}T_{\nu_i}}$.
We choose $\ell_{\rm min}\,{=}\,100$ and $\ell_{\rm max}\,{=}\,500$ and
compute the spectra in Eq.~\eqref{MC2} by cross-correlating half-mission maps
on the 60\,\% mask used to compute the $217\times217$ spectrum.
The resulting cleaning coefficients are $\alpha^T_{143} = 0.00194$
and $\alpha^T_{217} = 0.00765$; note that all of the
input maps are in units of thermodynamic temperature. The cleaning coefficients
are therefore optimized to remove Galactic dust at low multipoles,
though by using 545\,GHz as a dust template we find that the
cleaning coefficients are almost constant over the multipole range 50--2500.
We note, however, that this is not true if the 353- and 857-GHz maps are used
as dust templates, as discussed in Efstathiou et al.\ (in preparation).

The 545-GHz-cleaned spectra are shown by the red points in
Fig.~\ref{pgclean} and can be compared directly to the ``uncleaned''
spectra used in the \camspec\ likelihood (upper panels). As can be
seen, Galactic dust emission is removed to high accuracy and the
residual foreground contribution at high multipoles is strongly
suppressed in the $217\times217$ and $143\times217$
spectra. Nevertheless, there remains small foreground contributions
at high multipoles, which we model heuristically as power laws,
\begin{equation}
   {\cal \hat  D}_\ell = A \left (\ell \over 1500 \right )^{\epsilon},
   \label{MC3}
\end{equation}
with free amplitudes $A$ and spectral indices $\epsilon$. We
construct another \camspec\ cross-half-mission likelihood using
exactly the same sky masks as the 2015F(CHM) likelihood, but using
545-GHz-cleaned $217\times217$, $143\times217$, and $143\times143$ spectra.
We then use the simple model of Eq.~\eqref{MC3} in the likelihood to remove
residual unresolved foregrounds at high multipoles for each frequency combination. We do not
clean the $100\times100$ spectrum and so for this spectrum we use the standard
parametric foreground model in the likelihood.  The
lower panels in Fig.~\ref{pgclean} show the residuals with respect to
the best-fit base \LCDM\ model and foreground solution for the
uncleaned \camspec\ spectra (blue points) and for the 545-GHz-cleaned
spectra (red points). These residuals are almost identical, despite
the very different approaches to Galactic dust removal and foreground
modelling. The cosmological parameters from these two likelihoods are
also in very good agreement, typically to better than $0.1\,\sigma$,
with the exception of $\ns$, which is lower in the cleaned likelihood
by $0.26\,\sigma$. It is not surprising, given the heuristic nature of
the model (Eq.~\ref{MC3}), that $\ns$ shows the largest shift. We can also
remove the $100\times100$ spectrum from the likelihood entirely, with
very little impact on cosmological parameters.

Further tests of map-based cleaning are presented in
\cite{planck2014-a13}, which additionally describes another independently
written power-spectrum analysis pipeline (\mspec) tuned to map-cleaned
cross-spectrum analysis and using a more complex model for fitting
residual foregrounds than the heuristic model of Eq.~\eqref{MC3}.
\cite{planck2014-a13} also describes power spectrum analysis and cosmological
parameters derived from component-separated \Planck\ maps.  However,
the simple demonstration presented in this section shows that the
details of modelling residual dust contamination and other foregrounds
are under control in the 2015 \Planck\ likelihood. A further strong
argument that our $TT$ results are insensitive to foreground modelling
is presented in the next section, which compares the cosmological
parameters derived from the $TT$, $TE$, and $EE$ likelihoods.  Unresolved
foregrounds at high multipoles are completely negligible in the
polarization spectra and so the consistency of the parameters,
particularly from the $TE$ spectrum (which has higher signal-to-noise
than the $EE$ spectrum) provides an additional cross-check of the $TT$ results.

Finally, one can ask why we have not chosen to use a 545-GHz-cleaned
likelihood as the baseline for the 2015 \Planck\ parameter
analysis. Firstly, it would not make any difference to the results of
this paper had we chosen to do so. Secondly, we feel that the
parametric foreground model used in the baseline likelihood has a
sounder physical basis. This allows us to link the amplitudes of
the unresolved foregrounds across the various \Planck\ frequencies
 with the results from other ways of studying foregrounds, including the
higher resolution CMB experiments described in Sect.~\ref{sec:actspt}.

\subsection{The 2015 \Planck\ temperature and polarization spectra and
likelihood}
\label{subsec:planck_only3}

\begin{table*}[ht!]
\begingroup
\newdimen\tblskip \tblskip=5pt
\caption{Goodness-of-fit tests for the 2015 \planck\ temperature and
polarization spectra.  $\Delta \chi^2=\chi^2-N_{\rm dof}$ is the difference
from the mean assuming that the best-fit base \LCDM\ model (fitted to \planckTT) is correct
and $N_{\rm dof}$
is the number of degrees of freedom (set equal to the number of multipoles).
The sixth column expresses $\Delta \chi^2$ in units of
the expected dispersion, $\sqrt{2N_{\rm dof}}$, and the last column lists the
probability to exceed (PTE) the tabulated value of $\chi^2$.}
\label{tab:chi_squared}
\nointerlineskip
\vskip -2mm
%\footnotesize
%\scriptsize
\setbox\tablebox=\vbox{
   \newdimen\digitwidth
   \setbox0=\hbox{\rm 0}
   \digitwidth=\wd0
   \catcode`*=\active
   \def*{\kern\digitwidth}
   \newdimen\signwidth
   \setbox0=\hbox{+}
   \signwidth=\wd0
   \catcode`!=\active
   \def!{\kern\signwidth}
\halign{\tabskip 0pt#\hfil\tabskip 1.5em&
#\hfil&
\hfil#\hfil&
\hfil#\hfil&
\hfil#\hfil&
\hfil#\hfil&
\hfil#\hfil&
\hfil#\hfil\tabskip 0pt\cr                            % Template goes here.
\noalign{\doubleline}
\noalign{\vskip -3pt}
Likelihood& Frequency& Multipole range& $\chi^2$& $\chi^2/N_{\rm dof}$&
 $N_{\rm dof}$& $\Delta \chi^2/\sqrt{2N_{\rm dof}}$& PTE [\%]\cr
\noalign{\vskip 3pt\hrule\vskip 5pt}
   $TT$& 100$\times$100& *30--1197& 1234.37& 1.06& 1168& $!1.37$& *8.66\cr
\omit& 143$\times$143& *30--1996& 2034.45& 1.03& 1967& $!1.08$& 14.14\cr
\omit& 143$\times$217& *30--2508& 2566.74& 1.04& 2479& $!1.25$& 10.73\cr
\omit& 217$\times$217& *30--2508& 2549.66& 1.03& 2479& $!1.00$& 15.78\cr
\omit&       Combined& *30--2508& 2546.67& 1.03& 2479& $!0.96$& 16.81\cr
\noalign{\vskip 3pt\hrule\vskip 5pt}
   $TE$& 100$\times$100& *30--*999& 1088.78& 1.12& *970& $!2.70$& *0.45\cr
\omit& 100$\times$143& *30--*999& 1032.84& 1.06& *970& $!1.43$& *7.90\cr
\omit& 100$\times$217& 505--*999& *526.56& 1.06& *495& $!1.00$& 15.78\cr
\omit& 143$\times$143& *30--1996& 2028.43& 1.03& 1967& $!0.98$& 16.35\cr
\omit& 143$\times$217& 505--1996& 1606.25& 1.08& 1492& $!2.09$& *2.01\cr
\omit& 217$\times$217& 505--1996& 1431.52& 0.96& 1492& $-1.11$& 86.66\cr
\omit&       Combined& *30--1996& 2046.11& 1.04& 1967& $!1.26$& 10.47\cr
\noalign{\vskip 3pt\hrule\vskip 5pt}
   $EE$& 100$\times$100& *30--*999& 1027.89& 1.06& *970& $!1.31$& *9.61\cr
\omit& 100$\times$143& *30--*999& 1048.22& 1.08& *970& $!1.78$& *4.05\cr
\omit& 100$\times$217& 505--*999& *479.72& 0.97& *495& $-0.49$& 68.06\cr
\omit& 143$\times$143& *30--1996& 2000.90& 1.02& 1967& $!0.54$& 29.18\cr
\omit& 143$\times$217& 505--1996& 1431.16& 0.96& 1492& $-1.11$& 86.80\cr
\omit& 217$\times$217& 505--1996& 1409.58& 0.94& 1492& $-1.51$& 93.64\cr
\omit&       Combined& *30--1996& 1986.95& 1.01& 1967& $!0.32$& 37.16\cr
\noalign{\vskip 3pt\hrule\vskip 5pt}
}}
\endPlancktablewide
\endgroup
\end{table*}

The coadded 2015 \Planck\ temperature spectrum was introduced in
Fig.~\ref{pgTT_final}.  In this section, we present additional details
and consistency checks of the temperature likelihood and describe the
full mission \Planck\ $TE$ and $EE$ spectra and likelihood; preliminary
\Planck\ $TE$ and $EE$ spectra were presented in \paramsI. We then discuss
the consistency of the cosmological parameters for base
\LCDM\ measured separately from the $TT$, $TE$, and $EE$ spectra. For the
most part, the discussion given in this section is specific to the
\plik\ likelihood, which is used as the baseline in this paper. A more
complete discussion of the \plik\ and other likelihoods developed by the
\Planck\ team is given in \cite{planck2014-a13}.

\subsubsection{Temperature spectra and likelihood}

\paragraph{(1) Temperature masks.}
As in the 2013 analysis, the high-multipole $TT$ likelihood uses the
$100\times100$ , $143\times143$, $217\times217$, and $143\times 217$ spectra.
However, in contrast to the 2013 analysis, which used conservative sky masks to
reduce the effects of Galactic dust emission, we make more aggressive use of
sky in the 2015 analysis.
The 2015 analysis retains 80\,\%, 70\,\%, and 60\,\% of sky at 100\,GHz,
143\,GHz, and 217\,GHz, respectively, before apodization. We also apply
apodized point source masks to remove compact sources with a
signal-to-noise threshold $>5$ at each frequency (see \citealt{planck2014-a35}
for a description of the \Planck\ Catalogue of Compact Sources).  Apodized masks
are also applied to remove extended objects, and regions of high CO emission
were masked at 100\,GHz and 217\,GHz \citep[see][]{planck2014-a12}.
As an
estimate of the effective sky area, we compute the following sum over pixels:
\be
f^{\rm eff}_{\rm sky} = {1 \over 4 \pi} \sum w_i^2 \Omega_i, \label{plik1}
\ee
where $w_i$ is the weight of the apodized mask and $\Omega_i$ is the area
of pixel $i$. All input maps
are at \HEALpix\ \citep{gorski2005} resolution $N_{\rm side}=2048$.
Eq.~\eqref{plik1} gives $f^{\rm eff}_{\rm sky} = 66.3\,\%$ at 100\,GHz,
57.4\,\% at 143\,GHz, and 47.1\,\% at 217\,GHz.

\paragraph{(2) Galactic dust templates.}
With the increased sky coverage used in the 2015 analysis, we take a
slightly different approach to subtracting Galactic dust emission to
that described in \likeI\ and \paramsI.  The shape of the Galactic dust
template is determined from mask-differenced power spectra estimated
from the 545-GHz maps. The mask differencing removes the isotropic
contribution from the CIB and point sources. The resulting dust template has
a similar shape to the template used in the 2013 analysis, with
power-law behaviour $\mathcal{D}^{\rm dust}_\ell \propto \ell^{-0.63}$
at high multipoles, but with a ``bump'' at $\ell\approx 200$ (as shown in
Fig.~\ref{pgclean}). The absolute amplitude of the dust templates at
100, 143, and 217\,GHz is determined by cross-correlating the
temperature maps at these frequencies with the 545-GHz maps (with
minor corrections for the CIB and point source contributions).  This
allows us to generate priors on the dust template amplitudes, which
are treated as additional nuisance parameters when running MCMC chains
(unlike the 2013 analysis, in which we fixed the amplitudes of the
dust templates). The actual priors used in the \plik\ likelihood are
Gaussians on $\mathcal{D}^{\rm dust}_{\ell=200}$ with the following means
and dispersions: $(7\pm2)\,\muK^2$
for the $100\times100$ spectrum; $(9\pm2)\,\muK^2$ for $143\times143$;
$(21\pm8.5)\,\muK^2$ for $143\times217$; and $(80\pm20)\,\muK^2$ for
$217\times217$.  The MCMC solutions show small movements of the
best-fit dust template amplitudes, but always within statistically acceptable
ranges given the priors.

\paragraph{(3) Likelihood approximation and covariance matrices.}
The approximation to the likelihood function follows the methodology
described in \likeI\ and is based on a Gaussian likelihood assuming a
fiducial theoretical power spectrum (a fit to \plik\ TT with prior
$\tau=0.07\pm 0.02$). We have also included a number of small refinements to
the covariance matrices.  Foregrounds, including Galactic dust, are added to
the fiducial theoretical power spectrum, so that the additional small
variance associated with foregrounds is included, along with cosmic
variance of the CMB, under the assumption that the
foregrounds are Gaussian random fields. The 2013 analysis did not
include corrections to the covariance matrices arising from leakage of
low-multipole power to high multipoles via the point source
holes; these can introduce errors in the covariance matrices of a few percent
at $\ell \approx 300$, corresponding approximately to the first peak of the CMB
spectrum. In the 2015 analysis we apply corrections to the fiducial
theoretical power spectrum, based on Monte Carlo simulations, to
correct for this effect. We also apply Monte Carlo based corrections
to the analytic covariance matrices at multipoles $\ell \le 50$, where the
analytic approximations begin to become inaccurate even for large
effective sky areas \citep[see][]{Efstathiou:2004}. Finally, we add
the uncertainties on the beam shapes to the covariance matrix
following the methodology described in \likeI. The \planck\ beams are
much more accurately characterized in the 2015 analysis, and so the
beam corrections to the covariance matrices are extremely small. The
refinements to the covariance matrices described in this paragraph are
all relatively minor and have little impact on
cosmological parameters.

\paragraph{(4) Binning.}
The baseline \plik\ likelihood uses binned temperature and polarization spectra.
This is done because all frequency combinations of the $TE$ and $EE$ spectra
are used in the \plik\ likelihood, leading to a large data vector of length
$22\,865$ if the spectra are retained multipole-by-multipole. The baseline
\plik\ likelihood reduces the size of the data vector by binning these spectra.
The spectra are binned into bins of width $\Delta\ell=5$ for
$30 \le \ell \le 99$, $\Delta\ell=9$ for $100 \le \ell \le 1503$,
$\Delta\ell=17$ for $1504 \le \ell \le 2013$ and $\Delta\ell=33$ for
$2014 \le \ell \le 2508$, with a weighting of $C_\ell$ proportional to
$\ell(\ell+1)$ over the bin widths. The bins span an odd number of multipoles,
 since for approximately azimuthal masks we expect a nearly
symmetrical correlation function
around the central multipole. The binning does not affect
the determination of cosmological parameters in \LCDM-type models (which have
smooth power spectra), but significantly reduces the size of the
joint TT{,}TE{,}EE covariance matrix, speeding up the computation of the
likelihood. However, for
some specific purposes, e.g., searching for oscillatory features in the
$TT$ spectrum, or testing $\chi^2$ statistics, we produce blocks of the
likelihood multipole-by-multipole.

\paragraph{(5) Goodness of fit.}
The first five rows of Table~\ref{tab:chi_squared} list $\chi^2$ statistics for the $TT$ spectra (multipole-by-multipole)
relative to the \Planck\ best-fit base \LCDM\ model and foreground
parameters (fitted to \planckTT). The first four entries list the statistics
separately for each of the four spectra that form the $TT$ likelihood and the
fifth line (labelled ``Combined'') gives the $\chi^2$ value for the maximum
likelihood $TT$ spectrum plotted in Fig.~\ref{pgTT_final}.  Each of the
individual spectra provides an acceptable fit to the base
\LCDM\ model, as does the frequency-averaged spectrum plotted in
Fig.~\ref{pgTT_final}.  This demonstrates the excellent consistency of the
base \LCDM\ model across frequencies. More detailed consistency checks of
the \Planck\ spectra are presented in \cite{planck2014-a13}; however, as
indicated by Table~\ref{tab:chi_squared}, we find no evidence
for any inconsistencies between the foreground-corrected temperature power
spectra computed for different frequency combinations.
The temperature spectra are largely signal dominated over the multipole
ranges listed in Table~\ref{tab:chi_squared} and so the $\chi^2$ values are
insensitive to small errors in the
\Planck\ noise model used in the covariance matrices. As discussed in the
next subsection, this is not true for the $TE$ and
$EE$ spectra, which are noise dominated over much of the multipole range.

\begin{figure*}
\begin{center}
\includegraphics[width=141mm,angle=0]{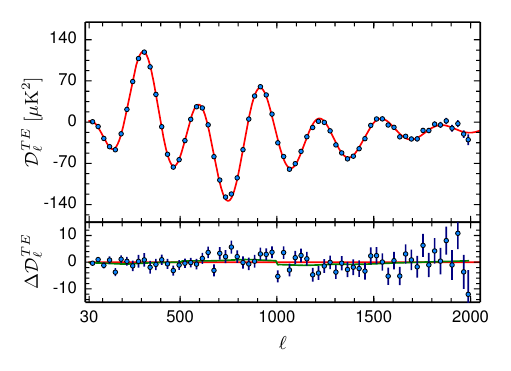} \\
\hspace{10mm}\includegraphics[width=144mm,angle=0]{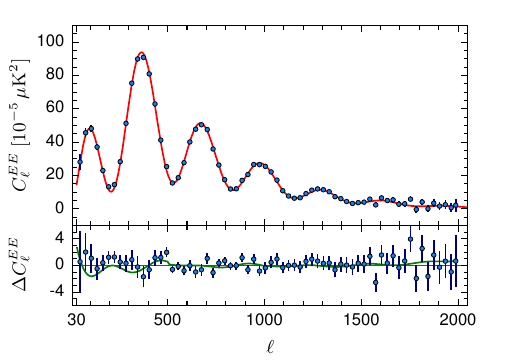}
\end{center}
\vspace{-1mm}
\caption {Frequency-averaged $TE$ and $EE$ spectra (without fitting for
  temperature-to-polarization leakage). The theoretical $TE$ and $EE$ spectra
  plotted in the upper panel of each plot are computed from the \planckTT\
  best-fit model
  of Fig.~\ref{pgTT_final}. Residuals with respect to this theoretical
  model are shown in the lower panel in each plot.  The error bars
  show $\pm 1\,\sigma$ errors. The green lines in the lower panels show the
  best-fit temperature-to-polarization leakage model of Eqs.~\eqref{PS0a} and
  \eqref{PS0b}, fitted separately to the $TE$ and $EE$ spectra. }
\label{pgTE+EE_final}
\end{figure*}

\begin{figure*}[t]
\begin{center}
\includegraphics[width=91mm,angle=0]{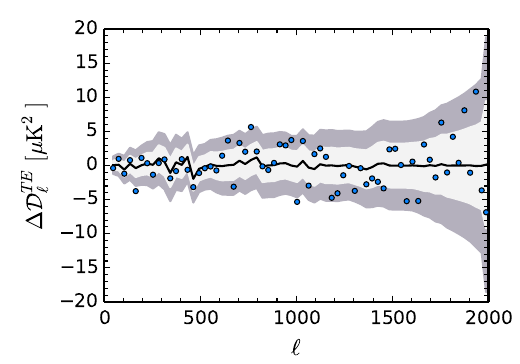}
\includegraphics[width=91mm,angle=0]{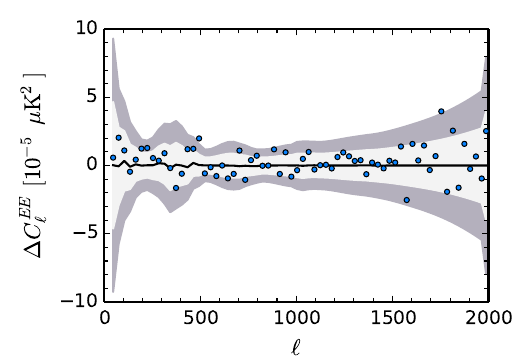}\\
\end{center}
\vspace{-5mm}
\caption
{Conditionals for the \plik\ $TE$ and $EE$ spectra, given the $TT$ data
computed from the \plik\ likelihood.  The black lines show the expected
$TE$ and $EE$ spectra {\it given the $TT$ data}. The shaded areas show the
$\pm1$ and $\pm2\,\sigma$ ranges computed from Eq.~\eqref{CVEC3}.
The blue points show the residuals for the measured $TE$ and $EE$ spectra.}
\label{pgTE+EE_cond}
\end{figure*}

\begin{figure*}
\begin{center}
\includegraphics[width=90mm,angle=0]{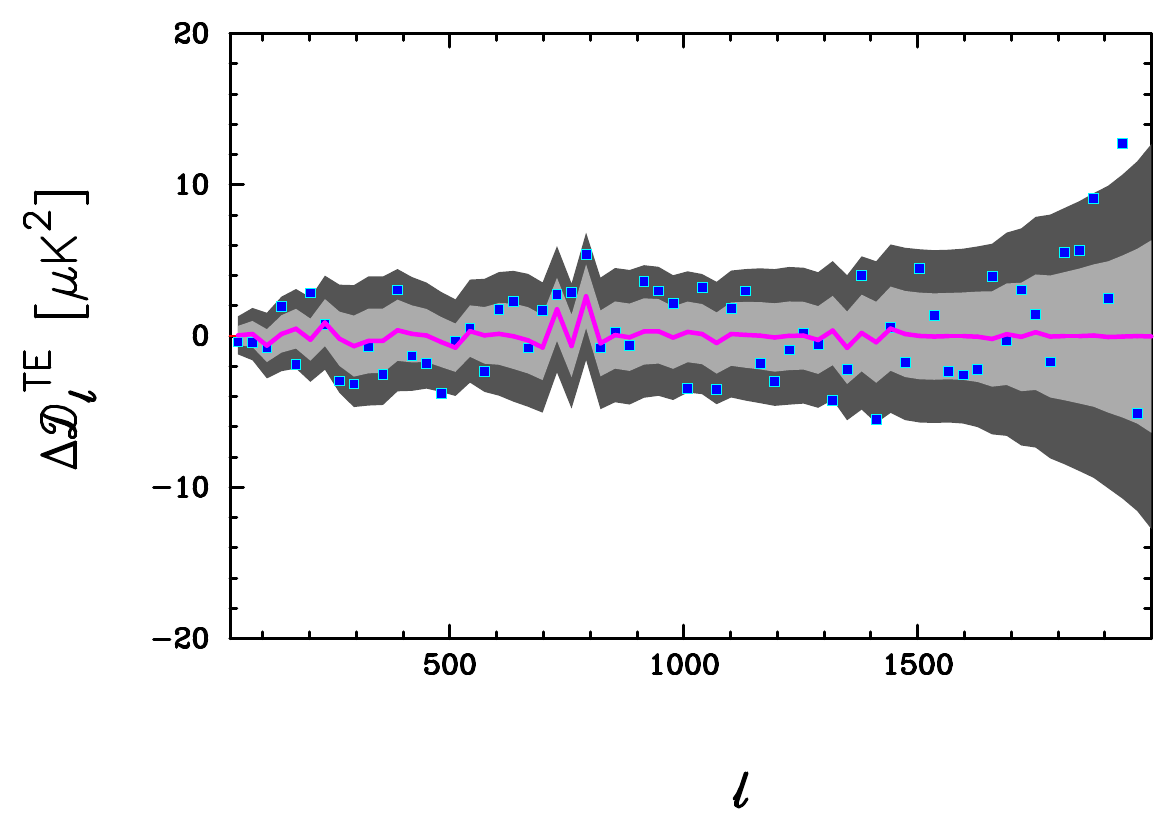}
\includegraphics[width=90mm,angle=0]{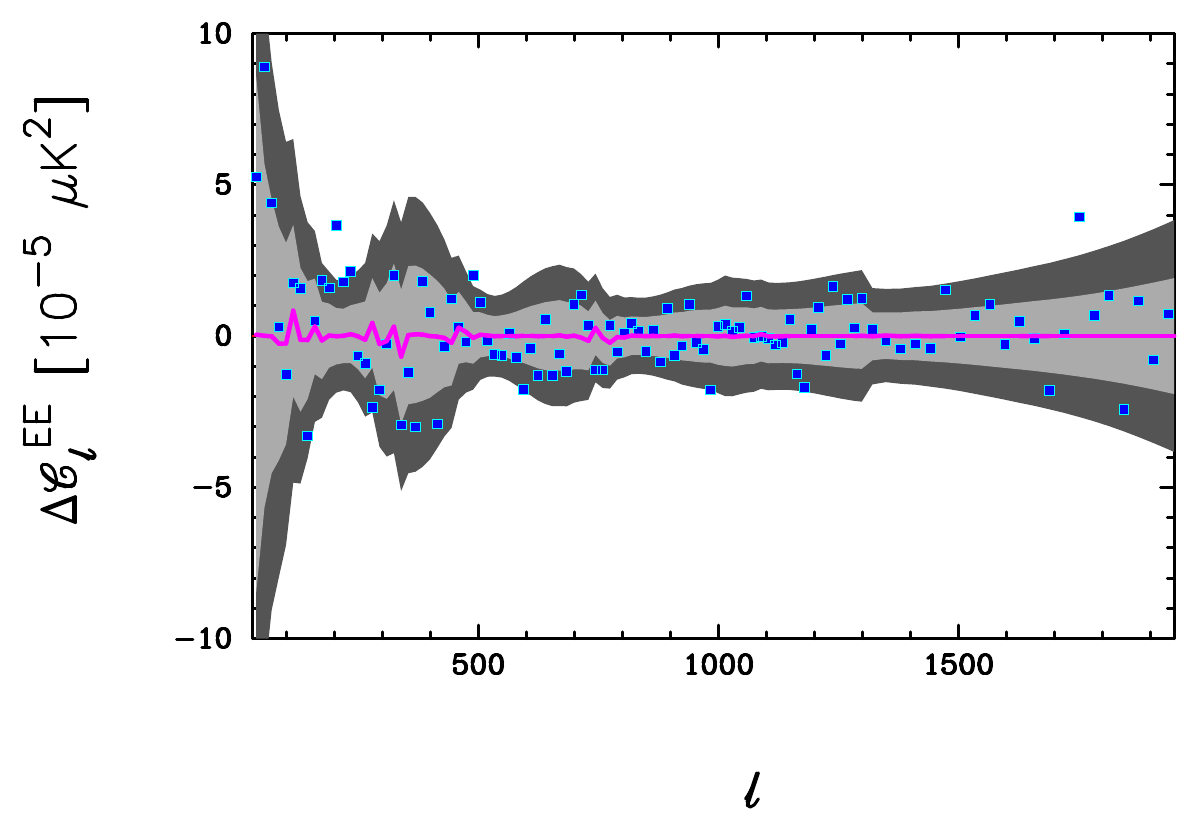}\\
\end{center}
\vspace{-3mm}
\caption {Conditionals for the \camspec\ $TE$ and $EE$ spectra, given the $TT$ data
computed from the \camspec\ likelihood. As in Fig.~\ref{pgTE+EE_cond},
the shaded areas show $\pm1$ and $\pm2\,\sigma$ ranges, computed
from Eq.~\eqref{CVEC3} and blue points show the residuals for the measured
$TE$ and $EE$ spectra.}
\label{pgTE+EE_condcamspec}
\end{figure*}

\subsubsection{Polarization spectra and likelihood}

In addition to the $TT$ spectra, the 2015 \Planck\ likelihood includes
the $TE$ and $EE$ spectra.  As discussed in Sect.~\ref{subsec:planck_only1},
the \Planck\ 2015 low-multipole polarization analysis is based on the
LFI 70-GHz data. Here we discuss the $TE$ and $EE$ spectra
that are used in the high-multipole likelihood, which are computed from
the HFI data at $100$, $143$ and $217$\,GHz.
 As summarized in \cite{planck2014-a13}, there is
no evidence for any unresolved foreground components at high multipoles in
the polarization spectra. We therefore include all frequency combinations in
computing the $TE$ and $EE$ spectra to maximize the signal-to-noise.\footnote{In
temperature, the $100 \times 143$ and $100\times217$
spectra are not included in the likelihood because the temperature spectra
are largely signal dominated. These spectra therefore add little new information
on the CMB, but would require additional nuisance parameters to correct for
unresolved foregrounds at high multipoles.}

\paragraph{(1) Masks and dust corrections.}
At low multipoles ($\ell \la 300$) polarized Galactic dust emission is
significant at all frequencies and is subtracted in a similar way to the dust
subtraction in temperature, i.e., by including additional nuisance parameters
quantifying the amplitudes of a power-law dust template
with a slope constrained to
$\mathcal{D}^{\rm dust}_\ell \propto \ell^{-0.40}$ for both $TE$ and $EE$
\citep{planck2014-XXX}. Polarized synchrotron emission (which has been shown
to be negligible at 100\,GHz and higher frequencies for \Planck\ noise levels,
\citealt{Fuskeland:2014}) is ignored.  Gaussian priors on the polarization
dust amplitudes are determined by cross-correlating the lower frequency
maps with the 353-GHz polarization maps (the highest frequency polarized
channel of the HFI) in a similar way to
the determination of temperature dust priors.
We use the temperature-based apodized masks in $Q$ and $U$ at each frequency,
retaining 70\,\%, 50\,\%, and 41\,\%
of the sky at 100, 143, and 217\,GHz, respectively, after apodization (slightly smaller
than the temperature masks at 143 and 217\,GHz).
However, we do not apply point source or CO masks to the $Q$ and $U$ maps.
The construction of the full TT{,}TE{,}EE likelihood is then a
straightforward extension of the $TT$ likelihood
using the analytic covariance matrices given by \cite{Efstathiou:2006}
and \cite{Hamimeche:2008}.

\paragraph{(2) Polarization spectra and residual systematics.}
Maximum likelihood frequency coadded $TE$ and $EE$ spectra are shown in
Fig.~\ref{pgTE+EE_final}. The theoretical curves plotted in these figures are
the $TE$ and $EE$ spectra computed from the best-fit base \LCDM\ model fitted
to the {\it temperature\/} spectra (\planckTT), as plotted in
Fig.~\ref{pgTT_final}. The lower panels in each figure show the residuals
with respect to this model. The theoretical model provides a very good fit
to the $TE$ and $EE$ spectra.
Table~\ref{tab:chi_squared} lists $\chi^2$ statistics for the $TE$ and $EE$
spectra for each frequency combination (with the $TE$ and $ET$ spectra
for each frequency combination coadded to form a single $TE$ spectrum).
Note that since the $TE$ and
$EE$ spectra are noisier than the $TT$ spectra, these values of $\chi^2$ are
sensitive to the procedure used to estimate \Planck\ noise
(see \citealt{planck2014-a13} for further details).

Some of these $\chi^2$ values are unusually high, for example the $100
\times 100$ and $143 \times 217$ $TE$ spectra and the $100\times 143$ $EE$
spectrum all have low PTEs.  The \Planck\ $TE$ and $EE$ spectra for
different frequency combinations are not as internally consistent as
the \Planck\ $TT$ spectra. Inter-comparison of the $TE$ and $EE$ spectra at
different frequencies is much more straightforward than for the
temperature spectra because unresolved foregrounds are unimportant in
polarization. The high $\chi^2$ values listed in Table~\ref{tab:chi_squared}
therefore provide clear evidence of residual instrumental systematics
in the $TE$ and $EE$ spectra.

With our present understanding of the \Planck\ polarization data, we believe
that the dominant source of systematic error in the polarization spectra is
caused by beam mismatch that generates leakage from temperature to
polarization (recalling that the HFI polarization maps are generated by
differencing signals between quadruplets of polarization sensitive
bolometers). In principle, with accurate knowledge of the beams this
leakage could be described by effective polarized beam window
functions. For the 2015 papers, we use the $TT$ beams rather than
polarized beams, and characterize temperature-to-polarization leakage using
a simplified model. The impact of beam mismatch on
the polarization spectra in this model is
\beglet
\begin{eqnarray}
 \Delta C_\ell^{TE} &=& \epsilon_\ell C_\ell^{TT}, \label{PS0a}\\
 \Delta C_\ell^{EE} &=& \epsilon_\ell^2C_\ell^{TT}
 + 2 \epsilon_\ell C_\ell^{TE}, \label{PS0b}
 \end{eqnarray}
\endlet
where $\epsilon_\ell$ is a polynomial in multipole.
As a consequence of the \Planck\ scanning strategy, pixels are visited
approximately every
six months, with a rotation of the focal plane by $180^\circ$, leading to a
weak coupling to beam modes $b_{\ell m}$ with odd values of $m$.
The dominant contributions are expected to come from modes with $m=2$ and
$4$, describing the beam ellipticity. We therefore fit the spectra using a
fourth-order polynomial,
\begin{equation}
\epsilon_\ell = a_0 + a_2 \ell^2 + a_4 \ell^4,   \label{PS1}
\end{equation}
treating the coefficients $a_0$, $a_2$, and $a_4$ as nuisance parameters
in the MCMC analysis.  We have ignored the odd coefficients of the polynomial,
which should be suppressed by our scanning strategy.
We do however include a constant term in the polynomial to account for small
deviations of the polarization efficiency from unity.

The fit is performed separately on the $TE$ and $EE$ spectra. A different
polynomial is used for each cross-frequency spectrum.  The coadded
corrections are shown in the lower panels of Fig.~\ref{pgTE+EE_final}.
Empirically, we find that temperature-to-polarization leakage
systematics tend to cancel in the coadded spectra.  Although the best-fit
leakage corrections to the coadded spectra are small, the corrections
for individual frequency cross-spectra can be up to 3 times
larger than those shown in Fig.~\ref{pgTE+EE_final}.
The model of Eqs.~(\ref{PS0a}) and (\ref{PS0b}) is clearly crude, but gives
us some idea of the impact of temperature-to-polarization leakage in the
coadded spectra.
With our present empirical understanding of leakage, we find a correlation
between the polarization spectra that have the highest expected
temperature-to-polarization leakage and those that display high $\chi^2$ in
Table~\ref{tab:chi_squared}. However, the characterization of this
leakage is not yet accurate enough to reduce the $\chi^2$
values for each frequency combination to acceptable levels.

As discussed in \paramsI, each \Planck\ data release and accompanying
set of papers should be viewed as a snapshot of the state of the \Planck\
analysis at the time of the release. For the 2015 release, we have a high
level of confidence in the temperature power spectra. However, we have
definite evidence for low-level systematics associated with
temperature-to-polarization
leakage in the polarization spectra.  The tests described
above suggest that these are at low levels of a few $(\mu {\rm K})^2$ in
$D_\ell$. However, temperature-to-polarization leakage can introduce
correlated features in the spectra, as shown by the $EE$ leakage model
plotted in Fig.~\ref{pgTE+EE_final}.  Until we have a more accurate
characterization of these systematics, we urge caution in the
interpretation of features in the $TE$ and $EE$ spectra. For some of the 2015
papers, we use the $TE$ and $EE$ spectra,
without leakage corrections. For most of the models considered in this paper,
the $TT$ spectra alone provide tight constraints and so we take a conservative
approach and usually quote the $TT$ results. However, as we will see, we find
a high level of consistency between the \Planck\ $TT$ and full \Planck\
$TT,TE,EE$ likelihoods.
Some models considered in Sect.~\ref{sec:grid} are, however, sensitive
to the polarization blocks of the likelihood. Examples include constraints
on isocurvature modes, dark matter annihilation, and non-standard recombination
histories. \Planck\ 2015 constraints on these models should be viewed as
preliminary, pending a more complete analysis of polarization systematics,
which will be presented in the next series of \Planck\ papers accompanying
a third data release.

\begin{table*}[tb]                 % table* is a two-column table.  Drop the * for one column.
\begingroup
\newdimen\tblskip \tblskip=5pt
\caption{%
Parameters of the base \lcdm\ cosmology computed from the 2015 baseline \Planck\
likelihoods, illustrating the consistency of parameters determined from the
temperature and polarization spectra at high multipoles.
Column [1] uses the $TT$ spectra at low and high multipoles and is the
same as column [6] of Table~\ref{paramtab1}.
Columns [2] and [3]
use only the $TE$ and $EE$ spectra at high multipoles, and only polarization
at low multipoles. Column [4] uses the full likelihood. The last
column lists the deviations of the cosmological parameters determined from
the \planckTT\ and \planckall\ likelihoods.
}                          % Caption goes here.
\label{paramtab2}                            % Label goes here.
\nointerlineskip
\vskip -3mm
%\footnotesize
%\scriptsize
\setbox\tablebox=\vbox{
   \newdimen\digitwidth
   \setbox0=\hbox{\rm 0}
   \digitwidth=\wd0
   \catcode`*=\active
   \def*{\kern\digitwidth}
   \newdimen\signwidth
   \setbox0=\hbox{+}
   \signwidth=\wd0
   \catcode`!=\active
   \def!{\kern\signwidth}
\halign{\hbox to 1.0in{#\leaderfil}\tabskip 1.2em&
\hfil#\hfil&
\hfil#\hfil&
\hfil#\hfil&
\hfil#\hfil&
\hfil#\hfil\tabskip 0pt\cr                            % Template goes here.
\noalign{\doubleline}
\omit\hfil Parameter\hfil& [1] \planckTT& [2] \Planck\ TE+\lowEB&
 [3] \Planck\ EE+\lowEB& [4] \planckall&$([1]-[4])/\sigma_{[1]}$\cr
\noalign{\vskip 3pt\hrule\vskip 5pt}

$\Omega_{\mathrm{b}}h^2$& $0.02222\pm0.00023$& $0.02228\pm0.00025$&
 $0.0240\pm0.0013$& $0.02225\pm0.00016$& $-0.1$\cr
$\Omega_{\mathrm{c}}h^2$& $0.1197\pm0.0022$& $0.1187\pm0.0021$&
 $0.1150^{+0.0048}_{-0.0055}$&$0.1198\pm 0.0015$& $!0.0$\cr
$100\theta_{\mathrm{MC}}$& $1.04085\pm0.00047$& $1.04094\pm0.00051$&
 $1.03988\pm 0.00094$&$1.04077\pm0.00032$& $!0.2$\cr
$\tau$&$0.078\pm0.019$& $0.053\pm0.019$& $0.059^{+0.022}_{-0.019}$&
 $0.079\pm 0.017$&$-0.1$\cr
$\ln(10^{10} A_\mathrm{s})$& $3.089\pm0.036$& $3.031\pm 0.041$&
 $3.066^{+0.046}_{-0.041}$& $3.094\pm0.034$& $-0.1$\cr
$n_\mathrm{s}$&$0.9655\pm0.0062$&$0.965\pm0.012$& $0.973\pm 0.016$&
 $0.9645\pm 0.0049$&$!0.2$\cr
$H_0$&$67.31\pm0.96*$& $67.73\pm0.92*$& $70.2\pm3.0*$& $67.27\pm0.66*$&$!0.0$\cr
$\Omm$& $0.315\pm0.013$& $0.300\pm0.012$& $0.286^{+0.027}_{-0.038}$&
 $0.3156\pm0.0091$& $!0.0$\cr
$\sigma_8$& $0.829\pm0.014$& $0.802\pm0.018$& $0.796\pm0.024$&
 $0.831\pm0.013$& $!0.0$\cr
$10^9A_{\mathrm{s}}e^{-2\tau}$& $1.880\pm0.014$& $1.865\pm0.019$&
 $1.907\pm0.027$& $1.882\pm0.012$& $-0.1$\cr
\noalign{\vskip 5pt\hrule\vskip 3pt}}}
%\endPlancktable                    % ends one-column \halign
\endPlancktablewide                 % ends two-column \halign
%\tablenote a Footnote a.\par
%\tablenote b Footnote b.\par
\endgroup
\end{table*}                        % table* is a two-column table.  Drop the * for one column.

\paragraph{(3) $TE$ and $EE$ conditionals.}
Given the best-fit base \LCDM\ cosmology and foreground parameters determined
from the temperature spectra, one can test whether the $TE$ and $EE$ spectra
are consistent with the $TT$ spectra by computing conditional
probabilities. Writing the data vector as
\begin{equation}
\vec{\hat C} = (\vec{\hat C}^{TT}, \vec{\hat C}^{TE},
 \vec{\hat C}^{EE})^{\sf T} =
 (\vec{\hat X}_T, \vec{\hat X}_P)^{\sf T}, \label{CVEC1}
\end{equation}
where the quantities $\vec{\hat C}^{TT}$, $\vec{\hat C}^{TE}$, and
 $\vec{\hat C}^{EE}$ are the maximum likelihood freqency co-added foreground-corrected spectra.
The covariance matrix of this vector can be partitioned as
\begin{equation}
\tens{\hat M} = \left ( \begin{array} {c|c}
  \tens{M}_T &  \tens{M}_{TP}  \\  \hline
  \tens{M}_{TP}^{\sf T} &   \tens{M}_P
              \end{array} \right ). \label{CSL1}
\end{equation}
The expected value of the polarization vector, given the observed temperature
vector $\vec{\hat X}_T$ is
\begin{equation}
\vec{\hat X}^{\rm cond}_P = \vec{\hat X}^{\rm theory}_P + \vec{M}^{\sf T}_{TP}
 \vec{M}^{-1}_T (\vec{\hat X}_T - \vec{\hat X}^{\rm theory}_T), \label{CVEC2}
\end{equation}
with covariance
\begin{equation}
\tens{\hat \Sigma}_P = \tens{M}_{P} - \tens{M}^{\sf T}_{TP} \tens{M}^{-1}_T
 \tens{M}_{TP}. \label{CVEC3}
\end{equation}
In Eq.~(\ref{CVEC2}), $\vec{X}^{\rm theory}_T$ and $\vec{X}^{\rm theory}_P$
are the theoretical temperature and polarization spectra
deduced from minimizing the \planckTT\ likelihood.  Equations~(\ref{CVEC2})
and (\ref{CVEC3}) give the expectation values and
distributions of the polarization spectra conditional on the observed
temperature spectra.  These are shown in
Fig.~\ref{pgTE+EE_cond}. Almost all of the data points sit within the
$\pm 2\,\sigma$ bands and in the case of the $TE$ spectra, the data
points track the fluctuations expected from the $TT$ spectra at
multipoles $\ell \la 1000$. Figure~\ref{pgTE+EE_cond} therefore
provides an important additional check of the consistency of the $TE$
and $EE$ spectra with the base \LCDM\ cosmology.

\paragraph{(4) Likelihood implementation.}
Section~\ref{subsec:planck_only1} showed good consistency between the
independently written \camspec\ and \plik\ codes in temperature. The
methodology used for the temperature likelihoods are very similar,
but the treatment of the polarization spectra in the two codes differs
substantially.   \camspec\ uses low-resolution CMB-subtracted 353-GHz
polarization maps thresholded by $P=(Q^2 + U^2)^{1/2}$ to define
diffuse Galactic polarization masks. The same apodized polarization
mask, with an effective sky
fraction $f^{\rm eff}_{\rm sky} = 48.8\,\%$ (as defined by
Eq.~\eqref{plik1}), is used for 100-, 143-, and 217-GHz $Q$ and $U$ maps.  Since there are no unresolved extragalactic
foregrounds detected in the $TE$ and $EE$ spectra, all of the different
frequency combinations of $TE$ and $EE$ spectra are compressed into
single $TE$ and $EE$ spectra (weighted by the inverse of the diagonals of the
appropriate covariance matrices), after foreground cleaning using the 353-GHz
maps\footnote{To reduce the impact of noise at 353\,GHz,
 the map-based cleaning of the $TE$ and $EE$ spectra is applied at
 $\ell \le 300$. At higher multipoles, the polarized dust corrections
 are small and are subtracted as power laws fitted to the Galactic
 dust spectra at lower multipoles.} (generalizing the map cleaning technique
described in Sect.~\ref{subsec:planck_only2} to polarization).
This allows the construction of a
full $TT,TE,EE$ likelihood with no binning of the spectra and with no
additional nuisance parameters in polarization. As noted in
Sect.~\ref{subsec:planck_only1} the consistency of results from the
polarization blocks of the \camspec\ and \plik\ likelihoods is not as
good as in temperature. Cosmological parameters from fits to the $TE$
and $EE$ \camspec\ and \plik\ likelihoods can differ by up to about
$1.5\,\sigma$, although no major
science conclusions would change had we chosen to use the
\camspec\ likelihood as the baseline in this paper. We will, however,
sometimes quote results from \camspec\ in addition to those from
\plik\ to give the reader an indication of the uncertainties in
polarization associated with different likelihood implementations.
Figure~\ref{pgTE+EE_condcamspec} shows the \camspec\ $TE$ and $EE$
residuals and error ranges conditional on the best-fit base \LCDM\ and
foreground model fitted to the \camspec\ temperature+lowP
likelihood. The residuals in both $TE$ and $EE$ are similar to those
from \plik. The main difference can be seen at low multipoles in the
$EE$ spectrum, where \camspec\ shows a higher dispersion, consistent with the
error model, though there are several high points at $\ell \approx 200$
corresponding to the minimum in the $EE$ spectrum, which may be caused
by small errors in the subtraction of polarized Galactic
emission using 353\,GHz as a foreground template (and there are also
differences in the covariance matrices at high multipoles caused by
differences in the methods used
in \camspec\ and \plik\ to estimate noise). Generally, cosmological
parameters determined from the \camspec\ likelihood have smaller formal
errors than those from \plik\ because there are no nuisance parameters
describing polarized Galactic foregrounds in \camspec.

\subsubsection{Consistency of cosmological parameters from the $TT$, $TE$,
and $EE$ spectra}
\label{subsec:consistency}

\begin{figure*}[tb]
\begin{center}
\includegraphics[width=18cm]{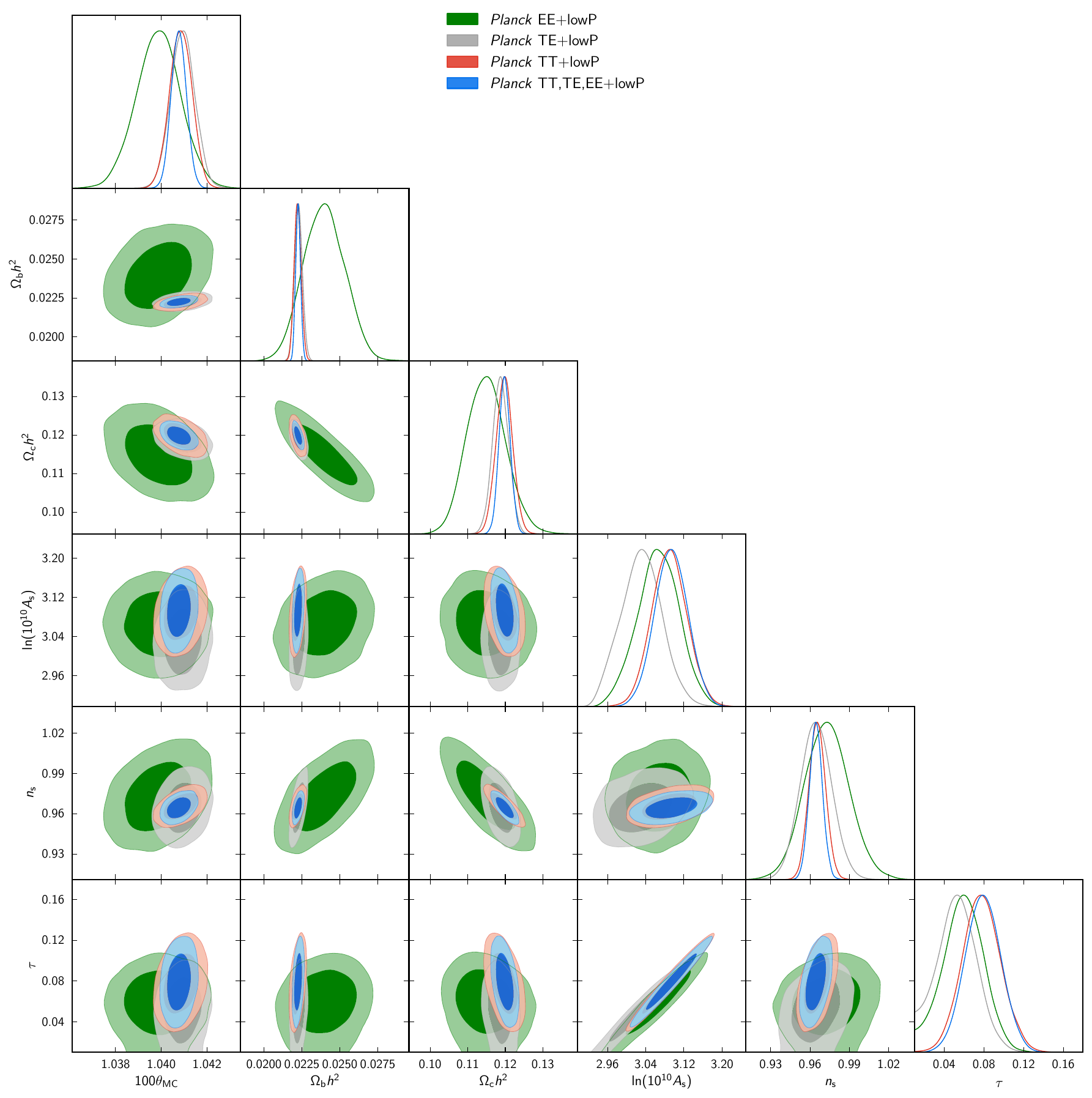}
\end{center}
\vspace{-3mm}
\caption {Comparison of the base \LCDM\ model parameter constraints from \planck\
temperature and polarization data. }
\label{fig:planckOnlyTriangle}
\end{figure*}

The consistency between parameters of the base \LCDM\ model determined
from the \plik\ temperature and polarization spectra are summarized in
Table~\ref{paramtab2} and in Fig.~\ref{fig:planckOnlyTriangle}.  As
pointed out by \cite{Zaldarriaga:1997ch} and \cite{Galli:2014kla},
precision measurements of the CMB polarization spectra have the
potential to constrain cosmological parameters to higher accuracy than
measurements of the $TT$ spectra because the acoustic peaks are narrower
in polarization and unresolved foreground contributions at high
multipoles are much lower in polarization than in temperature. The
entries in Table~\ref{paramtab2} show that cosmological parameters
that do not depend strongly on $\tau$ are consistent between the
$TT$ and $TE$ spectra, to within typically $0.5\,\sigma$ or better.
Furthermore, the cosmological parameters derived from the
$TE$ spectra have comparable errors to the $TT$
parameters.  None of the conclusions in this paper would change in any
significant way were we to use the $TE$ parameters in place of the $TT$
parameters. The consistency of the cosmological parameters for base
\LCDM\ between temperature and polarization therefore gives added
confidence that \Planck\ parameters are insensitive to the specific
details of the foreground model that we have used to correct the $TT$
spectra.  The $EE$ parameters are also typically within about $1\,\sigma$ of
the $TT$ parameters, though because the $EE$ spectra from \Planck\ are
noisier than the $TT$ spectra, the errors on the $EE$ parameters are
significantly larger than those from $TT$. However, both the $TE$ and $EE$
likelihoods give lower values of $\tau$, $\As$ and $\sigma_8$,
by over $1\,\sigma$ compared to the $TT$ solutions. Noticee that the $TE$ and
EE entries in Table~\ref{paramtab2} do not use any information from the
temperature in the low-multipole likelihood.  The tendency for higher
values of $\sigma_8$, $\As$, and $\tau$ in the \planckTT\ solution is
driven, in part, by the temperature power spectrum at low multipoles.

Columns [4] and [5] of Table~\ref{paramtab2} compare the parameters of
the \Planck\ $TT$ likelihood with the full \Planck\ $TT,TE,EE$ likelihood.
These are in agreement, shifting by less than $0.2\,\sigma$. Although we
have emphasized the presence of systematic effects in the \Planck\ polarization
spectra, which are not accounted for in the errors quoted in column [4]
of Table~\ref{paramtab2}, the consistency of the \Planck\ $TT$
and \Planck\ $TT,TE,EE$ parameters provides strong
evidence that residual systematics in the polarization spectra have
little impact on the scientific conclusions in this paper. The
consistency of the base \LCDM\ parameters from temperature and
polarization is illustrated graphically in
Fig.~\ref{fig:planckOnlyTriangle}. As a rough rule-of-thumb, for base
\LCDM, or extensions to \LCDM\ with spatially flat geometry, using the
full \Planck\ $TT,TE,EE$ likelihood produces improvements in cosmological
parameters of about the same size as adding BAO to the \planckTT\ likelihood.

\subsection{Constraints on the reionization optical depth parameter $\tau$}
\label{subsec:tau}

\begin{figure*}
%\vspace{1.5cm}
\begin{center}
\includegraphics[width=18.4cm]{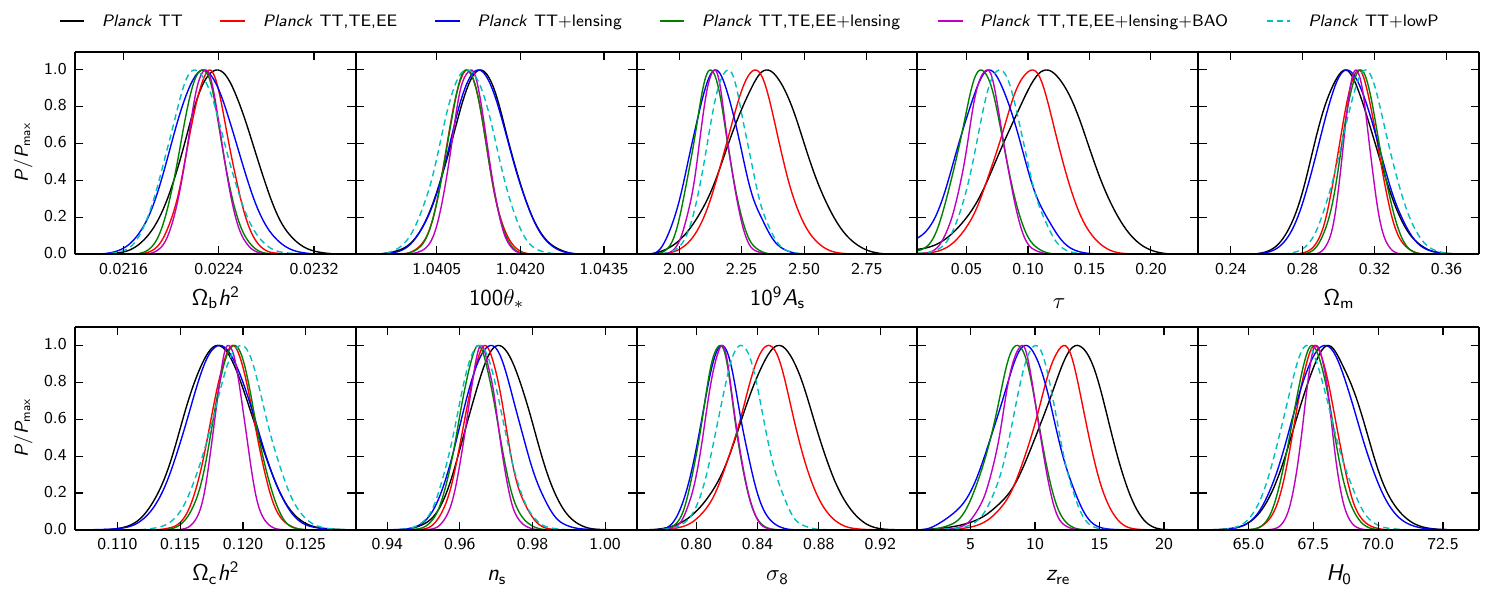}
\end{center}
\vspace{-6mm}
\caption {Marginalized constraints on parameters of the base \LCDM\ model
for various data combinations, excluding low-multipole polarization, compared to
the \planckTT\ constraints.}
%\vspace{0.5cm}
\label{fig:tau_compare}
\end{figure*}

The reionization optical depth parameter $\tau$ provides an important
constraint on models of early galaxy evolution and star formation. The
evolution of the inter-galactic Ly$\alpha$ opacity measured in the
spectra of quasars can be used to set limits on the epoch of
reionization \citep{Gunn:1965}.  The most recent measurements suggest
that the reionization of the inter-galactic medium was largely
complete by a redshift $z\approx6$ \citep{Fan:2006}.  The steep
decline in the space density of Ly$\alpha$-emitting galaxies over the
redshift range $6 \la z \la 8$ also implies a low redshift of
reionization \citep{Choudhury:2014}.  As a reference, for the
\Planck\ parameters listed in Table~\ref{paramtab2}, instantaneous
reionization at redshift $z = 7$ results in an optical depth of $\tau= 0.048$.

The optical depth $\tau$ can also be constrained from observations of
the CMB. The WMAP9 results of \citet{bennett2012} give $\tau=0.089 \pm
0.014$, corresponding to an instantaneous redshift of reionization
$\zre = 10.6 \pm 1.1$.  The WMAP constraint comes mainly from the $EE$
spectrum in the multipole range $\ell=2$--6. It has been argued
\citep[e.g.,][and references therein]{Robertson:2013} that the high
optical depth reported by WMAP cannot be produced by galaxies seen in
deep redshift surveys, even assuming high escape fractions for
ionizing photons, implying additional sources of photoionizing
radiation from still fainter objects. Evidently, it would be useful to
have an independent CMB measurement of $\tau$.

The $\tau$ measurement from CMB polarization is difficult because it
is a small signal, confined to low multipoles, requiring accurate
control of instrumental systematics and polarized foreground
emission. As discussed by \citet{komatsu2009}, uncertainties in
modelling polarized foreground emission are comparable to the
statistical error in the WMAP $\tau$ measurement. In particular, at
the time of the WMAP9 analysis there was very little information
available on polarized dust emission. This situation has been
partially rectified by the 353-GHz polarization maps from
\Planck\ \citep{planck2014-XXII, planck2014-XXX}.  In \likeI, we used
preliminary 353-GHz \planck\ polarization maps to
clean the WMAP Ka, Q, and V maps for polarized dust emission, using WMAP
K-band as a template for polarized synchrotron emission. This lowered
$\tau$ by about 1$\,\sigma$ to $\tau = 0.075 \pm 0.013$, compared to
$\tau = 0.089\pm 0.013$ using the WMAP dust model.\footnote{Neither
  of these error estimates reflect the true uncertainty in
  foreground removal.}  However, given the preliminary nature of the
\Planck\ polarization analysis we decided for the \Planck\ 2013 papers
to use the WMAP polarization likelihood, as produced by the WMAP team.

In the 2015 papers, we use \Planck\ polarization maps based on
low-resolution LFI 70-GHz maps, excluding Surveys~2 and 4. These maps
are foreground-cleaned using the LFI 30-GHz and HFI 353-GHz maps as
polarized synchrotron and dust templates, respectively. These cleaned
maps form the polarization part (``\lowTEB' ) of the low-multipole
\Planck\ pixel-based likelihood, as described in \cite{planck2014-a13}.
The temperature part of this likelihood is provided by the {\tt Commander}
component-separation algorithm.  The \Planck\ low-multipole likelihood
retains 46\,\% of the
sky in polarization and is completely independent of the WMAP
polarization likelihood. In combination with the \Planck\ high
multipole $TT$ likelihood, the \Planck\ low-multipole likelihood gives $\tau =
0.078\pm 0.019$. This constraint is somewhat higher than the
constraint $\tau = 0.067\pm 0.022$ derived from the
\Planck\ low-multipole likelihood alone (see \citealt{planck2014-a13} and also
Sect.~\ref{subsec:Alens}).

Following the 2013 analysis, we have used the 2015 HFI 353-GHz
polarization maps as a dust template, together with the WMAP K-band data as
a template for polarized synchrotron emission, to clean the
low-resolution WMAP Ka, Q, and V maps (see \citealt{planck2014-a13} for
further details).  For the purpose of cosmological parameter
estimation, this data set is masked using the WMAP P06 mask, which
retains 73\,\% of the sky.  The noise-weighted combination of the
\Planck\ 353-cleaned WMAP polarization maps yields $\tau = 0.071 \pm
0.013$ when combined with the \Planck\ $TT$ information in the range
$2 \le \ell \la 2508$, consistent with the value of $\tau$ obtained
from the LFI 70-GHz polarization maps. In fact, null tests described
in \cite{planck2014-a13} demonstrate that the LFI and WMAP
polarization data are statistically consistent. The HFI polarization
maps have higher signal-to-noise than the LFI and could, in principle,
provide a third cross-check. However, at the time of writing, we are
not yet confident that systematics in the HFI maps at low multipoles
($\ell \la 20$) are at negligible levels. A discussion of HFI
polarization at low multipoles will therefore be deferred to future
papers.\footnote{See \citet{planck2014-a10}, which has been
submitted since this paper was written.}

Given the difficulty of making accurate CMB polarization measurements
at low multipoles, it is useful to investigate other ways of
constraining $\tau$.  Measurements of the temperature power spectrum
provide a highly accurate measurement of the amplitude $\As
e^{-2\tau}$. However, as shown in {\paramsI} CMB {\it lensing\/}
breaks the degeneracy between $\tau$ and $\As$. The observed
\Planck\ $TT$ spectrum is, of course, lensed, so the degeneracy
between $\tau$ and $\As$ is partially broken when we fit models to the
\Planck\ $TT$ likelihood. However, the degeneracy breaking is much
stronger if we combine the \planckTTonly\ likelihood with the
\Planck\ lensing likelihood constructed from measurements of the power
spectrum of the lensing potential $C^{\phi\phi}_\ell$. The 2015
\planckTTonly\ and lensing likelihoods are statistically more powerful
than their $2013$ counterparts and the corresponding determination of
$\tau$ is more precise. The 2015 \Planck\ lensing likelihood (labelled
``lensing'') is
summarized in Sec.~\ref{sec:lensing} and discussed in more detail in
\citet{planck2014-a17}. The constraints on $\tau$ and
$\zre$\footnote{We use the same specific definition of $\zre$ as in
  the $2013$ papers, where reionization is assumed to be relatively
  sharp, with a mid-point parameterized by a redshift $\zre$ and width
  $\Delta\zre=0.5$.  Unless otherwise stated we impose a flat prior on
  the optical depth with $\tau>0.01$.} for various data combinations
{\it excluding\/} low-multipole polarization data from
\Planck\ are summarized in Fig.~\ref{fig:tau_compare} and
compared with the baseline \planckTT\ parameters. This figure
also shows the shifts of other parameters of the base
\LCDM\ cosmology, illustrating their sensitivity to changes in $\tau$.

The \planck\ constraints on $\tau$ and $\zre$ in the base \LCDM\ model for
various data combinations are:
\beglet
\begin{eqnarray}
\tau &=&  0.078^{+0.019}_{-0.019},  \ \zre = 9.9^{+1.8}_{-1.6}, \ \planckTT;
 \label{eq:tauconstraintsfirst} \\
\tau &=&  0.070^{+0.024}_{-0.024},  \ \zre = 9.0^{+2.5}_{-2.1},
 \ {\planckTTonly\dataplus\lensing}; \\
\tau &=&  0.066^{+0.016}_{-0.016},  \ \zre = 8.8^{+1.7}_{-1.4}, \ \planckTT\\
 \nonumber
     & &  \qquad \qquad \qquad \qquad  \qquad \qquad \qquad \qquad\qquad
  +\lensing; \\
\tau &=&  0.067^{+0.016}_{-0.016},  \  \zre = 8.9^{+1.7}_{-1.4}, \
 {\planckTTonly\dataplus\lensing} \\ \nonumber
     & &  \qquad \qquad \qquad \qquad  \qquad \qquad \qquad \qquad
 \qquad{\dataplus\BAO}; \\
\tau &=&  0.066^{+0.013}_{-0.013},  \  \zre = 8.8^{+1.3}_{-1.2}, \ \planckTT
 \label{eq:tauconstraintslast} \\ \nonumber
     & &  \qquad \qquad \qquad \qquad  \qquad \qquad \qquad
 \quad{\dataplus\lensing\dataplus\BAO}.
\end{eqnarray}
\label{eq:tauconstraintsall}
\endlet

The constraint from \planckTTonly+\lensing+BAO on $\tau$ {\it is
completely independent of low-multipole CMB polarization data\/} and
agrees well with the result from \planck\ polarization (and has
comparable precision).  These results all indicate a lower redshift of
reionization than the value $\zre=11.1 \pm 1.1$ derived in {\paramsI},
based on the WMAP9 polarization likelihood.  The low values of $\tau$
from \Planck\ are also consistent with the lower value of $\tau$
derived from the WMAP \Planck\ 353-GHz-cleaned polarization
likelihood, suggesting strongly that the WMAP9 value is biased
slightly high by residual polarized dust emission.

The \Planck\ results of Eqs.~(\ref{eq:tauconstraintsfirst})--(\ref{eq:tauconstraintslast}) provide evidence for a lower optical
depth and redshift of reionization than inferred from WMAP \citep{bennett2012}, partially alleviating the
difficulties in reionizing the intergalactic medium using starlight
from high-redshift galaxies. A key goal of the \Planck\ analysis over
the next year is to assess whether these results are consistent with
the HFI polarization data at low multipoles.

Given the consistency between the LFI and WMAP polarization maps when
both are cleaned with the HFI 353-GHz polarization maps, we have also
constructed a combined WMAP+\Planck\ low-multipole polarization
likelihood (denoted ``lowP+WP''). This likelihood uses 73\,\% of the sky and
is constructed from a noise-weighted combination of LFI 70-GHz and
WMAP Ka, Q, and V maps,
as summarized in Sect.~\ref{subsec:planck_only1} and discussed in more detail
in \cite{planck2014-a13}.  In combination with the \Planck\ high-multipole
$TT$ likelihood, the combined lowP+WP likelihood gives $\tau
= 0.074^{+0.011}_{-0.013}$, consistent with the individual LFI and WMAP
likelihoods to within about $0.5\,\sigma$.

\begin{figure}
\begin{center}
\includegraphics[width=\hsize]{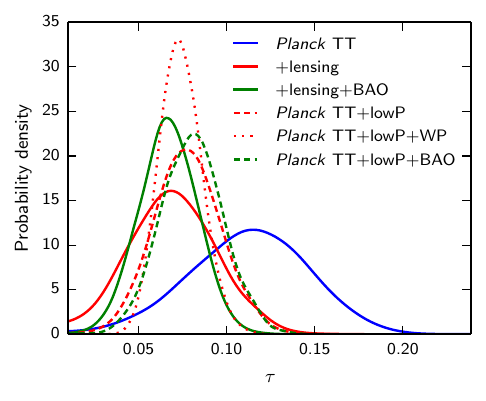}
\end{center}
\vspace{-5mm}
\caption {Marginalized constraints on the reionization optical depth in the
base \LCDM\ model for various data combinations. Solid lines do not
include low-multipole polarization; in these cases the optical depth is
constrained by \planck\ lensing. The dashed/dotted lines include
LFI polarization (+\lowTEB), or the combination of LFI and WMAP polarization
cleaned using 353\,GHz as a dust template (+\WMAPTEB). }
\label{fig:tau_constraints}
\end{figure}

The various \Planck\ and \Planck+WMAP constraints on $\tau$ are
summarized in Fig.~\ref{fig:tau_constraints}.  The tightest of these
constraints comes from the combined \WMAPTEB\ likelihood. It is
therefore reasonable to ask why we have chosen to use the
\lowTEB\ likelihood as the baseline in this paper, which gives a higher
statistical error on $\tau$. The principal reason is to produce a
\Planck\ analysis, utilizing the LFI polarization data, that is
independent of WMAP.  All of the constraints shown in
Fig.~\ref{fig:tau_constraints} are compatible with each other, and
insofar as other cosmological parameters are sensitive to small
changes in $\tau$, it would make very little difference to the results
in this paper had we chosen to use WMAP or \Planck+WMAP polarization
data at low multipoles.

\section{Comparison of the \Planck\ power spectrum with high-resolution experiments}\label{sec:actspt}

In {\paramsI} we combined \planck\ with the small-scale measurements of the
ground-based, high-resolution Atacama Cosmology Telescope (ACT) and
South Pole Telescope (SPT). The primary role of using ACT and SPT was to
set limits on foreground components that were poorly constrained by
\planck\ alone and to provide more accurate constraints on the damping
tail of the temperature power spectrum. In this paper, with the higher
signal-to-noise levels of the full mission \planck\ data, we have taken a
different approach, using the ACT and SPT data to impose a prior on the thermal
and kinetic SZ power spectrum parameters in the \planck\ foreground
model as described in Sect.~\ref{subsec:foreground}. In this section, we
check the consistency of the temperature power spectra measured by \planck,
ACT, and SPT, and test the effects of 
including the ACT and SPT data on the recovered CMB power spectrum.

We use the latest ACT temperature power spectra presented in
\cite{Das:2014}, with a revised binning described in
\citet{Calabrese:2013} and final beam estimates in
\citet{Hasselfield:2013}. As in {\paramsI} we use ACT data in the range
$1\,000<\ell<10\,000$ at 148\,GHz, and $1\,500<\ell<10\,000$ for the
$148\times218$ and 218-GHz spectra. We use SPT measurements in the
range $2\,000<\ell<13\,000$ from the complete $2\,540\,{\rm deg}^2$ SPT-SZ
survey at 95, 150, and 220\,GHz presented in \cite{George:2014}.

Each of these experiments uses a foreground model to describe the
multi-frequency power spectra. Here we implement a common foreground
model to combine \planck\ with the high-multipole data, following a similar
approach to {\paramsI} but with some refinements. Following the 2013 analysis,
 we solve for
common nuisance parameters describing the tSZ, kSZ, and tSZ$\times$CIB
components, extending the templates used for \planck\ to $\ell=13\,000$
to cover the full ACT and SPT multipole range. As in {\paramsI}, we use five
point-source amplitudes to fit for the total dusty and radio Poisson
power, namely $A^{\mathrm{PS,\,ACT}}_{148}$, $A^{\mathrm{PS,\,ACT}}_{218}$,
$A^{\mathrm{PS,\,SPT}}_{95}$, $A^{\mathrm{PS,\,SPT}}_{150}$, and
$A^{\mathrm{PS,\,SPT}}_{220}$. We rescale these amplitudes to
cross-frequency spectra using point-source correlation coefficients,
improving on the 2013 treatment by using different parameters for the
ACT and SPT correlations, $r^{\mathrm{PS, ACT}}_{148\times218}$ and
$r^{\mathrm{PS, SPT}}_{150\times220}$ (a single
$r^{\mathrm{PS}}_{150\times220}$ parameter was used in 2013). We vary
$r^{\mathrm{PS, SPT}}_{95\times150}$ and $r^{\mathrm{PS, SPT}}_{95\times220}$
as in 2013, and include dust amplitudes for
ACT, with Gaussian priors as in {\paramsI}.

As described in Sect.~\ref{subsec:foreground} we use a theoretically
motivated clustered CIB model fitted to \Planck+IRAS estimates of the
CIB. The model at all frequencies in the range 95--220\,GHz is
specified by a single amplitude $A^{\mathrm{CIB}}_{217}$.  The CIB
power is well constrained by \planck\ data at $\ell<2000$. At multipoles
$\ell \ga 3000$, the 1-halo component of the CIB model steepens and
becomes degenerate with the Poisson power. This causes an underestimate of the
Poisson levels for ACT and SPT, inconsistent with predictions from source
counts. We therefore use the \planck\ CIB template only in the range
$2<\ell<3000$, and extrapolate to higher multipoles using a power law
$\mathcal{D}_{\ell} \propto \ell^{\,0.8}$.  While this may
not be a completely accurate model for the clustered CIB spectrum at
high multipoles \citep[see, e.g.,][]{Viero2013,planck2013-pip56},
this extrapolation is consistent with the CIB model
used in the analysis of ACT and SPT.
We then need to extrapolate the \Planck\ 217-GHz CIB power to the ACT and
SPT frequencies. This requires converting the CIB measurement in the HFI
217-GHz channel to the ACT and SPT bandpasses assuming a spectral
energy distribution; we use the CIB
spectral energy distribution from \cite{bethermin2012}.
Combining this model with the ACT and SPT bandpasses, we find that
$A_{217}^{\rm CIB}$ has to be multiplied by 0.12 and 0.89 for ACT 148 and
218\,GHz, and by 0.026, 0.14, and 0.91 for SPT 95, 150, and 220\,GHz,
respectively.  With this model in place, the best-fit \Planck, ACT, and SPT
Poisson levels agree with those predicted from source counts, as 
discussed further in \cite{planck2014-a13}.

The nuisance model includes seven calibration parameters as in {\paramsI} (four
for ACT and three for SPT). The ACT spectra are internally calibrated
using the \WMAP\ 9-year maps, with 2\,\% and 7\,\% uncertainty at 148 and
218\,GHz, while SPT calibrates using the \planck\ 2013 143-GHz maps,
with 1.1\,\%, 1.2\,\%, and 2.2\,\% uncertainty at 95, 150, and 220\,GHz. To
account for the increased 2015 \planck\ absolute calibration (2\,\%
higher in power) we increase the mean of the SPT map-based calibrations from
1.00 to 1.01.

\begin{figure}
\centering
\includegraphics[width=\columnwidth]{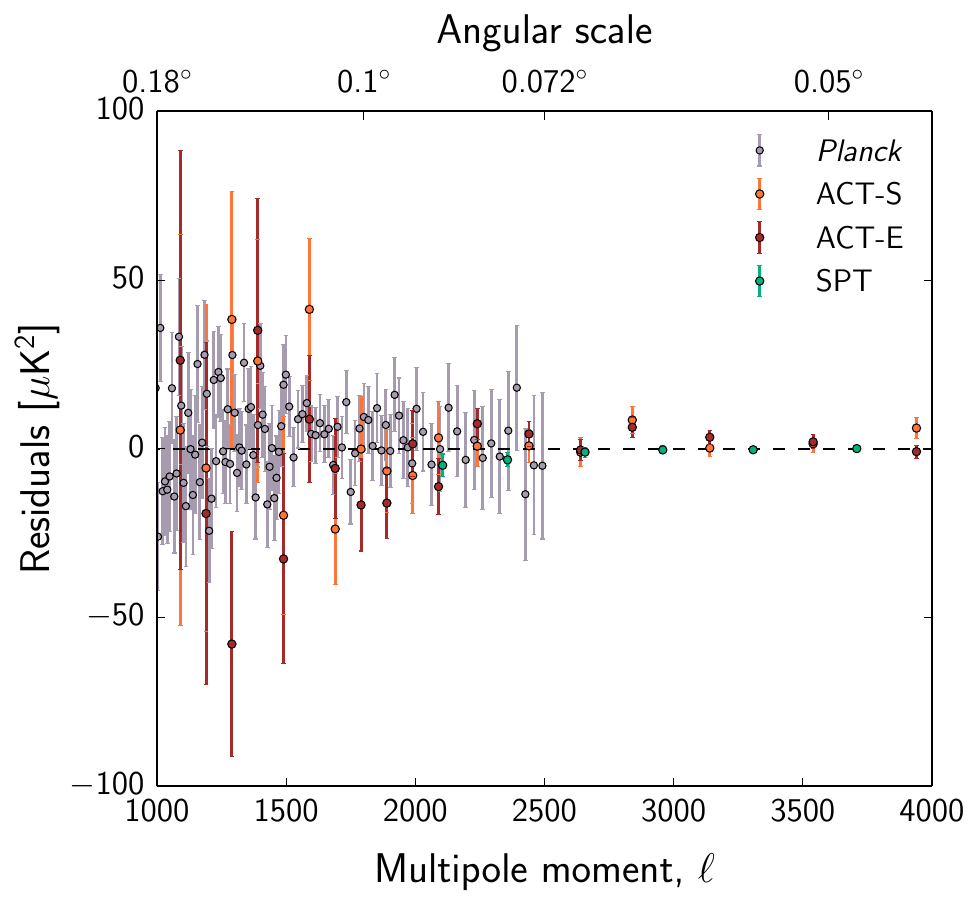}
\caption {Residual power with respect to the \planck\ TT+lowP
$\Lambda$CDM best-fit model for the \planck\ (grey), ACT south (orange),
ACT equatorial (red), and SPT (green) CMB bandpowers.
The ACT and SPT bandpowers are scaled by the best-fit calibration factors.}
\label{fig:resid_highL}
\end{figure}

This common foreground and calibration model fits the data well. We
first fix the cosmology to that of the best-fit \planckTT\ base-\lcdm\ model,
and estimate the foreground and calibration parameters, finding a
best-fitting $\chi^2$ of 734 for 731 degrees of freedom (reduced
$\chi^2=1.004$, PTE\,{=}\,0.46).  We then simultaneously estimate the
\planck, ACT- (S: south, E: equatorial) and SPT CMB bandpowers, $C_b$,
following the Gibbs sampling scheme of \citet{Dunkley:2013} and
\citet{Calabrese:2013}, marginalizing over the nuisance
parameters.

To simultaneously solve for the \planck, ACT, and SPT CMB spectra, we extend
the nuisance model described above, including the four \planck\ point
source amplitudes, the dust parameters and the \planck\ 100-GHz and
217-GHz calibration parameters (relative to 143\,GHz) with the same
priors as used in the \planck\ multi-frequency likelihood analysis. For ACT
and SPT, the calibration factors are defined for each frequency (rather
than relative to a central frequency). Following
\citet{Calabrese:2013}, we separate out the 148-GHz calibration for the
 ACT-(S,E) spectra and the 150-GHz calibration for SPT, estimating the CMB
bandpowers as $C_b/A_{\rm cal}$.\footnote{This means that the other calibration
factors (e.g., ACT 218\,GHz) are re-defined to be relative to 148\,GHz (or
150\,GHz for SPT) data.}  We impose Gaussian priors on $A_{\rm cal}$: $1.00 \pm
0.02$ for ACT-(S,E); and $1.010 \pm 0.012$ for SPT. The estimated CMB
spectrum will then have an overall calibration uncertainty for each of
the ACT-S, ACT-E, and SPT spectra.  We do not require the \planck\ CMB
bandpowers to be the same as those for ACT or SPT, so that we can
check for consistency between the three experiments.

In Fig.~\ref{fig:resid_highL} we show the residual CMB power with
respect to the \planck\ TT+lowP \lcdm\ best-fit model for the three
experiments. All of the data sets are consistent over the multipole range
plotted in this figure. For ACT-S, we find $\chi^2 = 17.54$
(18 data points, ${\rm PTE} = 0.49$); For ACT-E we find
$\chi^2 = 23.54$ (18 data points, ${\rm PTE} = 0.17$); and for SPT
$\chi^2 = 5.13$ (6 data points, ${\rm PTE} = 0.53$).

\begin{figure}
\centering
\includegraphics[width=\columnwidth]{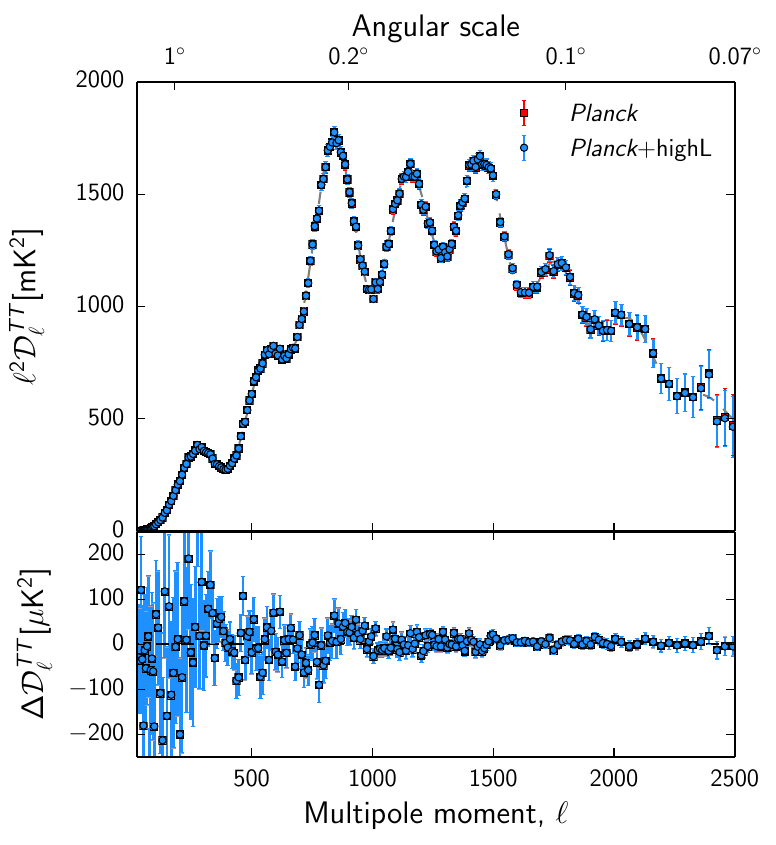}
\caption {\planck\ CMB power spectrum that is marginalized over foregrounds
(red), including a prior on the thermal and kinetic SZ power. The inclusion of
the full higher resolution ACT and SPT data (shown in blue) does not
significantly decrease the errors.}
\label{fig:planck_highL}
\end{figure}

Figure~\ref{fig:planck_highL} shows the effect of including ACT and SPT data
on the recovered \planck\ CMB spectrum. We find that including the ACT and SPT
data does not reduce the \planck\ errors significantly. This is expected
because the dominant small-scale foreground
contributions for \planck\ are the Poisson source amplitudes, which are treated
independently of the Poisson amplitudes for ACT and SPT. The
high-resolution experiments do help tighten the
CIB amplitude (which is reasonably well constrained by \Planck)
and the tSZ and kSZ amplitudes (which are sub-dominant
foregrounds for \Planck). The kSZ effect in particular is degenerate
with the CMB, since both have blackbody components; imposing a
prior on the allowed kSZ power (as discussed in Sect.~\ref{subsec:foreground}) breaks this degeneracy. The net effect is
that the errors on the recovered \planck\ CMB spectrum are only marginally
reduced with the inclusion of the ACT and SPT data. This motivates our choice
to include the information from ACT and SPT into the joint tSZ and kSZ prior
applied to \Planck. 

The Gibbs sampling technique recovers a best-fit CMB spectrum marginalized over
foregrounds and other nuisance parameters. The Gibbs samples can then be used
to form a fast CMB-only \Planck\ likelihood that depends on only one nuisance
parameter, the overall calibration $y_p$. MCMC chains run using the
CMB-only likelihood therefore converge much faster than using the full
multi-frequency \plik\ likelihood. The CMB-only likelihood is also extremely
accurate, even for extensions to the base \LCDM\ cosmology and is
discussed further in \citet{planck2014-a13}.

\section{Comparison of the \Planck\  base \lcdm\ model with other astrophysical
data sets}
\label{sec:datasets}

\subsection{CMB lensing measured by \Planck}
\label{sec:lensing}

Gravitational lensing by large-scale structure leaves imprints on the
CMB temperature and polarization that can be measured in high angular
resolution, low-noise observations, such as those from \planck. The
most relevant effects are a smoothing of the acoustic peaks and
troughs in the $TT$, $TE$, and $EE$ power spectra, the conversion
of $E$-mode polarization to $B$-modes, and the generation of
significant non-Gaussianity in the form of a non-zero connected
4-point function (see~\citealt{2006PhR...429....1L} for a review). The
latter is proportional to the power spectrum $C_\ell^{\phi\phi}$ of
the lensing potential $\phi$, and so one can estimate this power
spectrum from the CMB 4-point functions. In the 2013 \planck\ release,
we reported a $10\,\sigma$ detection of the lensing effect in the $TT$
power spectrum (see {\paramsI})
and a $25\,\sigma$ measurement of the amplitude of $C_\ell^{\phi\phi}$ from
the $TTTT$ 4-point function~\citep{planck2013-p12}. The power of such lensing
measurements is that they provide sensitivity to parameters that affect the
late-time expansion, geometry, and matter clustering (e.g., spatial curvature
and neutrino masses) \emph{from the CMB alone}.

Since the 2013 \planck\ release, there have been significant
developments in the field of CMB lensing. The SPT team have reported a
$7.7\,\sigma$ detection of lens-induced $B$-mode polarization based on
the $EB\phi^{\rm CIB}$ 3-point function, where $\phi^{\rm CIB}$ is a
proxy for the CMB lensing potential $\phi$ derived from CIB
measurements~\citep{2013PhRvL.111n1301H}. The POLARBEAR
collaboration~\citep{2014PhRvL.112m1302A} and the ACT
collaboration~\citep{vanEngelen:2014zlh} have performed similar
analyses at somewhat lower significance~\citep{2014PhRvL.112m1302A}.
In addition, the first detections of the polarization 4-point function
from lensing, at a significance of around $4\,\sigma$, have been
reported by the POLARBEAR~\citep{Ade:2013gez} and
SPT~\citep{Story:2014hni} collaborations, and the former have
also made a direct measurement of the $BB$ power spectrum due to
lensing on small angular scales with a significance around
$2\,\sigma$~\citep{2014ApJ...794..171T}. Finally, the $BB$ power
spectrum from lensing has also been detected on degree angular scales,
with similar significance, by the BICEP2
collaboration~\citep{BicepDetection}; see also~\BKP.

\subsubsection{The \Planck\ lensing likelihood}

Lensing results from the full-mission \planck\ data are discussed
in~\citet{planck2014-a17}.\footnote{In that paper we are careful
to highlight the 4-point function origin of the lensing power spectrum
reconstruction by using the index $L$; however, in this paper we use the
notation $\ell$.} With approximately twice the amount of
temperature data, and the inclusion of polarization, the noise levels
on the reconstructed $\phi$ are a factor of about 2 better than
in~\citet{planck2013-p12}. The broad-band amplitude of
$C_\ell^{\phi\phi}$ is now measured to better than $2.5\,\%$ accuracy,
the most significant measurement of CMB lensing to date. Moreover,
lensing $B$-modes are detected at $10\,\sigma$, both through a
correlation analysis with the CIB and via the $TTEB$ 4-point
function. Many of the results in this paper make use of the \planck\
measurements of $C_\ell^{\phi\phi}$. In particular, they provide an
alternative route to estimate the optical depth (as already discussed
in Sect.~\ref{subsec:tau}), and to tightly constrain spatial curvature
(Sect.~\ref{subsubsec:curv}).
\begin{figure*}
\begin{center}
\includegraphics[width=8.8cm,angle=0]{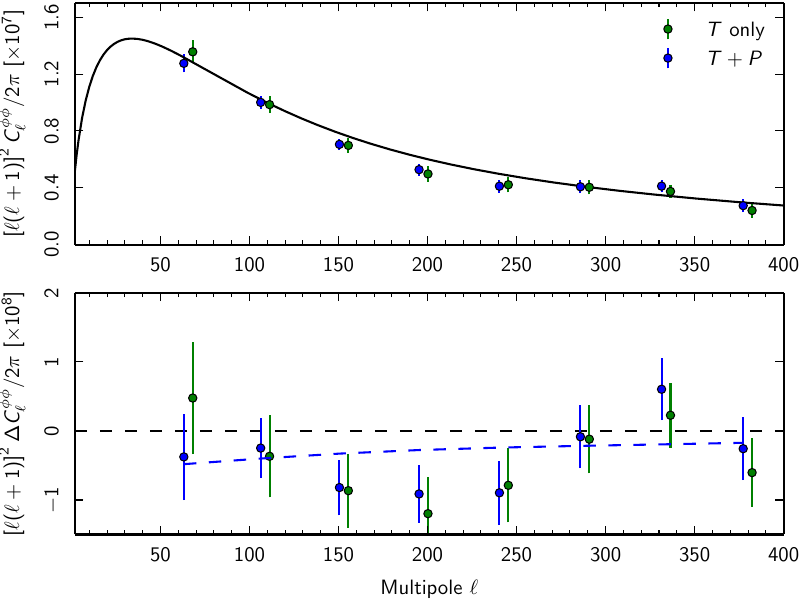}
\includegraphics[width=8.8cm,angle=0]{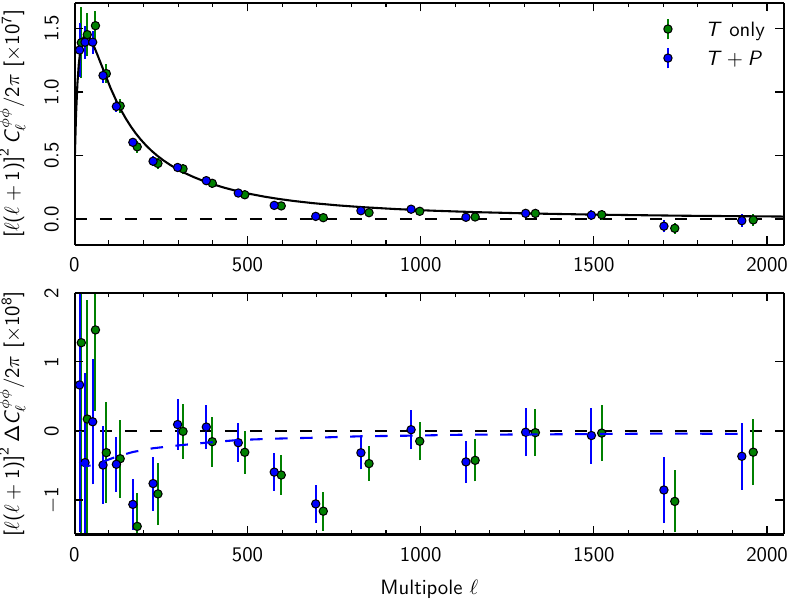}
\end{center}
 \caption{\Planck\ measurements of the lensing power spectrum compared to the
prediction for the best-fitting base \lcdm\ model to the \planckTT\ data.
\emph{Left}: the conservative cut of the \Planck\ lensing data used throughout
this paper, covering the multipole range $40 \leq \ell \leq 400$.
\emph{Right}: lensing data over the range $8 \leq \ell \leq 2048$,
demonstrating the general consistency with the \lcdm\ prediction over this
extended multipole range. In both cases, green points are the power from
lensing reconstructions using only temperature data, while blue points combine
temperature and polarization. They are offset in $\ell$ for clarity and
error bars are $\pm 1\,\sigma$. In the top panels the solid lines are the
best-fitting base \lcdm\ model to the \planckTT\ data with no renormalization
or $\delta N^{(1)}$ correction applied (see text for explanation).
The bottom panels show the
difference between the data and the renormalized and $\delta N^{(1)}$-corrected
theory bandpowers, which enter the likelihood. The mild preference of the
lensing measurements for lower lensing power around $\ell = 200$
pulls the theoretical prediction for $C_{\ell}^{\phi\phi}$ downwards at the
best-fitting parameters of a fit to the combined \planckTT+\lensing\ data,
shown by the dashed blue lines
(always for the conservative cut of the lensing data, including polarization).}
\label{fig:CMBlensing_LCDM}
\end{figure*}

The estimation of $C_\ell^{\phi\phi}$ from the \planck\ full-mission data is
discussed in detail in~\citet{planck2014-a17}. There are a number of
significant changes from the 2013 analysis that are worth noting here.
\begin{itemize}
\item The lensing potential power spectrum is now estimated from lens
reconstructions that use both temperature and polarization data in the
multipole range $100 \leq \ell \leq 2048$. The likelihood used here is based
on the power spectrum of a lens reconstruction derived from the
minimum-variance combination of five quadratic estimators
($TT$, $TE$, $EE$, $TB$, and $EB$). The power spectrum is therefore based on
15 different 4-point functions.
\item The results used here are derived from foreground-cleaned maps of the
CMB synthesized from all nine \planck\ frequency maps with the \smica\
algorithm, while the baseline 2013 results used a minimum-variance combination
of the 143-GHz and 217-GHz nominal-mission maps. After
masking the Galaxy and point-sources, $67.3\,\%$ of the sky is retained for
the lensing analysis.
\item The lensing power spectrum is estimated in the multipole range
$8 \leq \ell \leq 2048$.  Multipoles $\ell < 8$ have large mean-field
corrections due to survey anisotropy and are rather unstable to
analysis choices; they are therefore excluded from all lensing results.  Here,
we use only the range $40\leq \ell \leq 400$ (the same as used in the 2013
analysis), with eight bins each of width $\Delta \ell = 45$. This choice is
based on the extensive suite of null tests reported in~\citet{planck2014-a17}.
Nearly all tests are passed over the full multipole range
$8 \leq \ell \leq 2048$, with the exception of a slight excess of curl modes
in the $TT$ reconstruction around $\ell = 500$.  Given that the range
$40\leq \ell \leq 400$ contains most of the statistical power in the
reconstruction, we have conservatively adopted this range for use in the
\planck\ 2015 cosmology papers.
\item To normalize $C_\ell^{\phi\phi}$ from the measured 4-point functions
requires knowledge of the CMB power spectra. In practice, we normalize with
fiducial spectra, but then correct for changes in the true normalization at
each point in parameter space within the likelihood. The exact renormalization
scheme adopted in the 2013 analysis proved to be too slow for the extension to
polarization, so we now use a linearized approximation, based on pre-computed
response functions, which is very efficient within an MCMC analysis. Spot-checks
have confirmed the accuracy of this approach.
\item The measurement of $C_\ell^{\phi\phi}$ can be thought of as being derived
from an optimal combination of trispectrum configurations. In practice, the
expectation value of this combination at any multipole $\ell$ has a local part
proportional to $C_\ell^{\phi\phi}$, but also a non-local (``$N^{(1)}$ bias'')
part that couples to a broad range of multipoles in
$C_\ell^{\phi\phi}$~\citep{2003PhRvD..67l3507K};
this non-local part comes from non-primary trispectrum couplings.
In the \planck\ 2013 analysis we corrected for the $N^{(1)}$ bias by making a
fiducial correction, but this ignores its parameter dependence. We improve on
this in the 2015 analysis by correcting for errors in the fiducial $N^{(1)}$
bias at each point in parameter space within the lensing likelihood. As with
the renormalization above, we linearize this $\delta N^{(1)}$ correction for
efficiency. As a result, we no longer need to make an approximate correction
in the $C_\ell^{\phi\phi}$ covariance matrix to account for the cosmological
uncertainty in $N^{(1)}$.
\item Beam uncertainties are no longer included in the covariance matrix of
$C_\ell^{\phi\phi}$, since, with the improved knowledge of the beams, the
estimated uncertainties are negligible for the lensing analysis. The only
inter-bandpower correlations included in the $C_\ell^{\phi\phi}$ bandpower
covariance matrix are from the uncertainty in the correction applied for the
point-source 4-point function.
\end{itemize}
As in the 2013 analysis, we approximate the lensing likelihood as Gaussian in
the estimated bandpowers, with a fiducial covariance matrix. Following the
arguments in~\citet{Schmittfull:2013uea}, it is a good approximation to ignore
correlations between the 2- and 4-point functions; so, when combining the
\planck\ power spectra with \planck\ lensing, we simply multiply their
respective likelihoods.

It is also worth noting that the changes in absolute calibration of the
\planck\ power spectra (around $2\,\%$ between the 2013 and 2015 releases)
do not directly affect the lensing results. The CMB 4-point functions do, of
course, respond to any recalibration of the data, but in estimating
$C_\ell^{\phi\phi}$ this dependence is removed by normalizing with theory
spectra fit to the observed CMB spectra. The measured $C_\ell^{\phi\phi}$
bandpowers from the 2013 and current \planck\ releases can therefore be
directly compared, and are in good agreement~\citep{planck2014-a17}. Care is
needed, however, in comparing consistency of the lensing measurements across
data releases with the best-fitting model predictions. Changes in calibration
translate directly into changes in $\As e^{-2\tau}$, which, along with any
change in the best-fitting optical depth, alter $\As$, and hence the predicted
lensing power. These changes from 2013 to the current release go in opposite
directions, leading to a net decrease in $\As$ of $0.6\,\%$. This, combined with
a small ($0.15\,\%$) increase in $\theta_{\rm eq}$, reduces the
expected $C_\ell^{\phi\phi}$ by approximately $1.5\,\%$ for multipoles
$\ell > 60$.

The \planck\ measurements of $C_\ell^{\phi\phi}$, based on the temperature and
polarization 4-point functions, are plotted in Fig.~\ref{fig:CMBlensing_LCDM}
(with results of a temperature-only reconstruction included for comparison).
The measured $C_\ell^{\phi\phi}$ are compared with the predicted lensing power
from the best-fitting base \lcdm\ model to the \planckTT\ data in this figure. The bandpowers that
are used in the conservative lensing likelihood adopted in this paper are shown
in the left-hand plot, while bandpowers over the range
$8 \leq \ell \leq 2048$ are shown in the right-hand plot, to demonstrate the
general consistency with the \lcdm\ prediction over the full multipole range.
The difference between the measured bandpowers and the best-fit prediction are
shown in the bottom panels. Here, the theory predictions are corrected in the
same way as they are in the likelihood.\footnote{In detail, the theory spectrum
is binned in the same way as the data, renormalized to account for the
(very small) difference between the CMB spectra in the best-fit model and the
fiducial spectra used in the lensing analysis, and corrected for the difference
in $N^{(1)}$, calculated for the best-fit and fiducial models (around a
$4\,\%$ change in $N^{(1)}$, since the fiducial-model $C_\ell^{\phi\phi}$ is
higher by this amount than in the best-fit model).}

Figure~\ref{fig:CMBlensing_LCDM} suggests that the \planck\ measurements of
$C_\ell^{\phi\phi}$ are mildly in tension with the prediction of the
best-fitting \lcdm\ model. In particular, for the conservative multipole range
$40 \leq \ell \leq 400$, the temperature+polarization reconstruction has
$\chi^2 = 15.4$ (for eight degrees of freedom), with a PTE of $5.2\,\%$.
For reference, over the full multipole range $\chi^2 = 40.8$ for 19
degrees of freedom (PTE of $0.3\,\%$); the large $\chi^2$ is driven by a
single bandpower ($638 \leq \ell \leq 762$), and excluding this gives an
acceptable $\chi^2 = 26.8$ (PTE of $8\,\%$). We caution the reader that this
multipole range is where the lensing reconstruction shows a mild excess of
curl-modes~\citep{planck2014-a17}, and for this reason we adopt the
conservative multipole range for the lensing likelihood in this paper.

\begin{figure}
\centering
\includegraphics[width=88mm,angle=0]{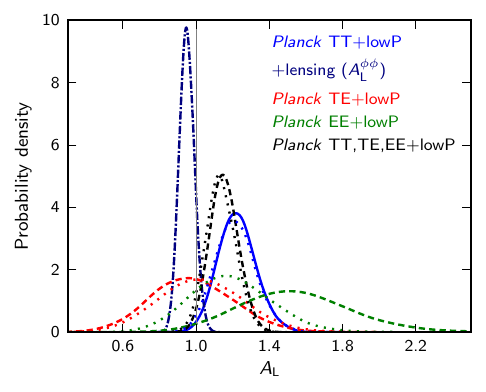}
\caption{Marginalized posterior distributions for measures of the lensing
power amplitude. The dark-blue (dot-dashed) line is the constraint on the
parameter $\Aphiphi$, which scales the amplitude of the lensing power spectrum
in the lensing likelihood for the \planckTT+\lensing\ data combination. The
other lines are for the $\Alens$ parameter, which scales the lensing power
spectrum used to lens the CMB spectra, for the data combinations \planckTT\
(blue, solid), \planck\ TE+\lowEB\ (red, dashed), \planck\ EE+\lowEB\
(green, dashed), and \planckall\ (black, dashed).
The dotted lines show the $\Alens$ constraints when the \plik\ likelihood is
replaced with \camspec, highlighting that the preference for high $\Alens$ in
the \planck\ EE+\lowEB\ data combination is not robust to the treatment of
polarization on intermediate and small scales.
}
\label{fig:alens}
\end{figure}

This simple $\chi^2$ test does not account for the uncertainty in the predicted
$C_\ell^{\phi\phi}$. In the \lcdm\ model, the dominant uncertainty in the
multipole range $40 \leq \ell \leq 400$ comes from that in $\As$
($1\,\sigma$ uncertainty of $3.7\,\%$ for \planckTT), which itself derives from the
uncertainty in the reionization optical depth, $\tau$. The predicted rms lensing deflection from
\planckTT\ data is $\langle d^2 \rangle^{1/2} = (2.50 \pm 0.05)\,
{\rm arcmin}$, corresponding to a $3.6\,\%$ uncertainty ($1\,\sigma$) in the
amplitude of $C_\ell^{\phi\phi}$ (which improves to $3.1\,\%$ uncertainty for
the combined \planck+WP likelihood). Note that this is larger
than the uncertainty on the measured amplitude, i.e., \emph{the lensing
measurement is more precise than the prediction from the CMB power spectra in
even the simplest \lcdm\ model}. This model uncertainty is reflected in a
scatter in the $\chi^2$ of the lensing data over the \planckTT\ chains,
$\chi^2_{\rm lens} = 17.9 \pm 9.0$, which is significantly larger than the
expected scatter in $\chi^2$ at the true model, due to the uncertainties in
the lensing bandpowers ($\sqrt{2 N_{\rm dof}} = 4$).
Following the treatment in {\paramsI},
we can assess consistency more carefully by introducing a parameter $\Aphiphi$
that scales the theory lensing trispectrum at every point in parameter space
in a joint analysis of the CMB spectra and the lensing spectrum. We find
\begin{equation}
\Aphiphi = 0.95 \pm 0.04 \quad \onesig{\planckTTlensing}, \label{CMBlens}
\end{equation}
in good agreement with the expected value of unity. The posterior for
$\Aphiphi$, and other lensing amplitude measures discussed below, is shown in
Fig.~\ref{fig:alens}.

Given the precision of the measured $C_\ell^{\phi\phi}$ compared to the
uncertainty in the predicted spectrum from fits to the \planckTT\ data, the
structure in the residuals seen in Fig.~\ref{fig:CMBlensing_LCDM} might be
expected to pull parameters in joint fits. As discussed
in~\citet{planck2014-a17} and~\citet{2014MNRAS.445.2941P}, the primary
parameter dependence of $C_\ell^{\phi\phi}$ at multipoles $\ell \ga 100$ is
through $\As$ and $\ell_{\rm eq}$ in \lcdm\ models. Here, $\ell_{\rm eq}
\propto 1/\theta_{\rm eq}$ is the angular multipole corresponding to the
horizon size at matter-radiation equality observed at a distance $\chi_{\ast}$.
The combination $\As \ell_{\rm eq}$ determines the mean-squared deflection
$\langle d^2 \rangle$, while $\ell_{\rm eq}$ controls the shape of
$C_\ell^{\phi\phi}$. For the parameter ranges of interest,
\begin{equation}
\delta C_\ell^{\phi\phi}/C_\ell^{\phi\phi} = \delta \As / \As + (n_\ell +1)
 \delta \ell_{\rm eq} / \ell_{\rm eq} ,
\label{eq:clppparams}
\end{equation}
where $n_\ell$ arises (mostly) from the strong wavenumber dependence of the transfer
function for the gravitational potential, with $n_{\ell} \approx 1.5$ around
$\ell = 200$.

In joint fits to \planckTT+\lensing, the main parameter changes from \planckTT\
alone are a $2.6\,\%$ reduction in the best-fit $\As$, with an accompanying
reduction in the best-fit $\tau$, to $0.067$ (around $0.6\,\sigma$; see
Sect.~\ref{subsec:tau}). There is also a $0.7\,\%$ reduction in
$\ell_{\rm eq}$, achieved at fixed $\thetastar$ by reducing $\omm$. These
combine to reduce $C_\ell^{\phi\phi}$ by approximately $4\,\%$ at $\ell = 200$,
consistent with Eq.~(\ref{eq:clppparams}). The difference between the theory
lensing spectrum at the best-fit parameters in the \planckTT\ and
\planckTT+\lensing\ fits are shown by the dashed blue lines in
Fig.~\ref{fig:CMBlensing_LCDM}. In the joint fit, the $\chi^2$ for the lensing
bandpowers improves by 6, while the $\chi^2$ for the \planckTT\ data degrades
by only $1.2$ ($2.8$ for the high-$\ell$ $TT$ data and $-1.6$ for the low-$\ell$
$TEB$ data).

The lower values of $\As$ and $\omm$ in the joint fit give a $2\,\%$ reduction
in $\sigma_8$, with
\be
\sigma_8 = 0.815\pm 0.009\quad\onesig{\planckTTlensing},
\ee
as shown in Fig.~\ref{fig:sigma8-omegam-planck}. The decrease in matter density
leads to a corresponding decrease in $\Omm$, and at fixed $\thetastar$
(approximately $\propto \Omm h^3$) a $0.5\,\sigma$ increase in $H_0$, giving
\twoonesig{H_0 &= (67.8 \pm 0.9) \, \Hunit}{\Omm &= 0.308 \pm 0.012}{\planckTTlensing. \label{lensing_H0}}
Joint \planck+lensing constraints on other parameters of the base \LCDM\ cosmology are given in
Table.~\ref{LCDMcompare}.

\citet{planck2014-a17} discusses the effect on parameters of extending the
lensing multipole range in joint fits with $\planckTT$. In the base \lcdm\
model, using the full multipole range $8 \leq \ell \leq 2048$, the parameter
combination $\sigma_8 \Omm^{1/4} \approx (\As \ell_{\rm eq}^{2.5})^{1/2}$
(which is well determined by the lensing measurements) is pulled around
$1\,\sigma$ lower that its value using the conservative lensing range, with a
negligible change in the uncertainty. Around half of this shift comes from the
$3.6\,\sigma$ outlying bandpower ($638 \leq \ell \leq 762$). In massive
neutrino models, the total mass is similarly pulled higher by around
$1\,\sigma$ when using the full lensing multipole range.

\subsubsection{Detection of lensing in the CMB power spectra}
\label{subsec:Alens}

The smoothing effect of lensing on the acoustic peaks and troughs of the $TT$
power spectrum is detected at high significance in the \planck\ data.
Following {\paramsI} (see also~\citealt{2008PhRvD..77l3531C}), we
introduce a parameter $\Alens$, which scales the theory
$C_\ell^{\phi\phi}$ power spectrum at each point in parameter space, and which
is used to lens the CMB spectra.\footnote{We emphasize the difference
  between the phenomenological parameters $\Alens$ and $\Aphiphi$
  (introduced earlier). The amplitude $\Alens$ multiplies
  $C_\ell^{\phi\phi}$ when calculating both the lensed CMB theory spectra and
  the lensing likelihood, while $\Aphiphi$ affects only the lensing
  likelihood by scaling the theory $C_\ell^{\phi\phi}$ when comparing with the
  power spectrum of the reconstructed lensing potential $\phi$.}
The expected value for base \LCDM\
is $\Alens = 1$. The results of such an analysis for models with variable
$\Alens$ is shown in Fig.~\ref{fig:alens}. The marginalized constraint on
$\Alens$ is
\begin{equation}
\Alens= 1.22 \pm 0.10 \quad \onesig{\planckTT} \, .
\end{equation}
This is very similar to the result from the 2013 \planck\ data reported in
{\paramsI}. The persistent preference for $\Alens > 1$ is discussed in detail
there. For the 2015 data, we find that $\Delta \chi^2 = -6.4$ between the
best-fitting \lcdm$+\Alens$ model and the best-fitting base \lcdm\ model.
There is roughly equal preference for high $\Alens$ from intermediate and high
multipoles (i.e., the \plik\ likelihood; $\Delta \chi^2 = -2.6$) and from the
low-$\ell$ likelihood ($\Delta \chi^2 = -3.1$), with a further small change
coming from the priors.

\begin{figure}
\centering
\includegraphics[width=88mm,angle=0]{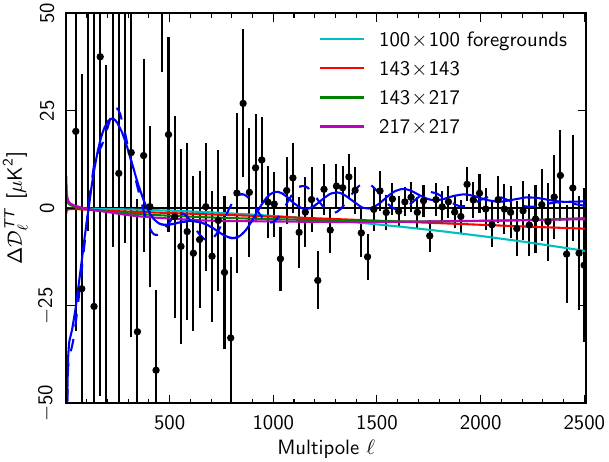}
\caption{Changes in the CMB $TT$ spectrum and foreground spectra, between the
best-fitting $\Alens$ model and the best-fitting base \lcdm\ model to the
\planckTT\ data.  The solid blue line shows the difference between the $\Alens$
model and \lcdm\, while the dashed line has the the same values of the
other cosmological parameters, but with $\Alens$ set to unity, to highlight
the changes in the spectrum arising from differences in the other
parameters.  Also shown are the changes in the best-fitting foreground
contributions to the four frequency cross-spectra between the $\Alens$ model
and the \lcdm\ model.
The data points (with $\pm 1\,\sigma$ errors) are the differences between the
high-$\ell$ maximum-likelihood frequency-averaged CMB spectrum and the
best-fitting \lcdm\ model to the \planckTT\ data (as in Fig.~\ref{pgTT_final}).
Note that the changes in the CMB spectrum and the foregrounds should be added
when comparing to the residuals in the data points.
}
\label{fig:Dls_base_Alens}
\end{figure}

Increases in $\Alens$ are accompanied by changes in all other parameters, with
the general effect being to reduce the predicted CMB power on large scales,
and in the region of the second acoustic peak, and to increase CMB power on small
scales (see Fig.~\ref{fig:Dls_base_Alens}). A reduction in the high-$\ell$
foreground power compensates the CMB increase on small scales. Specifically,
$\ns$ is increased by $1\,\%$ relative to the best-fitting base model and
$\As$ is reduced by $4\,\%$, both of which lower the large-scale power to
provide a better fit to the measured spectra around $\ell=20$ (see
Fig.~\ref{pgTT_final}).
The densities $\omb$ and $\omc$ respond to the change in $\ns$, following the
usual \lcdm\ acoustic degeneracy, and $\As e^{-2\tau}$ falls by $1\,\%$,
attempting to reduce power in the damping tail due to the increase in $\ns$ and
reduction in the diffusion angle $\theta_{\rm D}$ (which follows from the
reduction in $\omm$). The changes in $\As$ and $\As e^{-2\tau}$ lead to a
reduction in $\tau$ from $0.078$ to $0.060$. With these cosmological
parameters, the lensing power is lower than in the base model, which
additionally increases the CMB power in the acoustic peaks and reduces it in
the troughs. This provides a poor fit to the measured spectra around the
fourth and fifth peaks, but this can be mitigated by increasing $\Alens$
to give more smoothing from lensing than in the base model.
However, $\Alens$ further increases power in the damping tail, but this is
partly offset by reduction of the power in the high-$\ell$ foregrounds.

The trends in the $TT$ spectrum that favour high $\Alens$ have a similar pull
on parameters such as curvature (Sect.~\ref{subsubsec:curv}) and the dark
energy equation of state (Sect.~\ref{sec:dark_energy}) in extended models.
These parameters affect the late-time geometry and clustering and so alter the
lensing power, but their effect on the primary CMB fluctuations is degenerate
with changes in the Hubble constant (to preserve $\thetastar$). The same
parameter changes as those in $\Alens$ models are found in these extended
models, but with, for example, the increase in $\Alens$ replaced by a reduction
in $\Omk$. Adding external data, however, such as the \planck\ lensing data or
BAO (Sect.~\ref{sec:BAO}), pull these extended models back to base \lcdm.

Finally, we note that lensing is also detected at lower significance
in the polarization power spectra (see Fig.~\ref{fig:alens}):
\begin{subequations}
\begin{align}
\Alens &= 0.98^{+0.21}_{-0.24} \quad \onesig{\planck\ TE+\lowEB} \, ; \\
\Alens &= 1.54^{+0.28}_{-0.33} \quad \onesig{\planck\ EE+\lowEB} \, .
\label{eq:Alens}
\end{align}
\end{subequations}
These results use only polarization at low multipoles, i.e., with no temperature
data at multipoles $\ell < 30$.
 These are the first detections of lensing in the CMB polarization
spectra, and reach almost $5\,\sigma$ in $TE$. We caution the reader that the
$\Alens$ constraints from $EE$ and low-$\ell$ polarization are rather unstable
between high-$\ell$ likelihoods because of differences in the treatment of
the polarization data (see Fig.~\ref{fig:alens},
which compares constraints from the \plik\ and \camspec\ polarization
likelihoods).
The result of replacing \plik\ with the \camspec\ likelihood is $\Alens = 1.19^{+0.20}_{-0.24}$,
i.e., around $1\,\sigma$ lower than the result from \plik\ reported in
Eq.~\eqref{eq:Alens}.  If we additionally include the low-$\ell$ temperature
data, $\Alens$ from $TE$ increases:
\begin{equation}
\Alens= 1.13 \pm 0.2 \quad \onesig{\planck\ TE+lowT,P} \, .
\end{equation}
The pull to higher $\Alens$ in this case is due to the reduction in $TT$ power
in these models on large scales (as discussed above).

\subsection{Baryon acoustic oscillations}
\label{sec:BAO}

Baryon acoustic oscillation (BAO) measurements are geometric
and largely unaffected by uncertainties in the nonlinear
evolution of the matter density field and additional systematic errors
that may affect other types of astrophysical data. As in {\paramsI}, we
therefore use BAO as a primary astrophysical data set to break parameter
degeneracies from CMB measurements.

\begin{figure}
\begin{center}
\includegraphics[width=90mm,angle=0]{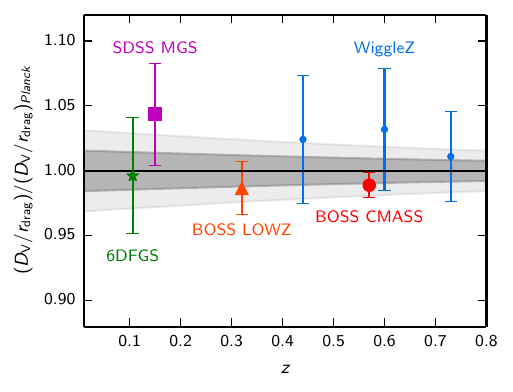}
\end{center}
\caption {Acoustic-scale distance ratio $\DVBAO(z)/\rdrag$ in the base \LCDM\
 model divided by the mean distance ratio from \planckTT+lensing.
 The points with $1\,\sigma$ errors are as follows: green star
 \citep[6dFGS, ][]{Beutler:11}; square \citep[SDSS MGS, ][]{Ross:14};
 red triangle and large circle
 \citep[BOSS ``LOWZ'' and CMASS surveys,][]{Anderson:14};
 and small blue circles \citep[WiggleZ, as analysed by][]{Kazin:14}.
 The grey bands show the $68\,\%$ and $95\,\%$ confidence ranges allowed by
 \planckTT+lensing.}
\label{BAO}
\end{figure}

Figure~\ref{BAO} shows an updated version of figure~15 from {\paramsI}.
The plot shows the acoustic-scale distance ratio $\DVBAO(z)/\rdrag$ measured
from a number of large-scale structure surveys with effective redshift $z$,
divided by the mean acoustic-scale ratio in the base \LCDM\ cosmology using
\planckTT+lensing. Here
$\rdrag$ is the comoving sound horizon at the end of the baryon drag epoch and
$\DVBAO$ is a combination of the angular diameter distance $\DA(z)$ and Hubble
parameter $H(z)$,
\begin{equation}
 \DVBAO(z) = \left [ (1+z)^2 \DA^2(z) {cz \over H(z)} \right ]^{1/3}.  \label{BAO1}
\end{equation}
The grey bands in the figure show the $\pm 1\,\sigma$ and $\pm 2\, \sigma$ ranges allowed by \Planck\
in the base \LCDM\ cosmology.

The changes to the data points compared to figure~15 of {\paramsI} are as
follows. We have replaced the SDSS DR7 measurements of \citet{Percival:2010}
with the recent analysis of the SDSS Main Galaxy Sample (MGS) of \citet{Ross:14}
at $z_{\rm eff}=0.15$, and by the \citet{Anderson:14} analysis of the Baryon
Oscillation Spectroscopic Survey (BOSS) ``LOWZ'' sample at $z_{\rm eff} = 0.32$.
Both of these analyses use peculiar velocity field reconstructions to
sharpen the BAO feature and reduce the errors on $\DVBAO/\rdrag$. The
blue points in Fig.~\ref{BAO} show a reanalysis of the WiggleZ redshift survey
by \citet{Kazin:14} that applyies peculiar velocity reconstructions.
These reconstructions cause small shifts in $\DVBAO/\rdrag$ compared to the
unreconstructed WiggleZ results of \citet{Blake:11} and lead to
reductions in the errors on the distance measurements at $z_{\rm eff} = 0.44$
and $z_{\rm eff} = 0.73$. The point labelled BOSS CMASS at $z_{\rm eff} = 0.57$
shows $\DVBAO/\rdrag$ from the analysis of \citet{Anderson:14}, updating the
BOSS-DR9 analysis of \citet{Anderson:12} used in {\paramsI}.

\begin{figure}
\begin{center}
\includegraphics[width=\hsize,angle=0]{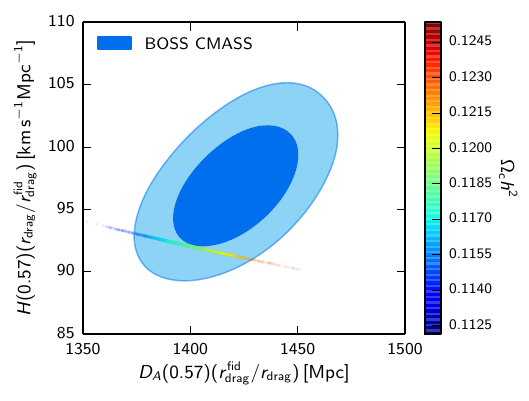}
\end{center}
\caption {68\,\% and 95\,\% constraints on the angular diameter distance
$\DA(z=0.57)$ and Hubble parameter $H(z=0.57)$ from the \citet{Anderson:14}
analysis of the BOSS CMASS-DR11 sample. The fiducial sound horizon adopted
by \citet{Anderson:14} is $\rdrag^{\rm fid} = 149.28\,{\rm Mpc}$.
Samples from the \planckTT+lensing chains are plotted coloured by their value of
$\Omc h^2$, showing consistency of the data, but also that the BAO measurement
can tighten the \Planck\ constraints on the matter density.}
\label{BOSSprob}
\end{figure}

In fact, the \citet{Anderson:14} analysis solves jointly for the positions of
the BAO feature in both the line-of-sight and transverse directions (the
distortion in the transverse direction caused by the background cosmology
is sometimes called the Alcock-Paczynski effect, \citealt{Alcock:79}), leading
to joint constraints on the angular diameter distance $\DA(z_{\rm eff})$
and the Hubble parameter $H(z_{\rm eff})$.
These constraints, using the tabulated likelihood included in the \COSMOMC\
module,\footnote{\url{http://www.sdss3.org/science/boss_publications.php}}
are plotted in Fig.~\ref{BOSSprob}.  Samples from the \planckTT+lensing
chains are shown for comparison, coloured by the value of $\Omc h^2$.
The length of the degeneracy line is set
by the allowed variation in $H_0$ (or equivalently $\Omm h^2$).
In the \planckTT+lensing \LCDM\ analysis the line is defined approximately by
\be
 \frac{\DA(0.57)/\rdrag}{ 9.384} \left(\frac{H(0.57)\rdrag/c}{0.4582 }
 \right)^{1.7} = 1.0000 \pm 0.0004,
\ee
which just grazes the BOSS CMASS 68\,\% error ellipse plotted in Fig.~\ref{BOSSprob}.
Evidently, the \Planck\ base \LCDM\ parameters are in good agreement with both the
isotropized $\DVBAO$ BAO measurements plotted in Fig.~\ref{BAO}, and with the anisotropic
constraints plotted in Fig.~\ref{BOSSprob}.

In this paper, we use the 6dFGS, SDSS-MGS, and BOSS-LOWZ BAO
measurements of $\DVBAO/\rdrag$ \citep{Beutler:11, Ross:14, Anderson:14} and
the CMASS-DR11 anisotropic BAO measurements of
\citet{Anderson:14}. Since the WiggleZ volume partially overlaps that of the
BOSS-CMASS sample, and the correlations have not been quantified, we do not
use the WiggleZ results in this paper. It is clear from Fig.~\ref{BAO} that
the combined BAO likelihood is dominated by the two BOSS measurements.

In the base \LCDM\ model, the \Planck\ data constrain
the Hubble constant $H_0$ and matter density $\Omm$ to high precision:
\twoonesig{H_0 &= (67.3 \pm 1.0) \, \Hunit}{\Omm &= 0.315 \pm 0.013}{\planckTT. \label{BAO_H01}}
With the addition of the BAO measurements, these constraints are strengthened
significantly to
\twoonesig{H_0 &= (67.6 \pm 0.6) \, \Hunit}{\Omm &= 0.310 \pm 0.008}{\planckTTBAO. \label{BAO_H02}}
These numbers are consistent with the \planck+lensing constraints of
Eq.~\eqref{lensing_H0}.
Section~\ref{sec:hubble} discusses the consistency of these estimates of $H_0$
with direct measurements.

Although low-redshift BAO measurements are in good agreement with \Planck\
for the base \LCDM\ cosmology, this may not be true at high redshifts.
Recently, BAO features have been measured in the flux-correlation function of
the Ly$\alpha$ forest of BOSS quasars \citep{Delubac:2014}
and in the cross-correlation of the Ly$\alpha$ forest with quasars
\citep{Font-Ribera:2014}. These observations give measurements of
$c/(H(z)\rdrag)$ and $\DA(z)/\rdrag$ (with somewhat lower precision) at
$z=2.34$ and $z=2.36$, respectively.  For example, from table~II of
\cite{Aubourg:2014} the two Ly$\alpha$ BAO measurements combined
give $c/(H(2.34)\rdrag) = 9.14 \pm 0.20$, compared to the predictions of the
base \Planck\ \LCDM\ cosmology of
$8.586\pm 0.021$, which are discrepant at the $2.7\,\sigma$ level. At present,
it is not clear whether this discrepancy is caused by systematics in the
Ly$\alpha$ BAO measurements (which are more complex and less mature than
galaxy BAO measurements) or an indicator of new physics. As
\cite{Aubourg:2014} discuss, it is difficult to find a physical
explanation for the Ly$\alpha$ BAO results without disrupting the consistency
with the much more precise galaxy BAO measurements at lower redshifts.

\subsection{Type Ia supernovae}
\label{sec:SNe}

Type Ia supernovae (SNe) are powerful probes of cosmology
\citep{Riess:1998,Perlmutter:1999} and particularly of the equation of
state of dark energy. In {\paramsI}, we used two samples of type Ia
SNe, the ``SNLS'' compilation \citep{Conley:2011} and the ``Union2.1''
compilation \citep{Suzuki:2012}. The SNLS sample was found to be in
mild tension, at about the $2\,\sigma$ level, with the 2013
\Planck\ base \LCDM\ cosmology favouring a value of $\Omm \approx
0.23$ compared to the \Planck\ value of $\Omm = 0.315 \pm
0.017$. Another consequence of this tension showed up in extensions to
the base \LCDM\ model, where the combination of \Planck\ and the SNLS
sample showed $2\,\sigma$ evidence for a ``phantom'' ($w<-1$) dark
energy equation of state.

Following the submission of {\paramsI}, \citet{Betoule:2013} reported
the results of an extensive campaign to improve the relative
photometric calibrations between the SNLS and SDSS supernova
surveys. The ``Joint Light-curve Analysis'' (JLA) sample, used in this
paper, is constructed from the SNLS and SDSS SNe data, together with
several samples of low redshift SNe.\footnote{A \COSMOMC\ likelihood
model for the JLA sample is available at
  \url{http://supernovae.in2p3.fr/sdss_snls_jla/ReadMe.html}.  The
  latest version in \COSMOMC\ includes numerical integration over the
  nuisance parameters for use when calculating joint constraints using
  importance sampling; this can give different $\chi^2$ values compared to
  parameter best fits.}. Cosmological constraints from the JLA sample
are discussed by \citet{Betoule:2014frx} and residual biases associated
with the photometry and light curve fitting are assessed by
\citet{Mosher:2014}. For the base \LCDM\ cosmology,
\citet{Betoule:2014frx} find $\Omm=0.295 \pm 0.034$, consistent with the
2013 and 2015 \Planck\ values for base \LCDM. This relieves the
tension between the SNLS and \Planck\ data reported in
{\paramsI}. Given the consistency between \Planck\ and the JLA sample
for base \LCDM, one can anticipate that the combination of these two
data sets will constrain the dark energy equation of state to be close
to $w=-1$ (see Sect. \ref{sec:dark_energy}).

Since the submission of {\paramsI}, first results from a sample of Type Ia SNe
discovered with the Pan-STARRS survey have been reported by \citet{Rest:2014}
and \citet{Scolnic:2014}. The Pan-STARRS sample is still relatively small
(consisting of $146$ spectroscopically confirmed Type Ia SNe) and is not
used in this paper.

\subsection{The Hubble constant}
\label{sec:hubble}

CMB experiments provide indirect and highly model-dependent estimates of the
Hubble constant.  It is therefore important to compare CMB estimates with
direct estimates of $H_0$, since any significant evidence of a tension could
indicate the need for new physics. In {\paramsI}, we used the
\citet[hereafter R11]{Riess:11} {\it Hubble\/} Space Telescope
(HST) Cepheid+SNe based estimate of
$H_0 = (73.8 \pm 2.4) \Hunit$ as a supplementary ``$H_0$-prior.''
This value was in tension at about the $2.5\,\sigma$ level with the
2013 \Planck\ base \LCDM\ value of $H_0$.

For the base \LCDM\ model, CMB and BAO experiments consistently find a value
of $H_0$ lower than the R11 value.
For example, the 9-year WMAP data \citep{bennett2012,hinshaw2012}
give:\footnote{These numbers are taken from our parameter grid, which includes
a neutrino mass of $0.06\,{\rm eV}$ and the same updated BAO compilation as
Eq.~\eqref{BAO_H02} (see Sect.~\ref{sec:BAO}).}
\beglet
\begin{eqnarray}
H_0 &=& (69.7 \pm 2.1) \, \Hunit, \qquad \ \ \datalabel{\rm WMAP9}, \\
H_0 &=& (68.0 \pm 0.7) \, \Hunit, \quad \ \ \datalabel{\rm WMAP9\dataplus\BAO}. \label{WMAPH0}
\end{eqnarray}
\endlet 
These numbers can be compared with the \Planck\ 2015 values
given in Eqs.~\eqref{BAO_H01} and (\ref{BAO_H02}).  The WMAP
constraints are driven towards the \Planck\ values by the addition
of the BAO data and so there is persuasive evidence for a low $H_0$ in
the base \LCDM\ cosmology {\it independently of the high-multipole CMB
results from \Planck}.  The 2015 \Planck\ TT+lowP value is entirely
consistent with the 2013 \Planck\ value and so the tension with the
R11 $H_0$ determination remains at about $2.4\,\sigma$.

The tight constraint on $H_0$ in Eq.~\eqref{WMAPH0} is an example of an
``inverse distance ladder,''  where the CMB primarily constrains the
sound horizon within a given cosmology, providing an absolute
calibration of the BAO acoustic-scale \citep[e.g.,][see also
{\paramsI}]{Percival:2010, Cuesta:2014, Aubourg:2014}.  In fact, in a
recent paper \citet{Aubourg:2014} use the 2013 \Planck\ constraints on
$\rs$ in combination with BAO and the JLA SNe data to find $H_0 = (67.3
\pm 1.1)\,\Hunit$, in excellent agreement with the 2015 \Planck\ value
for base \LCDM\ given in Eq.~\eqref{BAO_H01}, {\it which is based on the
\Planck\ temperature power spectrum}. Note that by adding SNe
data, the \citet{Aubourg:2014} estimate of $H_0$ is insensitive to
spatial curvature and to late time variations of the dark energy
equation of state. Evidently, there are a number of lines of evidence
that point to a lower value of $H_0$ than the direct determination of
R11.

The R11 Cepheid data have been reanalysed by
\citet[hereafter E14]{Efstathiou:14} using the revised geometric maser distance
to NGC 4258 of \citet{Humphreys:13}. Using NGC 4258 as a distance anchor,
E14 finds
\begin{equation}
H_0 = (70.6 \pm 3.3) \, \Hunit, \qquad {\rm NGC \ 4258}, \label{H0prior1}
\end{equation}
which is within $1\,\sigma$ of the \Planck\ TT estimate given in
Eq.~\eqref{BAO_H01}. In this paper we use Eq.~\eqref{H0prior1} as a
``conservative'' $H_0$ prior.

R11 also use Large Magellanic Cloud Cepheids and a small sample of Milky Way
Cepheids with parallax distances as alternative distance anchors to NGC4258.
The R11 $H_0$ prior used in {\paramsI} combines all
three distance anchors. Combining the LMC and MW distance anchors, E14 finds
\begin{equation}
H_0 = (73.9 \pm 2.7) \, \Hunit, \qquad {\rm LMC+MW}, \label{H0prior2}
\end{equation}
under the assumption that there is no metallicity variation of the
Cepheid period-luminosity relation.  This is discrepant with
Eq.~\eqref{BAO_H01} at about the $2.2\,\sigma$ level.  However,
neither the central value nor the error in Eq.~\eqref{H0prior2} is
reliable. The MW Cepheid sample is small and dominated by short period
($<10 \,{\rm day}$) objects.  The MW Cepheid sample therefore
 has very little overlap
with the period range of  SNe host galaxy Cepheids observed with
HST. As a result, the MW solutions for $H_0$ are unstable (see
Appendix A of E14). The LMC solution is sensitive to the metallicity
dependence of the Cepheid period-luminosity relation which is poorly
constrained by the R11 data. Furthermore, the estimate in Eq.~\eqref{H0prior1}
is based on a differential measurement, comparing HST
photometry of Cepheids in NGC 4258 with those in SNe host
galaxies. It is therefore less prone to
photometric systematics, such as crowding corrections, than is the
LMC+MW estimate of Eq.~\eqref{H0prior2}.  It is for these reasons that we
have adopted the prior of Eq.~\eqref{H0prior1} in preference
to using the LMC and MW distance anchors.\footnote{As this paper was
nearing completion, results from the Nearby Supernova Factory have been
presented that indicate a correlation between the peak brightness of Type Ia
SNe and the local star-formation rate \citep{Rigault:2014}. These
authors argue that this correlation introduces a systematic bias of around
$1.8 \Hunit$ in the SNe/Cepheid distance 
scale measurement of $H_0$ . For example,
according to these authors, the estimate of Eq.~(\ref{H0prior1}) should be
lowered to $H_0 = (68.8 \pm 3.3) \Hunit$, a downward shift of approximately
$0.5\,\sigma$. Clearly, further work needs to be done to assess the importance
of such a bias on the distance scale. It is ignored in the rest of this paper.}

Direct measurements of the Hubble constant have a long and sometimes
contentious history \citep[see, e.g.,][]{Tammann:08}. The controversy
continues to this day and in the literature one can find ``high'' values,
e.g., $H_0 = (74.3 \pm 2.6) \Hunit$ \citep{Freedman:12}, and ``low'' values,
e.g., $H_0 = (63.7 \pm 2.3) \Hunit$ \citep{Tammann:13}.
The key point that we wish to make is that
the \planck-only estimates of Eqs.~\eqref{lensing_H0} and \eqref{BAO_H01},  and
the  \Planck+BAO estimate of Eq.~\eqref{BAO_H02} all have small errors and
are consistent. If a persuasive case can be made that a direct
measurement of $H_0$ conflicts with these estimates, then this will
be strong evidence for additional physics beyond the base \LCDM\ model.

Finally, we note that in a recent analysis \citet{Bennett:2014} derive a
``concordance'' value of $H_0 = (69.6 \pm 0.7)\,\Hunit$ for base \LCDM\ by
combining WMAP9+SPT+ACT+BAO with a slightly revised version of the R11
$H_0$ value, $(73.0 \pm 2.4) \Hunit$. The \citet{Bennett:2014} central
value for $H_0$ differs from the \Planck\ value of Eq.~\eqref{BAO_H02} by
nearly $3\,\%$ (or $2.5\,\sigma$). The reason for this difference is that the
\Planck\ data are in tension with the \citet{Story:2013} SPT data
(as discussed in Appendix~B of {\paramsI};
note that the tension is increased with the \Planck\ full mission data) and
with the revised R11 $H_0$ determination. Both tensions drive the
\citet{Bennett:2014} value of $H_0$ away from the \Planck\ solution.

\subsection{Additional data}
\label{sec:additional_data}

\subsubsection{Redshift space distortions}
\label{subsec:RSD}

Transverse versus line-of-sight anisotropies in the redshift-space clustering
of galaxies induced by peculiar motions can, potentially, provide a
powerful way of constraining the growth rate of structure
\citep[e.g.,][]{PercivalW:2009}.  A number of studies of redshift-space
distortions (RSD) have been conducted to measure the parameter combination
$f\sigma_8(z)$, where for models with scale-independent growth
\begin{equation}
 f(z) = {d\ln D \over d\ln a}, \label{GR1}
\end{equation}
and $D$ is the linear growth rate of matter fluctuations.  Notice that the
parameter combination $f\sigma_8$ is insensitive to differences between the
clustering of galaxies and dark matter, i.e., to galaxy bias \citep{Song:09}.
In the base \LCDM\ cosmology, the
growth factor $f(z)$ is well approximated as $f(z) = \Omm(z)^{0.545}$.
More directly, in linear theory
the quadrupole of the redshift-space clustering anisotropy actually probes the
density-velocity correlation power spectrum, and we therefore define
\begin{equation}
 f\sigma_8(z) \equiv
  \frac{\left[\sigma_8^{(vd)}(z)\right]^2}{\sigma_8^{(dd)}(z)},
\end{equation}
as an approximate proxy for the quantity actually being measured.
Here $\sigma_8^{(vd)}$ measures the smoothed density-velocity correlation and
is defined analogously to $\sigma_8\equiv \sigma_8^{(dd)}$, but using the
correlation power spectrum $P_{vd}(k)$, where
$v = -\grad\cdot \vec{v}_{\rm N}/H$ and $\vec{v}_{\rm N}$ is the
Newtonian-gauge (peculiar) velocity of the baryons and dark matter, and $d$
is the total matter density perturbation.
This definition assumes that the observed galaxies follow the flow of the cold
matter, not including massive neutrino velocity effects. For models close to
\LCDM, where the growth is nearly scale independent, it is equivalent
to defining $f\sigma_8$ in terms of the growth of the baryon+CDM density
perturbations (excluding neutrinos).

\begin{figure}
\begin{center}
\includegraphics[width=90mm,angle=0]{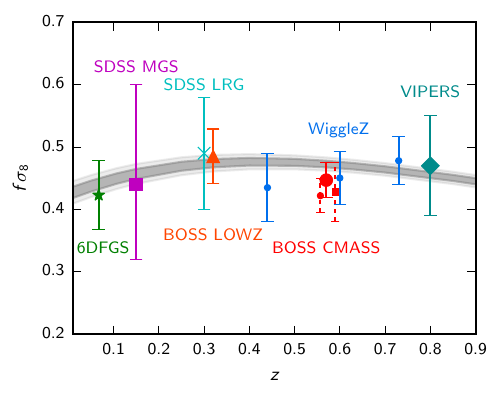}
\end{center}
\caption {Constraints on the growth rate of fluctuations from various redshift
surveys in the base \LCDM\ model: green star
\citep[6dFGRS, ][]{Beutler:12}; purple square
\citep[SDSS MGS, ][]{Howlett:14}; cyan cross \citep[SDSS LRG, ][]{Oka:14};
red triangle \citep[BOSS LOWZ survey,][]{Gil-Marin:2015sqa};
large red circle \citep[BOSS CMASS, as analysed by][]{Samushia:14};
blue circles \citep[WiggleZ,][]{Blake:12};
and green diamond \citep[VIPERS,][]{delaTorre:13}.  The points with dashed
red error bars correspond to alternative analyses of BOSS CMASS from
\citet[][small circle, offset for clarity]{Beutler:14} and
\citet[][small square]{Chuang:14}. Of the BOSS CMASS points, two are based on
the same DR11 data set~\citep{Samushia:14,Beutler:14}, while the third
is based on the more recent DR12~\citep{Chuang:14}, and are therefore not
independent.
The grey bands show the range allowed by \planckTT+lensing in the base
\LCDM\ model.  Where available (for SDSS MGS and BOSS), we have plotted
conditional constraints on $f\sigma_8$
assuming a \planck\ \LCDM\ background cosmology. The WiggleZ points are plotted conditional on the mean \Planck\
cosmology prediction for $F_{\rm AP}$ (evaluated using the covariance between
$f\sigma_8$ and $F_{\rm AP}$ given in~\citealt{Blake:12}).
The 6dFGS point is at sufficiently low redshift that it is insensitive to the
cosmology. }
\label{fsigma8}
\end{figure}

The use of RSD as a measure of the growth of structure is still under
active development and is considerably more difficult than measuring
the positions of BAO features. Firstly, adopting the wrong fiducial
cosmology can induce an anisotropy in the clustering of galaxies, via the
Alcock-Paczynski (AP) effect, which is strongly degenerate
with the anisotropy induced by peculiar motions. Secondly, much of the
RSD signal currently comes from scales where nonlinear effects and
galaxy bias are significant and must be accurately modelled in
order to relate the density and velocity fields \citep[see,
e.g., the discussions in][]{Bianchi:12,Okumura:12,Reid:14,White:14}.

Current constraints,\footnote{The constraint 
of \citet{Chuang:14} plotted in the original version of this paper was
subsequently shown to be in error. We therefore now show updated
BOSS data points for DR12 from \citet[][for CMASS]{Chuang:14} and
\citet[][for LOWZ]{Gil-Marin:2015sqa}.}
assuming a \planck\ base \LCDM\ model, are shown in
Fig.~\ref{fsigma8}.  Neglecting the AP effect can lead to biased measurements
of $f\sigma_8$ if the assumed cosmology differs, and to significant
underestimation of
the errors \citep{Howlett:14}.  The analyses summarized in Fig.~\ref{fsigma8}
solve simultaneously for RSD and the AP effect, except for the
6dFGS point (which is insensitive to cosmology) and the VIPERS point (which has a large error).
The grey bands show the range allowed by \planckTT+lensing in the base \lcdm\
model, and are consistent with the RSD data.
The tightest constraints on $f\sigma_8$ in this figure come from the BOSS
CMASS-DR11 analyses of \citet{Beutler:14} and \citet{Samushia:14}. The
\citet{Beutler:14} analysis is performed in Fourier space and shows a
small bias in $f\sigma_8$ compared to numerical simulations when
fitting over the wavenumber range 0.01--$0.20\, h {\rm Mpc}^{-1}$. The
\citet{Samushia:14} analysis is performed in configuration space and shows
no evidence of biases when compared to numerical simulations.  
The updated DR12 CMASS result from \citet{Chuang:14} marginalizes
over a polynomial model for systematic errors in the correlation function
monopole, and is consistent with these and the \planck\ constraints, with a
somewhat larger error bar.

\begin{figure}
\begin{center}
\includegraphics[width=90mm,angle=0]{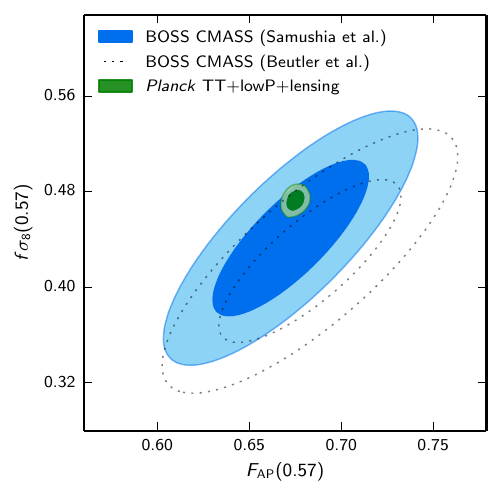}
\end{center}
\caption {68\,\% and 95\,\% contours in the $f\sigma_8$--$F_{\rm AP}$ plane
(marginalizing over $D_{\rm v}/r_{\rm s}$)
for the CMASS-DR11 sample as analysed by \citet[solid, our defult]{Samushia:14},
and \citet[dotted]{Beutler:14}. The green contours
show the constraint from \planckTT+lensing in the base \LCDM\ model.}
\label{Samushia}
\end{figure}

The \citet{Samushia:14} results are expressed as a $3\times3$ covariance
matrix for the three parameters $\DVBAO/\rdrag$, $F_{\rm AP}$ and $f\sigma_8$,
evaluated at an effective redshift of $z_{\rm eff}=0.57$, where $F_{\rm AP}$ is
the ``Alcock-Paczynski'' parameter
\begin{equation}
F_{\rm AP}(z) = (1+z) D_{\rm A} {H(z) \over c}.
\end{equation}
The principal degeneracy is between $f\sigma_8$ and $F_{\rm AP}$ and is
illustrated in Fig.~\ref{Samushia}, compared to the constraint from
\planckTT+lensing for the base \LCDM\ cosmology. The \Planck\ results sit
slightly high but overlap the $68\,\%$ contour from \citet{Samushia:14}.
The \Planck\ result lies about $1.5\,\sigma$ higher than the \citet{Beutler:14}
analysis of the BOSS CMASS sample.

RSD measurements are not used in combination with \Planck\ in this paper.
However, in the companion paper
exploring dark energy and modified gravity \citep{planck2014-a16}, the
RSD/BAO measurements of \citet{Samushia:14} are used together with \Planck.
Where this is done, we {\it exclude\/} the \citet{Anderson:14} BOSS-CMASS
results from the BAO likelihood. Since \citet{Samushia:14} do not apply a
density field reconstruction in their analysis,
the BAO constraints from BOSS-CMASS are then slightly weaker, though consistent,
with those of \citet{Anderson:14}.

\subsubsection{Weak gravitational lensing}
\label{subsec:WL}

Weak gravitational lensing offers a potentially powerful technique for
measuring the amplitude of the matter fluctuation spectrum at low redshifts.
Currently, the largest weak lensing data set is provided by the \CFHTLENS\
survey \citep{Heymans:2012, Erben:2013}. The first science results
from this survey appeared shortly before the completion of {\paramsI} and it
was not possible to do much more than offer a cursory comparison with the
\Planck\ 2013 results. As reported in {\paramsI}, at face value the results
from \CFHTLENS\ appeared to be in tension with the \Planck\ 2013 base \LCDM\
cosmology at about the 2--$3\,\sigma$ level. Since neither the \CFHTLENS\
results nor
the 2015 \Planck\ results have changed significantly from those in {\paramsI},
it is worth discussing this discrepancy in more detail in this paper.

Weak lensing data can be analysed in various ways.  For example, one can
compute two correlation functions from the ellipticities of pairs of images
separated by angle $\theta$,which are related to the convergence power
spectrum $P^\kappa(\ell)$ of the survey at multipole $\ell$ via
\begin{equation}
\xi_{\pm}(\theta) = {1 \over 2 \pi} \int d\ell \ell P^\kappa(\ell)
 J_{\pm}(\ell \theta), \label{WL1}
\end{equation}
where the Bessel functions in \eqref{WL1} are $J_+ \equiv J_0$ and
$J_- \equiv J_4$ \citep[see, e.g.,][]{Bartelmann:2001}. Much of the information
from the \CFHTLENS\ survey correlation function analyses comes from wavenumbers
at which the matter power spectrum is strongly nonlinear, complicating any
direct comparison with \Planck.

This can be circumventing by performing a 3D spherical harmonic
analysis of the shear field, allowing one to impose lower limits on the
wavenumbers that contribute to a weak lensing likelihood. This has
been done by \citet{Kitching:2014}. Including only wavenumbers with
$k \le 1.5\,h{\rm Mpc}^{-1}$, \cite{Kitching:2014}
find constraints in the $\sigma_8$--$\Omm$ plane that are consistent with the
results from \Planck. However, by excluding modes with higher wavenumbers,
the lensing constraints are weakened. When they increase the wavenumber
cut-off to $k = 5\,h{\rm Mpc}^{-1}$ some tension with \Planck\ begins to emerge
(which these authors argue may be an indication of the effects of baryonic
feedback in suppressing the matter power spectrum at small scales).
The large-scale properties of \CFHTLENS\ therefore seem broadly consistent
with \Planck\ and it is only as \CFHTLENS\ probes higher wavenumbers,
particular in the 2D and tomographic correlation function analyses
\citep{Heymans:2013, Kilbinger:2013, Fu:2014,MacCrann:2014wfa}, that
apparently strong discrepancies with \Planck\ appear.

\begin{figure}
\begin{center}
\includegraphics[width=90mm,angle=0]{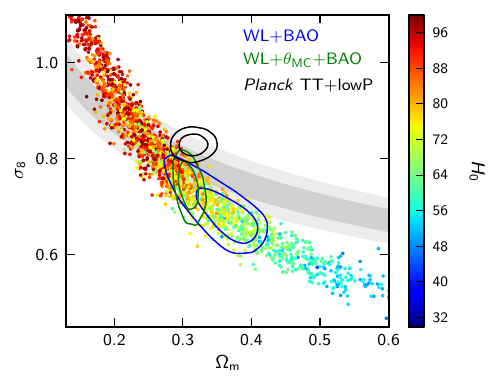}
\end{center}
\caption {Samples in the $\sigma_8$--$\Omm$ plane from the H13 \CFHTLENS\ data
(with angular cuts as discussed in the text), coloured by the value of the
Hubble parameter, compared to the joint constraints when the lensing data are
combined with BAO (blue), and BAO with the CMB acoustic scale parameter fixed
to $\thetaMC=1.0408$ (green).  For comparison, the \planckTT\ constraint
contours are shown in black. The grey bands show the constraint from \Planck\
CMB lensing.  We impose a weak prior on the primoridal
amplitude, $2<\ln(10^{10}A_{\rm s})<4$, which has some impact on the
distribution of \CFHTLENS-only samples.}
\label{WLonly}
\end{figure}

The situation is summarized in Fig.~\ref{WLonly}. The sample points show
parameter values in the $\sigma_8$--$\Omm$ plane for the \LCDM\ base model,
computed from the \citet[][hereafter H13]{Heymans:2013} tomographic
measurements of $\xi_{\pm}$.  These data consist of correlation function
measurements in six photometric redshift bins extending over the
redshift range 0.2--1.3. We use the blue galaxy sample, since H13
find that this sample shows no evidence for intrinsic galaxy
alignments (simplifying the comparison with theory) and we apply the
``conservative'' cuts of H13, intended to reduce sensitivity to the
nonlinear part of the power spectrum; these cuts eliminate measurements
with $\theta < 3^\prime$ for any redshift combination that involves the
lowest two redshift bins.  Here we have used the
\HALOFIT\ prescription of \cite{Takahashi:2012em} to model the
nonlinear power spectrum, but do not include any model of baryon feedback
or intrinsic alignments.  For the lensing-only constraint we also impose
additional priors in a similar way to the CMB lensing analysis
described in~\cite{planck2014-a17}, i.e., Gaussian priors
$\Omb h^2= 0.0223 \pm 0.0009$ and $\ns=0.96\pm 0.02$, where the exact values
(chosen to span reasonable ranges given CMB data) have little impact on the results. The sample
range shown also restricts the Hubble parameter to $0.2<h<1$; note that when
comparing with constraint contours, the location of the contours can change
significantly depending on the $H_0$ prior range assumed.
We also use a weak prior on the primoridal amplitude,
$2<\ln(10^{10}A_{\rm s})<4$,
which shows up the strong correlation between $\Omm$--$\sigma_8$--$H_0$ in
the region of parameter space relevant for comparison with \planck.
In Fig.~\ref{WLonly} we only show lensing contours after the samples have
been projected into the space allowed by the BAO data (blue contours), or also
additionally restricting to the reduced space where $\thetaMC$ is fixed to the
\Planck\ value, which is accurately measured.
The black contours show the constraints from \planckTT.

The lensing samples just overlap with \planck, and superficially one might
conclude that the two data sets are consistent. However, the weak lensing
constraints approximately define a 1D degeneracy in the
3D $\Omm$--$\sigma_8$--$H_0$ space, so consistency of the Hubble
parameter at each point in the projected space must also be considered
\citep[see appendix E1 of][]{planck2014-a17}.
Comparing the contours in Fig.~\ref{WLonly} (the regions
where the weak lensing constraints are consistent with BAO observations)
the \CFHTLENS\ data favour a lower value of $\sigma_8$ than the \Planck\
data (and much of the area of the blue contours also has higher $\Omm$).
However, even with the conservative angular cuts
applied by H13, the weak lensing constraints depend on the
nonlinear model of the power spectrum and on the possible
influence of baryonic feedback in reshaping the matter power spectrum at
small spatial scales \citep{Harnois-Deraps:2014sva,MacCrann:2014wfa}. The
importance of these effects can be reduced by imposing even more conservative
angular cuts on $\xi_{\pm}$, but of course, this weakens the statistical
power of the weak lensing data.  The \CFHTLENS\ data are not used in
combination with \Planck\ in this paper
(apart from specific cases in Sects.~\ref{sec:dark_energy}
and \ref{sec:neutrino_tension}) and, in any case, would have
little impact on most of the extended \LCDM\ constraints discussed in
Sect.~\ref{sec:grid}.  Weak lensing can, however, provide important constraints
on dark energy and modified gravity. The \CFHTLENS\ data are therefore
used in combination with \Planck\ in the companion paper \citep{planck2014-a16},
which explores several \HALOFIT\ prescriptions and the impact of
applying more conservative angular cuts to the H13 measurements.

\subsubsection{Planck cluster counts}

In 2013 we noted a possible tension between our primary CMB
constraints and those from the \Planck\ SZ cluster counts, with the
clusters preferring lower values of $\sigma_8$ in the base
$\Lambda$CDM model in some analyses \citep{planck2013-p15}.  The
comparison is interesting because the cluster counts directly measure
$\sigma_8$ at low redshift; any tension could signal the need for
extensions to the base model, such as non-minimal neutrino mass
(though see Sect.~\ref{sec:neutrino}).  However, limited knowledge of
the scaling relation between SZ signal and mass have hampered the
interpretation of this result.

With the full mission data we have created a larger catalogue of SZ
clusters with a more accurate characterization of its
completeness \citep{planck2014-a30}.  By fitting the counts in
redshift and signal-to-noise, we are able to simultaneously constrain
the slope of the SZ signal--mass scaling relation and the cosmological
parameters.  A major uncertainty, however, remains the overall mass
calibration, which in \citet{planck2013-p15} we quantified with a ``hydrostatic
bias'' parameter, $(1-b)$, with a fiducial value of $0.8$ and a range
$0.7<(1-b)<1$ (consistent with some other studies, e.g.,
\citealt{Simet:2015goa}).  In the base \LCDM\ model, the primary CMB
constraints prefer a normalization below the lower end of this range,
$(1-b) \approx 0.6$.  The recent, empirical normalization of the
relation by the Weighing the Giants lensing
programme \citep[WtG;][]{wtgplanck} gives $0.69\pm 0.07$ for the 22
clusters in common with the \Planck\ cluster sample.  This calibration
reduces the tension with the primary CMB constraints in base \LCDM.
In contrast, correlating the entire
\Planck\ 2015 SZ cosmology sample with \Planck\ CMB lensing gives
$1/(1-b)=1.0\pm 0.2$ \citep{planck2014-a30}, toward the upper end of the range
adopted in \citet{planck2013-p15}, although with a large uncertainty.
An alternative lensing calibration analysis by the Canadian Cluster Comparison
Project, which uses 37 clusters in common with the \Planck\ cluster
sample \citep{Hoekstra:2015} finds
$(1-b)=0.76\pm 0.05\ ({\rm stat.})\pm 0.06\ ({\rm syst.})$, which lies between
the other two mass calibrations.
These calibrations are not yet definitive and the situation will continue to
evolve with improvements in mass measurements from larger samples of clusters.

A recent analysis of cluster counts for an X-ray-selected sample
(REFLEX II) shows some tension with the \Planck\ base \LCDM\ cosmology
\citep{Bohringer:2014}.
However, an analysis of cluster counts of X-ray-selected clusters by the WtG
collaboration, incorporating the WtG weak lensing mass calibration, finds
$\sigma_8(\Omega_{\rm m}/0.3)^{0.17} = 0.81 \pm 0.03$, in good agreement
with the \planck\ CMB results for base \LCDM\ \citep{Mantz:2015}.
This raises the possibility that there may be systematic biases in the assumed
scaling relations for SZ-selected clusters compared to X-ray-selected clusters
(in addition to a possible mass calibration bias).
\citet{Mantz:2015} give a brief review of recent determinations of
$\sigma_8$ from X-ray, optically-selected, and SZ-selected samples,
to which we refer the reader.
More detailed discussion of constraints from combining \Planck\ cluster counts
with primary CMB anisotropies and other data sets can be found in
\citet{planck2014-a30}.

\subsection{Cosmic concordance?}
\label{subsec:data_summary}

\begin{figure}
\begin{center}
\includegraphics[width=\hsize]{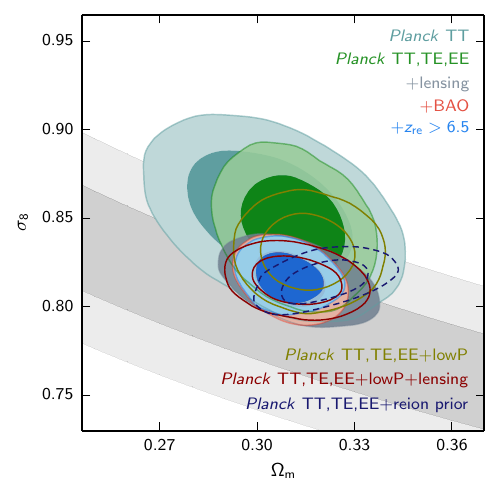}
\end{center}
\caption {
Marginalized constraints on parameters of the base \LCDM\ model without
low-$\ell$ $E$-mode polarization (filled contours), compared to the
constraints from using low-$\ell$ $E$-mode polarization (unfilled contours)
or assuming a strong prior that reionization was at $\zre = 7\pm 1$ and
$\zre > 6.5$ (``reion prior,'' dashed contours). Grey bands show the
constraint from CMB lensing alone.
}
\label{fig:sigma8-omegam-planck}
\end{figure}

Table~\ref{LCDMcompare} summarizes the cosmological parameters in the base
\LCDM\ for \Planck\ combined with
various data sets discussed in this section.  Although we have seen from the
survey presented above that base \LCDM\ is consistent with a wide range of
cosmological data, there are two areas of tension:

\begin{enumerate}
\item the Ly$\alpha$ BAO measurements at high redshift (Sect.~\ref{sec:BAO});
\item the \Planck\ CMB estimate of the amplitude of the fluctuation spectrum
and the lower values inferred from  weak lensing, and (possibly)
cluster counts and
redshift space distortions (Sect.~\ref{sec:additional_data}).
\end{enumerate}

The first point to note is that the astrophysical data in areas (1) and (2)
are complex and more difficult to interpret than most of the astrophysical
data sets discussed in this section. The interpretation of the data in
area (2) depends on nonlinear modelling of the power spectrum, and in the
case of clusters and weak lensing, on
uncertain baryonic physics. Understanding these effects more accurately
 sets a direction for future research.

It is, however, worth reviewing our findings on $\sigma_8$ and $\Omm$ from
\Planck\ assuming base \LCDM. These are summarized in
Fig.~\ref{fig:sigma8-omegam-planck} and the following constraints:
\beglet
\begin{eqnarray}
\sigma_8 &=& 0.829 \pm 0.014, \quad \datalabel{\planckTT}; \\
\sigma_8 &=& 0.815 \pm 0.009, \quad \datalabel{\planckTTlensing}; \\
\sigma_8 &=& 0.810 \pm 0.006, \quad \datalabel{\text{\planckTTonly+{\rm lensing}+$\zre$}}.
\label{DSsum1}
\end{eqnarray}
\endlet
The last line imposes a Gaussian prior of $\zre= 7 \pm 1$ with a limit
$\zre>6.5$ on the reionization redshift in place of the reionization
constraints from the \lowTEB\ likelihood. As discussed in Sect.~\ref{subsec:tau},
such a low redshift of reionization is close to
the lowest plausible value allowed by
astrophysical data (though such low values are
 not favoured by either the WMAP or LFI polarization data).  The addition of \Planck\ lensing
data pulls $\sigma_8$ down by about $1\,\sigma$ from the \planckTT\ value,
so Eq.~\eqref{DSsum1} is the lowest possible range allowed by the \Planck\
CMB data. As shown in Fig.~\ref{fig:sigma8-omegam-planck}, adding the $TE$
and $EE$ spectra at high multipoles does not change the \Planck\
constraints.  If a convincing case can be made that astrophysical data
conflict with the estimate of Eq.~\eqref{DSsum1}, then this will be powerful
evidence for new physics beyond base \LCDM\ with minimal-mass neutrinos.

A number of authors have interpreted the discrepancies in area (2) as
evidence for new physics in the neutrino sector
\citep[e.g.,][]{planck2013-p15, Hamann:2013iba,Battye:2013xqa,Battye:2014qga,Wyman:2013lza,
Beutler:2014yhv}.
They use various data combinations together with \Planck\ to argue for
massive neutrinos with mass $\sumnu \approx0.3\,{\rm eV}$ or for a single
sterile neutrino with somewhat higher mass.  The problem here is that
any evidence for new neutrino physics is driven mainly by the
additional astrophysical data, not by \Planck\ CMB anisotropy measurements.
In addition, the external data sets are
not entirely consistent, so tensions remain. As discussed in {\paramsI}
\citep[see also][]{Leistedt:2014sia,Battye:2014qga} \Planck\ data usually
favour base \LCDM\ over extended models.  Implications of the \Planck\ 2015
data for neutrino physics are discussed  in Sect.~\ref{sec:neutrino} and
tensions between \planck\ and external data in various extended neutrino
models are discussed further in Sect.~\ref{sec:neutrino_tension}.

As mentioned above, we do not use RSD or galaxy weak lensing
measurements for combined constraints in this paper (apart from
Sects. \ref{sec:dark_energy} and \ref{sec:neutrino_tension}, where we use the \CFHTLENS\ data) .
They are, however, used in the paper
exploring constraints on dark energy and modified gravity
\citep{planck2014-a16}. For some models discussed in that paper, the
combination of \Planck, RSD, and weak lensing data does prefer
extensions to the base \LCDM\ cosmology.

\begin{table*}
\begin{center}
%from cosmomcplanck,
%python python/makeTables.py main tables/LCDM_compare.tex --nobestfit --limit 1 --compare plikHM_TT_lowTEB plikHM_TT_lowTEB_lensing plikHM_TT_lowTEB_lensing_H070p6_JLA_BAO plikHM_TTTEEE_lowTEB plikHM_TTTEEE_lowTEB_lensing plikHM_TTTEEE_lowTEB_lensing_H070p6_JLA_BAO --paramtag='base' --forpaper --titles "\\shortTT;\\shortTT+\\lensing;\\shortTT+\\lensing+\\ext;\\shortall;\\shortall+\\lensing;\\shortall+\\lensing+\\ext" --blockEndParams "zeq;ns" --paramList batch2/outputs/cosmology.paramnames

\caption{Parameter $68\,\%$ confidence limits for the base \lcdm\ model from
\planck\ CMB power spectra, in combination with lensing reconstruction
(``\lensing'') and external data (``\ext'', BAO+JLA+$H_0$).
While we see no evidence that systematic effects in polarization are
biasing parameters in the base \LCDM\ model, a conservative choice would be to
use the parameter values listed in Column~3 (i.e., for TT+lowP+lensing).
Nuisance parameters are not listed here for brevity, but can be
found in the extensive tables on the Planck Legacy Archive,
\url{http://pla.esac.esa.int/pla}; however,
the last three parameters listed here give a summary measure of the total
foreground amplitude (in $\muK^2$) at $\ell=2000$ for the three high-$\ell$
temperature power spectra used by the likelihood.  In all cases the helium
mass fraction used is predicted by BBN from the baryon abundance
(posterior mean $\yhe\approx 0.2453$, with theoretical uncertainties in the
BBN predictions dominating over the \Planck\ error on $\Omb h^2$).
The Hubble constant is given in units of ${\rm km\,s}^{-1}\,{\rm Mpc}^{-1}$,
while $r_\ast$ is in Mpc and wavenumbers are in ${\rm Mpc}^{-1}$.
\label{LCDMcompare}
}
\vskip -3mm
\begingroup
\openup 5pt
\newdimen\tblskip \tblskip=5pt
\nointerlineskip
\vskip -3mm
\scriptsize
\setbox\tablebox=\vbox{
    \newdimen\digitwidth
    \setbox0=\hbox{\rm 0}
    \digitwidth=\wd0
    \catcode`*=\active
    \def*{\kern\digitwidth}
    \newdimen\signwidth
    \setbox0=\hbox{+}
    \signwidth=\wd0
    \catcode`!=\active
    \def!{\kern\signwidth}
\halign{
\hbox to 1.0in{$#$\leaderfil}\tabskip=0.6em&
\hfil$#$\hfil\tabskip=1.6em&
\hfil$#$\hfil&
\hfil$#$\hfil&
\hfil$#$\hfil\tabskip=1.1em&
\hfil$#$\hfil&
\hfil$#$\hfil\tabskip=0pt\cr
\noalign{\doubleline}
\multispan1\hfil \hfil&\multispan1\hfil \shortTT\hfil&\multispan1\hfil \shortTT+\lensing\hfil&\multispan1\hfil \shortTT+\lensing+\ext\hfil&\multispan1\hfil \shortall\hfil&\multispan1\hfil \shortall+\lensing\hfil&\multispan1\hfil \shortall+\lensing+\ext\hfil\cr
\noalign{\vskip -3pt}
\omit Parameter\hfil&\omit\hfil 68\,\% limits\hfil&\omit\hfil 68\,\% limits\hfil&\omit\hfil 68\,\% limits\hfil&\omit\hfil 68\,\% limits\hfil&\omit\hfil 68\,\% limits\hfil&\omit\hfil 68\,\% limits\hfil\cr
\noalign{\vskip 3pt\hrule\vskip 5pt}
\Omega_{\mathrm{b}} h^2&0.02222\pm 0.00023&0.02226\pm 0.00023&0.02227\pm 0.00020&0.02225\pm 0.00016&0.02226\pm 0.00016&0.02230\pm 0.00014\cr
\Omega_{\mathrm{c}} h^2&0.1197\pm 0.0022&0.1186\pm 0.0020&0.1184\pm 0.0012&0.1198\pm 0.0015&0.1193\pm 0.0014&0.1188\pm 0.0010\cr
100\theta_{\mathrm{MC}}&1.04085\pm 0.00047&1.04103\pm 0.00046&1.04106\pm 0.00041&1.04077\pm 0.00032&1.04087\pm 0.00032&1.04093\pm 0.00030\cr
\tau&0.078\pm 0.019&0.066\pm 0.016&0.067\pm 0.013&0.079\pm 0.017&0.063\pm 0.014&0.066\pm 0.012\cr
\ln(10^{10} A_\mathrm{s})&3.089\pm 0.036&3.062\pm 0.029&3.064\pm 0.024&3.094\pm 0.034&3.059\pm 0.025&3.064\pm 0.023\cr
n_\mathrm{s}&0.9655\pm 0.0062&0.9677\pm 0.0060&0.9681\pm 0.0044&0.9645\pm 0.0049&0.9653\pm 0.0048&0.9667\pm 0.0040\cr
\noalign{\vskip 5pt\hrule\vskip 3pt}
H_0&67.31\pm 0.96*&67.81\pm 0.92*&67.90\pm 0.55*&67.27\pm 0.66*&67.51\pm 0.64*&67.74\pm 0.46*\cr
\Omega_\Lambda&0.685\pm 0.013&0.692\pm 0.012&0.6935\pm 0.0072&0.6844\pm 0.0091&0.6879\pm 0.0087&0.6911\pm 0.0062\cr
\Omega_{\mathrm{m}}&0.315\pm 0.013&0.308\pm 0.012&0.3065\pm 0.0072&0.3156\pm 0.0091&0.3121\pm 0.0087&0.3089\pm 0.0062\cr
\Omega_{\mathrm{m}} h^2&0.1426\pm 0.0020&0.1415\pm 0.0019&0.1413\pm 0.0011&0.1427\pm 0.0014&0.1422\pm 0.0013&0.14170\pm 0.00097\cr
\Omega_{\mathrm{m}} h^3&0.09597\pm 0.00045&0.09591\pm 0.00045&0.09593\pm 0.00045&0.09601\pm 0.00029&0.09596\pm 0.00030&0.09598\pm 0.00029\cr
\sigma_8&0.829\pm 0.014&0.8149\pm 0.0093&0.8154\pm 0.0090&0.831\pm 0.013&0.8150\pm 0.0087&0.8159\pm 0.0086\cr
\sigma_8 \Omega_{\mathrm{m}}^{0.5}&0.466\pm 0.013&0.4521\pm 0.0088&0.4514\pm 0.0066&0.4668\pm 0.0098&0.4553\pm 0.0068&0.4535\pm 0.0059\cr
\sigma_8 \Omega_{\mathrm{m}}^{0.25}&0.621\pm 0.013&0.6069\pm 0.0076&0.6066\pm 0.0070&0.623\pm 0.011&0.6091\pm 0.0067&0.6083\pm 0.0066\cr
z_{\mathrm{re}}&9.9^{+1.8}_{-1.6}&8.8^{+1.7}_{-1.4}&8.9^{+1.3}_{-1.2}&10.0^{+1.7}_{-1.5}&8.5^{+1.4}_{-1.2}&8.8^{+1.2}_{-1.1}\cr
10^9 A_{\mathrm{s}}&2.198^{+0.076}_{-0.085}&2.139\pm 0.063&2.143\pm 0.051&2.207\pm 0.074&2.130\pm 0.053&2.142\pm 0.049\cr
10^9 A_{\mathrm{s}} e^{-2\tau}&1.880\pm 0.014&1.874\pm 0.013&1.873\pm 0.011&1.882\pm 0.012&1.878\pm 0.011&1.876\pm 0.011\cr
\mathrm{Age}/\mathrm{Gyr}&13.813\pm 0.038*&13.799\pm 0.038*&13.796\pm 0.029*&13.813\pm 0.026*&13.807\pm 0.026*&13.799\pm 0.021*\cr
z_\ast&1090.09\pm 0.42***&1089.94\pm 0.42***&1089.90\pm 0.30***&1090.06\pm 0.30***&1090.00\pm 0.29***&1089.90\pm 0.23***\cr
r_\ast&144.61\pm 0.49**&144.89\pm 0.44**&144.93\pm 0.30**&144.57\pm 0.32**&144.71\pm 0.31**&144.81\pm 0.24**\cr
100\theta_\ast&1.04105\pm 0.00046&1.04122\pm 0.00045&1.04126\pm 0.00041&1.04096\pm 0.00032&1.04106\pm 0.00031&1.04112\pm 0.00029\cr
z_{\mathrm{drag}}&1059.57\pm 0.46***&1059.57\pm 0.47***&1059.60\pm 0.44***&1059.65\pm 0.31***&1059.62\pm 0.31***&1059.68\pm 0.29***\cr
r_{\mathrm{drag}}&147.33\pm 0.49**&147.60\pm 0.43**&147.63\pm 0.32**&147.27\pm 0.31**&147.41\pm 0.30**&147.50\pm 0.24**\cr
k_{\mathrm{D}}&0.14050\pm 0.00052&0.14024\pm 0.00047&0.14022\pm 0.00042&0.14059\pm 0.00032&0.14044\pm 0.00032&0.14038\pm 0.00029\cr
z_{\mathrm{eq}}&3393\pm 49**&3365\pm 44**&3361\pm 27**&3395\pm 33**&3382\pm 32**&3371\pm 23**\cr
k_{\rm{eq}}&0.01035\pm 0.00015&0.01027\pm 0.00014&0.010258\pm 0.000083&0.01036\pm 0.00010&0.010322\pm 0.000096&0.010288\pm 0.000071\cr
100\theta_{\rm{s,eq}}&0.4502\pm 0.0047&0.4529\pm 0.0044&0.4533\pm 0.0026&0.4499\pm 0.0032&0.4512\pm 0.0031&0.4523\pm 0.0023\cr
\noalign{\vskip 5pt\hrule\vskip 3pt}
f_{2000}^{143}&29.9\pm 2.9*&30.4\pm 2.9*&30.3\pm 2.8*&29.5\pm 2.7*&30.2\pm 2.7*&30.0\pm 2.7*\cr
f_{2000}^{143\times217}&32.4\pm 2.1*&32.8\pm 2.1*&32.7\pm 2.0*&32.2\pm 1.9*&32.8\pm 1.9*&32.6\pm 1.9*\cr
f_{2000}^{217}&106.0\pm 2.0**&106.3\pm 2.0**&106.2\pm 2.0**&105.8\pm 1.9**&106.2\pm 1.9**&106.1\pm 1.8**\cr
\noalign{\vskip 5pt\hrule\vskip 3pt}
} % close halign
} % close vbox
\endPlancktable
\endgroup

\end{center}
\end{table*}

\section{Extensions to the base $\mathbf\Lambda$CDM model}
\label{sec:grid}

\begin{table*}
\begin{center}
\caption{Constraints on 1-parameter extensions to the base \lcdm\ model for
combinations of \planck\ power spectra, \planck\ lensing, and external data
(BAO+JLA+$H_0$, denoted ``\ext''). All limits and confidence
regions quoted here are 95\,\%.
\label{tab:grid_1paramext}
}
\begingroup
%\openup 5pt
\newdimen\tblskip \tblskip=5pt
\nointerlineskip
\vskip -4mm
\footnotesize
\setbox\tablebox=\vbox{
    \newdimen\digitwidth
    \setbox0=\hbox{\rm 0}
    \digitwidth=\wd0
    \catcode`"=\active
    \def"{\kern\digitwidth}
    \newdimen\signwidth
    \setbox0=\hbox{+}
    \signwidth=\wd0
    \catcode`!=\active
    \def!{\kern\signwidth}
\halign{
\hbox to 1.2in{$#$\leaderfil}\tabskip=1.5em&
\hfil$#$\hfil\tabskip=1.5em&
\hfil$#$\hfil\tabskip=1.5em&
\hfil$#$\hfil\tabskip=1.5em&
\hfil$#$\hfil\tabskip=1.0em&
\hfil$#$\hfil\tabskip=1.0em&
\hfil$#$\hfil\tabskip=0pt\cr
\noalign{\doubleline}
\omit\text{Parameter}
 \hfil& \hfil \quad \TT\quad\hfil& \hfil \TT{+}\lensing& \hfil \TT{+}\lensing{+}\ext \hfil& \hfil \TTTEEE \hfil& \hfil \TTTEEE{+}\lensing \hfil& \hfil \TTTEEE{+}\lensing{+}\ext\cr
\noalign{\vskip 3pt\hrule\vskip 5pt}

\Omega_K& -0.052^{+0.049}_{-0.055}& -0.005^{+0.016}_{-0.017}& -0.0001^{+0.0054}_{-0.0052}& -0.040^{+0.038}_{-0.041}& -0.004^{+0.015}_{-0.015}& 0.0008^{+0.0040}_{-0.0039}\cr
\Sigma m_\nu\,[\mathrm{eV}]& < 0.715& < 0.675& < 0.234& < 0.492& < 0.589& < 0.194\cr
N_{\mathrm{eff}}& 3.13^{+0.64}_{-0.63}& 3.13^{+0.62}_{-0.61}& 3.15^{+0.41}_{-0.40}& 2.99^{+0.41}_{-0.39}& 2.94^{+0.38}_{-0.38}& 3.04^{+0.33}_{-0.33}\cr
Y_{\mathrm{P}}& 0.252^{+0.041}_{-0.042}& 0.251^{+0.040}_{-0.039}& 0.251^{+0.035}_{-0.036}& 0.250^{+0.026}_{-0.027}& 0.247^{+0.026}_{-0.027}& 0.249^{+0.025}_{-0.026}\cr
\mathrm{d}n_{\mathrm{s}}/\mathrm{d}\ln k& -0.008^{+0.016}_{-0.016}& -0.003^{+0.015}_{-0.015}& -0.003^{+0.015}_{-0.014}& -0.006^{+0.014}_{-0.014}& -0.002^{+0.013}_{-0.013}& -0.002^{+0.013}_{-0.013}\cr
r_{0.002}& < 0.103& < 0.114& < 0.114& < 0.0987& < 0.112& < 0.113\cr
w& -1.54^{+0.62}_{-0.50}& -1.41^{+0.64}_{-0.56}& -1.006^{+0.085}_{-0.091}& -1.55^{+0.58}_{-0.48}& -1.42^{+0.62}_{-0.56}& -1.019^{+0.075}_{-0.080}\cr

\noalign{\vskip 5pt\hrule\vskip 3pt}
} % close halign
} % close vbox
\endPlancktable
\endgroup
%\hline
%\end{tabular}
\end{center}
\end{table*}

\subsection{ Grid of models}

\begin{figure*}
\centering
\includegraphics[width=18.5cm]{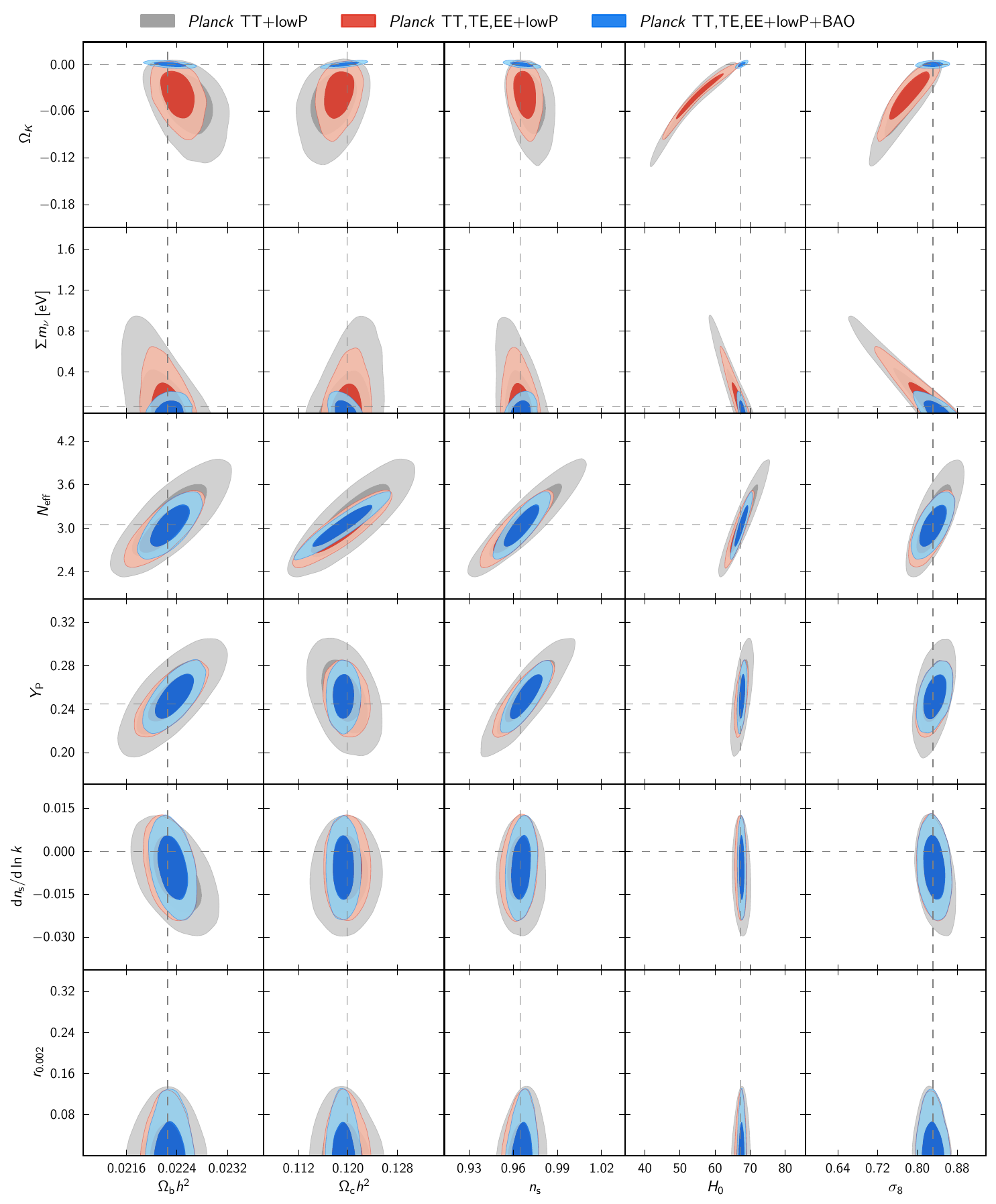}
\caption {68\,\% and 95\,\% confidence regions on 1-parameter extensions of
the base \LCDM\ model for \planckTT\ (grey), \planckall\ (red),
and \planckall+BAO (blue).
Horizontal dashed lines correspond to the parameter values assumed in the
base \LCDM\ cosmology, 
while vertical dashed lines show the mean posterior values in the base model
for \planckall+BAO.}
\label{fig:grid_1paramext}
\end{figure*}

The full grid results are available online.\footnote{See the Planck Legacy
Archive, \url{http://www.cosmos.esa.int/web/planck/pla}, which
contains considerably more detailed information than presented
in this paper.}
Figure~\ref{fig:grid_1paramext} and Table~\ref{tab:grid_1paramext} summarize
the constraints on one-parameter extensions to base \lcdm. As in {\paramsI}, we
find no strong evidence in favour of any of these
simple one-parameter extensions using \Planck\ or \Planck\ combined with BAO.
The entire grid has been run using both the \plik\ and \camspec\ likelihoods.
As noted in Sect.~\ref{sec:planck_only}, the parameters derived from these two
$TT$ likelihoods agree to better than $0.5\,\sigma$ for base \LCDM. This level
of agreement also holds for the extended models analysed in our grid.
In Sect.~\ref{sec:planck_only} we also pointed out that we have
definite evidence, by comparing spectra computed with different frequency
combinations, of residual systematics in the $TE$ and $EE$ spectra. These
systematics average down in the coadded $TE$ and $EE$ spectra, but the
remaining level of systematics in these coadded spectra are not yet well
quantified (though they are small). Thus, we urge the reader to treat
parameters computed from the TT{,}TE{,}EE likelihoods with some caution. In the
case of polarization, the agreement between the \plik\ and \camspec\ $TE$
and $EE$ likelihoods is less good, with shifts in parameters of up
to $1.5 \,\sigma$ (though such large shifts are unusual). In general, the behaviour of the
TT{,}TE{,}EE likelihoods is as shown in Fig.~\ref{fig:grid_1paramext}.
For extended models, the addition of the \Planck\ polarization
data at high multipoles reduces the errors on extended parameters compared
to the \Planck\ temperature data and pulls the parameters towards those of
base \LCDM. A similar behaviour is seen if the \Planck\ TT
(or \Planck\ TT{,}TE{,}EE) data are combined with BAO.

The rest of this section discusses the grid results in more detail and
also reports results on some additional models (specifically dark matter
annihilation, tests of the recombination history, and cosmic defects) that
are not included in our grid.

\subsection{Early-Universe physics}
\label{subsec:params_early}

Arguably the most important result from 2013 \Planck\ analysis was the finding
that simple single-field inflationary models, with a tilted scalar
spectrum $\ns \approx 0.96$, provide a very good fit to the
\Planck\ data. We found no evidence for a tensor component or running
of the scalar spectral index, no strong evidence for isocurvature
perturbations or features in the primordial power spectrum
\citep{planck2013-p17}, and no evidence for non-Gaussianity
\citep{planck2013-p09a}, cosmic strings or other topological defects
\citep{planck2013-p20}. On large angular scales, the \Planck\ data
showed some evidence for ``anomalies'' seen previously in the
\WMAP\ data \citep{bennett2010}.  These include a dip in the power
spectrum in the multipole range $20 \la \ell \la 30$ (see Fig.~\ref{pgTT_final}) and
some evidence for a departure from statistical isotropy on large angular
scales \citep{planck2013-p09}. However, the statistical significance
of these anomalies is not high enough to provide compelling evidence
for new physics beyond simple single-field inflation.

\begin{figure*}[t]
\centering
\includegraphics[width=88mm]{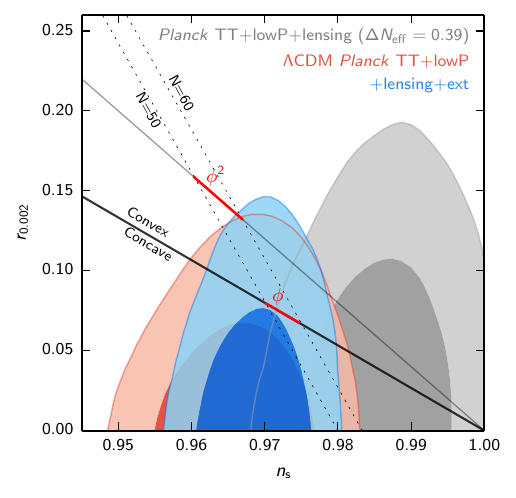}
\includegraphics[width=88mm]{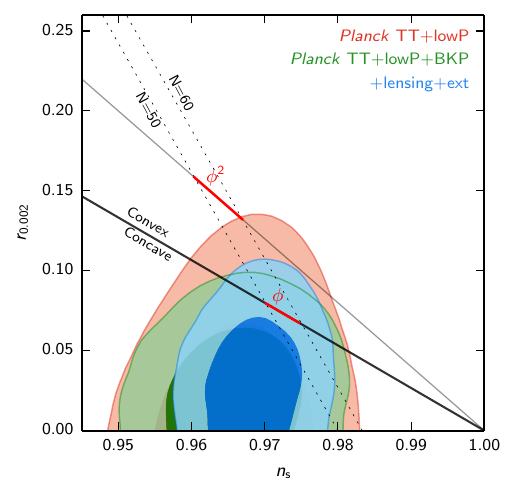}

\caption {
{\it Left}: Constraints on the tensor-to-scalar ratio $\rzerotwo$ in the
\lcdm\ model, using \planckTT\ and \planckTT+\lensing+BAO+JLA+$H_0$
(red and blue, respectively) assuming negligible running and the inflationary
consistency relation.  The result is model-dependent; for example, the grey
contours show how the results change if there were
additional relativistic degrees of freedom with $\Delta\nnu = 0.39$
(disfavoured, but not excluded, by \planck).
Dotted lines show loci of approximately constant $e$-folding number $N$,
assuming simple $V\propto (\phi/\mpl)^p$ single-field inflation. Solid lines
show the approximate $\ns$--$r$ relation for quadratic and linear
potentials, to first order in slow roll;
red lines show the approximate allowed range assuming $50<N<60$ and a power-law
potential for the duration of inflation.
The solid black line (corresponding to a linear potential) separates concave
and convex potentials.
{\it Right}: Equivalent constraints in the \lcdm\ model when adding $B$-mode
polarization results corresponding to the default configuration of the
{BICEP2/Keck Array+\Planck} (BKP) likelihood. These exclude the
quadratic potential at a higher level of significance compared to the
\planck-alone constraints.
\label{fig:ns_r_inflation}
}
\end{figure*}

The \Planck\ 2013 results led to renewed interest in the $R^2$
inflationary model, originally introduced by \cite{Starobinsky:1980},
and related inflationary models that have flat effective potentials of similar form
\citep[e.g.,][]{Kallosh:2013,Ferrara:2013,Buchmuller:2013,Ellis:2013}.
A characteristic of these models is that they produce a red tilted
scalar spectrum and a low tensor-to-scalar ratio. For reference, the
Starobinsky model predicts
\beglet
\begin{eqnarray}
 \ns & \approx & 1 - {2 \over N} \in (0.960, 0.967), \\
 r   & \approx & {12 \over N^2} \in (0.003 , 0.005),  \label{Inf1} \\
 \nrunfrac & \approx & - {2 \over N^2} \in (- 0.0008,-0.0006),
\end{eqnarray}
\endlet
where $N$ is the number of e-foldings between the end of inflation and
the time that our present day Hubble scale crossed the inflationary horizon,
and numerical values are for the range $50\le N\le 60$.

Although the \Planck\ 2013 results stimulated theoretical work on
inflationary models with low tensor-to-scalar ratios, the cosmological
landscape became more complicated following the detection of a $B$-mode
polarization anisotropy by the BICEP2 team \citep{BicepDetection}.
If the BICEP2 signal were primarily caused by primordial gravitational waves,
then the inferred tensor-to-scalar ratio would have been
$r_{0.01} \approx 0.2$,\footnote{The pivot scale quoted here is roughly
appropriate for the multipoles probed by BICEP2.} apparently in
conflict with the 2013 \Planck\ 95\,\% upper limit of $r_{0.002} < 0.11$, based
on fits to the temperature power spectrum. Since the
\Planck\ constraints on $r$ are highly model dependent (and fixed
mainly by lower $k$) it is possible to reconcile these
results by introducing additional parameters, such as large tilts or
strong running of the spectral indices.

The situation has been clarified following a joint analysis of
BICEP2/Keck observations and \Planck\ polarization data reported in
{\BKP}. This analysis shows that polarized dust emission contributes a
significant part of the BICEP2 signal. Correcting for polarized dust emission,
{\BKP} report a 95\,\% upper limit of $r_{0.05} < 0.12$ on scale-invariant
tensor modes, eliminating the tension between the BICEP2 and the \Planck\ 2013
results. There is therefore no evidence for inflationary tensor modes from
$B$-mode polarization measurements at this time (although the BKP
analysis leaves open the possibility of a much higher tensor-to-scalar
ratio than the prediction of Eq.~\ref{Inf1} for Starobinsky-type models).

The layout of the rest of this subsection is as follows. In
Sect.~\ref{subsubsec:tensors} we review the \Planck\ 2015 and \planck+BKP constraints
on $\ns$ and $r$. Constraints on the running of the scalar spectral
index are presented in Sect.~\ref{subsubsec:index}.  Polarization data
provide a powerful way of testing for isocurvature modes, as discussed
in Sect.~\ref{subsubsec:isocurvature}. Finally,
Sect.~\ref{subsubsec:curv} summarizes our results on spatial
curvature. A discussion of specific inflationary models and tests for
features in the primordial power spectrum can be found in
\citet{planck2014-a24}.

\subsubsection{Scalar spectral index and tensor fluctuations}
\label{subsubsec:tensors}

Primordial tensor fluctuations (gravitational waves) contribute to
both the CMB temperature and polarization power spectra. Gravitational
waves entering the horizon between recombination and the present day
generate a tensor contribution to the large-scale CMB temperature anisotropy. In
this data release, the strongest constraint on tensor modes from
\planck\ data still comes from the CMB temperature spectrum at
$\ell\la 100$. The corresponding
comoving wavenumbers probed by the \planck\ temperature spectrum have $k
\la 0.008\,\Mpc^{-1}$, with very little sensitivity to higher wavenumbers
because gravitational waves decay on sub-horizon scales.  The
precision of the \planck\ constraint is limited by cosmic
variance of the large-scale anisotropies (which are dominated by the scalar
component), and it is also model dependent.  In polarization, in addition to
$B$-modes, the $EE$ and $TE$ spectra also contain a signal from tensor modes
coming from the last-scattering and reionization epochs. However, in this
release the addition of \planck\ polarization constraints at $\ell \ge 30$ do
not significantly change the results from temperature and low-$\ell$
polarization (see Table~\ref{tab:grid_1paramext}).

Figure~\ref{fig:ns_r_inflation} shows the 2015 \planck\ constraint in the
$\ns$--$r$ plane, adding $r$ as a 1-parameter extension to base \LCDM.
For base \LCDM\ ($r=0$), the value of $\ns$ is
\begin{equation}
\ns = 0.9655 \pm 0.0062, \qquad \datalabel{\planckTT}. \label{Inf2}
\end{equation}
We highlight this number here since $\ns$, a key parameter for
inflationary cosmology, shows one of the largest shifts
of any parameter in base \LCDM\ between the \Planck\ 2013 and \Planck\ 2015
analyses (about $0.7\,\sigma$).
As explained in
Sect.~\ref{subsec:planck_only1}, part of this shift was caused by the
$\ell\approx1800$ systematic in the nominal-mission $217\times217$ spectrum
used in {\paramsI}.

The red contours in Fig.~\ref{fig:ns_r_inflation} show the constraints
from \planckTT.  These are similar to the constraints shown in Fig.~23
of {\paramsI}, but with $\ns$ shifted to slightly higher values.  The
addition of BAO or the \planck\ lensing data to \planckTT\ lowers the
value of $\Omc h^2$, which, at fixed $\thetastar$, increases the
small-scale CMB power. To maintain the fit to the \Planck\ temperature
power spectrum for models with $r=0$, these parameter shifts are
compensated by a change in the amplitude $\As$ and the tilt $\ns$ (by
about $0.4\,\sigma$). The increase in $\ns$ to match the observed
power on small scales leads to a decrease in the scalar power on large
scales, allowing room for a slightly larger contribution from tensor
modes. The constraints shown by the blue contours in
Fig.~\ref{fig:ns_r_inflation}, which combine \Planck\ lensing, BAO, and
other astrophysical data, are therefore tighter in the $\ns$ direction
and shifted to slightly higher values, but marginally weaker in the
$r$-direction. The 95\,\% limits on $r_{0.002}$ are
\beglet
\begin{eqnarray}
r_{0.002} &<& 0.10 , \quad \datalabel{\planckTT}, \label{Inf3} \\
r_{0.002} &<& 0.11 , \quad \datalabel{\planckTTlensext}, \label{Inf3b}
\end{eqnarray}
\endlet
consistent with the results reported in {\paramsI}. Here we assume
the second-order slow-roll consistency relation for the tensor spectral index.
The result in Eqs.~(\ref{Inf3}) and (\ref{Inf3b}) are mildly scale dependent,
with equivalent limits on $r_{0.05}$ being weaker by about $5\,\%$.

{\paramsI} noted a mismatch between the best-fit base \LCDM\ model and
the temperature power spectrum at multipoles $\ell \la 40$, partly
driven by the dip in the multipole range $20 \la \ell \la 30$. If this
mismatch is simply a statistical fluctuation of the \LCDM\ model (and
there is no compelling evidence to think otherwise), the strong
\planck\ limit (compared to forecasts) is the result of chance low
levels of scalar mode confusion. On the other hand, if the dip
represents a failure of the \LCDM\ model, the $95\,\%$ limits of
Eqs.~\eqref{Inf3} and \eqref{Inf3b} may be underestimates. These
issues are considered at greater length in \citet{planck2014-a24} and
will not be discussed further in this paper.

As mentioned above, the \Planck\ temperature constraints on $r$ are
model-dependent and extensions to \LCDM\ can give significantly
different results.  For example, extra relativistic degrees of freedom
increase the small-scale damping of the CMB anisotropies at a fixed
angular scale, which can be compensated by increasing $\ns$, allowing
a larger tensor mode. This is illustrated by the grey contours
in Fig.~\ref{fig:ns_r_inflation}, which show the constraints for a model
with $\Delta \nnu = 0.39$. Although this value of $\Delta \nnu$
is disfavoured by the \Planck\ data (see Sect.~\ref{sec:mnu}) it is not
excluded at a high significance level.

This example emphasizes the need for direct tests of tensor modes
based on measurements of a large-scale $B$-mode pattern in CMB
polarization.
\Planck\ $B$-mode constraints from the 100- and 143-GHz HFI channels, presented
in \citet{planck2014-a13}, give a
95\,\% upper limit of $r \la 0.27$.  However, at
present the tightest $B$-mode constraints on $r$ come
from the {\BKP} analysis of the BICEP2/Keck field, which covers approximately
$400\\,{\rm deg}^2$ centred on ${\rm RA}=0^{\rm h}$, ${\rm Dec}=-57\pdeg5$.
These measurements probe the peak of the $B$-mode power spectrum at
around $\ell=100$, corresponding to gravitational waves with $k\approx
0.01\,\Mpc^{-1}$ that enter the horizon during recombination (i.e.,
somewhat smaller than the scales that contribute to the
\Planck\ temperature constraints on $r$). The results of {\BKP}
give a posterior for $r$ that peaks at $r_{0.05}\approx0.05$, but is
consistent with $r_{0.05}=0$. Thus, at present there is no convincing evidence
of a primordial $B$-mode signal. At these low values of $r$, there is no
longer any tension with \Planck\ temperature constraints.

The analysis of {\BKP} constrains $r$ defined relative to a fixed
fiducial $B$-mode spectrum, and on its own does not
give a useful constraint on either the scalar amplitude or $\ns$.  A combined
analysis of the \planck\ CMB spectra and the BKP likelihood can,
self-consistently, give constraints in the $\ns$--$r$ plane,
as shown in the right-hand panel of Fig.~\ref{fig:ns_r_inflation}.
The BKP likelihood pulls the
contours to slightly non-zero values of $r$, with best fits of around
$r_{0.002}\approx 0.03$, but at very low levels of statistical
significance. The BKP likelihood also rules out the upper tail of $r$
values allowed by \planck\ alone. The joint \Planck+BKP likelihood analyses
give the $95\,\%$ upper limits
\beglet
\begin{eqnarray}
r_{0.002} &<& 0.08 , \quad \datalabel{\planckTT\dataplus{\rm BKP}}, \label{PBKPa}\\
r_{0.002} &<& 0.09 , \quad \datalabel{\planckTTlensext\dataplus{\rm BKP}}. \label{PBKPb}
\end{eqnarray}
\endlet
The exact values of these upper limits are weakly dependent on
the details of the foreground modelling applied in the BKP analysis
(see {\BKP} for further details).
The results given here are for the baseline 2-parameter
model, varying the $B$-mode dust amplitude and frequency scaling, using the lowest five $B$-mode bandpowers.

\begin{figure}[t]
\centering
\includegraphics[width=90mm]{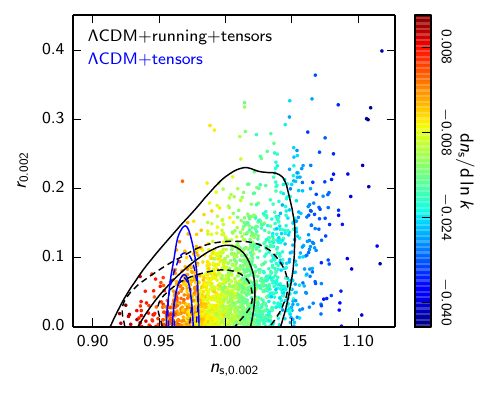}
\caption {
Constraints on the tensor-to-scalar ratio $\rzerotwo$ in the \lcdm\ model with
running, using \planckTT\ (samples, coloured by the running parameter), and
\planckTT+lensing+BAO (black contours).  Dashed contours show the
corresponding constraints also including the {\BKP} $B$-mode likelihood. These
are compared to the constraints when the running is
fixed to zero (blue contours).  Parameters are plotted at
$k=0.002\,\Mpc^{-1}$, which is approximately the scale at which \planck\
probes tensor fluctuations; however, the scalar tilt is only constrained
well on much
smaller scales. The inflationary slow-roll consistency relation is used here
for $\nt$ (though the range of running allowed is much larger than would be
expected in most slow-roll models).
\label{fig:ns_r_nrun}
}
\end{figure}

Allowing a running of the scalar spectral index as an additional
free parameter weakens the \Planck\ constraints on $r_{0.002}$, as shown in
Fig.~\ref{fig:ns_r_nrun}.  The coloured samples in
Fig.~\ref{fig:ns_r_nrun} illustrate how a negative running allows the
large-scale scalar spectral index $n_{{\rm s},0.002}$ to shift towards
higher values, lowering the scalar power on large scales relative to small
scales, thereby allowing a larger tensor contribution.  However
adding the BKP likelihood, which directly constrains the tensor
amplitude on smaller scales, significantly reduces the extent of this
degeneracy leading to a $95\,\%$ upper limit of $r_{0.002}< 0.10$ even in the
presence of running (i.e., similar to the results of Eqs.~\ref{PBKPa} and
\ref{PBKPb}).

The \planck+BKP joint analysis rules out a quadratic
inflationary potential ($V(\phi)\propto m^2\phi^2$, predicting
$r\approx 0.16$) at over 99\,\% confidence and reduces the allowed range of
the parameter space for models with convex potentials.  Starobinsky-type
models are an example of a wider class of inflationary theories in which
$\ns-1 = \clo(1/N)$ is not a coincidence, yet $r =
\clo(1/N^2)$
\citep{Roest:2013fha,Creminelli:2014nqa}.  These models have concave
potentials, and include a variety of string-inspired models with
exponential potentials.  Models with $r = \clo(1/N)$ are, however, still
allowed by the data, including a simple linear potential and
fractional-power monomials, as well as regions of parameter space in
between where $\ns-1 = \clo(1/N)$ is just a coincidence. Models that
have sub-Planckian field evolution, so satisfying the Lyth bound (\citealt{Lyth:1996im}; \citealt{Garcia-Bellido:2014wfa}), will typically have
$r\la 2\times 10^{-5}$ for $\ns\approx 0.96$, and are also consistent
with the tensor constraints shown in Fig.~\ref{fig:ns_r_inflation}.
For further discussion of the implications of the \Planck\ 2015 data
for a wide range of inflationary models see \citet{planck2014-a24}.

In summary, the \Planck\ limits on $r$ are consistent with the BKP
limits from $B$-mode measurements. Both data sets are consistent with
$r=0$; however, the combined data sets yield an upper limit to the
tensor-to-scalar ratio of $r\approx0.09$ at the 95\,\% level.
The \Planck\ temperature constraints on
$r$ are limited by cosmic variance. The only way of improving these limits,
or potentially detecting gravitational waves with $r\la0.09$, is through
direct $B$-mode detection. The \Planck\ 353-GHz polarization maps
\citep{planck2014-XXX} show that at frequencies of around 150\,GHz,
Galactic dust emission is an important contaminant at the $r\approx0.05$
level even in the cleanest regions of the sky. {\BKP} demonstrates further
that on small regions of the sky covering a few hundred square degrees
(typical of ground based $B$-mode experiments), the \Planck\ 353-GHz
maps are of limited use as monitors of polarized Galactic dust emission
because of their low signal-to-noise level. To achieve limits substantially
below $r\approx 0.05$ will require observations of comparably high sensitivity
over a range of frequencies, and with increased sky coverage.
The forthcoming measurements from Keck Array and BICEP3 at
95\,GHz and the Keck Array receivers at 220\,GHz should offer significant
improvements on the current constraints. A number of other ground-based and
sub-orbital experiments should also return high precision $B$-mode data within
the next few years \citep[see][for a review]{Abazajian:2015b}.

\subsubsection{Scale dependence of primordial fluctuations}
\label{subsubsec:index}

In simple single-field models of inflation, the running of the
spectral index is of second order in inflationary slow-roll parameters
and is typically small, $|\nrun| \approx (\ns-1)^2
\approx 10^{-3}$ \citep{Kosowsky:1995}. Nevertheless, it is possible to
construct models that produce a large running over a wavenumber range
accessible to CMB experiments, whilst simultaneously achieving enough
e-folds of inflation to solve the horizon problem.  Inflation with an
oscillatory potential of sufficiently long period, perhaps related to
axion monodromy, is an example \citep{Silverstein:2008,Meerburg:2014bpa,Czerny:2014wua,Minor:2014}.

\begin{figure}[t]
\centering
\includegraphics[width=92mm]{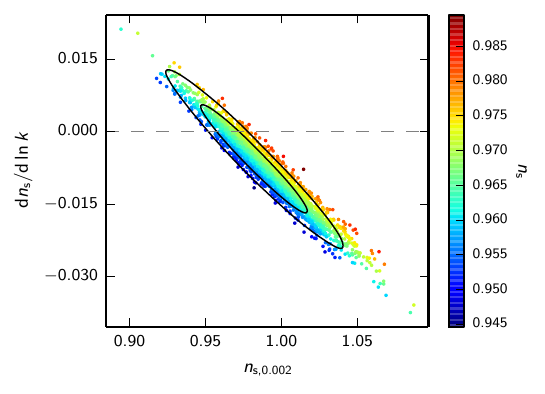}
\caption {
Constraints on the running of the scalar spectral index in the \lcdm\ model,
using \planckTT\ (samples, coloured by the spectral index at
$k=0.05\,\Mpc^{-1}$), and \planckall\ (black contours). The \planck\ data are
consistent with zero running, but also allow for significant negative
running, which gives a positive tilt on large scales and hence less
power on large scales.
\label{fig:ns_nrun}
}
\end{figure}

As reviewed in {\paramsI}, previous CMB experiments, either on their own or
in combination with other astrophysical data, have sometimes given hints of
a non-zero running at about the $2\,\sigma$ level
\citep{spergel2003, hinshaw2012,Hou:2014}. The results of {\paramsI} showed
a slight preference for negative running at the $1.4\,\sigma$ level,
driven almost entirely by the mismatch between the CMB temperature power
spectrum at high multipoles and the spectrum at multipoles $\ell \la 50$.

The 2015 \Planck\ results (Fig.~\ref{fig:ns_nrun}) are similar to those in
{\paramsI}.  Adding running as an additional parameter to base \LCDM\, with $r=0$, we find
\beglet
\begin{eqnarray}
\nrunfrac &=& -0.0084 \pm 0.0082, \quad \datalabel{\planckTT}, \label{Inf4a} \\
\nrunfrac &=& -0.0057 \pm 0.0071, \quad \datalabel{\planckall}. \label{Inf4b}
\end{eqnarray}
\endlet
There is a slight preference for negative running, which, as in {\paramsI},
is driven by the mismatch between the high and low multipoles
in the temperature power spectrum. However, in the 2015 \Planck\ data
the tension between high and low multipoles is reduced somewhat,
primarily because of changes to the HFI beams at multipoles $\ell \la 200$
(see Sect.~\ref{subsec:planck_only1}). A consequence of
this reduced tension can be seen in the 2015 constraints on models that
include tensor fluctuations in addition to running:
\beglet
\begin{eqnarray}
\nrunfrac &=& -0.0126^{+0.0098}_{-0.0087}, \quad \datalabel{\planckTT};
 \label{Inf5a} \\
\nrunfrac &=& -0.0085\pm 0.0076, \quad \datalabel{\planckall}; \label{Inf5b}\\
\nrunfrac &=& -0.0065\pm 0.0076, \quad \datalabel{\planckTT\dataplus\lensing}\nonumber\\
&&\qquad\qquad\qquad\qquad\qquad\qquad\datalabel{\dataplus\ext\dataplus{\rm BKP}}.
\end{eqnarray}
\endlet
{\paramsI} found an approximately $2\,\sigma$ pull towards negative running for
these models. This tension is reduced to about $1\,\sigma$ with the
2015 \Planck\ data, and to lower values when we include the {\BKP}
likelihood, which reduces the range of allowed tensor amplitudes.

In summary, the \Planck\ data are consistent with zero running of the scalar
spectral index. However, as illustrated in Fig.~\ref{fig:ns_nrun},
the \Planck\ data still allow running at roughly the $10^{-2}$ level, i.e.,
an order of magnitude higher than expected in simple inflationary models.
One way of potentially improving these constraints is to extend the wavenumber
range from CMB scales to smaller scales using additional astrophysical data,
for example by using measurements of the Ly$\alpha$ flux
power spectrum of high-redshift quasars \citep[as in the first year WMAP
analysis,][]{spergel2003}.
\citet{palanque:2014} have recently reported an analysis of a large
sample of quasar spectra from the SDSSIII/BOSS
survey. These authors find a low value of the scalar spectral index
$\ns =0.928 \pm 0.012 \ ({\rm stat.}) \pm (0.02) \ ({\rm syst.})$
on scales of $k \approx 1\,{\rm Mpc}^{-1}$. To extract
physical parameters, the Ly$\alpha$ power spectra need to
be calibrated against numerical hydrodynamical simulations.
The large systematic error in this spectral index determination is
dominated by the fidelity of the hydrodynamic simulations and by the
splicing used to achieve high resolution over large scales. These
uncertainties need to be reduced before addressing the consistency of
Ly$\alpha$ results with CMB measurements of the running of the spectral index.

\subsubsection{Isocurvature perturbations}
\label{subsubsec:isocurvature}

A key prediction of single-field inflation is that the primordial
perturbations are adiabatic.  More generally, the observed fluctuations will
be adiabatic in any model in which the curvature perturbations were the only
super-horizon perturbations left by the time that dark matter (and other
matter) first decoupled, or was produced by decay. The different matter
components then all have perturbations proportional to the curvature
perturbation, so there are no isocurvature perturbations. However,
it is possible to produce an observable amount of isocurvature modes
by having additional degrees of freedom present during inflation and through
reheating. For example, the curvaton model can generate correlated adiabatic
and isocurvature modes from a second field~\citep{Mollerach90,Lyth01}.

Isocurvature modes describe relative perturbations between the different
species~\citep{Bucher99}, with perhaps the simplest being a perturbation in
the baryonic or dark matter sector (relative to the radiation). However,
only one total matter isocurvature mode is observable in the linear CMB
(in the accurate approximation in which the baryons are pressureless);
a compensated mode (between the baryons and the cold dark matter)
with $\delta\rho_{\rm b} = -\delta\rho_{\rm c}$ has no net density
perturbation, and produces no CMB anisotropies~\citep{Gordon:2002gv,Grin:2011tf,Grin:2014}.
It is possible to generate isocurvature modes in the neutrino sector;
however, this requires interaction of an additional perturbed super-horizon
field with neutrinos after they have decoupled, and hence is harder to achieve.
Finally, neutrino velocity potential and vorticity modes are other possible
consistent perturbations to the photon-neutrino fluid after neutrino
decoupling. However, they are essentially impossible to excite,
since they consist of photon and neutrino fluids coherently moving in opposite
directions on super-horizon scales (although the relative
velocity would have been zero before neutrino decoupling).

\begin{figure}
\centering
\includegraphics[width=90mm]{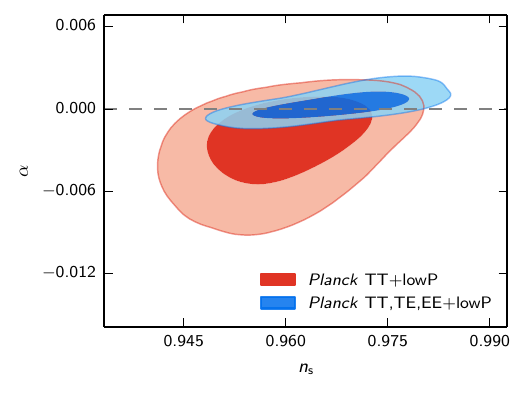}
\caption {Constraints on the correlated matter isocurvature mode amplitude
parameter $\alphaiso$, where $\alphaiso=0$ corresponds to purely adiabatic
perturbations. The \planck\ temperature data slightly favour negative values,
since this lowers the large-scale anisotropies; however, the polarization
signal from an isocurvature mode is distinctive and the \planck\ polarization
data significantly shrink the allowed region around the value $\alpha=0$
corresponding to adiabatic perturbations.
\label{fig:correlated_iso}
}
\end{figure}

\begin{figure*}
\centering
\includegraphics[width=3.6in]{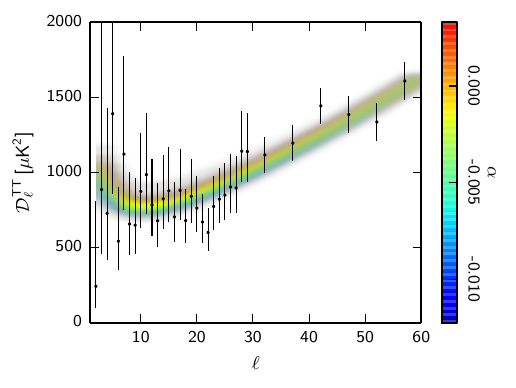}
\includegraphics[width=3.6in]{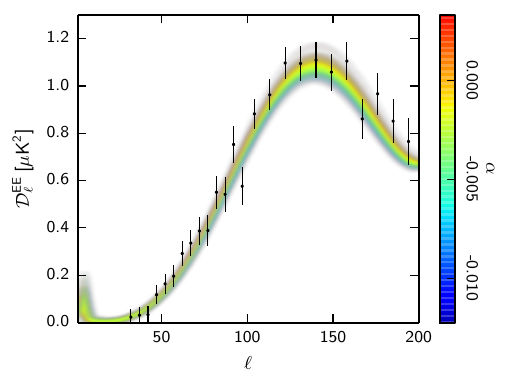}
\caption {Power spectra drawn from the \planckTT\ posterior for the correlated
matter isocurvature model, colour-coded by the value of the isocurvature
amplitude parameter $\alpha$, compared to the \planck\ data points. The
left-hand figure shows how the negatively-correlated modes lower the
large-scale temperature spectrum,
slightly improving the fit at low multipoles. Including polarization, the
negatively-correlated modes are disfavoured, as illustrated at the first
acoustic peak in $EE$ on the right-hand plot. Data points at $\ell<30$ are not
shown for polarization, as they are included with both the default temperature
(i.e., TT+lowP) and polarization (i.e., \TTTEEE+lowP) likelihood combinations.
\label{fig:isorainbows}
}
\end{figure*}

\cite{planck2013-p17} presented constraints on a variety of general
isocurvature models using the \planck\ temperature data, finding consistency
with adiabaticity, though with some mild preference for isocurvature models
that reduce the power at low multipoles to provide a better match to the
\Planck\ temperature spectrum at $\ell \la 50$.
For matter isocurvature perturbations, the photons are initially unperturbed
but perturbations develop as the Universe becomes more matter dominated. As a
result, the phase of the acoustic oscillations differs from adiabatic modes;
this is most clearly distinctive with the addition of polarization
data~\citep{Bucher:2000hy}

An extended analysis of isocurvature models is given in \cite{planck2014-a24}.
Here we focus on a simple illustrative case of a totally-correlated matter
isocurvature mode. We define an isocurvature amplitude parameter $\alphaiso$,
such that\footnote{\cite{planck2014-a24} gives equivalent one-tailed
constraints on $\beta_{\rm iso} = |\alpha|$, where the correlated and
anti-correlated cases are considered separately. }
\be
S_{\rm m} = {\rm sgn}(\alphaiso) \sqrt{\frac{|\alphaiso|}{1-|\alphaiso|}} \zeta ,
\ee
where $\zeta$ is the primordial curvature perturbation.
Here $S_{\rm m}$ is the total matter isocurvature mode, defined as the
observable sum of the baryon and CDM isocurvature modes, i.e.,
$S_{\rm m} = S_{\rm c} + S_{\rm b}(\rho_{\rm b}/\rho_{\rm c})$,
where
\be
S_i \equiv \frac{\delta\rho_i}{\rho_i}-\frac{3\delta\rho_\gamma}{4\rho_\gamma}.
\ee
All modes are assumed to have a power spectrum with the same spectral index
$\ns$, so that $\alphaiso$ is independent of scale.
For positive $\alphaiso$ this agrees with the definitions in
\cite{Bean:2006qz} and \cite{Larson:2010gs} for $\alpha_{-1}$, but also allows
for the correlation to have the opposite sign.  Approximately,
${\rm sgn}(\alphaiso)\,\alphaiso^2\approx B_{\rm c}$, where $B_{\rm c}$ is the
CDM version of the amplitude defined as in~\cite{Amendola:2001ni}. Note that
in our conventions, negative values of $\alphaiso$ \emph{lower} the
Sachs--Wolfe contribution to the large-scale $TT$ power spectrum. We caution
the reader that this convention differs from some others, e.g.,
\citet{Larson:2010gs}.

\planck\ constraints on the correlated isocurvature amplitude are shown in
Fig.~\ref{fig:correlated_iso}, with and without high-multipole polarization.
The corresponding marginalized limit from the temperature data is
\be
\alphaiso = -0.0025^{+0.0035}_{-0.0047} \quad\twosig{\planckTT}, \label{eq:isoalpha}
\ee
which is significantly tightened around zero when \Planck\ polarization
information is included at high multipoles:
\be
\alphaiso = 0.0003^{+0.0016}_{-0.0012} \quad\twosig{\planckall}. \label{eq:isoalphapol}
\ee
This strongly limits the isocurvature contribution to be less than about
3\,\% of the adiabatic modes.
Figure~\ref{fig:isorainbows} shows how models with negative correlation
parameter,  $\alpha$, fit the temperature data at low multipoles slightly better
than models with $\alpha=0$; however, these models are disfavoured from the
corresponding change in the polarization acoustic peaks.

In this model most of the gain in sensitivity comes from relatively large
scales, $\ell \la 300$, where the correlated isocurvature modes with delayed
phase change the first polarization acoustic peak ($\ell\approx 140$)
significantly more than in temperature~\citep{Bucher:2000hy}. The polarization
data are not entirely robust to systematics on these scales, but in this case
the result appears to be quite stable between the different likelihood
codes. However, it should be noted that a particularly low point in the
$TE$ spectrum at $\ell \approx 160$ (see Fig.~\ref{pgTE+EE_final}) pulls in
the direction of positive $\alpha$, and could be giving an artificially
strong constraint if this were caused by an unidentified systematic.

\subsubsection{Curvature}
\label{subsubsec:curv}

\begin{figure}[t]
\centering
\includegraphics[width=\hsize]{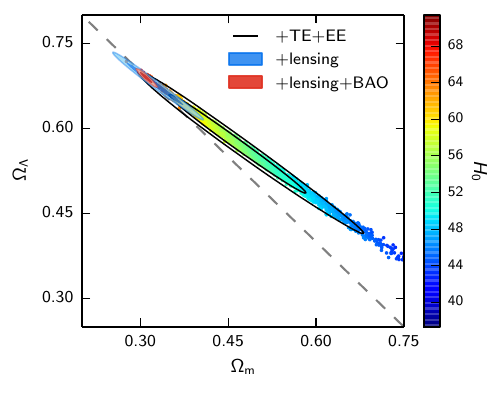}
\caption {Constraints in the $\Omega_{\rm m}$--$\Omega_\Lambda$ plane from the
\planckTT\ data (samples; colour-coded by the value of $H_0$)
and \planckall\ (solid contours). The geometric degeneracy
between $\Omega_{\rm m}$ and $\Omega_{\Lambda}$ is partially broken because
of the effect of lensing on the temperature and polarization power spectra.
These limits are improved significantly by the inclusion of the \Planck\
lensing reconstruction (blue contours) and BAO (solid red contours). The red
contours tightly constrain the geometry of our Universe to be nearly flat.}
\label{fig:geom}
\end{figure}

The simplifying assumptions of large-scale homogeneity and isotropy
lead to the familiar Friedman-Lema{\^\i}tre-Robertson-Walker (FLRW) metric that
appears to be an accurate description of our Universe. The base \LCDM\
cosmology assumes an FLRW metric with a flat 3-space. This is a very
restrictive assumption that needs to be tested empirically. In this subsection,
we investigate constraints on the parameter $\omegak$, where for \LCDM\ models
$\omegak \equiv 1 - \Omega_{\rm m} -\Omega_\Lambda$.
For FLRW models $\omegak >0$ corresponds to negatively-curved 3-geometries
while $\omegak < 0$ corresponds to positively-curved 3-geometries.
Even with perfect data within our past lightcone, our inference of
the curvature $\Omega_K$ is limited by the cosmic variance of curvature
perturbations that are still super-horizon at the present, since these cannot
be distinguished from background curvature within our observable volume.

The parameter $\omegak$ decreases exponentially with time during
inflation, but grows only as a power law during the radiation and
matter-dominated phases, so the standard inflationary prediction has
been that curvature should be unobservably small today.  Nevertheless,
by fine-tuning parameters it is possible to devise inflationary models
that generate open \citep[e.g.,][] {Bucher:1994gb,Linde:1999} or
closed universes
\citep[e.g.,][] {Linde:2003}. Even more speculatively, there has been 
interest recently in multiverse models, in which
topologically-open ``pocket universes'' form by bubble
nucleation~\citep[e.g.,][]{Coleman:1980aw,Gott:1982zf}
between different vacua of a ``string landscape'' \citep[e.g.,][]{Freivogel:2005vv, Bousso:2013}. Clearly, the detection of a significant
deviation from $\omegak=0$ would have profound consequences for inflation theory
and fundamental physics.

The \planck\ power spectra give the constraint
\be
\omegak= -0.052^{+0.049}_{-0.055} \quad \twosig{\planckTT}
 \label{omegak:planck}.
\ee
The well-known geometric degeneracy \citep{Bond:1997wr,Zaldarriaga:1997ch} allows for
the small-scale linear CMB spectrum to remain almost
unchanged if changes in $\omegak$ are compensated by changes in $H_0$ to obtain the same angular diameter distance to last scattering.
The \planck\ constraint is therefore mainly determined by the (wide)
priors on $H_0$, and the effect of lensing smoothing on the power
spectra. As discussed in Sect.~\ref{sec:lensing}, the \planck\ temperature
power spectra show a slight preference for more lensing than expected in the
base \LCDM\ cosmology, and since positive curvature increases the amplitude
of the lensing signal, this preference
also drives $\omegak$ towards negative values.

Taken at face value, Eq.~\eqref{omegak:planck} represents a detection
of positive curvature at just over $2\,\sigma$, largely via the impact of
lensing on the power spectra. One might wonder whether this is mainly
a parameter volume effect, but that is not the case, since the best fit
closed model has $\Delta\chi^2\approx 6$ relative to base \LCDM, and the fit is
improved over almost all the posterior volume, with the mean improvement
being $\langle \Delta\chi^2\rangle\approx 5$ (very similar
to the phenomenological case of \LCDM+$\Alens$). Addition of the \planck\
polarization spectra shifts $\omegak$ towards zero by $\Delta \omegak \approx
0.015$:
\be
\omegak= -0.040^{+0.038}_{-0.041} \quad \twosig{\planckall},
 \label{omegak:planckall}
\ee
but $\omegak$ remains negative at just over $2\,\sigma$.

What's more, the lensing reconstruction from \planck\ measures the lensing
amplitude directly and, as discussed in Sect.~\ref{sec:lensing}, this does
{\it not\/} prefer more lensing than base \LCDM. The
combined constraint shows impressive consistency with a flat universe:
\be
\omegak= -0.005^{+0.016}_{-0.017} \quad \twosig{\planckTTlensing}
 \label{omegak:plensing}.
\ee
The dramatic improvement in the error bar is another illustration of the power
of the lensing reconstruction from \planck.

The constraint can be sharpened further by adding external data that break
the main geometric degeneracy. Combining the \planck\ data with BAO, we find
\be
\omegak= 0.000\pm 0.005 \quad \twosig{ \planckTTlensing\dataplus\BAO}
 \label{omegak:planckall}.
\ee
This constraint is unchanged at the quoted precision if we add the JLA
supernovae data and the $H_0$ prior of Eq.~(\ref{H0prior1}).

Figure~\ref{fig:geom} illustrates these results in the $\Omm$--$\Oml$
plane.  We adopt Eq.~(\ref{omegak:planckall}) as our most reliable 
constraint on spatial curvature. Our Universe appears to be spatially
flat to a $1\,\sigma$ accuracy of 0.25\,\%.

\subsection{Dark energy\label{sec:dark_energy}}

\begin{figure}[!t]
\begin{center}
\includegraphics[width=0.5\textwidth]{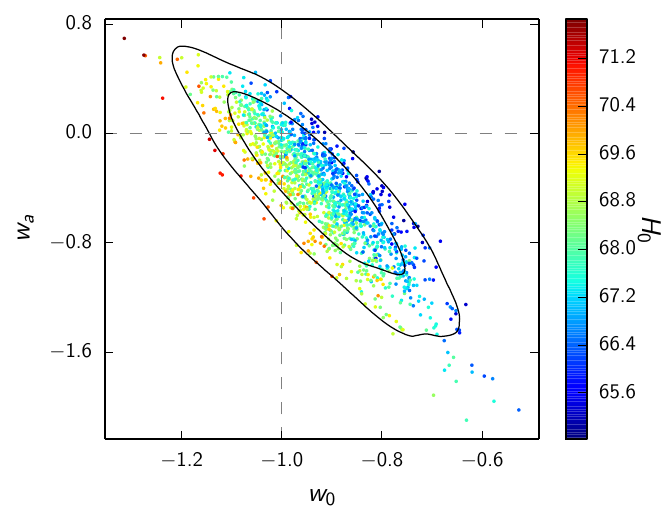}
\caption{Samples from the distribution of the dark energy parameters $w_0$
and $w_a$ using \planckTT+BAO+JLA data, colour-coded by the value of the Hubble
parameter $H_0$. Contours show the corresponding 68\,\% and 95\,\% limits.
Dashed grey lines intersect at the point in parameter space corresponding to
a cosmological constant.}
\label{2D_wwa}
\end{center}
\end{figure}

The physical explanation for the observed accelerated expansion of the
Universe is currently not known. In standard \LCDM\ the acceleration is
provided by a cosmological constant, i.e., an additional fluid
satisfying an equation of state
$w\equiv p_{\rm DE}/\rho_{\rm DE} = -1$. However, there are many possible
alternatives, typically described either in terms of extra degrees of freedom
associated with scalar fields or modifications of general relativity
on cosmological scales \citep[for reviews see,
e.g.,][]{Copeland:2006,Tsujikawa:2010}. A detailed study of these
models and the constraints imposed by \Planck\ and other data are
presented in a separate paper, \cite{planck2014-a16}.

Here we will limit ourselves to the most basic extensions of \LCDM, which
can be phenomenologically described in terms of the equation of state
parameter $w$ alone. Specifically we will use the
\CAMB\ implementation of the ``parameterized post-Friedmann'' (PPF)
framework of \cite{Hu:2007pj} and \cite{Fang:2008sn} to test whether there is
any evidence that $w$ varies with time. This framework aims to recover the
behaviour of canonical (i.e., those with a standard kinetic term)
scalar field cosmologies minimally coupled to
gravity when $w\ge -1$, and accurately approximates them for values
$w\approx -1$.  In these models the speed of sound is equal to the speed of
light, so that the clustering of the dark energy inside the horizon is
strongly suppressed. The advantage of using the PPF formalism is that
it is possible to study the phantom domain, $w < -1$, including
transitions across the ``phantom barrier,'' $w=-1$, which is
not possible for canonical scalar fields.

The CMB temperature data alone do not tightly constrain $w$, because of a
strong geometrical degeneracy, even for spatially-flat models. From \Planck\ we
find
\begin{equation}
w = -1.54^{+0.62}_{-0.50} \quad \twosig{\planckTT},
\label{eq:rawwlimit}
\end{equation}
i.e., almost a $2\,\sigma$ shift into the phantom domain.
This is partly, but not entirely, a parameter volume effect, with the average
effective $\chi^2$ improving by $\langle \Delta\chi^2 \rangle \approx 2$
compared to base \LCDM. This is consistent with the
preference for a higher lensing amplitude
discussed in Sect.~\ref{subsec:Alens},
improving the fit in the $w<-1$ region,
where the lensing smoothing amplitude becomes slightly larger.
However, the lower limit in Eq.~\eqref{eq:rawwlimit} is largely determined by
the (arbitrary) prior $H_0<100\,\Hunit$, chosen
for the Hubble parameter.  Much of the posterior volume in the phantom region
is associated with extreme values for cosmological parameters, which are
excluded by other astrophysical data. The mild tension with base \LCDM\
disappears as we add more data that break the geometrical degeneracy. Adding
\Planck\ lensing and BAO, JLA and
$H_0$ (``\ext'') gives the 95\,\% constraints
\beglet
\begin{eqnarray}
w &=& -1.023^{+0.091}_{-0.096} \quad\datalabel{\planckTT\dataplus\ext},
 \label{DE1a} \\
w &=& -1.006^{+0.085}_{-0.091} \quad\datalabel{\planckTTlensext},
 \label{DE1b} \\
w &=& -1.019^{+0.075}_{-0.080} \quad\datalabel{\planckalllensing\dataplus\ext}. \, \nonumber \\
 & & \quad \label{DE1c}
\end{eqnarray}
\endlet
The addition of \Planck\ lensing, or using the full
\Planck\ temperature+polarization likelihood together with the BAO, JLA, and
$H_0$ data does not substantially improve the constraint of Eq.~\eqref{DE1a}.
All of these data set combinations are compatible with the
base \LCDM\ value of $w=-1$. In {\paramsI}, we conservatively quoted
$w=-1.13^{+0.24}_{-0.25}$, based on combining \Planck\ with BAO, as our most
reliable limit on $w$. The errors in Eqs.~(\ref{DE1a})--(\ref{DE1c}) are
substantially smaller, mainly because of the addition of the JLA SNe data,
which offer a sensitive probe of the dark energy equation of state at
$z \la 1$. In {\paramsI}, the addition of the SNLS SNe data pulled $w$ into
the phantom domain at the $2\,\sigma$ level, reflecting the tension between
the SNLS sample and the \Planck\ 2013 base \LCDM\ parameters.
As noted in Sect.~\ref{sec:SNe}, this discrepancy is no longer present,
following improved photometric calibrations of the
SNe data in the JLA sample. One consequence of this is the tightening of the
errors in Eqs.~(\ref{DE1a})--(\ref{DE1c})
around the \LCDM\ value $w=-1$ when we combine the JLA sample with \Planck.

If $w$ differs from $-1$, it is likely to change with time. We consider here
the case of a Taylor expansion of $w$ at first order in the scale factor,
parameterized by
\begin{equation}
w = w_0 + (1-a) w_a. \label{DE2}
\end{equation}
More complex models of dynamical dark energy are discussed in
\cite{planck2014-a16}.  Figure~\ref{2D_wwa} shows the 2D marginalized
posterior distribution for $w_0$ and $w_a$ for the
combination \Planck+BAO+JLA.  The JLA SNe data are again crucial in breaking
the geometrical degeneracy at low redshift and with these data we find no
evidence for a departure from the base \LCDM\ cosmology. The points in
Fig.~\ref{2D_wwa} show samples from these chains colour-coded
by the value of $H_0$. From these MCMC chains, we find
$H_0 = (68.2 \pm 1.1)\Hunit$.
Much higher values of $H_0$ would favour the phantom regime, $w<-1$.

\begin{figure}[!t]
\begin{center}
\includegraphics[width=90mm]{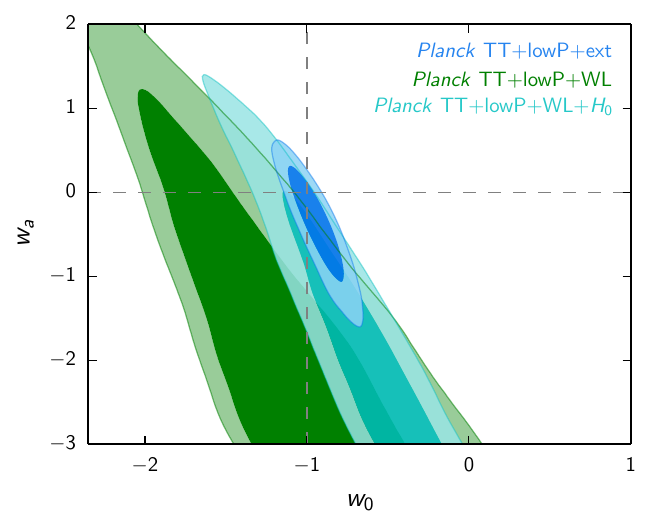}
\caption{Marginalized posterior distributions for ($w_0, w_a$) for various data
combinations. We show
$\planckTT$ in combination with BAO, JLA, $H_0$ (``ext''), and two data combinations
that add  the \CFHTLENS\ data with ultra-conservative cuts  as described in the text
(denoted ``WL''). Dashed grey lines show the parameter values corresponding to a
cosmological constant.}
\label{2D_wwa_separate_datasets}
\end{center}
\end{figure}

As pointed out in Sects.~\ref{subsec:WL} and \ref{subsec:data_summary}
the \CFHTLENS\ weak lensing data are in tension with the \Planck\ base
\LCDM\ parameters. Examples of this tension can be seen in
investigations of dark energy and modified gravity, since some of
these models can modify the growth rate of fluctuations from the base
\LCDM\ predictions. This tension can be seen even in the simple model
of Eq.~\eqref{DE2}. The green regions in Fig.~\ref{2D_wwa_separate_datasets}
show 68\,\% and 95\,\% contours in the
$w_0$--$w_a$ plane for {\planckTT} combined with the \CFHTLENS\ H13
data. In this example, we have applied ultra-conservative cuts,
excluding $\xi_{-}$ entirely and excluding measurements with $\theta <
17$\arcmin \  in $\xi_{+}$ for all tomographic redshift bins. As
discussed in \cite{planck2014-a16}, with these cuts the
\CFHTLENS\ data are insensitive to modelling the nonlinear evolution
of the power spectrum, but this reduction in sensitivity comes 
at the expense of reducing the statistical power
of the weak lensing data. Nevertheless,
Fig.~\ref{2D_wwa_separate_datasets} shows that the combination of
\planck+\CFHTLENS\ pulls the contours into the phantom domain and is
discrepant with base \LCDM\ at about the $2\,\sigma$ level.
The \planck+\CFHTLENS\
data also favour a high value of $H_0$. If we add the (relatively weak)
$H_0$ prior of Eq.~\eqref{H0prior1}, the contours (shown in cyan) in
Fig.~\ref{2D_wwa_separate_datasets} shift towards $w=-1$. It therefore seems
unlikely that the tension between \Planck\ and \CFHTLENS\ can be resolved by
allowing a time-variable equation of state for dark energy.

A much more extensive investigation of models of dark energy and also
models of modified gravity can be found in \cite{planck2014-a16}.  The
main conclusions of that analysis are as follows:
\begin{itemize}
\item an investigation of more general time-variations of the equation of
state shows a high degree of consistency with $w=-1$;
\item
a study of several dark energy and modified gravity models either finds
compatibility with base \LCDM,  or mild tensions, which are driven mainly by
external data sets.
\end{itemize}

\subsection{Neutrino physics and constraints on relativistic components}
\label{sec:neutrino}

In the following subsections, we update \Planck\ constraints on the
mass of standard (active) neutrinos, additional relativistic degrees
of freedom, models with a combination of the two, and models with
massive sterile neutrinos. In each subsection we emphasize the
\planck-only constraint, and the implications of the \planck\ result
for late-time cosmological parameters measured from  other
observations. We then give a brief discussion of tensions between
\planck\ and some discordant external data, and assess whether any of these
model extensions can help to resolve them.  Finally we provide
constraints on neutrino interactions.

\subsubsection{Constraints on the total mass of active neutrinos}
\label{sec:mnu}

Detection of neutrino oscillations has proved that neutrinos have mass
(see, e.g., \citealt{Lesgourgues:2006} and \citealt{NakamuraPetcov2014} for
reviews).  The \planck\ base
\LCDM\ model assumes a normal mass hierarchy with $\sumnu \approx
0.06\,\eV$ (dominated by the heaviest neutrino mass eigenstate) but
there are other possibilities, including a degenerate hierarchy with
$\sumnu\ga 0.1\,\eV$.  At this time there are no compelling
theoretical reasons to strongly prefer any of these possibilities, so
allowing for larger neutrino masses is perhaps one of the most well-motivated extensions to base \LCDM\ considered in this paper. There
has also been significant interest recently in larger neutrino masses
as a possible way to lower $\sigma_8$ (the late-time fluctuation
amplitude), and thereby reconcile \planck\ with weak lensing measurements and
the abundance of rich clusters (see Sects.~\ref{sec:additional_data}
and \ref{subsec:data_summary}).  Though model dependent, neutrino mass
constraints from cosmology are already significantly stronger than those
from tritium $\beta$-decay experiments \citep[see, e.g.,][]{Drexlin:2013}.

Here we give constraints assuming three species of degenerate massive
neutrinos, neglecting the small differences in mass expected from the
observed mass splittings. At the level of sensitivity of \Planck\ this
is an accurate approximation, but note that it does not quite match
continuously on to the base \LCDM\ model (which assumes two massless
and one massive neutrino with $\mnu=0.06\,\eV$). We assume that the
neutrino mass is constant, and that the distribution function is
Fermi-Dirac with zero chemical potential.

\begin{figure}
\centering
\includegraphics[width=92mm]{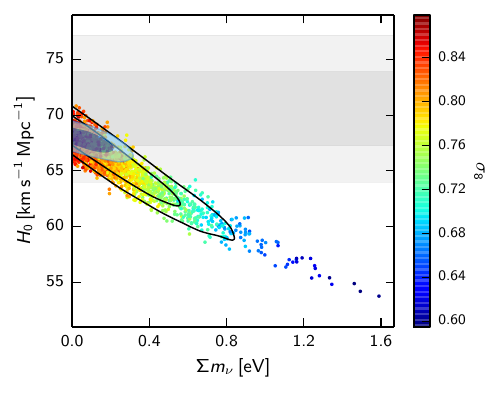}
\caption{Samples from the \planckTT\ posterior in the $\mnu$--$H_0$
 plane, colour-coded by $\sigma_8$. Higher $\mnu$ damps the matter
 fluctuation amplitude $\sigma_8$, but also decreases $H_0$.  The grey
 bands show the direct measurement, $H_0=(70.6\pm 3.3)\,\Hunit$,
 Eq.~\eqref{H0prior1}. Solid black contours show the constraint from
 \planckTT+\lensing\ (which mildly prefers larger masses), and filled
 contours show the  constraints from  \planckTT+\lensing+BAO.
\label{fig:mnu-H-sigma}
}
\end{figure}

Masses well below $1\,\eV$ have only a mild effect on the shape of the
CMB power spectra, since they became non-relativistic after
recombination.  The effect on the background cosmology can be
compensated by changes in $H_0$, to ensure the same observed acoustic
peak scale $\theta_*$. There is, however, some sensitivity of the CMB
anisotropies to neutrino masses as the neutrinos start to become less
relativistic at recombination (modifying the early ISW effect), and
from the late-time effect of lensing on the power spectrum. The
\planck\ power spectrum (95\,\%) constraints are
\beglet
\begin{eqnarray}
\sumnu &<& 0.72\,\eV \quad \datalabel{\planckTT}, \\
\sumnu &<& 0.21\,\eV \quad \datalabel{\planckTTBAO} \label{nu-no-lensing}, \\
\sumnu &<& 0.49\,\eV \quad \datalabel{\planckall}, \\
\sumnu &<& 0.17\,\eV \quad \datalabel{\planckallBAO}.
\label{nu-no-lensing1}
\end{eqnarray}
\endlet
The \planckTT\ constraint has a broad tail to high masses, as shown in
Fig.~\ref{fig:mnu-H-sigma}, which also illustrates the acoustic scale
degeneracy with $H_0$.  Larger masses imply a lower $\sigma_8$ through
the effects of neutrino free-streaming on structure formation, but the
larger masses also require a lower Hubble constant, leading to possible
tensions with direct measurements of $H_0$. Masses below about $0.4\,\eV$ can
provide an acceptable fit to the direct $H_0$ measurements, and adding the
BAO data helps to break the acoustic scale degeneracy and tightens the
constraint on $\sumnu$ substantially. Adding \Planck\ polarization data
at high multipoles produces a relatively small improvement to the \planckTT+BAO
constraint (and the improvement is even smaller with the alternative
\camspec\ likelihood), so we consider the $TT$ results to be  our most
reliable constraints.

The constraint of Eq.~\eqref{nu-no-lensing} is consistent with the 95\,\% limit
of $\sumnu < 0.23\,\eV$ reported in {\paramsI} for \Planck+BAO. The limits are
similar because the linear CMB is insensitive to the mass of neutrinos that
are relativistic at recombination.  There is little to be gained from improved
measurement of the CMB temperature power spectra, though improved external data
can help to break the geometric degeneracy to higher precision.
CMB lensing can also provide additional information at lower redshifts, and
future high-resolution CMB polarization measurements that accurately reconstruct
the lensing potential can probe much smaller masses
\citep[see, e.g.][]{Abazajian:2013oma}.

\begin{figure}
\centering
\includegraphics[width=\hsize]{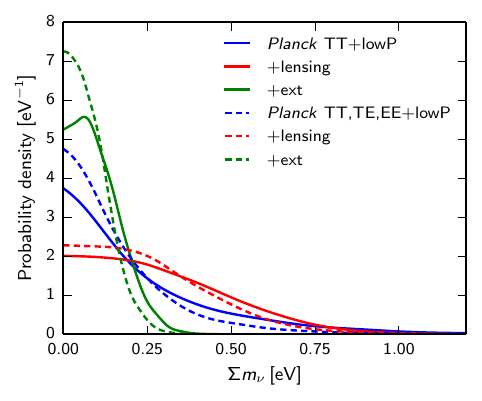}
\caption{Constraints on the sum of the neutrino masses for various data
combinations.
\label{fig:mnu-compare}
}
\end{figure}

As discussed in detail in {\paramsI} and Sect.~\ref{sec:lensing}, the
\planck\ CMB power spectra prefer somewhat more lensing smoothing than
predicted in \LCDM\ (allowing the lensing amplitude to vary gives
$\Alens>1$ at just over $2\,\sigma$). The neutrino mass constraint from
the power spectra is therefore quite tight, since increasing the
neutrino mass lowers the predicted smoothing even further compared to
base \LCDM. On the other hand the lensing reconstruction data, which
directly probes the lensing power, prefers lensing amplitudes slightly
below (but consistent with) the base \LCDM\ prediction
(Eq.~\ref{CMBlens}).  The \planck+lensing constraint therefore pulls
the constraints slightly away from zero towards higher neutrino
masses, as shown in Fig.~\ref{fig:mnu-compare}.  Although the
posterior has less weight at zero, the lensing data are
incompatible with very large neutrino masses so the
\planck+lensing 95\,\% limit is actually
tighter than the \planckTT\ result:
\be
\sumnu < 0.68\,\eV \quad \twosig{\planckTTlensing}.
\ee
Adding the polarization spectra improves this constraint slightly to
\be
\sumnu < 0.59\,\eV \quad \twosig{\planckalllensing}.
\ee
We take the combined constraint that further includes BAO, JLA, and $H_0$
(``ext'') as our best limit:
\twotwosig{\sumnu &< 0.23\,\eV}{\Omega_\nu h^2 &< 0.0025}
 {\planckTTlensext. \label{Mnu1}}
This is slightly weaker than the constraint from \planckall+lensing+BAO
(which is tighter in both the \camspec\ and \plik\ likelihoods), but is immune
to low level systematics that might affect the constraints from the \Planck\
polarization spectra.  Equation~\eqref{Mnu1} is therefore a conservative limit.
Marginalizing over the range of neutrino masses, the \planck\ constraints on
the late-time parameters are\footnote{To simplify the
displayed equations, $H_0$ is given in units of $\Hunit$ in this section.}
\twoonesig{H_0 = 67.7\pm 0.6}{\sigma_8= 0.810^{+0.015}_{-0.012}}
 {\planckTTlensext.}
For this restricted range of neutrino masses, the impact on the other
cosmological parameters is small and, in particular, low values of $\sigma_8$
will remain in tension with the parameter space preferred by \planck.

The constraint of Eq.~\eqref{Mnu1} is weaker than the constraint of
Eq.~\eqref{nu-no-lensing} excluding lensing, but there is no good
reason to disregard the \Planck\ lensing information while retaining other
astrophysical data. The CMB lensing signal probes very-nearly linear scales and
passes many consistency checks over the multipole range used in the
\Planck\ lensing likelihood (see Sect.~\ref{sec:lensing} and
\citealt{planck2014-a17}). The situation with galaxy weak lensing is
rather different, as discussed in Sect.~\ref{subsec:WL}. In addition to
possible observational systematics, the weak lensing data probe lower
redshifts than CMB lensing,  and smaller spatial scales, where uncertainties
in modelling nonlinearities in the matter power spectrum and baryonic
feedback become important~\citep{Harnois-Deraps:2014sva}.

A larger range of neutrino masses was found by~\cite{Beutler:2014yhv}
using a combination of RSD, BAO, and weak lensing information. The
tension between the RSD results and base \LCDM\ was subsequently
reduced following the analysis of~\cite{Samushia:14}, as shown in
Fig.~\ref{Samushia}. Galaxy weak lensing and some cluster constraints
remain in tension with base \LCDM, and we discuss possible neutrino
resolutions of  these problems in Sect.~\ref{sec:neutrino_tension}.

Another way of potentially improving neutrino mass constraints is to use measurements of the Ly$\alpha$ flux
power spectrum of high-redshift quasars. \citet{palanque:2014} have recently reported an
analysis of a large sample of quasar spectra from the SDSSIII/BOSS survey. When combining their results
with 2013 Planck data, these authors find a bound $\sumnu < 0.15\,\eV$  (95\,\% CL),
compatible with the results presented in this section.

An exciting future prospect is the possible direct detection of
non-relativistic cosmic neutrinos by capture on tritium, for example
with the PTOLEMY
experiment~\citep{Cocco:2007za,Betts:2013uya,Long:2014zva}. Unfortunately,
for the mass range $\sumnu < 0.23\,\eV$ preferred by \planck, detection
with the first generation experiment will be extremely difficult.

\subsubsection{Constraints on $\neff$}
\label{sec:Neff}

\begin{figure}[t]
\centering
\includegraphics[width=92mm]{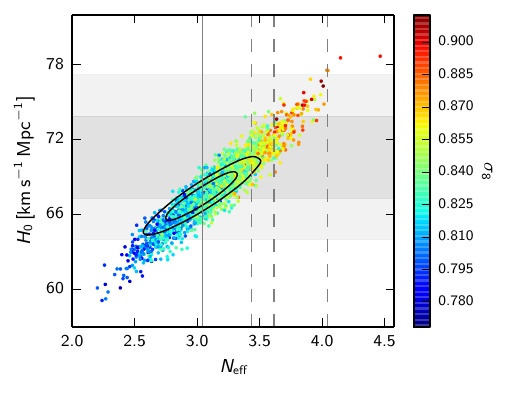}
\caption{Samples from \planckTT\ chains in the
$\neff$--$H_0$ plane, colour-coded by $\sigma_8$.  The grey bands show the
constraint $H_0=(70.6\pm 3.3)\,\Hunit$ of Eq.~\eqref{H0prior1}.
Notice that higher $\neff$ brings $H_0$ into better consistency with direct
measurements, but increases $\sigma_8$.
Solid black contours show the constraints from \planckall+BAO.
Models with $\neff < 3.046$ (left of the solid vertical line) require photon
heating after neutrino decoupling or incomplete thermalization. Dashed vertical
lines correspond to specific fully-thermalized particle models, for example
one additional massless boson that decoupled around the same time as the
neutrinos ($\Delta \nnu \approx 0.57$), or before muon annihilation
($\Delta \nnu \approx 0.39$), or an additional sterile neutrino that decoupled
around the same time as the active neutrinos ($\Delta \nnu \approx 1$).
\label{fig:nnu-H-sigma}
}
\end{figure}

Dark radiation density in the early Universe is usually parameterized
by  $\neff$, defined so that the total relativistic
energy density in neutrinos and any other dark radiation is given in
terms of the photon density $\rho_\gamma$ at $T\ll 1\,\MeV$ by
\be \rho
= \neff \frac{7}{8}\left(\frac{4}{11} \right)^{4/3} \rho_\gamma.
\ee
The numerical factors in this equation are included so that $\neff=3$
for three standard model neutrinos that were thermalized in the early
Universe and decoupled well before electron-positron annihilation.
The standard cosmological prediction is actually $\neff=3.046$, since
neutrinos are not completely decoupled at electron-positron
annihilation and are subsequently slightly
heated~\citep{Mangano:2001iu}.

In this section we focus on additional energy density from massless
particles. In addition to massless sterile neutrinos, a variety of
other particles could contribute to $\neff$. We assume that the
additional massless particles are produced well before recombination,
and neither interact nor decay, so that their energy density scales
with the expansion exactly like massless neutrinos. An additional
$\Delta \neff=1$ could correspond to a fully thermalized sterile
neutrino that decoupled at $T \la 100\,\MeV$; for example, any sterile
neutrino with mixing angles large enough to provide a potential
resolution to short-baseline reactor neutrino oscillation anomalies
would most likely thermalize rapidly  in the early Universe. However,
this solution to the neutrino oscillation anomalies requires approximately
1-eV sterile neutrinos, rather than the massless case considered in this
section; exploration of the two parameters $\neff$ and $ \sumnu$ is
reported in Sect.~\ref{subsec:joint_neff_mnu}. For a review of sterile
neutrinos see \cite{Abazajian:2012}.

More generally the additional radiation does not need to be fully
thermalized, for example there are many possible models of non-thermal
radiation production via particle decays \citep[see,
e.g.,][]{Hasenkamp:2012ii,Conlon:2013isa}. The radiation could also
be produced at temperatures $T> 100\,\MeV$, in which case typically
$\Delta \neff<1$ for each additional species, since heating by photon
production at muon annihilation (corresponding to $T\approx 100\,\MeV$)
decreases the fractional importance of the additional component at the later
times relevant for the CMB. For particles produced at $T\gg 100\,\MeV$ the
density would be diluted even more by numerous phase transitions and
particle annihilations, and give $\Delta \neff\ll 1$. Furthermore, if
the particle is not fermionic, the factors entering the entropy
conservation equation are different, and even thermalized particles
could give specific fractional values of $\Delta \neff$. For
example~\cite{Weinberg:2013kea} considers the case of a thermalized
massless boson, which contributes $\Delta\neff = 4/7\approx 0.57$ if
it decouples in the range $0.5\,\MeV < T<100\,\MeV$ like the neutrinos,
or $\Delta \neff \approx 0.39$ if it decouples at $T>100\,\MeV$ (before the
photon production at muon annihilation, hence undergoing
fractional dilution).

In this paper we follow the usual phenomenological approach, where one
constrains $\neff$ as a free parameter with a wide flat prior, although we
comment on a few discrete cases separately below. Values
of $\neff < 3.046$ are less well motivated, since they would require the
standard neutrinos to be incompletely thermalized or additional photon
production after neutrino decoupling, but we include this range for
completeness.

Figure~\ref{fig:nnu-H-sigma} shows that \planck\ is entirely consistent with
the standard value $\neff = 3.046$.  However, a significant density of
additional radiation is still allowed, with the (68\,\%) constraints
\beglet
\begin{eqnarray}
\nnu &=& 3.13 \pm 0.32 \quad \datalabel{\planckTT}, \label{Nu1}\\
\nnu &=& 3.15 \pm 0.23 \quad \datalabel{\planckTTBAO}, \label{Nu2}\\
\nnu &=& 2.99 \pm 0.20 \quad \datalabel{\planckall}, \label{Nu3}\\
\nnu &=& 3.04 \pm 0.18 \quad \datalabel{\planckallBAO}. \label{Nu4}
\end{eqnarray}
\endlet
Notice the significantly tighter constraint with the inclusion of \planck\
high-$\ell$ polarization, with $\Delta \nnu< 1$ at over $4\,\sigma$ from
\planck\ alone.  This constraint is not very stable between likelihoods, with
the  \camspec\ likelihood giving a roughly $0.8\,\sigma$
\emph{lower\/} value of $\nnu$. However, the strong limit from polarization is
also consistent with the joint \planckTT+BAO result, so Eq.~\eqref{Nu2} leads
to the robust conclusion that $\Delta \nnu< 1$ at over $3\,\sigma$.  The
addition of \Planck\ lensing has very little effect on this constraint.

For $\neff>3$, the \planck\ data favour higher values of the
Hubble parameter than the \Planck\ base \LCDM\ value, which as discussed
in Sect.~\ref{sec:hubble} may be in better agreement with some direct
measurements of $H_0$ . This is because \planck\ accurately measures the
acoustic scale $\rstar/\DAstar$; increasing $\neff$ means (via the Friedmann
equation) that the early Universe expands faster, so the sound horizon
at recombination, $\rstar$, is smaller and hence recombination has to be
closer (larger $H_0$ and hence smaller $\DAstar$) for it to subtend the
same angular size observed by \planck. However, models with $\neff>3$ and a
higher Hubble constant also have higher values of the fluctuation amplitude
$\sigma_8$, as shown by the coloured samples in Fig.~\ref{fig:nnu-H-sigma}.
As a result, these models {\it increase\/} the tensions
between the CMB measurements and astrophysical measurements of $\sigma_8$
discussed in Sect.~\ref{subsec:data_summary}. It therefore seems
unlikely that additional radiation alone can help to resolve tensions
with large-scale structure data.

The energy density in the early Universe can also be probed by the
predictions of BBN. In particular
$\Delta\neff>0$ increases the primordial expansion rate, leading to
earlier freeze-out, with a higher neutron density and hence a greater
abundance of helium and deuterium after BBN has completed. A detailed
discussion of the implications of \Planck\ for BBN is given in
Sect.~\ref{sec:bbn}. Observations of
both the primordial helium and deuterium abundance are compatible with the
predictions of standard BBN for the \Planck\ base \LCDM\ value of the baryon
density. The \Planck+BBN constraints on $\neff$
(Eqs.~\ref{neff_from_cmb_and_helium} and \ref{neff_from_cmb_and_deuterium})
are compatible, and slightly tighter than Eq.~\eqref{Nu2}.

Although there is a large continuous range of plausible $\nnu$ values,
it is worth mentioning briefly a few of the discrete values
from fully thermalized models. This serves as an indication of how strongly
\planck\ prefers base \LCDM, and also how the inferred values of other
cosmological parameters might be affected by this particular extension to base
\LCDM. As discussed above, one fully
thermalized neutrino ($\Delta\nnu\approx 1$) is ruled out at over
$3\,\sigma$, and is disfavoured by $\Delta \chi^2 \approx 8$ compared to
base \LCDM\ by \planckTT, and much more strongly in combination with \planck\
high-$\ell$ polarization or BAO data. The thermalized boson models that give
$\Delta \nnu = 0.39$ or $\Delta\nnu = 0.57$ are disfavoured by
$\Delta\chi^2\approx 1.5$ and $\Delta\chi^2\approx 3$, respectively, and are
therefore not strongly excluded.  We focus on the former, since it is
also consistent with the \planckTT+BAO constraint at $2\,\sigma$.  As shown
in Fig.~\ref{fig:nnu-H-sigma}, larger $\nnu$ corresponds to a region
of parameter space with significantly higher Hubble parameter,
\be H_0 = 70.6\pm 1.0 \quad\onesig{\planckTT;
$\Delta\nnu=0.39$}. \label{eq:nu39H0}
\ee
This can be compared to
the direct measurements of $H_0$ discussed in Sect.~\ref{sec:hubble}.
Evidently, Eq.~\eqref{eq:nu39H0} is consistent with the $H_0$ prior
adopted in this paper (Eq.~\ref{H0prior1}), but this example shows
that an accurate direct measurement of $H_0$ can potentially provide
evidence for new physics beyond that probed by \Planck.
As shown in Fig.~\ref{fig:nnu-H-sigma}, the $\Delta\nnu=0.39$ cosmology also
has a significantly higher small-scale fluctuation amplitude and the spectral
index $\ns$ is also bluer, with
\twoonesig{\sigma_8 = 0.850\pm 0.015}{\ns=0.983\pm0.006}
 {\planckTT; $\Delta\nnu=0.39$.}
The $\sigma_8$ range in this model is higher than preferred by the \planck\
lensing likelihood in base \LCDM. However, the fit to the \Planck\ lensing
likelihood is model dependent and the lensing degeneracy direction also
associates high $H_0$ and low $\Omm$ values with higher $\sigma_8$. The joint
\planckTT+\lensing\ constraint does pull $\sigma_8$ down slightly to
$\sigma_8=0.84\pm 0.01$ and provides an acceptable fit to the \Planck\ data.
For \planckTT+lensing, the difference in $\chi^2$ between the best
fit base \LCDM\ model and the extension with $\Delta\nnu=0.39$ is only
$\Delta \chi^2_{\rm CMB} \approx 2$.
The higher spectral index with $\Delta\nnu=0.39$ gives a decrease in
large-scale power, fitting the low $\ell<30$ \planck\ $TT$ spectrum better
by $\Delta\chi^2\approx 1$, but at the same time the high-$\ell$ data prefer
$\Delta\nnu\approx 0$. Correlations with other cosmological parameters can be
seen in Fig.~\ref{fig:grid_1paramext}.
Clearly, a very effective way of testing these models would be to obtain
reliable, accurate,  astrophysical measurements of $H_0$ and $\sigma_8$.

In summary, models with $\Delta \nnu=1$ are disfavoured by
\planck\ combined with BAO data at about the $3\,\sigma$ level. Models
with fractional changes of $\Delta \nnu \approx 0.39$ are mildly
disfavoured by \Planck, but require higher $H_0$ and $\sigma_8$
compared to base \LCDM.

\subsubsection{Simultaneous constraints on $\neff$ and neutrino mass}
\label{subsec:joint_neff_mnu}

\begin{figure}[t]
\centering
\includegraphics[width=92mm]{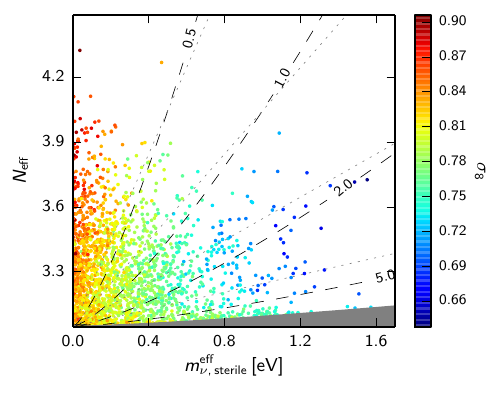}
\caption{Samples from \planckTT\ in the
$\neff$--$\mnusterile$ plane, colour-coded by $\sigma_8$, for models
with one massive sterile neutrino family, with effective mass $\mnusterile$,
and the three active neutrinos as in the base \lcdm\ model.
The physical mass of the sterile neutrino in the thermal scenario,
$m_{\rm sterile}^{\rm thermal}$, is constant along the grey dashed lines, with
the indicated mass in $\mathrm{eV}$; the grey shading shows the region
excluded by our prior $m_{\rm sterile}^{\rm thermal}< 10\,\eV$,
which cuts out most of the area where the neutrinos behave nearly like
dark matter.  The physical mass in the Dodelson-Widrow
scenario, $m_{\rm sterile}^{\rm DW}$, is constant along the dotted lines
(with the value indicated on the adjacent dashed lines).
\label{fig:meffsterile-sigma8}
}
\end{figure}

As discussed in the previous sections, neither a higher neutrino mass
nor additional radiation density alone can
resolve all of the tensions between \planck\ and other astrophysical data.
However, the presence of additional massive
particles, such as massive sterile neutrinos, could potentially
improve the situation by introducing enough freedom to allow
higher values of the Hubble constant \emph{and\/}
lower values of $\sigma_8$. As mentioned in Sect.~\ref{sec:Neff},
massive sterile neutrinos offer a possible solution to reactor neutrino
oscillation anomalies~\citep{Kopp:2013vaa,Giunti:2013aea} and this has led to
significant recent interest in this class of
models~\citep{Hamann:2013iba,Wyman:2013lza,Battye:2013xqa,Leistedt:2014sia,Bergstrom:2014fqa,MacCrann:2014wfa}. Alternatively,
active neutrinos could have significant degenerate masses above the
minimal baseline value together with additional massless particles contributing
to $\neff$. Many more complicated scenarios could also be envisaged.

\begin{figure*}[t]
\centering
\includegraphics[width=\hsize]{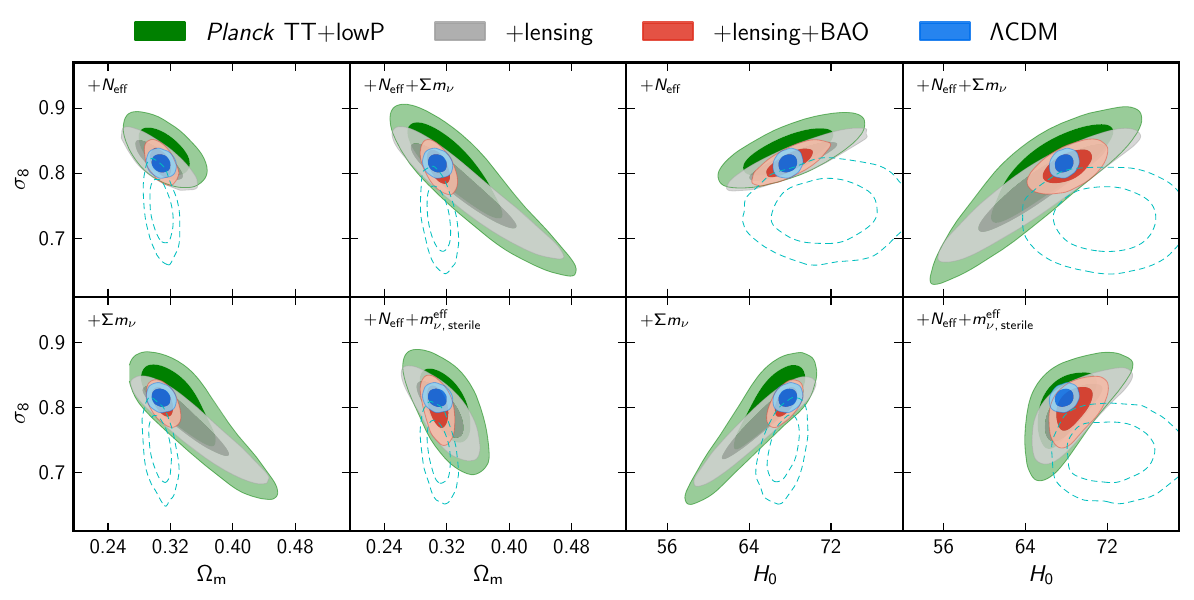}
\caption{68\,\% and 95\,\% constraints from \planckTT\ (green),
 \planckTT+lensing (grey), and \planckTT+lensing+BAO (red) on the
 late-Universe parameters $H_0$, $\sigma_8$, and $\Omm$ in various
 neutrino extensions of the base \LCDM\ model. The blue contours show
 the base \LCDM\ constraints from \planckTT+lensing+BAO.  The dashed
 cyan contours show  joint constraints from the H13 \CFHTLENS\ galaxy weak
 lensing likelihood (with angular cuts as in
 Fig.~\ref{WLonly}) at constant CMB acoustic scale $\thetaMC$ (fixed to
 the \planckTT\ \lcdm\ best fit) combined with BAO and the Hubble constant
 measurement of Eq.~\eqref{H0prior1}. These additional constraints break large
 parameter degeneracies in the weak lensing likelihood that would
 otherwise obscure the comparison with the \Planck\ contours.  Here priors
 on other parameters applied to the \CFHTLENS\ analysis are as described
 in Sect.~\ref{subsec:WL}.}
\label{fig:neutrinos_planes}
\end{figure*}

In the case of massless radiation density, the cosmological
predictions are independent of the actual form of the distribution function,
since all particles travel at the speed of light. However, for
massive particles the results are more model dependent. To formulate a
well-defined model, we follow {\paramsI} and consider the
case of one massive sterile neutrino parameterized by
$\mnusterile \equiv (94.1\, \Omega_{\nu,{\rm sterile}}h^2)\,\eV$,
in addition to the two approximately massless and one massive neutrino
of the baseline model.  For thermally-distributed sterile
neutrinos, $\mnusterile$ is related to the true mass via
\begin{equation}
\meffsterile = (\Tsterile/T_\nu)^3 m_{\rm sterile}^{\rm thermal}
 = (\Delta N_{\rm eff})^{3/4} \msthermal \,,
\end{equation}
and for the cosmologically-equivalent Dodelson-Widrow (DW) case
\citep{Dodelson:1994} the relation is given by
\begin{equation}
\meffsterile = \chi_{\rm s}\, \msDW\,,
\end{equation}
with $\Delta N_{\rm eff}=\chi_{\rm s}$.
We impose a prior on the physical thermal mass, $ \msthermal< 10\,\eV$,
when generating parameter chains, to exclude regions of parameter space in
which the particles are so massive that their effect on the CMB spectra
is identical to that of cold dark matter. Although we consider only
the specific case of one massive sterile neutrino with a thermal (or DW)
distribution, our constraints will be reasonably accurate for other models,
for example $\eV$-mass particles produced as non-thermal decay
products~\citep{Hasenkamp:2014hma}.

Figure~\ref{fig:meffsterile-sigma8} shows that although \planck\ is
perfectly consistent with no massive sterile neutrinos, a significant
region of parameter space with fractional $\Delta \neff$ is allowed,
where $\sigma_8$ is lower than in the base \lcdm\ model. This is also
the case for massless sterile neutrinos combined with massive active
neutrinos.  In the single massive sterile model, the combined
constraints are
\twotwosig{\nnu &< 3.7}{\meffsterile &< 0.52\,\eV}{\planckTT+\lensing+BAO.}
The upper tail of $\meffsterile$ is largely associated with high physical
masses near to the prior cutoff; if instead we restrict to the region where
$\msthermal < 2\,\eV$ the constraint is
\twotwosig{\nnu &< 3.7}{\meffsterile &< 0.38\,\eV}{\planckTT+\lensing+BAO.}
Massive sterile neutrinos with mixing angles large enough to help resolve the
reactor anomalies would typically imply full thermalization in the early
Universe, and hence give $\Delta \neff =1$ for each additional species. Such a
high value of $\neff$, especially combined with
$m_{\rm sterile} \approx 1\,\eV$, as required by reactor anomaly solutions,
were virtually ruled out by previous cosmological
data~\citep{Mirizzi:2013kva,Archidiacono:2013xxa,Gariazzo:2013gua}. This
conclusion is strengthened by the analysis presented here, since $\nnu=4$ is
excluded at greater than 99\,\% confidence. Unfortunately, there does not
appear to be a consistent resolution to the reactor anomalies, unless
thermalization of the massive neutrinos can be suppressed, for example,
by large lepton asymmetry, new interactions, or particle decay
\citep[see][and references therein]{Gariazzo:2014pja,Bergstrom:2014fqa}.

We have also considered the case of additional radiation and degenerate
massive active neutrinos, with the combined constraint
\twotwosig{\nnu&=3.2\pm 0.5}{\sumnu &< 0.32\,\eV}{\planckTT+\lensing+BAO.}
Again \planck\ shows no evidence for a deviation from the base \LCDM\ model.

\subsubsection{Neutrino models and tension with external data}
\label{sec:neutrino_tension}

The extended models discussed in this section allow \planck\ to be
consistent with a wider range of late-Universe parameters than in base
\LCDM.  Figure~\ref{fig:neutrinos_planes} summarizes the constraints on
$\Omm$, $\sigma_8$, and $H_0$ for the various models that we have
considered. The inferred Hubble parameter can increase or decrease, as
required to maintain the observed acoustic scale, depending on the
relative contribution of additional radiation (changing the sound
horizon) and neutrino mass (changing mainly the angular diameter
distance). However, all of the models follow similar degeneracy
directions in the $\Omm$--$\sigma_8$ and $H_0$--$\sigma_8$ planes,
so these models remain predictive: {\it large common areas of the parameter
space are excluded in all of these models}.  The two-parameter
extensions are required to fit substantially lower values of $\sigma_8$ without
also decreasing $H_0$ below the values determined from direct measurements,
but the scope for doing this is clearly limited.

External data sets need to be reanalysed consistently in extended
models, since the extensions change the growth of
structure, angular distances, and the matter-radiation equality
scale. For example, the dashed lines in
Fig.~\ref{fig:neutrinos_planes} show how different models affect the
\CFHTLENS\ galaxy weak lensing constraints from~\cite{Heymans:2013} (see
Sect.~\ref{subsec:WL}), when restricted to the region of parameter
space consistent with the \Planck\ acoustic scale measurements and the local
Hubble parameter. The filled green, grey, and red contours in
Fig.~\ref{fig:neutrinos_planes} show the CMB constraints on these models for
various data combinations. The tightest of these constraints comes from the
\planckTT+lensing+BAO combination. The blue contours show the constraints in
the base \LCDM\ cosmology. The red contours are broader than the
blue contours and there is greater overlap with the \CFHTLENS\ contours,
{\it but this offers only a marginal improvement compared to base \LCDM\/}
(compare with Fig.~\ref{WLonly}; see also the discussions in
\citealt{Leistedt:2014} and \citealt{Battye:2014qga}).
For each of these models, the \CFHTLENS\ results
prefer lower values of $\sigma_8$. Allowing for a higher neutrino mass lowers
$\sigma_8$ from \planck, but does not help alleviate the discrepancy with the
\CFHTLENS\ data, since the \planck\ data prefer a lower value of $H_0$. A joint
analysis of the \CFHTLENS\ likelihood with \planckTT\ shows a $\Delta\chi^2<1$
preference for the extended neutrino models compared to base \LCDM, and the
fits to \planckTT\ are worse in all cases. In base \LCDM\ the
\CFHTLENS\ data prefer a region of parameter space $\Delta\chi^2\approx 4$
away from the \planckTT+\CFHTLENS\ joint fit, indicative of the tension
between the data sets.  This is only slightly relieved to
$\Delta\chi^2\approx 3$ in the extended models.

In summary, modifications to the neutrino sector alone cannot easily
 explain the discrepancies between \Planck\ and other
astrophysical data described in Sect.~\ref{sec:additional_data}, including
the inference of a low value of $\sigma_8$ from rich cluster counts.

\subsubsection{Testing perturbations in the neutrino background}

As shown in the previous sections, the \Planck\ data provide evidence for a
cosmic neutrino background at a very high significance level. Neutrinos affect
the CMB anisotropies at the background level, by changing the expansion rate
before recombination and hence relevant quantities such as the sound horizon
and the damping scales. Neutrinos also affect the CMB anisotropies via their
perturbations. Perturbations in the neutrino background are coupled through
gravity to the perturbations in the photon background, and
can be described (for massless neutrinos) by the following set of
equations~\citep{Hu:1998kj, Hu:1998tk, Trotta:2004ty, Archidiacono:2011gq}:
\beglet
\begin{eqnarray}
\dot \delta_{\nu} &=&\frac{\dot a}{a} \left(1-3 c_{\rm eff}^2\right)
 \left(\delta_{\nu}+3\frac{\dot a}{a}
 \frac{q_{\nu}}{k}\right)-k\left(q_{\nu}+\frac{2}{3k}\dot h\right)\,;
  \label{NI1a} \\
\dot q_{\nu} &=& k \, c_{\rm eff}^2 \left(\delta_{\nu}+3\frac{\dot a}{a}
 \frac{q_{\nu}}{k}\right)-\frac{\dot a}{a} q_{\nu}-\frac{2}{3}k \pi_{\nu}\,;\\
\dot \pi_{\nu} &=& 3 k \, c_{\rm vis}^2 \left(\frac{2}{5} q_{\nu}
 + \frac{4}{15k}(\dot h + 6 \dot \eta) \right) - \frac{3}{5} k F_{\nu,3}\,;\\
\dot F_{\nu,\ell} &=& \frac{k}{2\ell+1} \left(\ell F_{\nu, \ell-1}
 -\left(\ell+1\right) F_{\nu, \ell+1}\right) , \quad (\ell\ge 3)\,. \label{NI1d}
\end{eqnarray}
\endlet
Here dots denote derivatives with respect to conformal time, $\delta_\nu$ is
the neutrino density contrast, $q_\nu$ is the neutrino velocity perturbation,
$\pi_\nu$ the anisotropic stress, $F_{\nu, \ell}$ are higher-order
moments of the neutrino distribution function, and $h$ and $\eta$ are the
scalar metric perturbations in the synchronous gauge.
In these equations,  $c_{\rm eff}^2$ is the neutrino sound speed in its own
reference frame and $c_{\rm vis}^2$ parameterizes the anisotropic stress.
For standard non-interacting massless neutrinos
$c_{\rm eff}^2=c_{\rm vis}^2=1/3$.  Any deviation from the expected values
could provide a hint of non-standard physics in the neutrino sector.

A greater (lower) neutrino sound speed would increase (decrease) the
neutrino pressure, leading to a lower (higher) perturbation
amplitude. On the other hand, changing $c_{\rm vis}^2$ alters the
viscosity of the neutrino fluid. For $c_{\rm vis}^2=0$, the
neutrinos act as a perfect fluid, supporting undamped acoustic oscillations.

\begin{figure}[t]
\includegraphics[width=90mm]{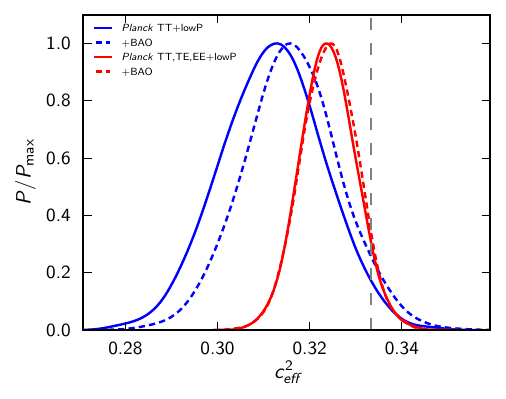}\\
\includegraphics[width=92mm]{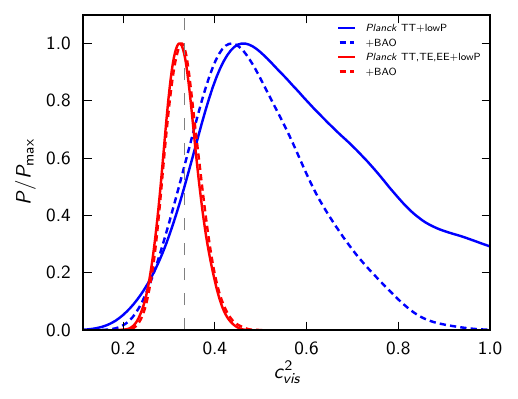}
\caption{\footnotesize 1D posterior distributions
for the neutrino
perturbation parameters $c_{\rm eff}^2$ (top) and $c_{\rm vis}^2$ (bottom).
Dashed vertical
lines indicate the conventional values $c_{\rm eff}^2=c_{\rm vis}^2=1/3$.}
\label{fig:cvisceff}
\end{figure}

Several previous studies have used this approach to constrain $c^2_{\rm eff}$
and $c^2_{\rm vis}$ using cosmological data
\citep[see, e.g.,][]{Trotta:2004ty, tristan,Archidiacono:2013lva,
Gerbino:2013ova,Audren:2014lsa},
with the motivation that deviations from the expected values could be a
hint of non-standard physics in the neutrino sector. Non-standard
interactions could involve, for example, neutrino coupling with light
scalar particles~\citep{Hannestad:2004qu, Beacom:2004yd, Bell:2005dr,
Sawyer:2006ju}. If neutrinos are strongly coupled at recombination,
this would result in a lower value for $c^2_{\rm vis}$ than in the
standard model.  Alternatively, the presence of early dark
energy that mimics a relativistic component at recombination could
possibly lead to a value for $c_{\rm eff}^2$ that differs from $1/3$
\citep[see, e.g.,][]{Calabrese:2010uf}.

In this analysis, for
simplicity, we assume $N_{\rm eff}=3.046$ and massless neutrinos. By using
an equivalent parameterization for massive neutrinos \citep{Audren:2014lsa} we
have checked that assuming one massive neutrino with
$\Sigma m_{\nu} \approx 0.06$\,eV, as in the base model used throughout this
paper, has no impact on the constraints on $c^2_{\rm eff}$ and $c^2_{\rm vis}$
reported in this section.\footnote{We also do not explore extended
cosmologies in this section, since no significant degeneracies are expected between
($\sum m_\nu$, $N_{\rm eff}$, $w$, $\nrun$) and ($c^2_{\rm eff}$,
$c^2_{\rm vis}$) ~\citep{Audren:2014lsa}.} We adopt a flat prior between zero
and unity for both $c^2_{\rm vis}$ and $c_{\rm eff}^2$.

The top and bottom panels of Fig.~\ref{fig:cvisceff} show the
posterior distributions of $c^2_{\rm eff}$ and $c^2_{\rm vis}$ from
{\planckTT}, {\planckTT}+BAO, {\planckall}, and {\planckall}+BAO.  The mean
values and 68\,\% errors on $c^2_{\rm eff}$ and $c^2_{\rm vis}$ are
\begin{subequations}
\twoonesig{c_{\rm eff}^2&=0.312 \pm 0.011\phantom{0}}
 {c_{\rm vis}^2&=0.47_{-0.12}^{+0.26}}{\planckTT,\phantom{,TE,EE+BAO}}
\twoonesig{c_{\rm eff}^2&=0.316 \pm 0.010\phantom{0}}
 {c_{\rm vis}^2&=0.44_{-0.10}^{+0.15}}{\planckTTBAO,\phantom{,TE,EE}}
\twoonesig{c_{\rm eff}^2&=0.3240 \pm 0.0060}{c_{\rm vis}^2&=0.327 \pm 0.037}
 {\planckall,\phantom{+BAO}}
\twoonesig{c_{\rm eff}^2&=0.3242 \pm 0.0059}{c_{\rm vis}^2&=0.331 \pm 0.037}
 {\planckallBAO.}
\end{subequations}

Constraints on these parameters are consistent with the conventional values
$c^2_{\rm eff} = c^2_{\rm vis}= 1/3$.  A vanishing value of $c_{\rm vis}^2$,
which might imply a strong interaction between neutrinos and other species, is
excluded at more than the $95\,\%$ level arising from the \Planck\ temperature
data.  This conclusion is greatly strengthened (to about $9\,\sigma$) when
\Planck\ polarization data are included.  As discussed in
\cite{Bashinsky:2003tk},
neutrino anisotropic stresses introduce a phase shift in the CMB angular
power spectra, which is more visible in polarization than temperature because
of the sharper acoustic peaks.  This explains why we see such a dramatic
reduction in the error on $c_{\rm vis}^2$ when including polarization data.

The precision of our results is consistent with the forecasts discussed
in~\cite{tristan}, and we find strong evidence, purely from CMB observations,
for neutrino anisotropies with the standard values
$c_{\rm vis}^2=1/3$ and $c_{\rm eff}^2=1/3$.

\subsection{Primordial nucleosynthesis}
\label{sec:bbn}

\subsubsection{Details of analysis approach}
 \label{sec:bbn_intro}

Standard big bang nucleosynthesis (BBN) predicts light element
abundances as a function of parameters relevant to the CMB, such as
the baryon-to-photon density ratio $\eta_\mathrm{b}\equiv
n_\mathrm{b}/n_\gamma$, the radiation density parameterized by
$N_\mathrm{eff}$, and the chemical potential of the electron
neutrinos. In {\paramsI}, we presented consistency checks between the
\Planck\ 2013 results, light element abundance data, and standard
BBN. The goal of Sect.~\ref{sec:bbn_consistency} below is to update these
results and to provide improved tests of the standard BBN model.  In
Sect.~\ref{sec:bbn_nuclear} we show how \Planck\ data can be used to
constrain nuclear reaction rates, and in Sect.~\ref{sec:bbn_helium} we
will present the most stringent CMB bounds to date on the primordial helium
fraction.

For simplicity, our analysis assumes a negligible leptonic asymmetry
in the electron neutrino sector. For a fixed photon temperature today
(which we take to be $T_{0}=2.7255$\,K), $\eta_\mathrm{b}$ can be related to
$\omega_{\mathrm{b}}\equiv \Omega_bh^2$, up to a small (and negligible)
uncertainty associated
with the primordial helium fraction. Standard BBN then predicts the abundance
of each light element as a function of only two parameters,
$\omega_\mathrm{b}$ and $\Delta N_\mathrm{eff} \equiv N_\mathrm{eff}-3.046$,
with a theoretical error coming mainly from uncertainties in the neutron
lifetime and a few nuclear reaction rates.

We will confine our discussion to BBN predictions for the primordial
abundances\footnote{BBN calculations usually refer to nucleon number density
fractions rather than mass fractions. To avoid any ambiguity with
the helium mass fraction $Y_\mathrm{P}$, normally used in CMB physics,
we use superscripts to distinguish between the two definitions
$Y_\mathrm{P}^\mathrm{CMB}$ and $Y_\mathrm{P}^\mathrm{BBN}$. Typically,
$Y_\mathrm{P}^\mathrm{BBN}$ is about 0.5\,\% higher
than $Y_\mathrm{P}^\mathrm{CMB}$.}
of $^4$He and deuterium, expressed, respectively as
$Y_\mathrm{P}^\mathrm{BBN} = 4 n_\mathrm{He}/n_\mathrm{b}$ and $y_\mathrm{DP} = 10^{5} n_\mathrm{D}/n_\mathrm{H}$.
We will not discuss other light elements, such as
tritium and lithium, because the observed abundance measurements
and their interpretation is more controversial \citep[see][for a recent
review]{Fields:2014uja}. As in {\paramsI}, the BBN predictions for
$Y_\mathrm{P}^\mathrm{BBN}(\omega_\mathrm{b},\Delta N_\mathrm{eff}$) and
$y_\mathrm{DP}(\omega_\mathrm{b},\Delta N_\mathrm{eff}$) are given by
Taylor expansions obtained with the {\tt PArthENoPE} code
\citep{Pisanti:2007hk}, similar to the ones presented in~\cite{Iocco:2008va},
but updated by the {\tt PArthENoPE} team with the latest observational data on
nuclear rates and on the neutron life-time:
\begin{align}
\begin{split}
Y_\mathrm{P}^\mathrm{BBN}&=
 0.2311+0.9502\omega_\mathrm{b}-11.27\omega_\mathrm{b}^2 \\
 \ \ &+ \Delta N_\mathrm{eff}\left(0.01356+0.008581\omega_\mathrm{b}
  -0.1810\omega_\mathrm{b}^2\right) \\
 \ \ &+\Delta N_\mathrm{eff}^2\left(-0.0009795-0.001370\omega_\mathrm{b}
  +0.01746\omega_\mathrm{b}^2\right)\,;
\end{split}
 \label{eq:fit_yp}
\\
\begin{split}
y_\mathrm{DP} &= 18.754-1534.4\omega_\mathrm{b}+48656\omega_\mathrm{b}^2
 -552670\omega_\mathrm{b}^3 \\
 \ \ &+ \Delta N_\mathrm{eff}\left(2.4914-208.11\omega_\mathrm{b}
 +6760.9\omega_\mathrm{b}^2-78007\omega_\mathrm{b}^3\right) \\
 \ \ &+ \Delta N_\mathrm{eff}^2\left(0.012907-1.3653\omega_\mathrm{b}
 +37.388\omega_\mathrm{b}^2-267.78\omega_\mathrm{b}^3\right).
\end{split}
\label{eq:fit_ydp}
\end{align}
By averaging over several
measurements, the Particle Data Group 2014 \citep{Agashe:2014kda}
estimates the neutron life-time to be $\tau_\neutron = (880.3\pm
1.1)\,{\rm s}$ at 68\,\%~CL.\footnote{However, the most recent individual
measurement by \cite{Yue:2013qrc} gives $\tau_\neutron=[887.8\pm 1.2
~\mathrm{(stat.)} \pm 1.9 ~\mathrm{(syst.)}]\,{\rm s}$, which is
discrepant at $3.3\,\sigma$ with the previous average (including only
statistical errors). Hence one should bear in mind that systematic effects
could be underestimated in the Particle Data Group result. Adopting the central
value of \cite{Yue:2013qrc} would shift our results by a small amount (by a
factor of 1.0062 for $Y_\mathrm{P}$ and 1.0036 for $y_\mathrm{DP}$).}
The expansions in Eqs.~(\ref{eq:fit_yp}) and (\ref{eq:fit_ydp}) are based on
this central value, and we assume that Eq.~\eqref{eq:fit_yp} predicts the
correct helium fraction
up to a standard error $\sigma(Y_\mathrm{P}^\mathrm{BBN})=0.0003$,
obtained by propagating the error on $\tau_\neutron$.

\begin{figure}
\centering
\includegraphics[width=9.1cm]{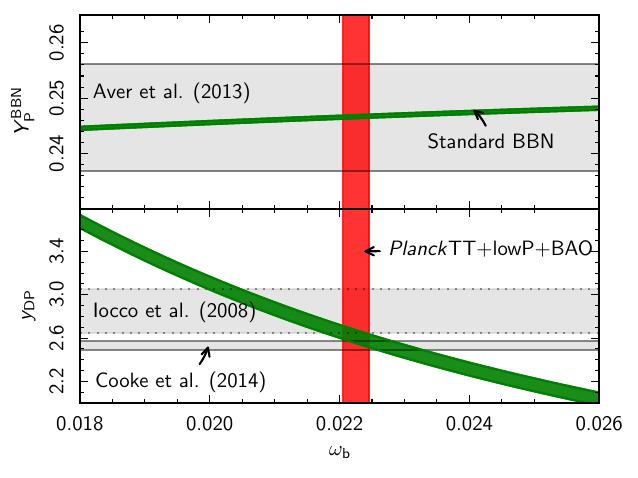}
\caption{
Predictions of standard BBN for the
primordial abundance of $^4$He (top) and deuterium (bottom), as a function of
the baryon density $\omb$. The width of the green stripes corresponds
to 68\,\% uncertainties on nuclear reaction rates and on the neutron lifetime.
The horizontal bands show observational bounds on primordial element abundances
compiled by various authors, and the red vertical band shows the
\planckTT+BAO bounds on $\omega_\mathrm{b}$ (all with 68\,\% errors). The BBN
predictions and CMB results shown here assume $N_\mathrm{eff}=3.046$ and no
significant lepton asymmetry.}
\label{fig:bbn_a}
\end{figure}

The uncertainty on the deuterium fraction is dominated by that on the rate
of the reaction ${\rm d}({\rm p},\gamma)^3 \mbox{He}$. For that rate,
in {\paramsI} we
relied on the result of~\cite{Serpico:2004gx}, obtained by fitting several
experiments. The expansions of Eqs.~(\ref{eq:fit_yp}) and (\ref{eq:fit_ydp})
now adopt the latest experimental determination by \cite{Adelberger:2010qa}
and use the best-fit expression in their equation~(29). We also rely on the
uncertainty quoted in \cite{Adelberger:2010qa}
and propagate it to the deuterium fraction. This gives
a standard error $\sigma(y_\mathrm{DP})=0.06$, which is more conservative
than the error adopted in {\paramsI}.

\subsubsection{Primordial abundances from \Planck\ data and standard BBN
 \label{sec:bbn_consistency}}

We first investigate the consistency of standard BBN and the CMB by fixing the
radiation density to its standard value, i.e., $N_\mathrm{eff}=3.046$, based
on the assumption of standard neutrino decoupling and no extra light relics.
We can then use \Planck\ data to measure $\omega_\mathrm{b}$, assuming base
\LCDM, and test for consistency with experimental abundance measurements.
The 95\,\%~CL bounds obtained for the base $\Lambda$CDM model for various
data combinations are
\begin{equation}
\omega_\mathrm{b} =
\left\{
\mathrm{
\begin{tabular}{ll}
$0.02222^{+0.00045}_{-0.00043}$ & \datalabel{\planckTT}, \\
$0.02226^{+0.00040}_{-0.00039}$ & \datalabel{\planckTTBAO}, \\
$0.02225^{+0.00032}_{-0.00030}$ & \datalabel{\planckall},\\
$0.02229^{+0.00029}_{-0.00027}$ & \datalabel{\planckallBAO},
\end{tabular}
}
\right.
\end{equation}
\noindent
corresponding to a predicted primordial $^4$He number density fraction
(95\,\%~CL) of
\begin{equation}
Y_\mathrm{P}^\mathrm{BBN} \!\! = \!\!
\left\{
\!\!
\mathrm{
\begin{tabular}{ll}
$0.24665^{+(0.00020)\ 0.00063}_{-(0.00019) \ 0.00063}$ & \!\!\!\! \datalabel{\planckTT},\\
$0.24667^{+(0.00018)\ 0.00063}_{-(0.00018)\ 0.00063}$ & \!\!\!\! \datalabel{\planckTTBAO},\\
$0.24667^{+(0.00014)\ 0.00062}_{-(0.00014)\ 0.00062}$ & \!\!\!\! \datalabel{\planckall},\\
$0.24668^{+(0.00013)\ 0.00061}_{-(0.00013)\ 0.00061}$ & \!\!\!\! \datalabel{\planckallBAO},
\end{tabular}
}
\right. \label{yp_from_cmb}
\end{equation}
\noindent
and deuterium fraction (95\,\%~CL)
\begin{equation}
y_\mathrm{DP} =
\left\{
\mathrm{
\begin{tabular}{ll}
$2.620^{+(0.083)\ 0.15}_{-(0.085)\ 0.15}$ & \datalabel{\planckTT}, \\
$2.612^{+(0.075)\ 0.14}_{-(0.074)\ 0.14}$ & \datalabel{\planckTTBAO}, \\
$2.614^{+(0.057)\ 0.13}_{-(0.060)\ 0.13}$ & \datalabel{\planckall}, \\
$2.606^{+(0.051)\ 0.13}_{-(0.054)\ 0.13}$ & \datalabel{\planckallBAO}.
\end{tabular}
}
\right. \label{ydp_from_cmb}
\end{equation}
The first set of error bars (in parentheses) in Eqs.~\eqref{yp_from_cmb}
and \eqref{ydp_from_cmb} reflect only the uncertainty on $\omega_\mathrm{b}$.
The second set includes the theoretical uncertainty on the BBN predictions, added in
quadrature to the errors from $\omega_\mathrm{b}$. The total errors in the
predicted helium abundances are dominated by the BBN uncertainty, as in
{\paramsI}. For deuterium, the \Planck\ 2015 results improve the determination
of $\omega_\mathrm{b}$ to the point where the theoretical errors
are comparable or larger than the errors from the CMB. In other words, for
base \LCDM\ the predicted abundances cannot be improved substantially by
further measurements of the CMB. This also means that \Planck\ results
can, in principle, be used to investigate nuclear reaction rates that dominate
the theoretical uncertainty (see Sect.~\ref{sec:bbn_nuclear}).

\begin{figure}
\centering
\includegraphics[width=9.1cm]{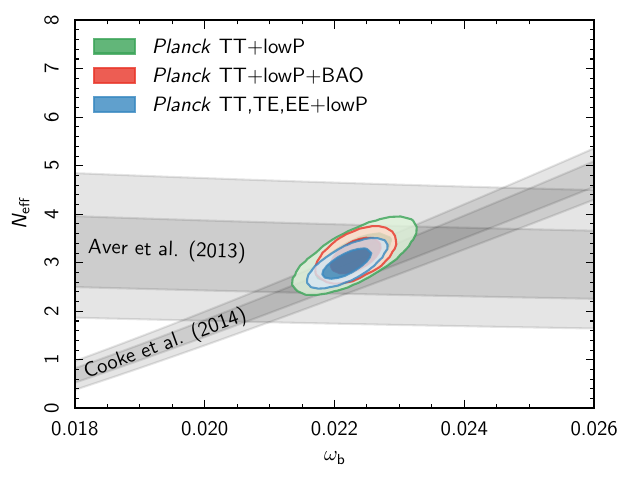}
\caption{
Constraints in the $\omb$--$\neff$ plane from \Planck\ and \Planck+BAO data
(68\,\% and 95\,\% contours) compared to the predictions of BBN, given
primordial element abundance measurements.
We show the 68\,\% and 95\,\% confidence regions derived from $^4$He bounds
compiled by \cite{Aver:2013wba} and from deuterium bounds compiled by
\cite{Cooke:2013cba}.  In the CMB analysis, $N_\mathrm{eff}$ is allowed to vary
as an additional parameter to base \LCDM, with $Y_{\rm P}$ fixed as
a function of $\omega_\mathrm{b}$ and $N_\mathrm{eff}$, according to BBN
predictions. These constraints assume no significant
lepton asymmetry.}
\label{fig:bbn_c}
\end{figure}

The results of Eqs.~\eqref{yp_from_cmb} and \eqref{ydp_from_cmb} are well
within the ranges indicated by the latest measurements of primordial
abundances, as illustrated in Fig.~\ref{fig:bbn_a}. The helium data
compilation of \cite{Aver:2013wba} gives $Y_\mathrm{P}^\mathrm{BBN} =
0.2465 \pm 0.0097$ (68\,\%~CL), and the \Planck\ prediction is near the
middle of this range.\footnote{A substantial part of this error comes
from the regression to zero metallicity. The mean of the 17
measurements analysed by \cite{Aver:2013wba} is $\langle
Y_\mathrm{P}^\mathrm{BBN} \rangle = 0.2535 \pm 0.0036$, i.e., about
$1.7\,\sigma$ higher than the \Planck\ predictions of
Eq.~\eqref{yp_from_cmb}.} As summarized by
\citet{Aver:2013wba} and \citet{Peimbert:2008}, helium abundance measurements
derived from emission lines in low-metallicity \ion{H}{ii} regions are
notoriously difficult and prone to systematic errors. As a result, many
discrepant helium abundance measurements can be found in the
literature.  \citet{Izotov:2014fga} have reported
$Y_\mathrm{P}^\mathrm{BBN} =0.2551 \pm 0.0022$, which is discrepant
with the base \LCDM\ predictions by $3.4\,\sigma$. Such a high
helium fraction could be accommodated by increasing $N_{\rm eff}$ (see
Fig.~\ref{fig:bbn_c} and Sect.~\ref{sec:bbn_helium}); however, at
present it is not clear whether the error quoted by
\citet{Izotov:2014fga} accurately reflects systematic uncertainties,
including in particular the error in extrapolating to zero metallicity.

Historically, deuterium abundance measurements have shown excess scatter over
that expected from statistical errors, indicating the presence of systematic
uncertainties in the observations. Figure~\ref{fig:bbn_a} shows
the data compilation of \cite{Iocco:2008va}, $y_\mathrm{DP}=2.87 \pm
0.22$ (68\,\%~CL), which includes measurements based on damped
Ly$\alpha$ and Lyman limit systems.  We also show the more recent
results by \cite{Cooke:2013cba} \citep[see also][]{Pettini-Cooke} based
on their observations of low-metallicity damped Ly$\alpha$
absorption systems in two quasars (SDSS J1358+6522, $z_{\rm abs}=
3.06726$ and SDSS J1419+0829, $z_{\rm abs}= 3.04973$) and a reanalysis of
archival spectra of damped Ly$\alpha$ systems in three further
quasars that satisfy strict selection criteria. The
\cite{Cooke:2013cba} analysis gives $y_\mathrm{DP}=2.53 \pm 0.04$
(68\,\%~CL), somewhat lower than the central \cite{Iocco:2008va} value,
and with a much smaller error.  The \cite{Cooke:2013cba} value is almost
certainly the more reliable measurement, as evidenced by the consistency of
the deuterium abundances of the five systems in their analysis.
The \Planck\ base \LCDM\ predictions of Eq.~\eqref{ydp_from_cmb} lie within
1$\,\sigma$ of the \cite{Cooke:2013cba} result. This is a remarkable success
for the standard theory of BBN.

It is worth noting that the \Planck\ data are so accurate that $\omb$ is
insensitive to the underlying cosmological model. In our grid of extensions
to base \LCDM\ the largest degradation of the error in $\omb$ is in models
that allow $N_{\rm eff}$ to vary. In these models, the mean value of $\omb$
is almost identical to that for base \LCDM, but the error on $\omb$
increases by about 30\,\%. The value of $\omb$ is stable to even more
radical changes to the cosmology, for example, adding general isocurvature
modes \citep{planck2014-a24}.

If we relax the assumption that $N_\mathrm{eff}=3.046$ (but adhere to
the hypothesis that electron neutrinos have a standard distribution,
with a negligible chemical potential), BBN predictions depend on both
parameters $(\omega_\mathrm{b}$ and $N_\mathrm{eff}$. Following
the same methodology as in Sect.~6.4.4 of
{\paramsI}, we can identify the region of the
$\omega_\mathrm{b}$--$N_\mathrm{eff}$ parameter space that is
compatible with direct measurements of the primordial helium and
deuterium abundances, including the BBN theoretical errors.
This is illustrated in Fig.~\ref{fig:bbn_c} for the
 $N_\mathrm{eff}$ extension to base \LCDM. The region preferred by CMB observations lies at the
intersection between the helium and deuterium abundance 68\,\%~CL preferred
regions and is compatible with the standard value of $N_{\rm eff}=3.046$.
This confirms the beautiful agreement between CMB and BBN
physics.  Figure~\ref{fig:bbn_c} also shows that the \planck\
polarization data help in reducing the degeneracy between
$\omega_\mathrm{b}$ and $N_\mathrm{eff}$.

We can actually make a more precise statement by combining the posterior
distribution on $\omega_\mathrm{b}$ and $N_\mathrm{eff})$ obtained for \Planck\
with that inferred from helium and deuterium abundance, including
observational and theoretical errors. This provides joint CMB+BBN predictions
on these parameters.  After marginalizing over $\omega_\mathrm{b}$, the
95\,\% CL preferred ranges for $N_\mathrm{eff}$ are
\begin{equation}
N_\mathrm{eff} =
\left\{
\mathrm{
\begin{tabular}{ll}
$3.11_{-0.57}^{+0.59}$ & \datalabel{He+\planckTT}, \\
$3.14_{-0.43}^{+0.44}$ & \datalabel{He+\planckTTBAO}, \\
$2.99_{-0.39}^{+0.39}$ & \datalabel{He+\planckall},
\label{neff_from_cmb_and_helium}
\end{tabular}
}
\right.
\end{equation}
when combining \Planck\ with the helium abundance estimated by~\cite{Aver:2013wba}, or
\begin{equation}
N_\mathrm{eff} =
\left\{
\mathrm{
\begin{tabular}{ll}
$2.95_{-0.52}^{+0.52}$ & \datalabel{D+\planckTT}, \\
$3.01_{-0.37}^{+0.38}$ & \datalabel{D+\planckTTBAO}, \\
$2.91_{-0.37}^{+0.37}$ & \datalabel{D+\planckall},
\end{tabular}
}
\right. \label{neff_from_cmb_and_deuterium}
\end{equation}
when combining with the deuterium abundance measured by~\cite{Cooke:2013cba}.
These bounds represent the best currently-available estimates of
$N_\mathrm{eff} $ and are remarkably consistent with the standard
model prediction.

The allowed region in $\omega_\mathrm{b}$--$N_\mathrm{eff}$ space
does not increase significantly when other parameters are allowed to
vary at the same time. From our grid of extended models, we have checked that
this conclusion holds in models with neutrino masses, tensor
fluctuations, or running of the scalar spectral index, for example.

\subsubsection{Constraints from \Planck\ and deuterium observations on nuclear
reaction rates \label{sec:bbn_nuclear}}

We have seen that primordial element abundances estimated from direct
observations are consistent with those inferred from \Planck\ data
under the assumption of standard BBN. However, the
\Planck\ determination of $\omega_\mathrm{b}$ is so precise that the
theoretical errors in the BBN predictions are now a dominant source of
uncertainty. As noted by ~\cite{Cooke:2013cba}, one can begin to
think about using CMB measurements together with accurate deuterium
abundance measurements to learn about the underlying BBN physics.

While for helium the theoretical error comes mainly from the uncertainties in
the neutron lifetime, for deuterium it is dominated by uncertainties in the
radiative capture process ${\rm d}({\rm p},\gamma)^3 \mbox{He}$, converting
deuterium into helium. The present experimental uncertainty for the
$S$-factor at low energy (relevant for BBN), is in the range
6--10\,\% \citep{Ma:1997}. However, as noted by several authors
\citep[see, e.g.,][]{Nollett:2011aa, divalentino} the best-fit value
of $S(E)$ inferred from experimental data in the range
$30\,{\rm keV}\,{\leq}\,E\,{\leq}\,300\,{\rm keV}$
is about 5--10\,\% lower than theoretical
expectations \citep{Viviani:1999us,Marcucci:2005zc}. The
\texttt{PArthENoPE} BBN code assumes the lower experimental value
for ${\rm d}({\rm p},\gamma)^3 \mbox{He}$, and this might explain why the
deuterium abundance determined by \cite{Cooke:2013cba} is slightly lower than
the value inferred by \Planck.

To investigate this further, following the methodology of \cite{divalentino},
we perform a combined analysis of \Planck\ and deuterium observations,
to constrain the value of the ${\rm d}({\rm p},\gamma)^3 \mbox{He}$
reaction rate.
As in \cite{divalentino}, we parameterize the thermal rate $R_2(T)$ of the
${\rm d}({\rm p},\gamma)^3 \mbox{He}$ process in the
\texttt{PArthENoPE} code by introducing a rescaling factor $A_2$ of the
experimental rate $R_2^{\,\rm ex}(T)$, i.e., $R_2(T) = A_2 \, R_2^{\rm ex}(T)$,
and solve for $A_2$ using various \Planck+BAO data combinations, given
the \cite{Cooke:2013cba} deuterium abundance measurements.

\begin{figure}[htb!]
\includegraphics[width=9.0cm,]{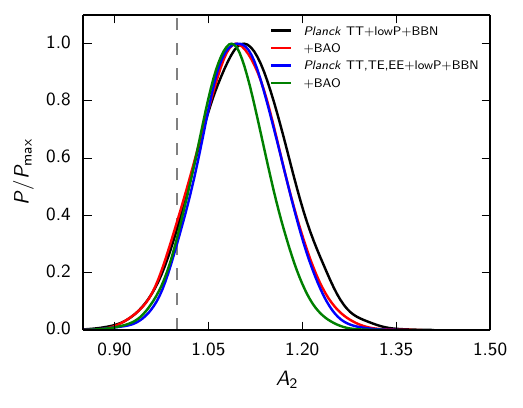}
\caption{Posteriors for the $A_2$ reaction rate parameter for
various data combinations. The vertical dashed line shows the value $A_2=1$
that corresponds to the current experimental estimate of the
${\rm d}({\rm p},\gamma)^3 \mbox{He}$ rate used in the \texttt{PArthENoPE}
BBN code.}
\label{fig.A2}
\end{figure}

Assuming the base \LCDM\ model we find (68\,\% CL)
\beglet
\begin{eqnarray}
A_2 &=& 1.106 \pm 0.071 \quad \datalabel{\planckTT}\,, \\
A_2 &=& 1.098 \pm 0.067 \quad \datalabel{\planckTTBAO}\,, \\
A_2 &=& 1.110 \pm 0.062 \quad \datalabel{\planckall}\,, \\
A_2 &=& 1.109 \pm 0.058 \quad \datalabel{\planckallBAO}\,.
\end{eqnarray}
\endlet
The posteriors for $A_2$ are shown in Fig.~\ref{fig.A2}.  These results
suggest that the ${\rm d}({\rm p},\gamma)^3 \mbox{He}$ reaction rate may
be have been underestimated by about 10\,\%. Evidently, tests of the
standard BBN picture appear to have reached the point where they are
limited by uncertainties in nuclear reaction rates. There is therefore
a strong case to improve the precision of experimental measurements
\citep[e.g.,][]{Anders:2014} and theoretical computations of key
nuclear reaction rates relevant for BBN.

\subsubsection{Model-independent bounds on the helium fraction from
 \Planck\label{sec:bbn_helium}}

\begin{figure}[t]
\centering
\includegraphics[width=9.0cm]{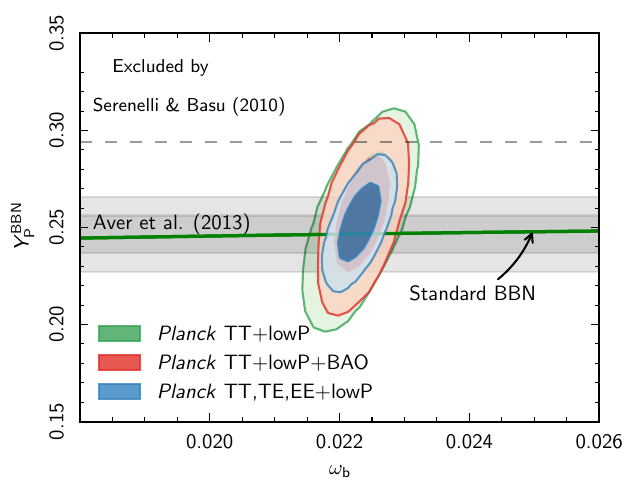}
\caption{
Constraints in the $\omb$--$Y_{\rm P}^\mathrm{BBN}$ plane from \planck\ and
\planck+BAO, compared to helium abundance measurements.  Here 68\,\% and 95\,\%
contours are plotted for the CMB(+BAO) data combinations when
$Y_{\rm P}^\mathrm{BBN}$ is allowed to vary as an additional parameter to
base \LCDM. The horizontal band shows observational bounds on $^4$He compiled
by~\cite{Aver:2013wba} with 68\,\% and 95\,\% errors, while the dashed line at
the top of the figure delineates the conservative 95\,\% upper bound inferred
from the Solar helium abundance by~\citet{serenelli10}. The green stripe shows
the predictions of standard BBN for the
primordial abundance of $^4$He as a function of the baryon density. Both BBN
predictions and CMB results assume $N_\mathrm{eff}=3.046$ and no
significant lepton asymmetry.}
\label{fig:bbn_b}
\end{figure}

\begin{figure}
\centering
\includegraphics[width=9.0cm]{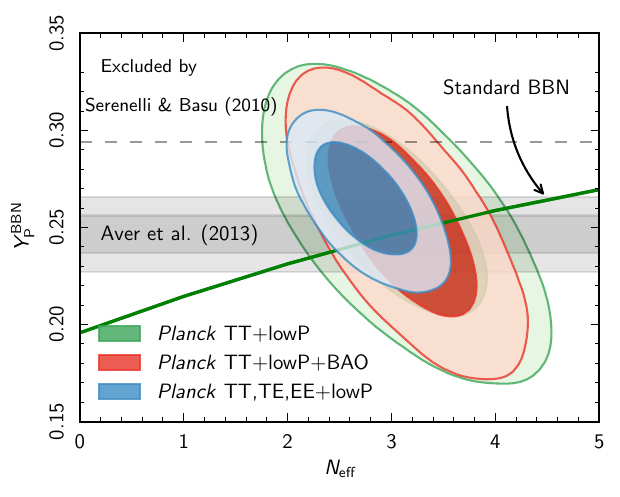}
\caption{As in Fig.~\ref{fig:bbn_b}, but now allowing
$Y_{\rm P}^\mathrm{BBN}$ and $N_\mathrm{eff}$ to vary as parameter extensions
to base \LCDM.}
\label{fig:bbn_d}
\end{figure}

Instead of inferring the primordial helium abundance from BBN codes
using $(\omega_\mathrm{b},N_\mathrm{eff})$ constraints from \Planck,
we can measure it directly, since variations in $Y_\mathrm{P}^\mathrm{BBN}$
modify the density of free electrons between helium and hydrogen recombination
and therefore affect the damping tail of the CMB anisotropies.

If we allow $Y_\mathrm{P}^\mathrm{CMB}$ to vary as an additional parameter to
base \LCDM, we find the following constraints (at 95\,\%~CL):
\begin{equation}
Y_\mathrm{P}^\mathrm{BBN} =
\left\{
\mathrm{
\begin{tabular}{ll}
$0.253^{+0.041}_{-0.042}$ & \datalabel{\planckTT}\,;\\
$0.255^{+0.036}_{-0.038}$ & \datalabel{\planckTTBAO}\,;\\
$0.251^{+0.026}_{-0.027}$ & \datalabel{\planckall}\,;\\
$0.253^{+0.025}_{-0.026}$ & \datalabel{\planckallBAO}\,.\\
\end{tabular}
}
\right. \label{yp_from_cmb_only}
\end{equation}
Joint constraints on $Y_\mathrm{P}^\mathrm{BBN}$ and $\omega_\mathrm{b}$ are
shown in Fig.~\ref{fig:bbn_b}. The addition of \Planck\ polarization
measurements results in a substantial reduction in the uncertainty on the
helium fraction. In fact, the standard deviation on
$Y_\mathrm{P}^\mathrm{BBN}$ in the case of \planckall\ is only 30\,\% larger
than the observational error quoted by \cite{Aver:2013wba}. As emphasized
throughout this paper, the systematic effects in the \Planck\ polarization
spectra, although at low levels, have not been accurately characterized at this
time.  Readers should therefore treat the polarization constraints with some
caution.  Nevertheless, as shown in Fig.~\ref{fig:bbn_b}, all three data
combinations agree well with the observed helium abundance measurements and
with the predictions of standard BBN.

There is a well-known parameter degeneracy between $Y_\mathrm{P}$ and the
radiation density (see the discussion in {\paramsI}).
Helium abundance predictions from the CMB are therefore particularly sensitive
to the addition of the parameter $N_\mathrm{eff}$ to base \LCDM. Allowing
both $Y_\mathrm{P}^\mathrm{BBN}$ and $N_\mathrm{eff}$ to vary
we find the following constraints (at 95\,\%~CL):
\begin{equation}
Y_\mathrm{P}^\mathrm{BBN} =
\left\{
\mathrm{
\begin{tabular}{ll}
$0.252^{+0.058}_{-0.065}$ & \datalabel{\planckTT}\,;\\
$0.251^{+0.058}_{-0.064}$ & \datalabel{\planckTTBAO}\,;\\
$0.263^{+0.034}_{-0.037}$ & \datalabel{\planckall}\,;\\
$0.262^{+0.035}_{-0.037}$ & \datalabel{\planckallBAO}\,.\\
\end{tabular}
}
\right. \label{fig:yp_from_cmb_only_free_neff}
\end{equation}
Contours in the $Y_\mathrm{P}^\mathrm{BBN}$--$N_\mathrm{eff}$ plane are shown
in Fig.~\ref{fig:bbn_d}.  Here again, the impact of \Planck\ polarization data
is important, and helps to substantially reduce the degeneracy
between these two parameters.  The \planckall\ contours are in
very good agreement with standard BBN and $N_\mathrm{eff}=3.046$.
However, even if we relax the assumption of standard BBN, the
CMB does not allow high values of $N_\mathrm{eff}$. It is
therefore difficult to accommodate an extra thermalized relativistic species,
even if the standard BBN prior on the helium fraction is relaxed.

\subsection{Dark matter annihilation}
\label{sec:DM}
\newcommand{\pann}{\ifmmode p_{\rm ann}\else $p_{\rm ann}$\fi}

Energy injection from dark matter (DM) annihilation can alter the
recombination history, leading to changes in the temperature and
polarization power spectra of the CMB
\citep[e.g.,][]{Chen:2003gz,Padmanabhan:2005es}.  As demonstrated in
several papers \citep[e.g.,][]{Galli:2009zc,Slatyer:2009yq,
 Finkbeiner:2011dx}, CMB anisotropies offer an opportunity to
constrain the nature of DM.  Furthermore, CMB
experiments such as \Planck\ can achieve limits on the annihilation
cross-section that are relevant for the interpretation of the rise in
the cosmic-ray positron fraction at energies $\ga 10\,{\rm GeV}$
observed by PAMELA, {\it Fermi}, and AMS
\citep{Adriani:2009,Ackermann:2012, Aguilar:2014}. The CMB constraints
are complementary to those determined from other astrophysical
probes, such as the $\gamma$-ray observations of dwarf galaxies
by the {\it Fermi\/} satellite \citep{Ackermann:2013yva}.

The way in which DM annihilations heat and ionize the gaseous
background depends on the nature of the cascade of particles produced
following annihilation and, in particular, on the production of
${\rm e}^{\pm}$ pairs and photons that couple to the gas.
The fraction of the rest mass energy that is injected into the gas
can be modelled by an ``efficiency factor,'' $f(z)$, which
is typically in the range 0.01--1 and depends on
redshift.\footnote{To maintain consistency with other papers on dark
  matter annihilation, we retain the notation $f(z)$ for the
  efficiency factor in this section; it should not be confused with
  the growth rate factor introduced in Eq.~\eqref{GR1}.}
Computations of $f(z)$ for various annihilation channels can be found
in \citet{Slatyer:2009yq}, \citet{Huetsi2009}, and
\citet{Evoli:2012qh}.  The rate of energy release per unit volume by
annihilating DM can therefore be written as
\begin{equation}
\label{enrateselfDM}
\frac{dE}{dtdV}(z)= 2\, g\, \rho^2_{\rm crit} c^2 \Omega^2_{\rm c}
 (1+z)^6 \pann(z),
\end{equation}
where $\pann$ is defined as
\begin{equation}
\label{pann}
\pann (z) \equiv f(z) \frac{\langle\sigma \varv \rangle}{m_\chi}.
\end{equation}
Here $\rho_{\rm crit}$ the critical density of the Universe today, $m_\chi$ is
the mass of the DM particle,
and $\langle\sigma \varv \rangle$ is the thermally-averaged annihilation
cross-section times the velocity (explicitly the so-called
M$\o$ller velocity); we will refer to this quantity
loosely as the ``cross-section'' hereafter. In Eq.~\eqref{enrateselfDM},
$g$ is a degeneracy factor that is equal to $1/2$ for Majorana
particles and $1/4$ for Dirac particles. In this paper, the
constraints will refer to Majorana particles. Note that to produce
the observed dark matter density from thermal DM relics requires an
 annihilation cross-section of $\langle \sigma \varv \rangle \approx 3
\times 10^{-26}\,{\rm cm}^{3}\,{\rm s}^{-1}$ (assuming s-wave annihilation)
at the time of freeze-out
\citep[see, e.g., the review by][]{Profumo:2013}.

Both the amplitude and redshift dependence of the efficiency factor
$f(z)$ depend on the details of the annihilation process
\citep[e.g.,][]{Slatyer:2009yq}.  The functional shape of $f(z)$ can
be taken into account using generalized parameterizations or principal
components \citep{Finkbeiner:2011dx,Hutsi:2011vx}, similar to the
analysis of the recombination history presented in
Sect.~\ref{sec:eXeM}. However, as shown in \cite{Galli:2011rz},
\cite{Giesen:2012rp}, and \cite{Finkbeiner:2011dx}, to a first
approximation the redshift dependence of $f(z)$ can be ignored, since
current CMB data (including \Planck) are sensitive to energy injection over
a relatively narrow range of redshift, typically $z\approx1000$--600.
The effects of DM annihilation can therefore be reasonably well parameterized
by a single constant parameter, $\pann$ (with $f(z)$ set to a constant
$f_{\rm eff}$), which encodes the dependence on the properties of the DM
particles. In the following, we calculate constraints on the $\pann$
parameter, assuming that it is constant, and then project these constraints
on to a particular dark matter
model assuming $f_{\rm eff}\equiv f(z=600)$, since the effect of dark matter
annihilation peaks at $z\approx600$ \citep[see][]{Finkbeiner:2011dx}. The
$f(z)$ functions used here are those calculated in \cite{Slatyer:2009yq},
with the updates described in \cite{Galli:2013dna} and
\cite{Madhavacheril:2013cna}.  Finally, we estimate the fractions of injected
energy that affect the gaseous background, from heating,
ionizations, or Ly$\alpha$ excitations, using the updated calculations
described in \citet{Galli:2013dna} and
\citet{Valdes:2009cq}, following \citet{Shull1985}.

We compute the theoretical angular power spectrum in the presence of DM
annihilations by modifying the \RECFAST\ routine \citep{SeagerRecfast1999} in
the \CAMB\ code as in \cite{Galli:2011rz}.\footnote{We checked that we obtain
similar results using either the \HYREC\ code \citep{AliHaimoud:2010dx}, as
detailed in \cite{Giesen:2012rp}, or
\COSMOREC\ \citep{Chluba2010b}, instead of \RECFAST.}
We then add $\pann$ as an additional parameter to those of the base \LCDM\
cosmology.  Table~\ref{tab:pannconstraints} shows the constraints for various
data combinations.

\begin{table}[htbp]
\begingroup
\newdimen\tblskip \tblskip=5pt
\caption{Constraints on $\pann$ in units of
${\rm cm}^3\,{\rm s}^{-1}\,{\rm GeV}^{-1}$.}
\label{tab:pannconstraints}
\vskip -3mm
%\footnotesize
\setbox\tablebox=\vbox{
 \newdimen\digitwidth
 \setbox0=\hbox{\rm 0}
 \digitwidth=\wd0
 \catcode`*=\active
 \def*{\kern\digitwidth}
 \newdimen\signwidth
 \setbox0=\hbox{+}
 \signwidth=\wd0
 \catcode`!=\active
 \def!{\kern\signwidth}
 \halign{\tabskip=0pt\hbox to 1.75in{#\leaderfil}\tabskip=0.5em&
 \hfil#\hfil\tabskip=0pt\cr
\noalign{\doubleline}
\omit Data combinations\hfil& \pann \ (95\,\% upper limits)\cr
\noalign{\vskip 3pt\hrule\vskip 5pt}
\shortTT& $< 5.7\times10^{-27}$\cr
EE\dataplus\lowEB& $< 1.4\times10^{-27}$\cr
TE\dataplus\lowEB& $< 5.9\times10^{-28}$\cr
\shortTT\dataplus\lensing& $< 4.4\times10^{-27}$\cr
\noalign{\vskip 2mm}
\shortall& $< 4.1\times10^{-28}$\cr
\shortall\dataplus\lensing& $< 3.4\times10^{-28}$\cr
\shortall\dataplus\ext& $< 3.5\times10^{-28}$\cr
\noalign{\vskip 3pt\hrule\vskip 5pt}
}}
\endPlancktable
\endgroup
\end{table}

\begin{figure}[t]
%left bottom right top
\centering
 \includegraphics[width=9.1cm,angle=0]{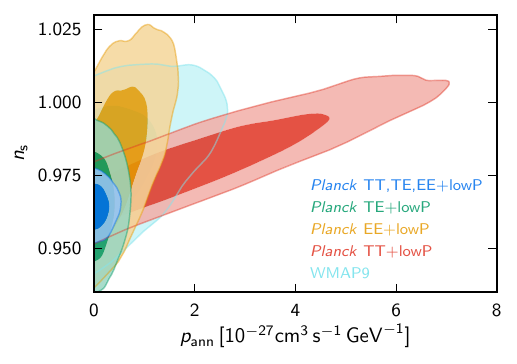}
\caption{2D marginal distributions in the $p_{\rm ann}$--$n_{\rm s}$ plane
for \planckTT\ (red), \Planck\ EE+\lowTEB\ (yellow), \Planck\ TE+\lowTEB\
(green), and \planckall\ (blue) data combinations. We also show the
constraints obtained using WMAP9 data (light blue).}
\label{fig:pann2D}
\end{figure}

The constraints on \pann\ from the \planckTT\ spectra are about 3
times weaker than the 95\,\% limit of $\pann < 2.1 \times 10^{-27}
\,{\rm cm^3}\,{\rm s}^{-1}\,{\rm GeV}^{-1}$ derived from WMAP9, which
includes WMAP polarization data at low multipoles.  On the other hand, the
\Planck\ $TE$ or $EE$ spectra improve the constraints on \pann\ by about
an order of magnitude compared to those from \Planck\ $TT$ alone. This is
because the main effect of dark matter annihilation is to increase the
width of last scattering, leading to a suppression of the amplitude of
the peaks, both in temperature and polarization. As a result, the
effects of DM annihilation on the power spectra at high multipole are
degenerate with other parameters of base \LCDM, such as $\ns$ and $\As$
\citep{Chen:2003gz,Padmanabhan:2005es}.  At large angular scales
($\ell\lesssim 200$), however, dark matter annihilation can produce an
enhancement in polarization, caused by the increased ionization
fraction in the freeze-out tail following recombination.  As a result,
large-angle polarization information is crucial for breaking the
degeneracies between parameters, as illustrated in
Fig.~\ref{fig:pann2D}. The strongest constraints on \pann\ therefore
come from the full \Planck\ temperature and polarization likelihood
and there is little improvement if other astrophysical data, or
\Planck\ lensing, are added.\footnote{It is interesting to note that
the constraint derived from \planckall\ is consistent with the
forecast given in \cite{Galli:2009zc}, $\pann<3\times 10^{-28}
\,{\rm cm^3}\,{\rm s}^{-1}\,{\rm GeV}^{-1}$.}

We verified the robustness of the \planckall\ constraint by also
allowing other parameter extensions of base $\Lambda$CDM ($\nnu$, $\nrun$,
or $\yhe$) to vary together with $\pann$. We found that the constraint
is weakened by up to 20\,\%. Furthermore, we have verified that we
obtain consistent results
when relaxing the priors on the amplitudes of the Galactic dust templates
or if we use the \camspec\ likelihood instead of the baseline \plik\
likelihood.

\begin{figure}
\centering
\includegraphics[angle=0,width=91mm,page=1]{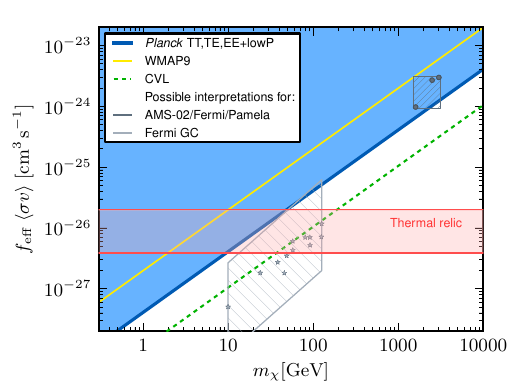}
\caption{Constraints on the self-annihilation cross-section at recombination,
$\langle \sigma \varv \rangle_{z_\ast}$, times the efficiency parameter, $f_{\rm eff}$
(Eq.~\ref{pann}). The blue area shows the parameter space excluded by the
\Planck\ TT{,}TE{,}EE+\lowTEB\ data at 95\,\% CL. The yellow line indicates
the constraint using WMAP9 data. The dashed green line delineates the region
ultimately accessible to a cosmic-variance-limited experiment with angular
resolution comparable to that of \Planck. The horizontal red band includes
the values of the
thermal-relic cross-section multiplied by the appropriate $f_{\rm eff}$ for
different DM annihilation channels. The dark grey circles show the best-fit
DM models for the PAMELA/AMS-02/{\it Fermi\/} cosmic-ray excesses, as
calculated in \citet{Cholis:2013psa} (caption of their figure~6). The light
grey stars show the best-fit DM models for the {\it Fermi\/} Galactic centre
$\gamma$-ray excess, as calculated by \citet{Calore:2014nla} (their tables~I,
II, and III), with the light grey area indicating the astrophysical
uncertainties on the best-fit cross-sections.}
\label{fig:pannconstraints}
\end{figure}

Figure~\ref{fig:pannconstraints} shows the constraints from WMAP9,
\planckall, and a forecast for a cosmic-variance-limited experiment
with similar angular resolution to \Planck.\footnote{We assumed here that the
  cosmic-variance-limited experiment would measure the angular power spectra
  up to a maximum multipole of $\ell_\mathrm{max}=2500$, observing a sky
  fraction $f_{\rm sky}=0.65$.}
The horizontal red band includes the values of the thermal-relic cross-section
multiplied by the appropriate $f_{\rm eff}$ for different DM annihilation
channels. For example, the upper red line corresponds to $f_{\rm eff}=0.67$,
which is appropriate for a DM particle of mass $m_\chi=10\,{\rm GeV}$
annihilating into ${\rm e}^+{\rm e}^-$, while the lower red line corresponds to
$f_{\rm eff}=0.13$,
for a DM particle annihilating into $2 \pi^+\pi^-$ through an intermediate
mediator \citep[see, e.g.,][]{ArkaniHamed:2008qn}. The \Planck\ data exclude at
95\,\% confidence level a thermal relic cross-section for DM particles of mass
$m_\chi\la 44\,{\rm Gev}$ annihilating into ${\rm e}^+{\rm e}^-$
($f_{\rm eff}\approx 0.6$), $m_\chi\la 16\,{\rm GeV}$ annihilating into
$\mu^+\mu^-$ or ${\rm b}\bar{\rm b}$ ($f_{\rm eff}\approx 0.2$),
and $m_\chi\la 11\,{\rm GeV}$ annihilating into $\tau^+\tau^-$
($f_{\rm eff}\approx 0.15$).

The dark grey shaded area in Fig.~\ref{fig:pannconstraints} shows the
approximate allowed region of parameter space, as calculated by
\cite{Cholis:2013psa} on the assumption that the PAMELA, AMS, and {\it Fermi\/}
cosmic-ray excesses are caused by DM annihilation; the dark grey dots
indicate the best-fit dark matter models described in that paper
\citep[for a recent discussion on best-fitting models, see
also][]{Boudaud:2014dta}.
The favoured value of the cross-section is about two orders of magnitude
higher than the thermal relic cross-section ($\approx 3\times10^{-26}
\,{\rm cm^3}\,{\rm s}^{-1}$). Attempts to reconcile such a high
cross-section with the relic abundance of DM include a Sommerfeld
enhanced cross-section (that may saturate at $\langle \sigma \varv \rangle
\approx 10^{-24}\,{\rm cm^3}\,{\rm s}^{-1}$) or non-thermal
production of DM \citep[see, e.g., the discussion by][]
{Madhavacheril:2013cna}. Both of these possibilities are strongly
disfavoured by the \Planck\ data. We cannot, however, exclude more
exotic possibilities, such as DM annihilation through a p-wave channel
with a cross-section that scales as $\varv^2$
\citep{Diamanti:2013bia}. Since the relative velocity of DM particles
at recombination is many orders of magnitude smaller than in the
Galactic halo, such a model cannot be constrained using CMB data.

Observations from the {\it Fermi\/} Large Area Telescope of extended
$\gamma$-ray emission towards the centre of the Milky Way, peaking at energies
of around 1--3\,GeV, have been interpreted as evidence for annihilating DM
\citep[e.g.,][]{Goodenough:2009,Gordon:2013,Daylan:2014,Abazajian:2014fta,
Lacroix:2014eea}. The light grey
stars in Fig.~\ref{fig:pannconstraints} show specific models of DM
annihilation designed to fit the {\it Fermi\/} $\gamma$-ray excess
\citep{Calore:2014nla}, while the light grey box shows the uncertainties of
the best-fit cross-sections due to imprecise knowledge of the Galactic DM
halo profile. Although the interpretation of the {\it Fermi\/}
excess remains controversial (because of uncertainties in the
astrophysical backgrounds), DM annihilation remains a possible
explanation. The best-fit models of \citet{Calore:2014nla} are
consistent with the \Planck\ constraints on DM annihilation.

\subsection{Testing recombination physics with \Planck\ }
\newcommand{\COSMORECvII}{{\tt CosmoRec v2.0}}
\newcommand{\changeJ}[1]{{\textcolor{red}{#1}}}
\newcommand{\changeJII}[1]{{\textcolor{green}{#1}}}
\label{sec:recombination}
The cosmological recombination process determines how CMB photons
decoupled from baryons around redshift $z \approx 10^3$, when the Universe
was about 400\,000 years old. The importance of this transition on
the CMB anisotropies has long been recognized
\citep{Sunyaev1970, Peebles1970}. The most advanced
computations of the ionization history \citep[e.g.,][]{Yacine2010,Chluba2010b,
 AliHaimoud:2010dx, Chluba2012HeRec} account for many subtle atomic
physics and radiative transfer effects that were not included in the
earliest calculations \citep{Zeldovich68, Peebles68}.

With precision data from \Planck, we are sensitive to sub-percent
variations of the free electron fraction around last-scattering
\citep[e.g.,][]{Hu1995, Seager2000, Seljak2003}. Quantifying the impact of
uncertainties in
the ionization history around the maximum of the Thomson visibility
function on predictions of the CMB power spectra is thus crucial
for the scientific interpretation of data from \Planck. In particular, for tests
of models of inflation and extensions to \LCDM, the interpretation of the
CMB data can be significantly compromised by inaccuracies in the recombination
calculation \citep[e.g.,][]{Wong2008,Jose2010, Shaw2011}. This problem can be
approached in two ways, either by using modified recombination models with a
specific physical process (or parameter) in mind, or in a semi-blind,
model-independent way. Both approaches provide useful insights in
assessing the robustness of the results from \Planck.

Model-dependent limits on {varying fundamental constants}
\citep{Kaplinghat1999, Scoccola2009, Galli2009b}, {annihilating} or
{decaying particles}
\citep[e.g.,][and Sect.~\ref{sec:DM}]{Chen:2003gz,Padmanabhan:2005es,Zhang2006fr},
or more general sources of {extra ionization} and {excitation photons}
\citep{Peebles2000, Doroshkevich2003, Galli2008}, have been discussed
extensively in the literature.

As already discussed in \paramsI, the choice for \Planck\ has been to use the
rapid calculations of the \RECFAST\ code, modified using corrections calculated
with the more precise codes.  To start this sub-section we
quantify the effect on the analysis of \Planck\ data of the remaining
uncertainties in the standard recombination history obtained with different
recombination codes
(Sect.~\ref{sec:standard_rec}). We also derive CMB anisotropy-based
measurements of the hydrogen $2s$--$1s$ two-photon decay rate,
$A_{2s\rightarrow 1s}$ (Sect.~\ref{sec:A2s1s}), and the average CMB
temperature, $T_0$ derived at the last-scattering epoch
(Sect.~\ref{sec:T0}). These two parameters strongly
affect the recombination history but are usually kept fixed when
fitting models to CMB data
(as in the analyses described in previous sections).
Section~\ref{sec:eXeM} describes model-independent constraints on perturbed
recombination scenarios. A discussion of these
cases provides both a test of the consistency of the CMB data with the
standard recombination scenario and also a demonstration of the
impressive sensitivity of \Planck\ to small variations in the
ionization history at $z\approx 1100$.

\subsubsection{Comparison of different recombination codes}
\label{sec:standard_rec}
Even for pre-\Planck\ data, it was realized that the early recombination
calculations of \cite{Zeldovich68} and \cite{Peebles68} had to be improved.
This led to the development of the widely-used and computationally quick
\RECFAST\ code
\citep{SeagerRecfast1999, Seager2000}.  However, for \Planck, the recombination
model of \RECFAST\ in its original form is not accurate enough.
Percent-level corrections, due to detailed radiative transfer and atomic
physics effects have to be taken into account.  Ignoring these effects can bias
the inferred cosmological parameters, some by as much as a few
standard deviations.

The recombination problem was solved as a common effort of several
groups \citep{Dubrovich2005, Kholupenko2007, Chluba2006,Jose2006,Karshenboim2008,Wong2007,Switzer2007I,Grin2009,Yacine2010}. This work
was undertaken, to a large extent, in preparation for the precision data from
\Planck. Both \COSMOREC\ \citep{Chluba2010b} and
\HYREC\ \citep{AliHaimoud:2010dx}
allow fast and precise computations of the ionization history,
explicitly capturing the physics of the recombination problem. For the
standard cosmology, the ionization histories obtained from these two
codes in their default settings agree to within 0.05\,\% for hydrogen
recombination ($600\la z\la 1600$) and 0.35\,\%
during helium recombination\footnote{Helium recombination is treated
 in more detail by \COSMOREC\ \citep[e.g.,][]{Jose2008,
 Chluba2012HeRec}, which explains most of the difference.}
($1600\la z\la 3000$). The effect of these
small differences on the CMB power spectra is $\la 0.1\,\%$ at
$\ell<4000$ and so has a negligible impact on the interpretation
of precision CMB data; for the standard six parameters of base \LCDM, we find
that the largest effect is a bias in $\ln(10^{10} A_{\rm s})$ at the level of
$0.04\,\sigma \approx 0.0012$ for \planckallBAO.

For \Planck\ analyses, the recombination model of \RECFAST\ is used by
default. In \RECFAST, the precise dynamics of recombination is not
modelled physically, but approximated with fitting-functions
calibrated against the full
recombination calculations assuming
a reference cosmology \citep{SeagerRecfast1999, Seager2000, Wong2008}.
At the level of precision required for
\Planck, the \RECFAST\ approach is sufficiently accurate,
provided that the cosmologies are close to base \LCDM\
\citep{Jose2010, Shaw2011}. Comparing the latest version of
\RECFAST\ (\CAMB\ version) with \COSMOREC, we find agreement to within
0.2\,\% for hydrogen recombination ($600\la z\la 1600$)
and 0.2\,\% during helium recombination for the
standard ionization history. The effect on the CMB power spectra
is $\la 0.15\,\%$ at $\ell<4000$, although with
slightly more pronounced shifts in the peak positions than when
comparing \COSMOREC\ and \HYREC. For the base \LCDM\ model, we
find that the largest bias is on $\ns$, at the level of $0.15\,\sigma$
($\approx 0.0006$) for \planckall+BAO. Although this is about 5 times larger
than the difference in $\ns$ between \COSMOREC\ and \HYREC, this bias is
nevertheless
unimportant at the current level of precision (and smaller than the differences
seen from different likelihoods, see Sect.~\ref{subsec:planck_only1}).

Finally we compare \COSMOREC\ with \RECFAST\ in its original form
(i.e., before recalibrating the fitting-functions on refined recombination
calculations).  For base \LCDM, we expect to see biases of
$\Delta \Omb h^2 \approx -2.1\,\sigma \approx -0.00028$ and
$\Delta \ns \approx -3.3\,\sigma \approx -0.012$ \citep{Shaw2011}.
Using the actual data (\planckall+BAO) we find biases of
$\Delta \Omb h^2 \approx -1.8\,\sigma \approx -0.00024$ and
$\Delta \ns \approx -2.6\,\sigma \approx-0.010$, very close
to the expected values.
This illustrates explicitly the importance of the improvements of
\COSMOREC\ and \HYREC\ over the original version of \RECFAST\
for the interpretation of \Planck\ data.  However, \COSMOREC\ and
\HYREC\ themselves are much more computationally intensive than the
modified \RECFAST, which is why we use \RECFAST\ in most \Planck\ cosmological
analyses.

\subsubsection{Measuring $A_{2s\rightarrow 1s}$ with \Planck\ }
\label{sec:A2s1s}
The crucial role of the $2s$--$1s$ two-photon decay channel for the
dynamics of hydrogen recombination has been appreciated since the early days of
CMB research \citep{Zeldovich68, Peebles68}. Recombination is an
out-of-equilibrium process and energetic photons emitted in the far
Wien tail of the CMB by Lyman continuum and series transitions keep
the primordial plasma ionized for a much longer period than expected from
simple equilibrium recombination physics. Direct recombinations to the ground
state of hydrogen are prohibited,
causing a modification of the free electron number density, $N_{\rm e}$,
by only $\Delta N_{\rm e}/N_{\rm e}\approx 10^{-6}$ around
$z\approx 10^3$ \citep{Chluba2007b}. Similarly, the slow escape of
photons from the Ly$\alpha$ resonance reduces the effective
Ly-$\alpha$ transition rate to $A^\ast_{2p\rightarrow1s}\approx
1$--10\,s$^{-1}$ (by more than seven orders of
magnitude), making it comparable to the vacuum $2s$--$1s$ two-photon
decay rate of $A_{2s\rightarrow 1s}\approx 8.22\,{\rm s^{-1}}$. About
$57\,\%$ of all hydrogen atoms in the Universe became neutral through
the $2s$--$1s$ channel \citep[e.g.,][]{Wong2006,Chluba2006b},
and subtle effects, such as
the induced $2s$--$1s$ two-photon decay and Ly$\alpha$
re-absorption, need to be considered in precision recombination calculations
 \citep{Chluba2006, Kholu2006, Hirata2008}.

The high sensitivity of the recombination process to the $2s$--$1s$
two-photon transition rate also implies that instead of simply
adopting a value for $A_{2s\rightarrow 1s}$ from theoretical
computations \citep{Breit1940, Spitzer1951,Goldman1989}
one can directly determine it
with CMB data. From the theoretical point of view it would be
surprising to find a value that deviates significantly from
$A_{2s\rightarrow 1s}=8.2206\,{\rm s^{-1}}$, derived from the most detailed
computation \citep{Labzowsky2005}. However, laboratory
measurements of this transition rate are extremely challenging
\citep{Connell1975, Krueger1975, Cesar1996}.
The most stringent limit is for the differential decay
rate, $A_{2s\rightarrow 1s}(\lambda)\,{\rm d} \lambda=(1.5 \pm
0.65)\,{\rm s}^{-1}$ (a $43\,\%$ error) at wavelengths
$\lambda=255.4$--$232.0\,{\rm nm}$, consistent with the
theoretical value of $A_{2s\rightarrow 1s}(\lambda)\,{\rm d}
\lambda=1.02\,{\rm s}^{-1}$ in the same wavelength range
\citep{Krueger1975}. With precision data from \Planck\ we are in a
position to perform the best measurement to date, using cosmological
data to inform us about atomic transition rates at last scattering \citep[as
also emphasized by][]{Mukhanov2012}.

The $2s$--$1s$ two-photon rate affects the CMB anisotropies only
through its effect on the recombination history. A larger value of
$A_{2s\rightarrow 1s}$, accelerates recombination, allowing photons and
baryons to decouple earlier, an effect that shifts the acoustic peaks
towards smaller scales. In addition, slightly less damping occurs, as
in the case of the stimulated $2s$--$1s$ two-photon decays
\citep{Chluba2006}. This implies that for flat cosmologies, variations
of $A_{2s\rightarrow 1s}$ correlate with $\Omc h^2$ and $H_0$ (which
affect the distance to the last scattering surface), while
$A_{2s\rightarrow 1s}$ anti-correlates with $\Omb h^2$ and $\ns$
(which modify the slope of the damping tail). Despite these degeneracies,
one expects that \Planck\ will provide a measurement of $A_{2s\rightarrow 1s}$
to within $\pm 0.5\,{\rm s}^{-1}$, corresponding to an approximately 6\,\%
uncertainty \citep{Mukhanov2012}.

\begin{figure}
\centering
\includegraphics[angle=0,width=90mm,page=1]{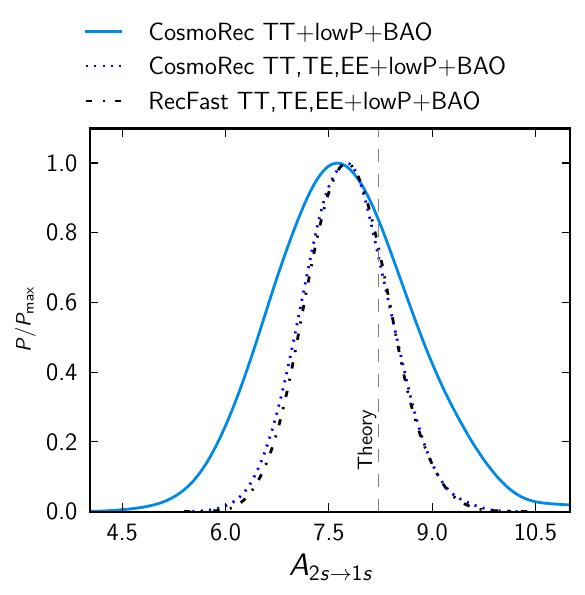}
\vspace{-3mm}
\caption{Marginalized posterior for $A_{2s\rightarrow 1s}$, obtained using
\COSMOREC, with and without small-scale polarization data.
We find good agreement with the theoretical value of
$A_{2s\rightarrow 1s}=8.2206\,{\rm s^{-1}}$. For comparison, we also
show the result for \planckall+BAO obtained with \RECFAST, emphasizing the
consistency of different treatments.}
\label{fig:A2s1s}
\end{figure}

In Fig.~\ref{fig:A2s1s}, we show the marginalized posterior for
$A_{2s\rightarrow 1s}$ from \Planck\ and for \Planck\ combined with BAO.
Using \COSMOREC\ to compute the recombination history, we find
\beglet
\begin{eqnarray}
A_{2s\rightarrow 1s}&\!=&\!7.70\pm 1.01\,{\rm s}^{-1} \quad
 \datalabel{\planckTT}, \\
A_{2s\rightarrow 1s}&\!=&\!7.72\pm 0.60\,{\rm s}^{-1} \quad
 \datalabel{\planckall}, \\
A_{2s\rightarrow 1s}&\!=&\!7.71\pm 0.99\,{\rm s}^{-1} \quad
 \datalabel{\planckTTBAO}, \\
A_{2s\rightarrow 1s}&\!=&\!7.75\pm 0.61\,{\rm s}^{-1} \quad
 \datalabel{\planckall}\nonumber\\
 &&\qquad\qquad\qquad\qquad\qquad\qquad\qquad\qquad {\dataplus\BAO}.
\end{eqnarray}
\endlet
These results are in very good agreement with the theoretical value,
$A_{2s\rightarrow 1s}=8.2206\,{\rm s^{-1}}$. For \planckall+BAO, approximately
8\,\% precision is reached using cosmological data. These constraints
are not sensitive to the addition of BAO, or other external data (JLA+$H_0$).
The slight shift away from the theoretical value is accompanied by
small (fractions of a $\sigma$) shifts in $\ns$, $\Omc h^2$, and
$H_0$, to compensate for the effects of $A_{2s\rightarrow 1s}$
on the distance to the last scattering surface and damping tail.
This indicates that
additional constraints on the acoustic scale are required to fully
break degeneracies between these parameters and their effects on the
CMB power spectrum, a task that could be achieved in the future using
large-scale structure surveys and next generation CMB experiments.

The values for $A_{2s\rightarrow 1s}$ quoted above were obtained using
\COSMOREC. When varying $A_{2s\rightarrow 1s}$, the range of
cosmologies becomes large enough to potentially introduce a mismatch of the
\RECFAST\ fitting-functions that could affect the posterior. In
particular, with \RECFAST\ the $2s$--$1s$ two-photon and
Ly$\alpha$ channels are not treated separately, so that changes
specific to the $2s$--$1s$ decay channel propagate
inconsistently.\footnote{One effect is that by increasing
 $A_{2s\rightarrow 1s}$ fewer Ly$\alpha$ photons are
 produced. This reduces the Ly$\alpha$ feedback correction to the
 $2s$--$1s$ channel, which further accelerates recombination, an
 effect that is not captured with \RECFAST\ in the current
 implementation.} However, repeating the analysis with \RECFAST, we find
$A_{2s\rightarrow 1s}=7.78 \pm 0.58\,{\rm s}^{-1}$ (see
Fig.~\ref{fig:A2s1s}), for \planckall+BAO, which is in excellent agreement
with \COSMOREC, showing that these effects can be neglected.

\subsubsection{Measuring $T_0$ at last-scattering with \Planck\ }
\label{sec:T0}
Our best constraint on the CMB monopole temperature comes from the
measurements of the CMB spectrum with COBE/FIRAS, giving a $0.02\,\%$
determination of $T_0$ \citep{Fixsen1996, Fixsen2009}. Other
constraints from molecular lines typically reach $1\,\%$ precision
\citep[see table 2 in][for an overview]{Fixsen2009}, while independent
BBN constraints provide 5--10\,\% limits \citep{Simha2008,
 Jeong2014}.

The CMB anisotropies provide additional ways of determining
the value of $T_0$ (for fixed values of $\neff$ and $\yhe$).
One is through the energy distribution of the CMB anisotropies
\citep{Fixsen1996, Fixsen2003, Chluba2014TRR} and another through their
power spectra \citep{Opher2004, Opher2005, Chluba2014TRR}. Even small changes
in $T_0$, compatible with the COBE/FIRAS error, affect the ionization history
at the 0.5\,\% level around last-scattering, propagating to a roughly 0.1\,\%
uncertainty in the CMB power spectrum \citep{Chluba2008T0}. Overall, the
effect of this uncertainty on the parameters of \LCDM\ models is small
\citep{Hamann2008}; however, without prior knowledge of $T_0$ from the
COBE/FIRAS measurement, the situation would change significantly.

The CMB monopole affects the CMB anisotropies in several ways. Most
importantly, for larger $T_0$, photons decouple from baryons at lower
redshift, since more ionizing photons are present in the Wien-tail of
the CMB. This effect is amplified because of the exponential dependence
of the atomic level populations on the ratio of the ionization
potentials and CMB temperature. In addition, increasing $T_0$ lowers
the expansion timescale of the Universe and the redshift of matter-radiation
equality, while increasing the photon sound speed. Some of
these effects are also produced by varying $\neff$; however, the
effects of $T_0$ on the ionization history and photon sound speed are
distinct.

\begin{figure}
\centering
\includegraphics[angle=0,width=90mm,page=1]{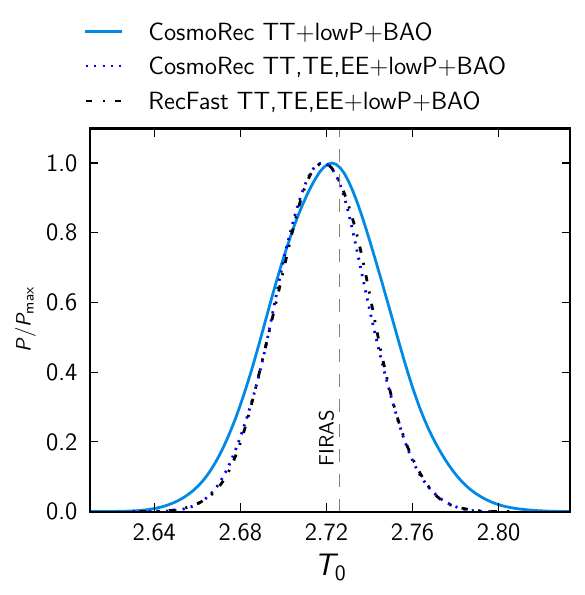}
\vspace{-3mm}
\caption{Marginalized posterior for $T_0$. We find excellent agreement with
the COBE/FIRAS measurement. For comparison, we show the result for
\planckall+BAO obtained using both \COSMOREC\ and \RECFAST, emphasizing the
consistency of different treatments.}
\label{fig:T0}
\end{figure}

With CMB data alone, the determination of $T_0$ is degenerate with other
parameters, but the addition of other
data sets breaks this degeneracy.
Marginalized posterior distributions for $T_0$ are shown in Fig.~\ref{fig:T0}.
Using \COSMOREC, we find
\beglet
\begin{eqnarray}
T_0&=&2.722\pm 0.027\,{\rm K} \quad\datalabel{\planckTTBAO},
\\
T_0&=&2.718 \pm 0.021\,{\rm K} \quad\datalabel{\planckallBAO},
\end{eqnarray}
\endlet
and similar results are obtained with \RECFAST.
This is in excellent agreement with the COBE/FIRAS measurement,
$T_0=2.7255\pm 0.0006\,{\rm K}$ \citep{Fixsen1996, Fixsen2009}.
These measurements of $T_0$
reach a precision that is comparable to the accuracy obtained with
interstellar molecules. Since the systematics of these independent
methods are very different, this result demonstrates the consistency of
all these data. Allowing $T_0$ to vary causes the errors of the
other cosmological parameters to increase. The strongest effect is
on $\theta_{\rm MC}$, which is highly degenerate with $T_0$. The error on
$\theta_{\rm MC}$ increases by a factor of roughly $25$ if $T_0$ is allowed
to vary. The error
on $\Omb h^2$ increases by a factor of about 4, while the errors on $\ns$
and $\Omc h^2$ increase by factors of 1.5--2, and the other cosmological
parameters are largely unaffected by variations in $T_0$. Because of the
strong degeneracy with $\theta_{\rm MC}$, no constraint on $T_0$ can be
obtained using \Planck\ data alone.  External data, such as BAO, are
therefore required to break this geometric degeneracy.

It is important to emphasize that the CMB measures the temperature at a
redshift of $z\approx 1100$, so the comparison with measurements of $T_0$
at the present day is effectively a test of the constancy of $aT_{\rm CMB}$,
where $a \approx 1/1100$ is the scale-factor at the time
of last-scattering. It is remarkable that we are able to test
the constancy of $aT_{\rm CMB}\equiv T_0$ over such a large dynamic range in
redshift.  Of course, if we did find that $aT_{\rm CMB}$ around recombination
were discrepant with $T_0$ now, then we would need to invent a finely-tuned
late-time photon injection mechanism\footnote{Pure energy release in the form
of heating of ordinary matter would leave a Compton $y$-distortion
\citep{Zeldovich1969} at these late times
\citep{Burigana1991, Hu1993, Chluba2011therm}.} to explain the anomaly.
Fortunately, the data are consistent with the standard
$T_{\rm CMB}\propto (1+z)$ scaling of the CMB temperature.

Another approach to measuring $a T_{\rm CMB}$ is through the thermal
Sunyaev-Zeldovich effect in rich clusters of galaxies at various redshifts
\citep{Fabbri1978, Rephaeli1980}, although it is unclear how one would
interpret a failure of this test without an explicit model.
In practice this approach
is consistent with a scaling $aT_{\rm CMB}= {\rm constant}$, but with lower
precision than obtained here from \Planck\
\citep[e.g.,][]{Battistelli2002, Luzzi2009, Saro2013, Hurier2014}.
A simple $T_{\rm CMB}=T_0(1+z)^{1-\beta}$ modification to the standard
temperature redshift relation is frequently discussed in the literature
\citep[though this case is not justified by any physical model and is
difficult to realize without creating a CMB spectral distortion,
see][]{Chluba2014TRR}. For this parameterization we find
\beglet
\begin{eqnarray}
\beta&=& (0.2\pm 1.4)\times 10^{-3} \,\ \ \datalabel{\planckTTBAO},
\\
\beta&=& (0.4\pm 1.1)\times 10^{-3} \,\ \ \datalabel{\planckallBAO},
\end{eqnarray}
\endlet
where we have adopted a recombination redshift of
$z_\ast =1100$.\footnote{The test depends on the logarithm of the
redshift and so is insensitive to the precise value adopted for
$z_\ast$.} Because of the long lever-arm in redshift afforded by the CMB,
this is an improvement over earlier constraints by more than an order of
magnitude \citep[e.g.,][]{Hurier2014}.

In a self-consistent picture, changes of $T_0$ would also affect the
BBN era. We might therefore consider a simultaneous variation of
$\neff$ and $\yhe$ to reflect the variation of the neutrino energy
density accompanying a putative variation in the photon energy
density. Since we find $aT_{\rm CMB}$ at recombination to be highly consistent
with the observed CMB temperature from COBE/FIRAS, considering this
extra variation seems unnecessary. Instead, we may view the $aT_{\rm CMB}$
variation investigated here as further support for the limits
discussed in Sects.~\ref{sec:neutrino} and \ref{sec:bbn}.

\subsubsection{Semi-blind perturbed recombination analysis}
\label{sec:eXeM}

The high sensitivity of small-scale CMB anisotropies to the ionization
history of the Universe around the epoch of recombination allows us to
constrain possible deviations from the standard recombination scenario
in a model-independent way \citep{Farhang2011, Farhang2013}. The
method relies on an eigen-analysis (often referred to as a {principle
component analysis}) of perturbations in the free electron fraction,
$X_{\rm e}(z)=N_{\rm e}/N_{\rm H}$, where $N_{\rm H}$ denotes the
number density of hydrogen nuclei. The eigenmodes selected are specific to the
data used in the analysis. Similar approaches
have been used to constrain deviations of the reionization
history from the simplest models \citep{Mortonson2008} and
annihilating dark matter scenarios \citep{Finkbeiner:2011dx}, both
with the prior assumption that the standard recombination physics is
fully understood, as well as for constraining
trajectories in inflation \cite{planck2014-a24} and
dark energy \cite{planck2014-a16} parameterizations.

Here, we use \Planck\ data to find preferred ionization fraction
trajectories $X_{\rm e}(z)$ composed of low-order perturbation
eigenmodes to the standard history ($X_{\rm e}$-modes).  The $X_{\rm e}$-modes
are constructed through the eigen-decomposition of the
inverse of the Fisher information matrix for base \LCDM\ (the six
cosmological parameters and the nuisance parameters) and
recombination perturbation parameters \citep[see][for
 details]{Farhang2011}. This procedure allows us to estimate the
errors on the eigenmode amplitudes, $\mu_i$, providing a rank
ordering of the $X_{\rm e}$-modes and their information content.

\begin{figure}[t]
\centering
\includegraphics[width=96mm]{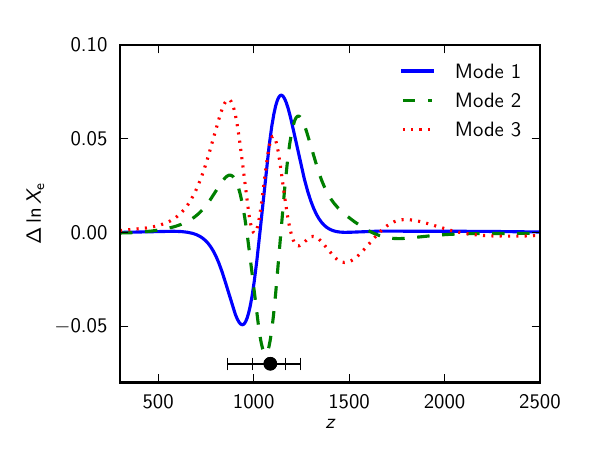} \\
\includegraphics[width=96mm]{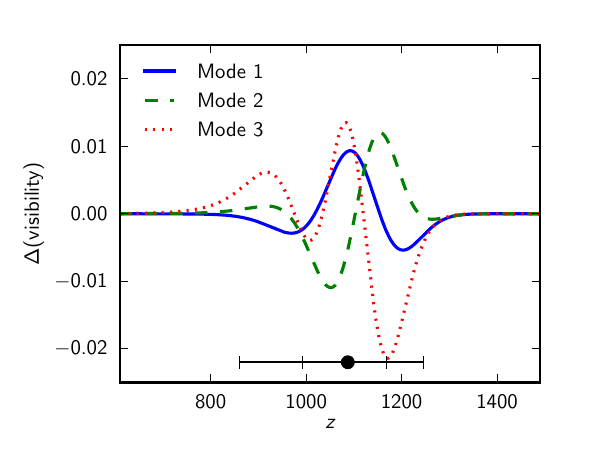}
\vspace{-3mm}
\caption{Eigen-modes of the recombination history, marginalized over the
standard six cosmological and \Planck\ nuisance parameters.
The upper panel shows the first three $X_{\rm e}$-modes constructed for
\planckall\ data.  The lower panel show changes in the differential visibility
corresponding to $1\,\sigma$ deviations from the standard recombination
scenario for the first three $X_{\rm e}$-modes. The maximum of the Thomson
visibility function and width are indicated in both figures.}
\label{modes-Xe-vis}
\end{figure}

\begin{figure}[t]
\centering
\includegraphics[width=96mm]{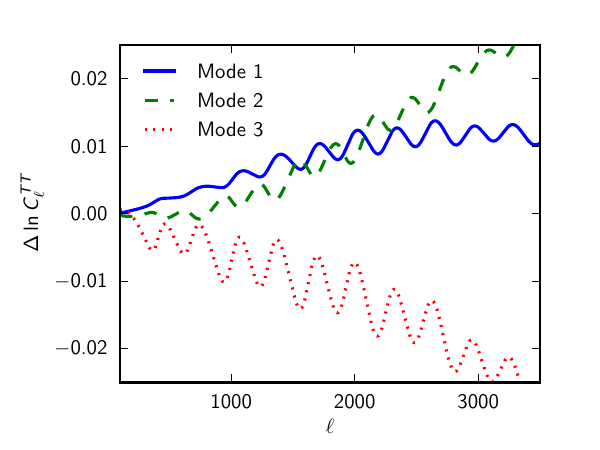} \\
\includegraphics[width=96mm]{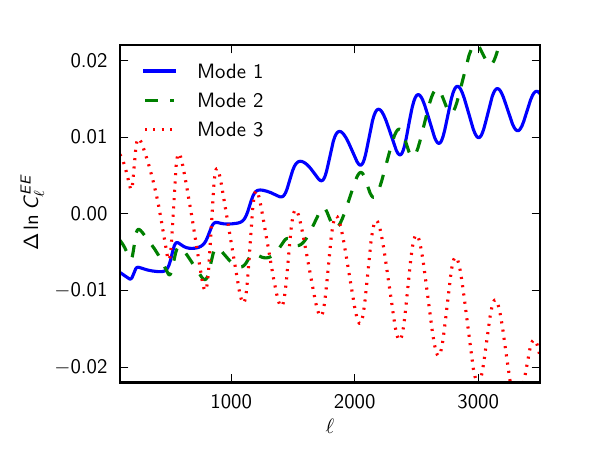}
\vspace{-3mm}
\caption{Changes in the $TT$ (upper panel) and $EE$ (lower panel) power
spectra caused by a $1\,\sigma$ deviation from the standard recombination
scenario for the first three $X_{\rm e}$-modes (see Fig.~\ref{modes-Xe-vis}).}
\label{modes-cls}
\end{figure}

The first three $X_{\rm e}$-modes for \planckall\ are illustrated in
Fig.~\ref{modes-Xe-vis}, together with their impact on
 the differential visibility function. Figure~\ref{modes-cls} shows
the response of the CMB temperature and polarization power spectra to these
eigenmodes. The first mode mainly leads to a
change in the width and height of the Thomson visibility function
(bottom panel of Fig.~\ref{modes-Xe-vis}). This implies less diffusion
damping, which is also reflected in the modifications to the CMB power
spectra (as shown in Fig.~\ref{modes-cls}). The second mode causes the
visibility maximum to shift towards higher redshifts for $\mu_2>0$,
which leads to a shift of the
CMB extrema to smaller scales; however, for roughly constant width of
the visibility function it also introduces less damping at small
scales. The third mode causes a combination
of changes in both the position and width of the visibility function,
with a pronounced effect on the location of the acoustic peaks.
For the analysis of \planck\ data
combinations, we only use $X_{\rm e}$-modes that are optimized for \planckall.

We modified \COSMOMC\ to estimate the mode amplitudes. The results for
\planckall+BAO are presented in Table~\ref{xe-mcmc-2}. Although all
mode amplitudes are consistent with standard recombination, adding the
second $X_{\rm e}$-mode causes mild shifts in $H_0$ and $\tau$. For
\planckTT, we find $\mu_1= -0.11 \pm 0.51$ and $\mu_2=-0.23 \pm 0.50$,
using the \planckall\ eigenmodes, again consistent with the standard
recombination scenario.  Adding the
polarization data improves the errors by more than a factor of 2.
However, the mode amplitudes are insensitive to the addition of
external data.

\begin{table}[tb]
\begingroup
\newdimen\tblskip \tblskip=5pt
\caption{%
Standard parameters and the first three $X_{\rm e}$-modes, as determined for
\planckall+BAO.
}
\label{xe-mcmc-2}
\nointerlineskip
\vskip -3mm
\small
\setbox\tablebox=\vbox{
   \newdimen\digitwidth
   \setbox0=\hbox{\rm 0}
   \digitwidth=\wd0
   \catcode`*=\active
   \def*{\kern\digitwidth}
   \newdimen\signwidth
   \setbox0=\hbox{+}
   \signwidth=\wd0
   \catcode`!=\active
   \def!{\kern\signwidth}
\halign{
\hbox to 0.5in{#\leaderfil}
\tabskip -1.0em&
\hfil#\hfil\tabskip -0.5em&
\hfil#\hfil\tabskip 0em&
\hfil#\hfil\tabskip 0pt\cr
\noalign{\doubleline}
\noalign{\vskip -2pt}
\omit Parameter\hfil& *+ 1 mode& *+ 2 modes& *+ 3 modes\cr
\noalign{\vskip 3pt\hrule\vskip 5pt}
                                    % Body of table goes here.
$\Omb h^2$&
 $!*0.02229\pm0.00017$& $!*0.02237\pm0.00018$& $!0.02237\pm0.00019$\cr
$\Omc h^2$&
 $!**0.1190\pm0.0010*$& $!**0.1186\pm0.0011*$& $!*0.1187\pm0.0012*$\cr
$H_0$&
 $!***67.64\pm0.48***$& $!***67.80\pm0.51***$& $!**67.80\pm0.56***$\cr
$\tau$&
 $!***0.065\pm0.012**$& $!***0.068\pm0.013**$& $!**0.068\pm0.013**$\cr
$\ns$&
 $!**0.9667\pm0.0053*$& $!**0.9677\pm0.0055*$& $!*0.9678\pm0.0067*$\cr
$\ln(10^{10}A_{\rm s})$&
 $!***3.062\pm0.023**$& $!***3.066\pm0.024**$& $!**3.066\pm0.024**$\cr
$\mu_1$&
 $****-0.03\pm0.12***$& $!****0.03\pm0.14***$& $!***0.02\pm0.15***$\cr
$\mu_2$&
               !*\dots& $****-0.17\pm0.18***$& $***-0.18\pm0.19***$\cr
$\mu_3$&
               !*\dots&               !*\dots& $***-0.02\pm0.88***$\cr
\noalign{\vskip 3pt\hrule\vskip 3pt}}}
\endPlancktablewide
\endgroup
\end{table}

With pre-\Planck\ data, only the amplitude, $\mu_1$, of the first
eigenmode could be constrained. The corresponding change in the
ionization history translates mainly into a change in the slope of the
CMB damping tail, with this mode resembling the first mode determined using
\Planck\ data
(Fig.~\ref{modes-Xe-vis}). The WMAP9+SPT data gave a non-zero value
for the first eigenmode at about $2\,\sigma$, $\mu^{\rm SPT}_1=-0.80\pm0.37$.
However, the WMAP9+ACT data gave $\mu^{\rm ACT}_1=0.14\pm0.45$
and the combined pre-\Planck\ data
(WMAP+ACT+SPT) gave $\mu^{\rm pre}_1=-0.44\pm0.33$, both consistent
with the standard recombination scenario \citep{Calabrese:2013}.
The variation among these results is another manifestation of the
tensions between different pre-\Planck\ CMB data, as discussed in \paramsI.

Although not optimal for \Planck\ data, we also compute the
amplitudes of the first three $X_{\rm e}$-modes constructed for the
WMAP9+SPT data set.  This provides a more direct comparison with the
pre-\Planck\ constraints. For \planckall+BAO we obtain
$\mu^{\rm SPT}_1=-0.10 \pm 0.13$ and $\mu^{\rm SPT}_2=-0.13 \pm 0.18$.
The mild tension of the pre-\Planck\ data with
the standard recombination scenario disappears when using \Planck\ data. This
is especially impressive, since the errors have improved by more than a factor
of 2.
By projecting onto the \Planck\ modes, we find that the first two SPT
modes can be expressed as $\mu^{\rm SPT}_1\approx 0.69 \mu_1 + 0.66
\mu_2\approx -0.09$ and $\mu^{\rm SPT}_2\approx -0.70 \mu_1 + 0.64
\mu_2\approx -0.13$, which emphasizes the consistency of the
results. Adding the first three SPT modes, we obtain $\mu^{\rm
 SPT}_1=-0.09 \pm 0.13 $, $\mu^{\rm SPT}_2=-0.14 \pm 0.21$, and
$\mu^{\rm SPT}_3=-0.12 \pm 0.86$, which again is consistent with the standard
model of recombination.  The small changes in the mode
amplitudes when adding the third mode arise because the SPT modes are
not optimal for \Planck\ and so are correlated.

\subsection{Cosmic defects\label{sec:defects}}

Topological defects are a generic by-product of symmetry-breaking
phase transitions and a common phenomenon in condensed matter
systems. Cosmic defects of various types can be formed in phase
transitions in the early Universe \citep{Kibble:1976sj}.  In
particular, cosmic strings can be produced in some supersymmetric and
grand-unified theories at the end of inflation
\citep{Jeannerot:2003qv}, as well as in higher-dimensional theories
\citep[e.g.,][]{Polchinski:2004hb}.  Constraints on the abundance of
cosmic strings and other defects therefore place limits on a range of
models of the early Universe.  More discussion on the formation, evolution, and
cosmological role of topological defects can be found, for example, in
the reviews by \cite{vilenkin2000cosmic}, \cite{Hindmarsh:1994re}, and
\cite{Copeland:2009ga}.

In this section we revisit the power spectrum-based constraints
on the abundance of cosmic strings and other topological defects
using the 2015
\Planck\ data, including \Planck\ polarization measurements.
The general approach follows that described in the \Planck\ 2013
analysis of cosmic defects \citep{planck2013-p20},
so here we focus on the updated constraints rather than on details of
the methodology.

\begin{figure}[!t]
\begin{center}
\includegraphics[width=86mm]{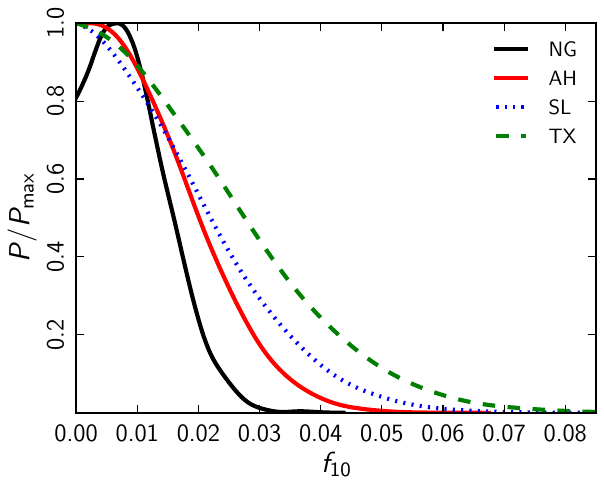}
\includegraphics[width=86mm]{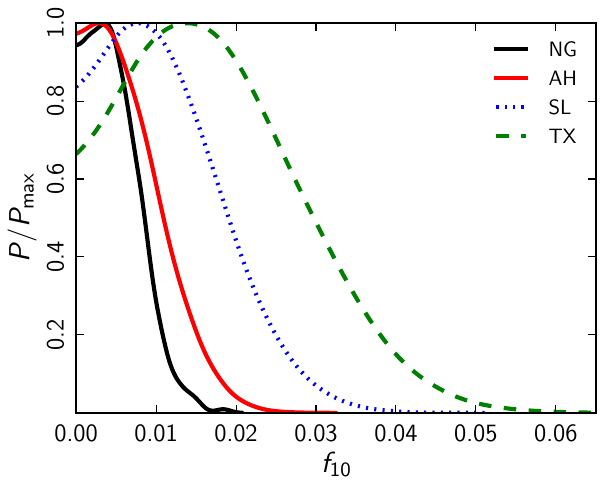}
\caption{Marginalized posterior distributions for the fractional
contribution, $f_{10}$, of the defect contribution to the temperature power
spectrum at $\ell=10$ (see the text for the precise definition).
Here we show the constraints for the Nambu-Goto cosmic strings
(NG, solid black), field-theory simulations of Abelian-Higgs cosmic strings
(AH, solid red), semi-local strings (SL, dotted blue), and global textures
(TX, dashed green).
The upper panel shows the 1D posterior for the \Planck+lowP data, while
constraints shown in the lower panel additionally use the $TE$ and $EE$ data.}
\label{fig:def_field_1D}
\end{center}
\end{figure}

Topological defects are non-perturbative excitations of the underlying
field theory and their study requires numerical simulations.
Unfortunately, since the Hubble scale, $c/H_0$, is over 50
orders of magnitude greater that the thickness of a GUT-scale string, approximately
$(\hbar/\mu c)^{1/2}$ with $\mu$ the mass per unit length of
the string, it is impractical to simulate the string
dynamics exactly in the late Universe. For this reason one needs to make
approximations. One approach considers the limit of an infinitely thin
string, which corresponds to using the Nambu-Goto (``NG'') action for the
string dynamics.  In an alternative approach, the actual field dynamics for a
given model are solved on a lattice. In this case it is necessary to
resolve the
string core, which generally requires more computationally intensive simulations than in the NG
approach. Lattice simulations, however, can include additional physics,
such as field radiation that is not present in NG simulations. Here we will use
field-theory simulations of the Abelian-Higgs action (``AH''); details of these
simulations are discussed in \cite{Bevis:2006mj,Bevis:2010gj}.

The field-theory approach also allows one to simulate theories in which
the defects are not cosmic strings and so cannot be described by the NG
action. Examples include semi-local strings
\citep[``SL'',][]{Urrestilla:2007sf} and global defects. Here we will
specifically consider the breaking of a global $O(4)$ symmetry resulting in
texture defects (``TX'').

For the field-theory defects, we measure the energy-momentum tensor
from the simulations and insert it as an additional constituent into a
modified version of the {\tt CMBEASY} Boltzmann code
\citep{Doran:2003sy} to predict the defect contribution to
the CMB temperature and polarization power spectra \citep[see,
e.g.,][]{Durrer:2001cg}. The same approach can be applied to NG
strings, but rather than using simulations directly, we model the strings
using the unconnected segment model
\citep[``USM'',][]{Albrecht:1997mz,Pogosian:1999np}. In this model, strings
are represented by a set of uncorrelated straight segments, with scaling
properties chosen to match those determined from numerical simulations. In this
case, the string energy-momentum tensor can be computed analytically
and used as an active source in a modified Boltzmann code. For this
analysis we use {\tt CMBACT} version
4,\footnote{\url{http://www.sfu.ca/~levon/cmbact.html}}
whereas
\cite{planck2013-p20} used version 3. There have been several
improvements to the code since the 2013 analysis, including a correction to the
normalization of vector mode spectra. However, the largest change comes
from an improved treatment of the scaling properties. The string
correlation length and velocity are described by an updated
velocity-dependent one-scale model~\citep{Martins:2000cs}, which
provides better agreement with numerical simulations.
Small-scale structure of the string, which was previously a free
parameter, is accounted for by the one-scale model.

The CMB power spectra from defects are proportional to $(G\mu/c^2)^2$. We scale the computed template CMB spectra, and add these to the inflationary and
foreground power spectra, to form the theory spectra that enter the likelihood. In practice, we parameterize the defects with their relative contribution to the $TT$ spectrum at multipole $\ell=10$, $f_{10} \equiv
C^{TT{\rm(defect)}}_{10}/C^{TT{\rm(total)}}_{10}$. We vary $f_{10}$ and the standard six parameters of the base \LCDM\ model, using \COSMOMC. We also report our results in terms of the derived parameter $G\mu/c^2$.

\begin{table}[!bt]
\begingroup
\newdimen\tblskip \tblskip=5pt
\caption{95\,\% upper limits on the parameter $f_{10}$
and on the derived parameter $G\mu/c^2$ for the defect models discussed in
the text. We show results for \Planck\ TT+lowP data as well as for
\Planck\ TT{,}TE{,}EE+lowP. }
\label{tab:defect_constraints}
\nointerlineskip
\vskip -3mm
\setbox\tablebox=\vbox{
   \newdimen\digitwidth
   \setbox0=\hbox{\rm 0}
   \digitwidth=\wd0
   \catcode`*=\active
   \def*{\kern\digitwidth}
   \newdimen\signwidth
   \setbox0=\hbox{+}
   \signwidth=\wd0
   \catcode`!=\active
   \def!{\kern\signwidth}
\halign{\hbox to 0.75in{#\leaderfil}\tabskip=1em&
        \hfil#\hfil&
        \hfil#\hfil&
        \hfil#\hfil&
        \hfil#\hfil\tabskip=0pt\cr                % Template goes here.
\noalign{\doubleline}
\omit& \multispan2\hfil{TT+lowP}\hfil& \multispan2\hfil{TT,TE,EE+lowP}\hfil\cr
\noalign{\vskip 2pt}
\omit Defect type\hfil& $f_{10}$& $G\mu/c^2$& $f_{10}$& $G\mu/c^2$\cr
\noalign{\vskip 3pt\hrule\vskip 5pt}
NG& $<0.020$& $<*1.8 \times 10^{-7}$& $<0.011$& $<1.3 \times 10^{-7}$\cr
AH& $<0.030$& $<*3.3 \times 10^{-7}$& $<0.015$& $<2.4 \times 10^{-7}$\cr   
SL& $<0.039$& $<10.6 \times 10^{-7}$& $<0.024$& $<8.5 \times 10^{-7}$\cr  
TX& $<0.047$& $<*9.8 \times 10^{-7}$& $<0.036$& $<8.6 \times 10^{-7}$\cr 
\noalign{\vskip 3pt\hrule\vskip 4pt}}}
\endPlancktable                    % ends one-column \halign
\endgroup
\end{table}

The constraints on $f_{10}$ and the inferred limits on $G\mu/c^2$ are
summarized in Table~\ref{tab:defect_constraints}. The marginalized 1D
posterior distribution functions are shown in
Fig.~\ref{fig:def_field_1D}. For \planckTT\ we find that the
results are similar to the \Planck +WP constraints reported in
\citet{planck2013-p20}, for the AH model, or somewhat better for SL
and TX.  However, the addition of the \Planck\ high-$\ell$ $TE$ and $EE$
polarization data leads to a significant improvement compared to the
2013 constraints. 

For the NG string model, the results based on
\planckTT\ are slightly weaker than the 2013
\Planck +WP constraints. This is caused by a
difference in the updated defect spectrum from the USM model, which has a
less pronounced peak and shifts towards the AH spectrum. With the
inclusion of polarization, \planckall\ improves the
upper limit on $f_{10}$ by a factor of 2, as for the AH
model.  The differences between the AH and NG results quoted here can be
regarded as a rough indication of the uncertainty in the theoretical
string power spectra.

In summary, we find no evidence for cosmic defects from the \Planck\ 2015 data,
with tighter limits than before.

\section[Conclusions]{Conclusions\footnote{As in the abstract, here we quote
68\,\% confidence limits on measured parameters and 95\,\% upper limits on
other parameters.}}  \label{sec:conclusions}

\paragraph{(1)}
The six-parameter base \LCDM\ model continues to
provide a very good match to the more extensive 2015 \Planck\ data, including
polarization.
This is the most important conclusion of this paper.

\paragraph{(2)}
The 2015 \Planck\ $TT$, $TE$, $EE$, and lensing spectra are consistent with
each other under the assumption of the base \LCDM\ cosmology.
However, when comparing the $TE$ and $EE$ spectra computed for different
frequency combinations, we find evidence for systematic effects caused by
temperature-to-polarization leakage. These effects are at low
levels and have little impact on the science conclusions of this paper.

\paragraph{(3)}
We have presented the first results on polarization from the LFI at
low multipoles. The LFI polarization data, together with
\Planck\ lensing and high-multipole temperature data, gives a
reionization optical depth of $\tau = 0.066 \pm 0.016$ and a
reionization redshift of $\zre = 8.8^{+1.7}_{-1.4}$. These numbers are
in good agreement with those inferred from the WMAP9 polarization data
cleaned for polarized dust emission using HFI 353-GHz maps.
They are also in good agreement with results from \Planck\ temperature
and lensing data, i.e., excluding any information from polarization at
low multipoles.

\paragraph{(4)}
The absolute calibration of the \Planck\ 2015 HFI spectra is higher by
$2\,\%$ (in power) compared to 2013, largely resolving the calibration
difference noted in {\paramsI} between WMAP and \Planck. In addition, there
have been a number of small changes to the low-level
\Planck\ processing and more accurate calibrations of the HFI
beams. The 2015 \Planck\ likelihood also makes more aggressive use of
sky than in {\paramsI} and incorporates some refinements to the modelling
of unresolved foregrounds. Apart from differences in $\tau$ (caused by
switching to the LFI low-multipole polarization likelihood, as described in
item 3 above) and the amplitude-$\tau$ combination
$A_{\rm s} {\rm e}^{-2 \tau}$ (caused by the change in absolute
calibration), the 2015 parameters for base \LCDM\ are in good
agreement with those reported in {\paramsI}.

\paragraph{(5)}
The \Planck\ $TT$, $TE$, and $EE$ spectra are accurately described by a
purely adiabatic spectrum of fluctuations with a spectral tilt $\ns =
0.968 \pm 0.006$, consistent with the predictions of single-field
inflationary models. Combining \Planck\ data with BAO, we find tight
limits on the spatial curvature of the Universe, $\vert \Omega_K\vert
< 0.005$, again consistent with the inflationary prediction of
a spatially-flat Universe.

\paragraph{(6)}
The \Planck\ data show no evidence for tensor modes. Adding a tensor
amplitude as a one-parameter extension to base \LCDM, we derive a 95\,\%
upper limit of $r_{0.002} < 0.11$. This is consistent with the
$B$-mode polarization analysis reported in {\BKP}, resolving the
apparent discrepancy between the \Planck\ constraints on $r$ and the
BICEP2 results reported by \citet{BicepDetection}. In fact, by combining
the \Planck\ and BKP likelihoods, we find an even tighter constraint,
$r_{0.002} < 0.09$, strongly disfavouring inflationary models with a
$V(\phi) \propto \phi^2$ potential.

\paragraph{(7)}
The \Planck\ data show no evidence for any significant running of
the spectral index. We also set strong limits on a possible departure from
a purely adiabatic spectrum, either through an admixture of
fully-correlated isocurvature modes or from cosmic defects.

\paragraph{(8)}
The \Planck\ best-fit base \LCDM\ cosmology (we quote numbers for
\planckTT+lensing here) is in good agreement with results from BAO
surveys, and with the recent JLA sample of Type Ia SNe. The Hubble
constant in this cosmology is $H_0 = (67.8 \pm 0.9)\Hunit$,
consistent with the direct measurement of $H_0$ of Eq.~(\ref{H0prior1})
used as an $H_0$ prior in this paper. The \Planck\ base
\LCDM\ cosmology is also consistent with the recent analysis of
redshift-space distortions of the BOSS CMASS-DR11 data by
\citet{Samushia:14} and \citet{Beutler:14}.  The amplitude of the
present-day fluctuation spectrum, $\sigma_8$, of the \Planck\ base
\LCDM\ cosmology is higher than inferred from weak lensing measurements
from the \CFHTLENS\ survey \citep{Heymans:2012, Erben:2013} and, possibly,
from counts of rich clusters of galaxies (including \Planck\ cluster counts
reported in \citealt{planck2014-a30}). The \Planck\ base \LCDM\ cosmology is
also discordant with Ly$\alpha$ BAO measurements at $z\approx 2.35$
\citep{Delubac:2014, Font-Ribera:2014}. At present, the reasons for these
tensions are unclear.

\paragraph{(9)}
By combining the \planckTT+lensing data with other astrophysical data,
including the JLA supernovae, the equation of state for dark energy is
constrained to $w = -1.006 \pm 0.045$ and is therefore compatible with a
cosmological constant, as assumed in the base \LCDM\ cosmology.

\paragraph{(10)}
We have presented a detailed analysis of possible extensions to the
neutrino sector of the base \LCDM\ model.  Combining \planckTT+lensing with BAO 
we find $N_{\rm eff} = 3.15\pm 0.23$ for the effective number of relativistic degrees of freedom,
consistent with the value $N_{\rm eff} = 3.046$ of the standard
model. The sum of neutrino masses is constrained to $\sum m_\nu < 0.23
\,{\rm eV}$. The \Planck\ data strongly disfavour fully thermalized
sterile neutrinos with $m_{\rm sterile} \approx 1\,{\rm eV}$ that have
been proposed as a solution to reactor neutrino oscillation
anomalies. From \Planck, we find no evidence for new neutrino physics.
Standard neutrinos with masses larger than those in the minimal mass hierarchy are still allowed, 
and could be detectable in combination with future astrophysical and CMB lensing data.

\paragraph{(11)}
The standard theory of big bang nucleosynthesis, with $N_{\rm eff} =
3.046$ and negligible leptonic asymmetry in the electron neutrino
sector, is in excellent agreement with \Planck\ data and observations
of primordial light element abundances. This agreement is particularly
striking for deuterium, for which accurate primordial abundance measurements
have been reported recently \citep{Cooke:2013cba}. The BBN theoretical
predictions for deuterium are now dominated by uncertainties in nuclear
reaction rates (principally the ${\rm d}({\rm p},\gamma)^3 \mbox{He}$
radiative capture process), rather than from \Planck\ uncertainties in the
physical baryon density $\omb \equiv \Omb h^2$.

\paragraph{(12)}
We have investigated the temperature and polarization signatures associated with annihilating dark matter and
possible deviations from the standard recombination history. Again, we find no evidence for new physics from
the \Planck\ data.

\medskip

In summary, the \Planck\ temperature and polarization spectra
presented in Figs.~\ref{pgTT_final} and \ref{pgTE+EE_final} are more
precise (and accurate) than those from any previous CMB experiment,
and improve on the 2013 spectra presented in {\paramsI}. Yet we find
no signs for any significant deviation from the base
\LCDM\ cosmology. Similarly, the analysis of 2015 \Planck\ data
reported in \citet{planck2014-a19} sets unprecedentedly tight limits
on primordial non-Gaussianity.  The \Planck\ results offer powerful
evidence in favour of simple inflationary models, which provide an
attractive mechanism for generating the slightly tilted spectrum of
(nearly) Gaussian adiabatic perturbations that match our data
to such high precision. In addition, the \Planck\ data show that the
neutrino sector of the theory is consistent with the assumptions of
the base \LCDM\ model and that the dark energy is compatible with a
cosmological constant.  If there is new physics beyond base \LCDM,
then the corresponding observational signatures in the CMB are weak
and difficult to detect. This is the legacy of the \Planck\ mission
for cosmology.

\vspace{0.7cm}

%\clearpage

\begin{acknowledgements}
The Planck Collaboration acknowledges the support of: ESA; CNES and CNRS/INSU-IN2P3-INP (France); ASI, CNR, and INAF (Italy); NASA and DoE (USA); STFC and UKSA (UK); CSIC, MINECO, JA, and RES (Spain); Tekes, AoF, and CSC (Finland); DLR and MPG (Germany); CSA (Canada); DTU Space (Denmark); SER/SSO (Switzerland); RCN (Norway); SFI (Ireland); FCT/MCTES (Portugal); ERC and PRACE (EU). A description of the Planck Collaboration and a list of its members, indicating which technical or scientific activities they have been involved in, can be found at \href{http://www.cosmos.esa.int/web/planck/planck%2dcollaboration}{\texttt{http://www.cosmos.esa.int/web/planck/planck-collaboration}}.
Some of the results in this paper have been derived using the {\tt HEALPix}
package.
The research leading to these results has received funding from the European
Research Council under the European Union's Seventh Framework Programme
(FP/2007–2013) / ERC Grant Agreement No. [616170] and from the UK Science and
Technology Facilities Council [grant number  ST/L000652/1].
Part of this work was undertaken on the STFC DiRAC HPC Facilities at the
University of Cambridge, funded by UK BIS National E-infrastructure capital
grants, and on the Andromeda cluster of the University of Geneva.
A large set of cosmological parameter constraints from different
data combinations, and including many separate extensions to the 6-parameter
base \LCDM\ model, are available at
\href{http://pla.esac.esa.int/pla/\#cosmology}{\texttt{http://pla.esac.esa.int/pla/\#cosmology}}.
\end{acknowledgements}

%Put these back into two lines as tex editor doesn't detect long line sets, arg.
\bibliography{Planck_bib,model_and_params,planck_only,section4,introduction,Planck_only_comparison,grid,grid_neutrino,grid_bbn,datasets_additional,datasets_CMBlensing,datasets_H0,datasets_BAO,datasets_SNe,datasets_clusters,grid_Recomb,grid_neutrino_pert,grid_early,grid_DM,grid_DE,grid_defects,cosmomc,grid_early_curv,grid_iso}{}
\bibliographystyle{aat}

\raggedright

\end{document}